\documentclass[a4paper,twoside]{book}
\newif\ifdraft \global\drafttrue
\def\production{\global\draftfalse}

\production
\usepackage[T1]{fontenc}
\usepackage{a4wide}
\usepackage{times}
\usepackage{graphicx}
\usepackage{theorem}
\usepackage{index} 
\usepackage[colorlinks=true,hyperindex=true]{hyperref}
\usepackage{framed}
\usepackage{tikz}
\usepackage{color}
\usepackage{fancyhdr}
\usepackage{amsmath}
\usepackage{bbm}
\usepackage{amsfonts}
\usepackage{enumerate}
\usepackage{cancel}
\ifdraft
\proofmodetrue
\usepackage{showkeys}
\fi
\newindex{default}{idx}{ind}{Index}
\newindex{notation}{ndx}{nnd}{Notations}

\let\myindex\index
\renewcommand{\index}[1]{\myindex{#1|idx}}
\newcommand{\bindex}[1]{\myindex{#1|bidx}}
\newcommand{\nindex}[3]{\myindex[notation]{1 #1@#2: #3|bidx}}
\newcommand{\appendixsection}[1]{%
\phantomsection
\chapter*{#1}
\addcontentsline{toc}{chapter}{#1}
}
\newcommand{\appendixsubsection}[1]{%
\phantomsection
\section*{#1}
\addcontentsline{toc}{section}{#1}}
\newtheorem{theorem}{Theorem}[chapter]
\newtheorem{proposition}[theorem]{Proposition}
\newtheorem{lemma}[theorem]{Lemma}
\newtheorem{definition}[theorem]{Definition}
\newtheorem{corollary}[theorem]{Corollary}
\theoremstyle{plain}
\theorembodyfont{\upshape}
\newtheorem{example}{Example}[chapter]

\newcounter{smallarabics}
\newenvironment{arabicenumerate}
{\begin{list}{{\normalfont\textrm{(\arabic{smallarabics})}}}
{\usecounter{smallarabics}\setlength{\itemindent}{0cm}
\setlength{\leftmargin}{5ex}\setlength{\labelwidth}{4ex}
\setlength{\topsep}{0.75\parsep}\setlength{\partopsep}{0ex}
\setlength{\itemsep}{0ex}}}
{\end{list}}
\newcounter{exercise}[chapter]
\renewcommand\theexercise{\arabic{chapter}.\arabic{exercise}}
\newlength{\exoskip}
\newlength{\exopskip}
\newcounter{exopcnt}[exercise]
\newcommand{\exop}{
\stepcounter{exopcnt}
\ifnum\value{exopcnt}>0\vskip\exopskip\fi
\noindent\arabic{exopcnt}.\ }
\definecolor{shadecolor}{gray}{0.95}

\newenvironment{exo}
{\refstepcounter{exercise}%
\fboxsep=1.2\FrameSep
\MakeFramed{\advance\hsize-20pt\FrameRestore}%
\noindent\textbf{Exercise~\theexercise. }
\ignorespaces
}{\endMakeFramed}
\def\bel{\begin{lemma}}
\def\eel{\end{lemma}}
\def\bec{\begin{corollary}}
\def\eec{\end{corollary}}
\def\bet{\begin{theorem}}
\def\eet{\end{theorem}}
\def\bed{\begin{definition}}
\def\eed{\end{definition}}
\def\bep{\begin{proposition}}
\def\eep{\end{proposition}}
\def\ben{\begin{arabicenumerate}}  
\def\een{\end{arabicenumerate}}
\def\beq{\begin{equation}}
\def\eeq{\end{equation}}  
\def\demo{{\noindent\bf Proof.\ }}
\def\qed{\hfill$\square$}
\newcommand{\e}{\mathrm{e}}
\let\oldi\i
\newcommand{\myi}{\oldi}
\renewcommand{\i}{\mathrm{i}}

\renewcommand{\d}{\mathrm{d}}
\def\nn{{\mathbb N}}
\def\zz{{\mathbb Z}}

\def\rr{{\mathbb R}}
\def\cc{{\mathbb C}}
\def\one{{\mathbbm 1}}
\def\cL{\mathcal{ L}}
\def\cC{\mathcal{C}}
\def\cS{\mathcal{S}}
\def\cQ{\mathcal{Q}}
\def\cK{\mathcal{K}}
\def\cH{\mathcal{H}}
\def\cR{\mathcal{R}}
\def\cO{\mathcal{O}}
\def\bPt{\bar P^{\kern1pt t}}
\def\bQt{\bar Q^{\kern1pt t}}
\def\balpha{{\boldsymbol\alpha}}
\newcommand{\ie}{{\sl i.e.,\ }}
\newcommand{\eg}{{\sl e.g.,\ }}
\newcommand{\etal}{{\sl et al.\ }}
\def\Im{\mathrm{Im}\,}
\def\Re{\mathrm{Re}\,}
\def\Ran{\mathrm{Ran}\,}
\def\Ker{\mathrm{Ker}\,}
\def\ac{\mathrm{ac}}
\def\sp{\mathrm{sp}}
\def\rh{\mathrm{e}}
\def\rc{\mathrm{c}}
\def\eq{\mathrm{eq}}
\def\rE{\mathrm{E}}
\def\s{\mathrm{s}}
\def\tr{\mathrm{tr}}
\def\interior{\mathrm{int}}
\def\closure{\mathrm{cl}}
\def\Os{\mathcal{O}_{\mathrm{self}}}
\def\abso{|}
\def\norm{\|}
\newcommand{\slim}{\mathop{\mathrm{s-lim}}\limits}
\def\bar{\overline}
\def\ubar{\underline}
\def\cal{\mathcal}
\renewcommand{\atop}[2]{\genfrac{}{}{0pt}{1}{#1}{#2}}
\begin{document}
\thispagestyle{empty}
\title{\bf\Huge Entropic Fluctuations in\\ Quantum Statistical Mechanics\\\vskip0.3cm An Introduction\vskip1cm}
\author{\sc V. Jak\v{s}i\'c$^{a}$, Y. Ogata$^{b}$, Y. Pautrat$^{c}$, C.-A. Pillet$^{d}$\\ \\ \\
\\ \\ \\ 
$^a$Department of Mathematics and Statistics\\ 
McGill University\\
805 Sherbrooke Street West \\
Montreal,  QC,  H3A 2K6, Canada
 \\ \\
$^b$Department of Mathematical Sciences\\
University of Tokyo\\
Komaba, Tokyo, 153-8914\\
Japan
 \\ \\
$^c$Laboratoire de Math\'ematiques\\
Universit\'e Paris-Sud\\
91405 Orsay Cedex, France
\\ \\
$^d$Centre de Physique Th\'eorique\footnote{%
UMR 6207:Universit\'e de Provence, Universit\'e de la M\'editerran\'ee, %
Universit\'e de Toulon et CNRS, FRUMAM}\\
Universit\'e du Sud Toulon-Var,  B.P. 20132 \\
F-83957 La Garde Cedex, France
}
\def\today{}
\maketitle
\tableofcontents

\chapter*{Introduction}

These lecture notes are the second instalment in a series of papers dealing with 
entropic fluctuations in non-equilibrium statistical mechanics. The first instalment 
\cite{JPR} concerned classical statistical mechanics. This one deals with the quantum 
case and is an introduction to the results of \cite{JOPP}. 
Although these  lecture notes could be read
independently of \cite{JPR}, a reader who wishes to get a proper grasp of the material is strongly 
encouraged  to consult \cite{JPR} for the classical analogs of the results presented here. In fact, to 
emphasize the link between the mathematical structure of classical and quantum theory of entropic 
fluctuations, we shall start the lectures with a {\em classical} example: a thermally driven harmonic 
chain. This example will serve as a prologue for the rest of the lecture notes.

The mathematical theory of entropic fluctuations developed in \cite{JPR,JOPP} is axiomatic 
in nature. Starting with a general classical/quantum dynamical system, the basic objects of the 
theory---entropy production observable, finite time entropic functionals, finite time fluctuation theorems 
and relations, finite time linear response theory---are introduced/derived at a great level of generality. 
The  axioms concern the large time limit $t\rightarrow \infty$, \ie the existence and  the regularity 
properties of the limiting entropic functionals. The introduced axioms are natural and minimal (\ie
necessary to have a meaningful theory), ergodic in nature, and typically difficult to verify in physically 
interesting models. Some of the quantum models for which the axioms have been verified 
(Spin-Fermion model, Electronic Black Box model) are  described in Chapter \ref{chap:FERMION}.

However, apart for Chapter \ref{chap:TD-limit}, we shall not discuss the axiomatic approach of
\cite{JOPP} here. The main body of the lecture notes is  devoted  to a pedagogical 
self-contained introduction to the finite time entropic functionals  and  fluctuation relations for 
{\em finite} quantum systems. A typical example the reader should have in mind is a quantum spin 
system or a Fermi gas with finite configuration space $\Lambda\subset{\mathbb Z}^d$. After the 
theory is developed, one proceeds by taking first the thermodynamic limit 
($\Lambda \rightarrow {\mathbb Z}^d$), and then the large time limit $t\rightarrow \infty$. The 
thermodynamic limit of the finite time/finite volume theory is typically an easy exercise 
in the techniques developed in the 70's (the two volumes monograph of Bratteli and Robinson provides
a good introduction to this subject). 
On the other hand, the large time limit, as to be expected, is typically a very difficult ergodic-type  
problem. In these notes we shall discuss the thermodynamic and 
the large time limits only in Chapter \ref{chap:TD-limit}. 
This section is intended for more advanced readers who are familiar 
with our previous works and lectures notes. It may be entirely skipped, although even technically 
less prepared readers my benefit from Sections \ref{sect:TDoverview} and \ref{sect:LThypo} up to
and including the proof of Theorem \ref {EBBQHTthm}.

Let us comment on our choice of the topic. From a mathematical point of view, there is a complete 
parallel between  classical and quantum theory of entropic fluctuations. The quantum theory applied 
to commutative structures (algebras) reduces to the classical theory, \ie  the classical theory is a  
special case of the quantum one. There is, however, a big difference in mathematical tools needed  
to describe the respective theories.  Only basic results of measure theory are needed for the finite 
time theory in classical statistical mechanics. In the non-commutative setting  these familiar tools 
are replaced by  the Tomita-Takesaki modular theory of von Neumann algebras. For example, 
Connes cocycles  and relative modular operators replace Radon-Nikodym derivatives. The quantum 
transfer operators act on Araki-Masuda non-commutative $L^p$-spaces which replace the 
familiar $L^p$-spaces of measure theory on which Ruelle-Perron-Frobenius (classical) transfer 
operators act, etc. The remarkably beautiful and powerful modular theory needed to describe 
quantum theory of entropic fluctuations has been  developed in 1970's and 80's, primarily by 
Araki, Connes and Haagerup. Although modular theory has played a key role in the mathematical
development of non-equilibrium quantum statistical mechanics over the last decade, 
the extent of its application to quantum theory of entropic fluctuations is somewhat striking.
Practically all  fundamental results  of modular  theory play a role. Some of them, like  the
Araki-Masuda theory of non-commutative $L^p$-spaces, have found in this context their first 
application to quantum statistical mechanics. 

The power of modular theory is somewhat shadowed by its technical aspects. Out of necessity, a 
reader of \cite{JOPP} must be familiar with the full machinery of algebraic quantum statistical mechanics 
and modular theory. Finite quantum systems, \ie quantum systems described by finite 
dimensional Hilbert spaces, are special since all the structures and results of this machinery can be 
described by elementary tools. The purpose of these lecture notes is to provide a self-contained 
pedagogical introduction to the algebraic structure of quantum statistical mechanics, finite time 
entropic functionals, and finite time fluctuation relations for {\em finite quantum systems}. 
For most part, the lecture notes should be easily accessible to an undergraduate student with basic 
training in linear algebra and analysis. Apart from occasional remarks/exercises and Chapter 
\ref{chap:TD-limit}, more advanced tools
enter only in the computations of the thermodynamic limit and the large time limit of the examples in 
Chapters \ref{chap:HarmoChain} and \ref{chap:FERMION}. A student who has taken a course in 
quantum mechanics and/or operator theory should have no difficulties with those tools either. 

Apart from from a few comments in Chapter \ref{chap:TD-limit} we shall not discuss here the 
Gallavotti-Cohen fluctuation theorem and the principle of regular entropic fluctuations. 
These  important topics concern non-equilibrium steady states  and require a technical machinery
not covered in these notes.
\index{theorem!Gallavotti-Cohen fluctuation}
\index{principle of regular entropic fluctuations}

The lecture notes are organized as follows. In the Prologue, Chapter \ref{chap:HarmoChain}, we 
describe the classical theory of entropic fluctuations on the example of a classical harmonic chain. 
The rest of the notes can be read independently of this section. Chapter \ref{chap:QSMfin} is devoted 
to the algebraic quantum statistical mechanics of finite quantum systems. In Chapters
\ref{chap:FINthermo} and \ref{chap:OQS} this algebraic structure is applied to the study of entropic 
functionals and fluctuation relations of finite quantum systems. In Chapter \ref{chap:FERMION} we 
illustrate the results of Chapters \ref{chap:FINthermo} and \ref{chap:OQS} on examples of fermionic 
systems. Large deviation theory and the G\"artner-Ellis  theorem play a key role in entropic 
fluctuation theorems and for this reason we review the G\"artner-Ellis theorem 
in Appendix \ref{appx:LDP}.  Another tool, a convergence result based on Vitali's theorem, will be often used in the 
lecture notes, and we provide its proof in Appendix \ref{appx:Vitali}.

\bigskip
{\noindent\bf Acknowledgment.} The research  of V.J. was partly supported by NSERC. 
The research of Y.O. was supported by JSPS Grant-in-Aid for Young Scientists (B), 
Hayashi Memorial Foundation for Female Natural Scientists, Sumitomo Foundation, and 
Inoue Foundation. The research of C.-A.P. was partly supported by ANR (grant 09-BLAN-0098).
A part of the lecture notes was written during the stay at the first author at IHES. V.J. wishes to thank 
D.~Ruelle  for hospitality and useful discussions. Various parts of the lecture notes have been 
presented by its authors in mini-courses at University of Cergy-Pontoise,
Erwin Schr\"odinger Institute (Vienna), Centre de Physique Th\'eorique (Marseille and Toulon), 
University of British Columbia (Vancouver), Ecole Polytechnique (Paris), 
Institut Henri Poincar\'e (Paris) and Ecole de Physique des Houches.
The lecture notes  have gained a lot from these presentations and we wish to thank 
the  respective institutions and F.~Germinet, J.~Yngvanson, R.~Froese,
S.~Kuksin, G.~Stoltz, J.~Fr\"ohlich  for making these mini-courses possible.

\chapter{Prologue: A thermally driven classical harmonic chain}
\label{chap:HarmoChain}

In this section we will discuss a very simple {\sl classical} example: a finite harmonic chain $\cC$ 
coupled at its left and right ends to two harmonic heat reservoirs $\cR_L$, $\cR_R$. This model is 
exactly solvable and  allows for a transparent review of the classical theory of entropic fluctuations 
developed in \cite{JPR}. Needless to say, models of this type have a long history in the physics 
literature and we refer the reader to Lebowitz and Spohn \cite{LS1} for references and
additional information. The reader should compare Chapter \ref{chap:OQS}, which deals with the 
non-equilibrium statistical mechanics of open quantum systems, with the  example of open classical 
system described here. The same remark applies to Section \ref{sect:EBBM}, where we study
the non-equilibrium statistical mechanics of ideal Fermi gases.

\section{The finite harmonic chain}
\label{sect:HarmoChainDef}

We start with the description of an isolated harmonic chain on the finite
1D-lattice $\Lambda=[A,B]\subset\zz$ (see Fig. \ref{fig:chain} below).
Its phase space is
$$
\Gamma_\Lambda=\{(p,q)=(\{p_x\}_{x\in\Lambda},\{q_x\}_{x\in\Lambda})
\,|\,p_x,q_x\in\rr\}=\rr^\Lambda\oplus\rr^\Lambda,
$$
and its Hamiltonian is given by
$$
H_\Lambda(p,q)=\sum_{x\in\zz}\left(\frac{p_x^2}2+\frac{q_x^2}2
+\frac{(q_x-q_{x-1})^2}2
\right),
$$
where we set $p_x=q_x=0$ for $x\not\in\Lambda$.

\begin{figure}[htbp]
\begin{center}
    \includegraphics[width=\textwidth]{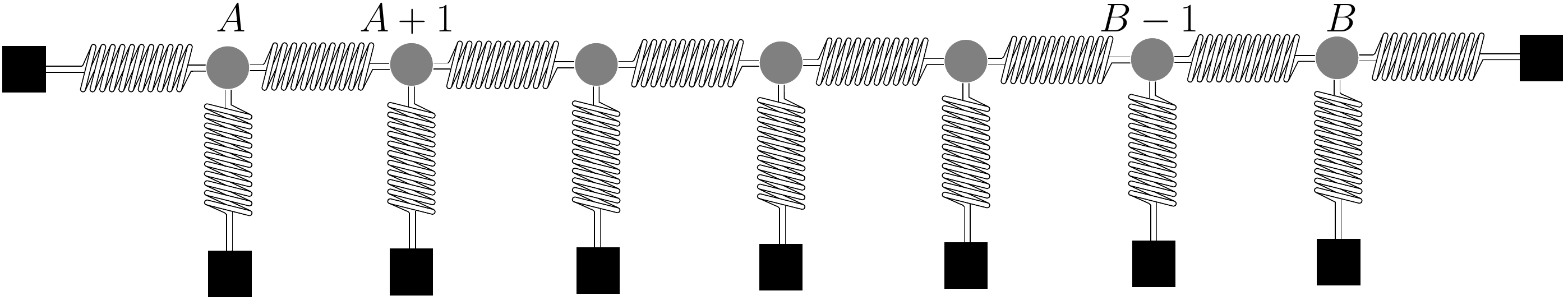}
    \caption{The finite harmonic chain on $\Lambda=[A,B]$.}
    \label{fig:chain}
\end{center}
\end{figure}

Thus, w.r.t.\ the natural Euclidian structure of $\Gamma_\Lambda$, the function
$2H_\Lambda(p,q)$ is the quadratic form associated to the symmetric matrix
$$
h_\Lambda=\left[\begin{array}{cc}
\one&0\\0&\one-\Delta_\Lambda
\end{array}
\right],
$$
where $\Delta_\Lambda$ denotes the discrete Laplacian on $\Lambda=[A,B]$ 
with Dirichlet boundary conditions
\begin{equation}
(-\Delta_\Lambda u)_x=\left\{\begin{array}{ll}
2u_{A}-u_{A+1}&\text{ for } x=A;\\[2pt]
2u_x-u_{x-1}-u_{x+1}&\text{ for } x\in]A,B[;\\[2pt]
2u_B-u_{B-1}&\text{ for } x=B.
\end{array}
\right.
\label{DeltaLambda}
\end{equation}\nindex{Delta Lambda}{$\Delta_\Lambda$}{discrete Dirichlet Laplacian}
\bindex{Laplacian!discrete Dirichlet ($\Delta_\Lambda$)}
The equations of motion of the chain,
$$
\dot p=-(\one-\Delta_\Lambda)q,\qquad\dot q=p,
$$
define a Hamiltonian flow on $\Gamma_\Lambda$, the one-parameter group 
$\e^{t\cL_\Lambda}$ generated by
$$
\cL_\Lambda=jh_\Lambda,\qquad j=\left[\begin{array}{cc}
0&-\one\\\one&0
\end{array}
\right].
$$
This flow has two important properties:

\begin{enumerate}[(i)]
\item Energy conservation: 
$\e^{t\cL_\Lambda^\ast}h_\Lambda\,\e^{t\cL_\Lambda}=h_\Lambda$.
\item Liouville's theorem: $\det\left(\e^{t\cL_\Lambda}\right)=\e^{t\,\tr (\cL_\Lambda)}=1$.
\end{enumerate}

An observable of the harmonic chain is a real (or vector) valued function on its
phase space $\Gamma_\Lambda$ and a state is a probability measure on $\Gamma_\Lambda$.
If $f$ is an observable and $\omega$ a state, we denote by
$$
\omega(f)=\int_{\Gamma_\Lambda}f(p,q)\,\d\omega(p,q),
$$
the expectation of $f$ w.r.t. $\omega$. Under the flow of the Hamiltonian $H_\Lambda$ the 
observables evolve as
$$
f_t=f\circ\e^{t\cal L_\Lambda}.
$$
In terms of the Poisson bracket
$$
\{f,g\}=\nabla_pf\cdot\nabla_qg-\nabla_qf\cdot\nabla_pg,
$$
the evolution of an observable $f$ satisfies
$$
\partial_tf_t=\{H_\Lambda,f_t\}=\{H_\Lambda,f\}_t.
$$
The evolution of a state $\omega$ is given by duality
$$
\omega_t(f)=\omega(f_t),
$$
and satisfies
$$
\partial_t\omega_t(f)=\omega_t(\{H_\Lambda,f\}).
$$
$\omega$ is called steady state or stationary state if it is
invariant under this evolution, \ie $\omega_t=\omega$ for all $t$. 
If $\omega$ has a density w.r.t. Liouville's measure
on $\Gamma_\Lambda$, \ie $\d\omega(p,q)=\rho(p,q)\,\d p\d q$, then Liouville's theorem
yields
\begin{align*}
\omega_t(f)&=\int_{\Gamma_\Lambda}f\circ\e^{t\cal L_\Lambda}(p,q)\rho(p,q)\,\d p\d q\\
&=\int_{\Gamma_\Lambda}f(p,q)\rho\circ\e^{-t\cal L_\Lambda}(p,q)\det\left(\e^{-t\cal L_\Lambda}\right)\,\d p\d q\\
&=\int_{\Gamma_\Lambda}f(p,q)\rho\circ\e^{-t\cal L_\Lambda}(p,q)\,\d p\d q,
\end{align*}
and so  $\omega_t$ also has a density w.r.t. Liouville's measure
given by $\rho\circ\e^{-t\cal L_\Lambda}$. If $D$ is a positive definite matrix on
$\Gamma_{\Lambda}$ and $\omega$ is 
the centered Gaussian measure with covariance $D$,
$$
\d\omega(p,q)=\det\left(2\pi D\right)^{-1/2}\e^{-D^{-1}[p,q]/2}\,\d p\d q,
$$
where $D^{-1}[p,q]$ denotes the quadratic form associated to $D^{-1}$, then $\omega_t$
is the centered Gaussian measure with covariance 
$D_t=\e^{t\cal L_\Lambda}D\e^{t\cal L_\Lambda^\ast}$.

The thermal equilibrium state of the chain at inverse temperature $\beta$ 
is the Gaussian measure with covariance $(\beta h_\Lambda)^{-1}$,
$$
\d\omega_{\Lambda\beta}(p,q)
=\sqrt{\det\left(\frac{\beta h_\Lambda}{2\pi}\right)}\,
\e^{-\beta H_\Lambda(p,q)}\d p\d q.
$$
Thermal equilibrium states are invariant under
the Hamiltonian flow of $H_\Lambda$.

\section{Coupling to the reservoirs}
\label{sect:CouplingRes}

As a small system, we consider the harmonic chain $\cal C$ on $\Lambda=[-N,N]$.
The left and right reservoirs are harmonic chains $\cR_L$ and $\cR_R$ on
$\Lambda_L=[-M,-N-1]$ and $\Lambda_R=[N+1,M]$ respectively.
In our discussion we shall keep $N$ fixed, but eventually let $M\to\infty$. 
In any case, the reader  should always have in mind that $M\gg N$.

The Hamiltonian of the joint but decoupled  system is 
$$
H_0(p,q)=H_{\Lambda}(p,q)+H_{\Lambda_L}(p,q)+H_{\Lambda_R}(p,q).
$$
The Hamiltonian  of the coupled system is
$$
H(p,q)=H_{\Lambda_L\cup\Lambda\cup\Lambda_R}(p,q)=H_0(p,q)+V_L(p,q)+V_R(p,q),
$$
where $V_L(p,q)=-q_{-N-1}q_{-N}$ and $V_R(p,q)=-q_Nq_{N+1}$.

\begin{figure}[htbp]
\begin{center}
    \includegraphics[width=\textwidth]{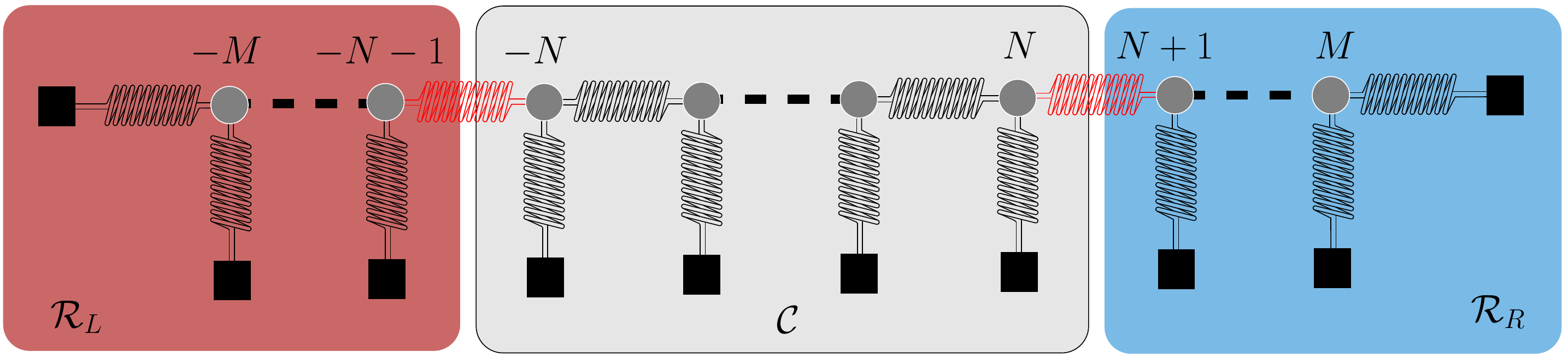}
    \caption{The chain $\cC$ coupled at its left and right ends to the reservoirs $\cR_L$ and $\cR_R$.}
    \label{fig:chain_coupled}
\end{center}
\end{figure}

We denote by $h_0$, $h_L$, $h_R$ and $h$ the symmetric matrices associated to the quadratic
forms $2H_0$, $2H_L$, $2H_R$ and $2H$ and by $\cal L_0=jh_0$ and $\cal L=jh$ the generators 
of the corresponding Hamiltonian flows. We also set $v=v_L+v_R=h-h_0$ where $v_L$ and $v_R$ 
are associated to $2V_L$ and $2V_R$ respectively.

\section{Non-equilibrium reference measure}
\label{sect:NonEquRefState}

We shall assume that  initially each  subsystem is in  thermal equilibrium, the reservoirs  at  
temperatures $T_{L/R}=1/\beta_{L/R}$, and  the small system at temperature   $T=1/\beta$. 
The initial (reference) state is therefore
\begin{equation}
\d\omega_{\Lambda_L\beta_L}\otimes\d\omega_{\Lambda\beta}\otimes
\d\omega_{\Lambda_R\beta_R}(p,q)
=Z^{-1}\,{\e^{-(\beta_LH_{\Lambda_L}(p,q)+\beta H_\Lambda(p,q)
+\beta_RH_{\Lambda_R}(p,q))}\,\d p\d q}.
\label{FirstGuess}
\end{equation}
If the temperatures of the reservoirs are different, the system is initially out of equilibrium. We set
$X_L=\beta-\beta_L$, $X_R=\beta-\beta_R$ and $X=(X_L,X_R)$. We call $X$ the thermodynamic
force acting on the chain  $\cal C$. $X_L$ and $X_R$ are sometimes called affinities in 
non-equilibrium thermodynamics (see, \eg \cite{dGM}). When $X=0$,
one has $\beta_L=\beta_R=\beta$ and  the joint system is in equilibrium at inverse temperature 
$\beta$ for the decoupled dynamics generated by $H_0$. 

In view of the coupled dynamics generated by $H$, it will be more convenient to use a  slightly 
modified initial state
\begin{align*}
\d\omega_X(p,q)&=Z_X^{-1}\,
\e^{-(\beta_LH_{\Lambda_L}(p,q)+\beta H_\Lambda(p,q)+\beta_RH_{\Lambda_R}(p,q)+\beta V(p,q))}
\d p\d q\\[5pt]
&=Z_X^{-1}\,
\e^{-(\beta H(p,q)-X_L H_{\Lambda_L}(p,q)-X_RH_{\Lambda_R}(p,q))}
\d p\d q,
\end{align*}
which, for $X=0$, reduces to the thermal equilibrium state at inverse temperature $\beta$
of the joint system under the coupled dynamics. Note that $\omega_X$ is the Gaussian measure
with covariance 
$$
D_X=(\beta h-k(X))^{-1},\qquad k(X)=X_Lh_L+X_Rh_R,
$$
whereas \eqref{FirstGuess} is Gaussian with covariance $(\beta h_0-k(X))^{-1}$. Since
$h-h_0=v$ is a rank $4$ matrix which is well localized at the boundary of $\Lambda$,
these two states describe the same thermodynamics.

\section{Comparing states}
\label{sect:StateCompare}

Under the Hamiltonian flow of $H$, the state $\omega_X$ evolves into $\omega_{X,t}$,
the Gaussian measure with covariance
$$
D_{X,t}=\e^{t\cal L}D_X\e^{t\cal L^\ast}=\left(\beta h-\e^{-t\cal L^\ast}k(X)\e^{-t\cal L}\right)^{-1}.
$$
As time goes on, the state $\omega_{X,t}$ diverges from the initial state $\omega_X$.
In order to quantify this divergence, we  need  a way to describe the ``rate of change"
of the state, \ie a concept of ``distance" between states. Classical information theory provides 
several candidates for such a distance. In this section, we introduce two of them
and explore their physical meaning.

Let $\nu$ and $\omega$ be two states. Recall that $\nu$ is said to be absolutely continuous w.r.t. 
$\omega$, written $\nu\ll\omega$,  if there exists a density, a non-negative function
$\rho$ satisfying $\omega(\rho)=1$, such that $\nu(f)=\omega(\rho f)$ for all observables $f$.
The function $\rho$ is called Radon-Nikodym derivative \index{Radon-Nikodym derivative} of $\nu$
w.r.t. $\omega$ and is denoted ${\d\nu}/{\d\omega}$.

The relative entropy of $\nu$ w.r.t. $\omega$ is defined by\index{entropy!relative}
\begin{equation}
S(\nu|\omega)=\left\{
\begin{array}{ll}
\displaystyle\nu\left(-\log \frac{\d\nu}{\d\omega}\right)&\text{ if } \nu\ll\omega,\\[10pt]
-\infty&\text{ otherwise.}
\end{array}
\right.
\label{RelatEntDef}
\end{equation}\index{entropy!relative}

\medskip
\begin{exo}
\label{Exo:first}
\exop Show that $\log(x^{-1})\le x^{-1}-1$ for $x>0$, where equality holds iff $x=1$.

\exop Using the previous inequality, show that $S(\nu|\omega)\le0$ with equality iff $\nu=\omega$.
This justifies the use of relative
entropy (or rather of $-S(\nu|\omega)$) as a measure of the ``distance" between $\nu$ and $\omega$. 
Note however that $-S(\nu|\omega)$ is not a metric in the usual sense since it is not symmetric and
does not satisfy the triangle inequality.
\end{exo}

\medskip
Applying Definition \eqref{RelatEntDef} to $\omega_{X,t}$ and $\omega_X$, we get
\begin{equation}
-\log\left(\frac{\d\omega_{X,t}}{\d\omega_X}\right)
=\displaystyle X_L(H_{\Lambda_L}-H_{\Lambda_L,-t})
+X_R(H_{\Lambda_R}-H_{\Lambda_R,-t}),
\label{prerelatent}
\end{equation}
and hence
\begin{align*}
S(\omega_{X,t}|\omega_X)
&=\omega_{X,t}\left(
\displaystyle X_L(H_{\Lambda_L}-H_{\Lambda_L,-t})
+X_R(H_{\Lambda_R}-H_{\Lambda_R,-t})\right)\\
&=X_L\omega_X\left(\displaystyle H_{\Lambda_L,t}-H_{\Lambda_L}\right)
+X_R\omega_X\left(H_{\Lambda_R,t}-H_{\Lambda_R})\right).
\end{align*}
Since the observable $H_{\Lambda_R,t}-H_{\Lambda_R}$ measures the increase of the energy in the 
right reservoir during the time interval $[0,t]$ and
$$
H_{\Lambda_R,t}-H_{\Lambda_R}
=\int_0^t\frac{\d\ }{\d s}H_{\Lambda_R,s}\,\d s
=\int_0^t\{H,H_{\Lambda_R}\}_s\,\d s,
$$
we interpret
$$
\Phi_R=-\{H,H_{\Lambda_R}\}=\{H_{\Lambda_R},V_R\}=-p_{N+1}q_{N},
$$
as the energy flux out of the right reservoir. Similarly,\index{flux}
$$
\Phi_L=-\{H,H_{\Lambda_L}\}=\{H_{\Lambda_L},V_L\}=-p_{-N-1}q_{-N},
$$
is the energy flux out of the left reservoir.

\medskip
\begin{exo}
Compare the equation of motion of the isolated reservoir $\cR_R$ with that of the same reservoir 
coupled to $\cal C$. Deduce that the force exerted on the reservoir by the system $\cal C$ is
given by $q_{N}$ and therefore that $q_Np_{N+1}$ is the power dissipated into the right
reservoir. 
\end{exo}

\medskip
In terms of fluxes, we have obtained the following {\sl entropy balance relation}\index{entropy!balance}
\begin{equation}
S(\omega_{X,t}|\omega_X)=-\int_0^t\omega_{X}(\sigma_{X,s})\,\d s,
\label{ClassicEbalance}
\end{equation}
where
$$
\sigma_X=X_L\Phi_L+X_R\Phi_R.
$$
This bilinear expression in the thermodynamic forces and the corresponding fluxes
has precisely the form of entropy production as derived in phenomenological non-equilibrium
thermodynamics (see, \eg Section IV.3 of \cite{dGM}). For this reason,
we shall call $\sigma_X$ the {\sl entropy production} observable and\index{entropy!production}
\begin{equation}
\Sigma^t=\frac1t\int_0^t\sigma_{X,s}\,\d s,
\label{SigmatDef}
\end{equation}
the {\sl mean entropy production rate}\footnote{Various other names are commonly used in the 
literature for  the observable $\sigma_X$: phase space contraction rate, dissipation function, etc.} 
over the time interval $[0,t]$. The important fact is that  the mean entropy production rate 
has  non-negative expectation for $t>0$:
\begin{equation}
\omega_X(\Sigma^t)=\frac1t\int_0^t\omega_X(\sigma_{X,s})\,\d s
=-\frac1tS(\omega_{X,t}|\omega_X)\ge0.
\label{PositivSigma}
\end{equation}

Another widely used measure of the discrepancy between two states $\omega$ and $\nu$
is R\'enyi relative $\alpha$-entropy, defined for any $\alpha\in\rr$ by\index{entropy!R\'enyi}
$$
S_\alpha(\nu|\omega)=\left\{
\begin{array}{ll}
\displaystyle\log\omega\left(\left(\frac{\d\nu}{\d\omega}\right)^\alpha\right)&\text{ if } \nu\ll\omega,\\[10pt]
-\infty&\text{ otherwise.}
\end{array}
\right.
$$
Starting from Equ. \eqref{prerelatent} one easily derives the formula
\begin{equation}
\log\frac{\d\omega_{X,t}}{\d\omega_X}=\int_0^{-t}\sigma_{X,s}\,\d s=t\Sigma^{-t},
\label{SigmatMeans}
\end{equation}
so that
\begin{equation}
e_t(\alpha)=
S_\alpha(\omega_{X,t}|\omega_X)
=\log\omega_X\left(\left(\frac{\d\omega_{X,t}}{\d\omega_X}\right)^\alpha\right)
=\log\omega_X\left(\e^{\alpha t\Sigma^{-t}}\right).
\label{SalphaFirst}
\end{equation}

\medskip
\begin{exo}
\exop Assuming $\nu\ll\omega$ and using H\"older's inequality, show that 
$\alpha\mapsto S_\alpha(\nu|\omega)$
is convex.

\exop Show that $S_0(\nu|\omega)=S_1(\nu|\omega)=0$ and conclude that
$S_\alpha(\nu|\omega)$ is non-positive for $\alpha\in]0,1[$ and non-negative for
$\alpha\not\in]0,1[$.

\exop Assuming also $\omega\ll\nu$, show that $S_{1-\alpha}(\nu|\omega)=S_\alpha(\omega|\nu)$.
\end{exo}

\section{Time reversal invariance}
\label{sect:TRIClassic}

Our dynamical system is time reversal invariant: the map $\vartheta(p,q)=(-p,q)$  is an
anti-symplectic involution, \ie $\{f\circ\vartheta,g\circ\vartheta\}=-\{f,g\}\circ\vartheta$ and 
$\vartheta\circ\vartheta=\mathrm{Id}$. Since $H\circ\vartheta=H$, it satisfies
$$
\e^{t\cal L}\circ\vartheta=\vartheta\circ\e^{-t\cal L},
$$
and leaves our reference state $\omega_X$ invariant,
\index{time reversal invariance}
$$
\omega_X(f\circ\vartheta)=\omega_X(f).
$$
It follows that $\omega_{X,t}(f\circ\vartheta)=\omega_{X,-t}(f)$, $\Phi_{L/R}\circ\vartheta=-\Phi_{L/R}$
and $\sigma_X\circ\vartheta=-\sigma_X$. Note in particular that $\omega_X(\Phi_{L/R})=0$ and
$\omega_X(\sigma_X)=0$. Applying time reversal to Definition \eqref{SigmatDef} we further get
\begin{align}
\Sigma^t\circ\vartheta&=\frac1t\int_0^t\sigma_X\circ\e^{s\cal L}\circ\vartheta\,\d s
=\frac1t\int_0^t\sigma_X\circ\vartheta\circ\e^{-s\cal L}\,\d s\nonumber\\
&=-\frac1t\int_0^t\sigma_X\circ\e^{-s\cal L}\,\d s
=\frac1t\int_0^{-t}\sigma_X\circ\e^{s\cal L}\,\d s\label{SigmatTRI}\\
&=-\Sigma^{-t},\nonumber
\end{align}
and Equ. \eqref{SalphaFirst} becomes
\begin{equation}
e_t(\alpha)
=\log\omega_X\left(\e^{\alpha t\Sigma^{-t}\circ\vartheta}\right)
=\log\omega_X\left(\e^{-\alpha t\Sigma^{t}}\right).
\label{SalphaSecond}
\end{equation}
Thus, $\alpha\mapsto t^{-1}e_t(\alpha)$ is the cumulant generating function
of the observable $-t\Sigma^t$ in the state $\omega_X$, and in particular
\begin{align*}
\left.\frac{\d\ }{\d\alpha}t^{-1}e_t(\alpha)\right|_{\alpha=0}
&=-\omega_X\left(\frac1t\int_0^t\sigma_{X,s}\,\d s\right),\\[5pt]
\left.\frac{\d^2\ }{\d\alpha^2}t^{-1}e_t(\alpha)\right|_{\alpha=0}
&=\omega_X\left(\left(\frac1{\sqrt t}\int_0^t
\left(\sigma_{X,s}-\omega_X(\sigma_{X,s})\right)\d s\right)^2\right).
\end{align*}

\section{A universal symmetry}
\label{sect:univsym}

Let us look more closely at the positivity property \eqref{PositivSigma}. To this end, we introduce
the distribution of the observable $\Sigma^t$ induced by the state $\omega_X$, \ie the 
probability measure $P^t$ defined by
$$
P^t(f)=\omega_X(f(\Sigma^t)).
$$
To comply with \eqref{PositivSigma}, this distribution should be asymmetric and give more weight to 
positive values than to negative ones. Thus, let us compare $P^t$ with the distribution
$\bPt(f)=\omega_X(f(-\Sigma^t))$ of $-\Sigma^t$. Observing that
\begin{align}
\Sigma^{-t}&=-\frac1t\int_0^{-t}\sigma_X\circ\e^{s\cal L}\,\d s
=\frac1t\int_0^{t}\sigma_X\circ\e^{-s\cal L}\,\d s\nonumber\\
&=\left(\frac1t\int_0^t\sigma_X\circ\e^{(t-s)\cal L}\,\d s\right)\circ\e^{-t\cal L}
=\Sigma^t\circ\e^{-t\cal L},\label{SigmatTRIagain}
\end{align}
we obtain, using \eqref{SigmatMeans} and \eqref{SigmatTRI}
\begin{align*}
\bPt(f)&=\omega_X(f(-\Sigma^t))=\omega_X(f(\Sigma^{-t}\circ\vartheta))
=\omega_X(f(\Sigma^{-t}))=\omega_{X}(f(\Sigma^t\circ\e^{-t\cal L}))\\
&=\omega_{X,-t}(f(\Sigma^t))
=\omega_X\left(\frac{\d\omega_{X,-t}}{\d\omega_X}f(\Sigma^t)\right)
=\omega_X\left(\e^{-t\Sigma^t}f(\Sigma^t)\right),
\end{align*}
from which we conclude that $\bPt\ll P^t$ and
\begin{equation}
\frac{\d\bPt}{\d P^t}(s)=\e^{-ts}.
\label{ESRelation}
\end{equation}
This relation shows that negative values of $\Sigma^t$ are exponentially suppressed as $t\to\infty$.
One easily deduces from \eqref{ESRelation} that
$$
-s-\delta\le\frac1t\log
\frac{\omega_X(\{\Sigma^t\in[-s-\delta,-s+\delta]\})}{\omega_X(\{\Sigma^t\in[s-\delta,s+\delta]\})}
\le -s+\delta,
$$
for $t,\delta>0$ and any $s\in\rr$. Such a property was discovered in numerical experiments
on shear flows by Evans \etal \cite{ECM}. Evans and Searles \cite{ES} were the first 
to provide a theoretical analysis of the underlying mechanism. Since then, a large body of theoretical 
and experimental literature has been devoted to similar ``fluctuation relations" or ``fluctuation theorems".
They have been derived for various types of systems: Hamiltonian and non-Hamiltonian mechanical
systems, discrete and continuous time dynamical systems, Markov processes, ...
We refer the reader to the review by Rondoni and Me\'\myi ja-Monasterio \cite{RM} for historical 
perspective and references and to \cite{JPR} for a more mathematically 
oriented presentation.\index{relation!Evans-Searles}\index{theorem!Evans-Searles fluctuation}

We can rewrite Equ. \eqref{SalphaSecond} in terms of the Laplace transform of the measure $P^t$, 
$$
e_t(\alpha)=\log\int\e^{-\alpha ts}\,\d P^t(s).
$$
Relation \eqref{ESRelation} is equivalent to
$$
\int\e^{-(1-\alpha) ts}\,\d P^t(s)=\int\e^{\alpha ts}\,\d \bPt(s)=
\int\e^{-\alpha ts}\,\d P^t(s),
$$
and therefore can be expressed in the form 
\begin{equation}
e_t(1-\alpha)=e_t(\alpha).
\label{FTESsym}
\end{equation}
We shall  call the last relation the  {\sl finite time Evans-Searles symmetry} of the function $e_t(\alpha)$. 
The above derivation directly extends to a general  time-reversal invariant dynamical system, see 
\cite{JPR}.\index{symmetry!Evans-Searles}

\section{A generalized Evans-Searles symmetry}
\label{sect:GESym}

Relation \eqref{ESRelation} deals with the mean entropy production rate $\Sigma^t$.
It can be generalized to the mean energy flux, the vector valued observable
$$
{\mathbf\Phi}^t=\frac1t\int_0^t\left(\Phi_L\circ\e^{s\cal L},\Phi_R\circ\e^{s\cal L}\right)\,\d s.
$$

\medskip
\begin{exo}
\label{Exo:Qt}
Denote by $Q^t$ (respectively $\bQt$) the distribution of ${\mathbf\Phi}^t$ (respectively
$-{\mathbf\Phi}^t$) induced by the state $\omega_X$, \ie
$Q^t(f)=\omega_X(f({\mathbf\Phi}^t))$ and $\bQt(f)=\omega_X(f(-{\mathbf\Phi}^t))$.
Using the fact that $X\cdot{\mathbf\Phi}^t=\Sigma^t$ and mimicking the proof of \eqref{ESRelation}
show that
\begin{equation}
\frac{\d\bQt}{\d Q^t}({\mathbf s})=\e^{-tX\cdot{\mathbf s}}.
\label{GESRelation}
\end{equation}
Again, this derivation can be extended to an arbitrary time-reversal invariant 
dynamical system, see \cite{JPR}.
\end{exo} 

\medskip
Introducing the cumulant generating function
\begin{equation}
\label{gXYdef}
g_t(X,Y)=\log\omega_X\left(\e^{-tY\cdot{\mathbf\Phi}^t}\right),
\end{equation}
and proceeding as in the previous section, we see that Relation \eqref{GESRelation} is equivalent
to
$$
\int\e^{-t(X-Y)\cdot{\mathbf s}}\,\d Q^t({\mathbf s})
=\int\e^{tY\cdot{\mathbf s}}\,\d \bQt({\mathbf s})
=\int\e^{-tY\cdot{\mathbf s}}\,\d Q^t({\mathbf s}),
$$
which leads to the {\sl generalized finite time Evans-Searles symmetry}\index{symmetry!Evans-Searles!generalized}
\begin{equation}
g_t(X,X-Y)=g_t(X,Y).
\label{FTGESsym}
\end{equation}

\medskip
\begin{exo}
Check that
\begin{equation}
g_t(X,Y)=-\frac12\log\det\left(
\one-D_X\left(\e^{t\cal L^\ast}k(Y)\e^{t\cal L}-k(Y)\right)\right),
\label{GESFormOne}
\end{equation}
where we adopt the convention that $\log x=-\infty$ whenever $x\le0$. Using this formula 
verify directly  Relation \eqref{FTGESsym}.
\end{exo}

\medskip

\section{Thermodynamic limit}
\label{sect:ClassicTD}
\begin{figure}[htbp]
\begin{center}
    \includegraphics[scale=0.7]{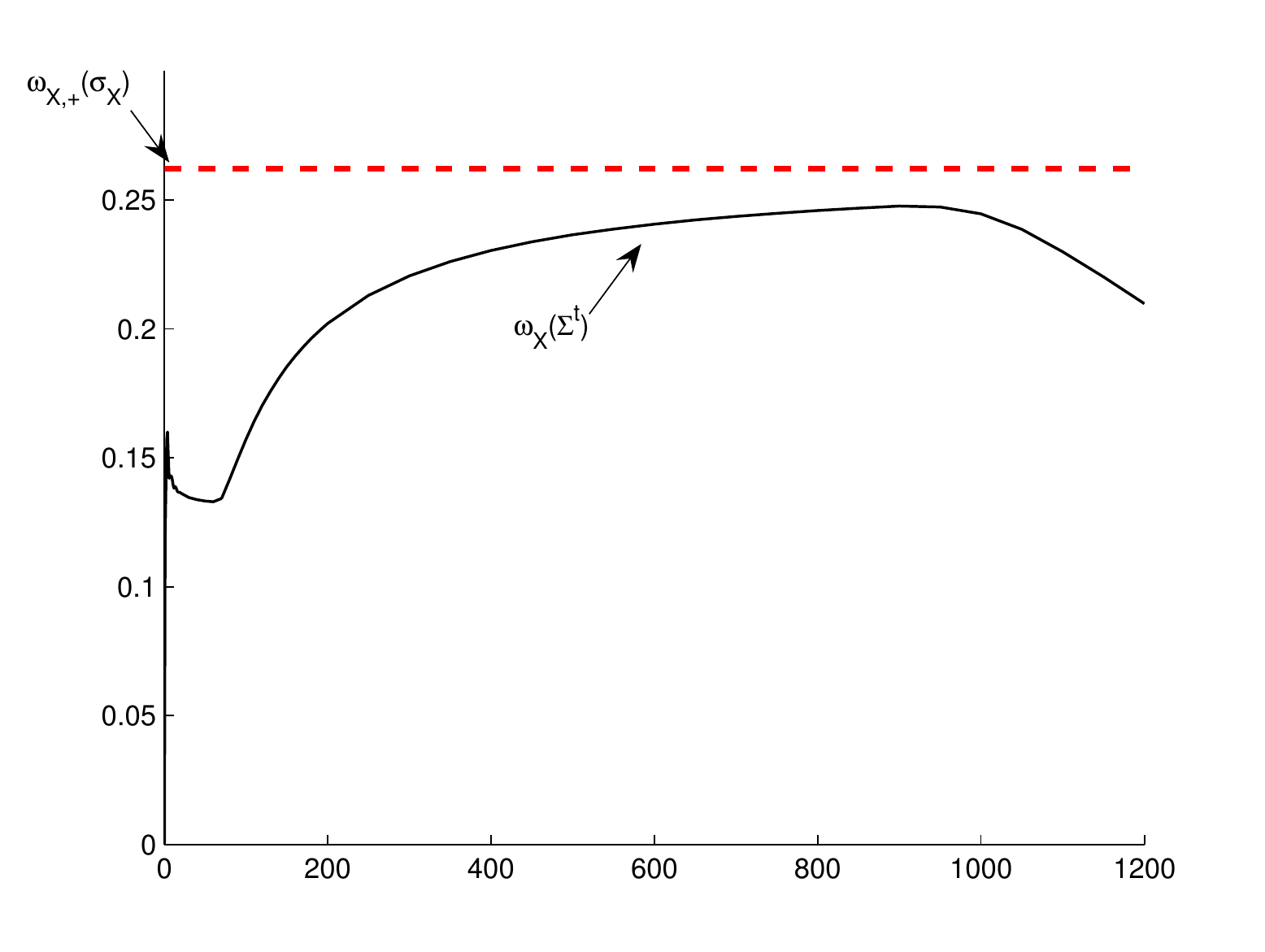}
    \caption{The typical behavior of the mean entropy production rate $t\mapsto\omega_X(\Sigma^t)$
     for a finite system ($N=20$, $M=300$). The dashed line represent the steady state value
     $\omega_{X,+}(\sigma_X)=\lim_{t\to\infty}\omega_X(\Sigma^t)$ for the same finite chain
     ($N=20$) coupled to two infinite reservoirs.}
    \label{fig:sigma}
\end{center}
\end{figure}

\index{thermodynamic limit}
So far we were dealing with a finite dimensional harmonic system. Its Hamiltonian flow $\e^{t\cal L}$ is
quasi-periodic and it is therefore not a surprise that entropy production vanishes in the large
time limit,
$$
\lim_{t\to\infty}\omega_X(\Sigma^t)
=\lim_{t\to\infty}\frac1{2t}\tr\,\left(
D_X\left(k(X)-\e^{t\cal L^\ast}k(X)\e^{t\cal L}
\right)\right)=0,
$$
see also Figure \ref{fig:sigma}.
To achieve a strictly positive entropy production rate in the asymptotic regime $t\to\infty$, the 
thermodynamic limit of the reservoirs must be taken prior to the large time limit.

To take  $M\to\infty$ while  keeping $N$ fixed we observe that the phase space $\Gamma_{[-M,M]}$
is naturally embedded in the real Hilbert space $\Gamma=\ell_\rr^2(\zz)\oplus\ell_\rr^2(\zz)$ and that
$h_0$, $h_L$, $h_R$ and $h$ are uniformly bounded and strongly convergent as operators on
this space. For example
$$
\slim_{M\to\infty}\,h_0=
\left[
\begin{array}{cc}
\one&0\\
0&\one-\Delta_0
\end{array}
\right],
$$
where $\Delta_0=\Delta_{]-\infty,-N-1]}\oplus\Delta_{[-N,N]}\oplus\Delta_{[N+1,\infty[}$ is the discrete
Laplacian on $\zz$ with Dirichlet decoupling at $\pm N$ and
$$
\slim_{M\to\infty}\,h=
\left[
\begin{array}{cc}
\one&0\\
0&\one-\Delta
\end{array}
\right],
$$
where $\Delta=\Delta_\zz$ is the discrete Laplacian on $\zz$. It follows that
$\cal L_0=jh_0$ and $\cal L=jh$ are also strongly convergent. Hence, the Hamiltonian flows 
$\e^{t\cal L_0}$ and $\e^{t\cal L}$ converge strongly and uniformly on compact time intervals
to the uniformly bounded, norm continuous groups on $\Gamma$ generated
by the strong limits of $\cal L_0$ and $\cal L$. Finally, since the covariance 
$D_X=(\beta h-k(X))^{-1}$ of the state $\omega_X$ converges strongly, the state $\omega_X$ 
converges weakly to the Gaussian measure with the limiting covariance. In the following, we shall
use the same notation for these objects after the limit $M\to\infty$, \ie
$h$, $h_0$, $\cal L$, $\cal L_0$, $k(X)$, $\omega_X$, ...  denote the thermodynamic limits of the
corresponding finite volume objects.

After the thermodynamic limit, we are left with a linear dynamical system on the
$L^2$-space of the Gaussian measure $\omega_X$. Denoting by $\phi_{L/R}$ the
finite rank operators corresponding to the flux observable $2\Phi_{L/R}$ and setting
$\phi(Y)=Y_L\phi_L+Y_R\phi_R$, we can write
\begin{equation}
\e^{t\cal L^\ast}k(Y)\e^{t\cal L}-k(Y)=-\int_0^t\e^{s\cal L^\ast}\phi(Y)\e^{s\cal L}\,\d s.
\label{eq:kevol}
\end{equation}
Since the right hand side of this identity is trace class for every $Y\in\rr^2$ and $t\in\rr$,
we conclude from
\begin{equation}
D_{X,t}^{-1}-D_X^{-1}=\e^{-t\cal L^\ast}k(X)\e^{-t\cal L}-k(X),
\label{eq:kevol2}
\end{equation}
and the Feldman-Hajek-Shale theorem (see, \eg \cite{Si}) that the Gaussian measure 
$\omega_{X,t}$ and $\omega_X$ are equivalent and that Relation \eqref{prerelatent} still holds 
in the following form
$$
-\log\left(\frac{\d\omega_{X,t}}{\d\omega_X}\right)
=\int_0^{-t}\sigma_{X,s}\,\d s.
$$
For the same reason, Equ. \eqref{GESFormOne} for the generalized Evans-Searles functional
$g_t(X,Y)$ remains valid in the thermodynamic limit.

\section{Large time limit I: Scattering theory}
\label{sect:ClassicScattering1}

Taking the limit $t\to\infty$ in \eqref{eq:kevol}, \eqref{eq:kevol2} we obtain the formal result
$$
D_{X,+}^{-1}=\lim_{t\to\infty}D_{X,t}^{-1}=D_X^{-1}
+\int_{-\infty}^0\e^{s\cal L^\ast}\phi(X)\e^{s\cal L}\,\d s,
$$
which we can interpret in the following way: the state $\omega_{X,t}$, Gaussian with
covariance $D_{X,t}$, converges as $t\to\infty$ towards a non-equilibrium steady state (NESS) 
$\omega_{X,+}$, Gaussian with covariance $D_{X,+}$, which formally writes
\begin{align*}
&\d\omega_{X,+}(p,q)=\\[8pt]
&\frac1{Z_{X,+}}\e^{-\left(\beta H(p,q)-X_LH_{\Lambda_L}(p,q)-X_RH_{\Lambda_R}(p,q)
+\int_{-\infty}^0\left(X_L\Phi_{L,s}(p,q)+X_R\Phi_{R,s}(p,q)\right)\,\d s\right)}\d p\d q.
\end{align*}
This formal expression  is a special case of the McLennan-Zubarev non-equilibrium ensemble
(see \cite{McL,Zu1,Zu2}). In this and the following sections we shall turn this 
formal argument into a rigorous construction.\index{McLennan-Zubarev ensemble}

The study of the limit $t\to\infty$ in our infinite dimensional harmonic system reduces to an application of 
trace class  scattering theory. We refer to \cite{RS3} for basic facts about 
scattering theory. We start with a few simple remarks:
\begin{enumerate}[(i)]
\item We denote by $\cal H=\ell^2_\cc(\zz)\oplus\ell^2_\cc(\zz)\simeq\ell^2_\cc(\zz)\otimes\cc^2$
the complexified phase space and extend all operators on $\Gamma$ to $\cal H$ by $\cc$-linearity.
The inner product on the complex Hilbert space $\cal H$ is written $\langle\phi|\psi\rangle$.
\item $h-h_0=v$ is finite rank and hence   trace class. Since $h_0\ge \one$ and
$h\ge \one$,  $h^{1/2}-h_0^{1/2}$ is also  trace class.
\item $h^{1/2}h_0^{-1/2}-\one=(h^{1/2}-h_0^{1/2})h_0^{-1/2}$ is  trace class. The same is true for
$h_0^{1/2}h^{-1/2}-\one$, $h_0^{-1/2}h^{1/2}-\one$ and $h^{-1/2}h_0^{1/2}-\one$.
\item $L_0=\i h_0^{1/2}jh_0^{1/2}$ and $L=\i h^{1/2}jh^{1/2}$ are self-adjoint, $L-L_0$ is
trace class and 
$$
\e^{-\i tL_0}=h_0^{1/2}\e^{t\cal L_0}h_0^{-1/2},\qquad \e^{-\i tL}=h^{1/2}\e^{t\cal L}h^{-1/2}.
$$
Note that $\i L_0$ (respectively $\i L$) acting on $\cal H$ is unitarily equivalent to
$\cal L_0$ (respectively $\cal L$) acting on the ``energy" Hilbert space 
$\ell^2_\cc(\zz)\oplus\ell^2_\cc(\zz)$ equipped with the inner product
$\langle\phi|\psi\rangle_{h_0}=\langle\phi|h_0|\psi\rangle$ (respectively 
$\langle\phi|\psi\rangle_{h}=\langle\phi|h|\psi\rangle$).
\item $L$ has purely absolutely continuous spectrum.
\item The Hilbert space $\cal H$ has a direct decomposition into three parts, 
$\cal H=\cal H_L\oplus\cal H_{\cal C} \oplus\cal H_R$,
corresponding to the three subsystems $\cR_L$, $\cal C$ and $\cR_R$.
We denote by $P_L$, $P_{\cal C}$ and $P_R$ the corresponding orthogonal
projections.
\item This decomposition reduces $L_0$ so that $L_0=L_L\oplus L_{\cal C}\oplus L_R$. The
operators $L_L$ and $L_R$ have purely absolutely continuous spectrum and $L_{\cal C}$ has purely 
discrete spectrum. In particular, $P_L+P_R$ is the spectral projection of $L_0$ onto its absolutely 
continuous part.
\end{enumerate}

\bigskip\noindent
By Kato-Birman theory, the wave operators \index{wave operator}
$$
W_\pm=\slim_{t\to\pm\infty}\e^{\i tL}\e^{-\i tL_0}(P_L+P_R)
$$
exists and are complete, \ie
$$
W_\pm^\ast=\slim_{t\to\pm\infty}\e^{\i tL_0}\e^{-\i tL},
$$
also exists and satisfy $W_\pm^\ast W_\pm=P_L+P_R$, $W_\pm W_\pm^\ast=\one$.
The scattering matrix $S=W_+^\ast W_-$ \index {scattering matrix}
is unitary on $\cal H_L\oplus\cal H_R$. A few more remarks are needed to actually compute $S$:

\bigskip
\begin{enumerate}[(i)]
\setcounter{enumi}{7}
\item One has
$$
U^\ast L_0U=\left[\begin{array}{cc}
\Omega_0&0\\0&-\Omega_0
\end{array}\right],\qquad
U^\ast LU=\left[\begin{array}{cc}
\Omega&0\\0&-\Omega
\end{array}\right],
$$
where $\Omega=\sqrt{\one-\Delta}$ and $\Omega_0=\sqrt{\one-\Delta_0}$ are discrete Klein-Gordon 
operators and $U$ is the unitary
$$
U=\frac1{\sqrt2}\left[\begin{array}{cc}
1&\i\\\i&1
\end{array}\right].
$$
\item It follows that
$$
W_\pm=U\left[\begin{array}{cc}
w_\pm&0\\0&w_\mp
\end{array}\right]U^\ast,
$$
where
$$
w_\pm=\slim_{t\to\pm\infty}\e^{\i t\Omega}\e^{-\i t\Omega_0}(P_L+P_R).
$$
In particular, one has
\begin{equation}
S=U\left[\begin{array}{cc}
w_+^\ast w_-&0\\0&w_-^\ast w_+
\end{array}\right]U^\ast.
\label{SForm}
\end{equation}
\item By the invariance principle for wave operators, we have
\begin{align*}
w_\pm
&=\slim_{t\to\pm\infty}\e^{\i t\Omega^2}\e^{-\i t\Omega_0^2}P_\ac(\Omega_0^2)\\
&=\slim_{t\to\pm\infty}\e^{\i t(-\Delta)}\e^{-\i t(-\Delta_0)}P_\ac(-\Delta_0).
\end{align*}
\end{enumerate}
\noindent
We proceed to  compute the scattering matrix. 
A complete set of (properly normalized) generalized eigenfunctions for the absolutely 
continuous part of $-\Delta_0$ is given by
$$
\phi_{\sigma,k}(x)=\sqrt{\frac2\pi}\,\theta(\sigma x-N)\sin k|\sigma x- N|,
\qquad (\sigma,k)\in\{-,+\}\times [0,\pi],
$$
where $\theta$ denotes the Heaviside step function and 
$-\Delta_0\phi_{\sigma,k}=2(1-\cos k)\phi_{\sigma,k}$. For the operator $-\Delta$, 
such a set is given by
$$
\chi_{\sigma,k}(x)=\frac1{\sqrt{2\pi}}\,\e^{\i\sigma kx},
\qquad (\sigma,k)\in\{-,+\}\times [0,\pi].
$$
Since 
$$
w_\pm\phi_{\sigma,k}=\mp\sigma\i\e^{\mp\i kN}\chi_{\pm\sigma,k},
$$
we deduce that
\begin{equation}
w_\pm^\ast w_\mp\phi_{\sigma,k}=\e^{\pm2\i kN}\phi_{-\sigma,k}.
\label{smallsform}
\end{equation}
We shall denote by $\mathfrak h_{k\pm}$ the 2-dimensional generalized eigenspace of $L_0$ 
to the ``eigenvalue" $\pm\varepsilon(k)=\pm\sqrt{3-2\cos k}$. The space $\mathfrak h_{k+}$
is spanned by the two basis vectors
$$
\psi_{\sigma,k,+}=
U\left[\begin{array}{c}
\phi_{\sigma,k}\\0
\end{array}\right],\qquad
\sigma\in\{-,+\},
$$
and $\mathfrak h_{k-}$ is the span of
$$
\psi_{\sigma,k,-}=
U\left[\begin{array}{c}
0\\\phi_{\sigma,k}
\end{array}\right],\qquad
\sigma\in\{-,+\}.
$$
In the direct integral representation
$$
\cal H_L\oplus\cal H_R
=\bigoplus_{\mu=\pm}\left(\int^\oplus_{[0,\pi]}\mathfrak h_{k\mu}\,\d k\right),
$$
the scattering matrix is given by
$$
S=\bigoplus_{\mu=\pm}\left(\int^\oplus_{[0,\pi]}S_\mu(k)\,\d k\right),
$$
where, thanks to \eqref{SForm} and \eqref{smallsform}, the on-shell $S$-matrix $S_\mu(k)$ is 
given by
\begin{equation}
S_\pm(k)=S|_{\mathfrak h_{k\pm}}=\e^{\pm2\i kN}\left[\begin{array}{cc}
0&1\\1&0
\end{array}\right].
\label{SAction}
\end{equation}

\section{Large time limit II: Non-equilibrium steady state}
\label{sect:ClassicNESS}

We shall now use scattering theory to compute the weak limit, as $t\to\infty$, of the state
$\omega_{X,t}$. Setting $\widehat X=X_LP_L+X_RP_R$ for $X=(X_L,X_R)\in\rr^2$, one has
$$
k(X)=X_Lh_L+X_Rh_R=h_0^{1/2}\widehat X h_0^{1/2}.
$$
Energy conservation yields $\e^{-t\cal L_0^\ast}k(X)\e^{-t\cal L_0}=k(X)$ and 
\begin{align*}
\e^{t\cal L^\ast}k(X)\e^{t\cal L}
&=\e^{t\cal L^\ast}\e^{-t\cal L_0^\ast}h_0^{1/2}\widehat X h_0^{1/2}\e^{-t\cal L_0}\e^{t\cal L}\\
&=\e^{t\cal L^\ast}h_0^{1/2}\e^{-\i tL_0}\widehat X \e^{\i tL_0}h_0^{1/2}\e^{t\cal L}\\
&=\e^{t\cal L^\ast}h^{1/2}h^{-1/2}h_0^{1/2}\e^{-\i tL_0}\widehat X
\e^{\i tL_0}h_0^{1/2}h^{-1/2}h^{1/2}\e^{t\cal L}\\
&=h^{1/2}\e^{\i tL}h^{-1/2}h_0^{1/2}\e^{-\i tL_0}\widehat X
\e^{\i tL_0}h_0^{1/2}h^{-1/2}\e^{-\i tL}h^{1/2}.
\end{align*}
By Property (ii) of the previous section, one has
\begin{gather*}
\slim_{t\to\pm\infty}\,\e^{\i tL}h^{-1/2}h_0^{1/2}\e^{-\i tL_0}(P_L+P_R)=W_\pm,\\
\slim_{t\to\pm\infty}\,\e^{\i tL_0}h_0^{1/2}h^{-1/2}\e^{-\i tL}=W_\pm^\ast,
\end{gather*}
and so 
\begin{equation}
\slim_{t\to\pm\infty}\,\e^{t\cal L^\ast}k(X)\e^{t\cal L}=h^{1/2}W_\pm\widehat X W_\pm^\ast h^{1/2}.
\label{eq:klimit}
\end{equation}
It follows that
\begin{align}
\slim_{t\to\infty}\, D_{X,t}&=\slim_{t\to\infty}\, (\beta h-\e^{-t\cal L^\ast}k(X)\e^{-t\cal L})^{-1}\nonumber\\
&=(\beta h-h^{1/2}W_-\widehat X W_-^\ast h^{1/2})^{-1}\label{eq:Dlimit}\\
&=h^{-1/2}W_-(\beta-\widehat X)^{-1} W_-^\ast h^{-1/2}=D_{X,+},\nonumber
\end{align}
which implies that the state $\omega_{X,t}$ converges weakly to the Gaussian measure
$\omega_{X,+}$ with covariance $D_{X,+}$. The state $\omega_{X,+}$ is invariant under the Hamiltonian flow
$\e^{t\cal L}$ and is called the non-equilibrium steady state (NESS) associated to the reference state
$\omega_X$. Note that in the equilibrium case $\beta_L=\beta_R$ the operator 
$\widehat X$ is a multiple of the identity and
\index{state!non-equilibrium steady}
\myindex{NESS|see{state, non-equilibrium steady}}
$$
D_{X,+}=(\beta_L h)^{-1},
$$
which means that the stationary state $\omega_{X,+}$ is the thermal equilibrium state of
the coupled system at inverse temperature $\beta_L=\beta_R$.

\medskip
\begin{exo}
If  $X_L\not=X_R$ then $\omega_{X,+}$ is singular w.r.t. $\omega_X$, \ie
$$
D_{X,+}^{-1}-D_X^{-1}=h_0^{1/2}\widehat X h_0^{1/2}-h^{1/2}W_-\widehat X W_-^\ast h^{1/2},
$$
is not Hilbert-Schmidt. Prove this fact by  deriving  explicit formulas for $W_-P_{L/R}W_-^\ast$ .
\end{exo}
\begin{exo}
\label{Exo:SteadyFlux}
Compute $\omega_{X,+}(\Phi_{L/R})=\frac12\tr ( D_{X,+}\phi_{L/R})$ and show that
$$
\omega_{X,+}(\Phi_L)=-\omega_{X,+}(\Phi_R)=\kappa(T_L-T_R),
$$
where $T_{L/R}=\beta_{L/R}^{-1}$ is the temperature of the left/right reservoir and
$$
\kappa=\frac{\sqrt5-1}{2\pi}.
$$
Note in particular that $\omega_{X,+}(\Phi_L)+\omega_{X,+}(\Phi_R)=0$. What is the physical
origin of this fact ? Show that, more generally, if $\omega$ is a stationary state such that 
$\omega(p_x^2+q_x^2)<\infty$ for all $x\in\zz$, then $\omega(\Phi_L)+\omega(\Phi_R)=0$.
\end{exo}

\medskip
Using the result of Exercise \ref{Exo:SteadyFlux} we conclude that
\begin{align*}
\omega_{X,+}(\sigma_X)&=X_L\omega_{X,+}(\Phi_L)+X_R\omega_{X,+}(\Phi_R)\\
&=(X_L-X_R)\omega_{X,+}(\Phi_L)\\
&=\kappa\frac{(T_L-T_R)^2}{T_LT_R}>0,
\end{align*}
provided $T_L\not=T_R$. This implies that the mean entropy production rate in the state $\omega_X$
is strictly positive in the asymptotic regime\footnote{Recall  that if $\lim_{t\to+\infty}f(t)=a$ exists then
it coincide with the Ces\' aro limit of $f$ at $+\infty$, $\lim_{T\to+\infty}T^{-1}\int_0^Tf(t)\,\d t=a$, 
and with its Abel limit, $\lim_{\eta\downarrow0}\eta\int_0^\infty\e^{-\eta t}f(t)\,\d t=a$.\label{GenLim}},
$$
\lim_{t\to\infty}\omega_X(\Sigma^t)
=\lim_{t\to\infty}\frac1t\int_0^t\omega_X(\sigma_{X,s})\,\d s
=\lim_{t\to\infty}\omega_{X,t}(\sigma_X)=\omega_{X,+}(\sigma_X)>0,
$$
and that it is constant and strictly positive in the NESS $\omega_{X,+}$,
$$
\omega_{X,+}(\Sigma^t)=\frac1t\int_0^t\omega_{X,+}(\sigma_{X,s})\,\d s=\omega_{X,+}(\sigma_X)>0.
$$ 
\index{entropy!production}

\section{Large time limit III: Generating functions}
\label{sect:GF}

In this section we  use scattering theory to study the large time asymptotic of the
Evans-Searles functional $e_t(\alpha)$ (Equ. \eqref{SalphaSecond}) and 
the generalized Evans-Searles functional $g_t(X,Y)$ (Equ. \eqref{gXYdef}).

Starting from Equ. \eqref{GESFormOne}, and using \eqref{eq:kevol} to write
$$
T_t=-D_X\left(\e^{t\cal L^\ast}k(Y)\e^{t\cal L}-k(Y)\right)
=\int_0^t D_X\e^{s\cal L^\ast}\phi(Y)\e^{s\cal L}\,\d s,
$$
we get
\begin{align*}
\frac1t g_t(X,Y)
&=-\frac1{2t}\log\det\left(\one+T_t\right)\\
&=-\frac1{2t}\tr\log\left(\one+T_t\right)\\
&=-\frac1{2t}\int_0^1\frac{\d\ }{\d u}\tr\log\left(\one+uT_t\right)\d u.
\end{align*}
Using the result of Exercise \ref{Exo:Dlog}, we further get
\begin{align*}
\frac1t g_t(X,Y)
&=-\frac1{2t}\int_0^1\tr\left(\left(\one+uT_t\right)^{-1}T_t\right)\d u\\
&=-\frac1{2t}\int_0^1\int_0^t\tr\left(\left(\one+uT_t\right)^{-1}D_X\e^{s\cal L^\ast}\phi(Y)\e^{s\cal L}\right)\d s\,\d u\\
&=-\frac1{2t}\int_0^1\int_0^t\tr\left[\left(D_X^{-1}-u\left(\e^{t\cal L^\ast}k(Y)\e^{t\cal L}-k(Y)\right)\right)^{-1}
\e^{s\cal L^\ast}\phi(Y)\e^{s\cal L}\right]\d s\,\d u\\
&=-\frac1{2}\int_0^1\int_0^1\tr\left[\e^{st\cal L}
     \left(D_X^{-1}-u\left(\e^{t\cal L^\ast}k(Y)\e^{t\cal L}-k(Y)\right)\right)^{-1}\e^{st\cal L^\ast}
     \phi(Y)\right]\d s\,\d u.
\end{align*}
Writing
\begin{align*}
\e^{st\cal L}&\left(D_X^{-1}-u\left(\e^{t\cal L^\ast}k(Y)\e^{t\cal L}-k(Y)\right)\right)^{-1}\e^{st\cal L^\ast}\\
&=\left(\e^{-st\cal L^\ast}D_X^{-1}\e^{-st\cal L}
-u\,\e^{-st\cal L^\ast}\left(\e^{t\cal L^\ast}k(Y)\e^{t\cal L}-k(Y)\right)\e^{-st\cal L}\right)^{-1}\\
&=\left(D_{X,st}^{-1}
-u\left(\e^{(1-s)t\cal L^\ast}k(Y)\e^{(1-s)t\cal L}-\e^{-st\cal L^\ast}k(Y)\e^{-st\cal L}\right)\right)^{-1},
\end{align*}
and using \eqref{eq:klimit} and \eqref{eq:Dlimit}, we obtain
\begin{align*}
\slim_{t\to\infty}\,\e^{st\cal L}&
\left(D_X^{-1}-u\left(\e^{t\cal L^\ast}k(Y)\e^{t\cal L}-k(Y)\right)\right)^{-1}\e^{st\cal L^\ast}\\
&=\left(D_{X,+}^{-1}
-uh^{1/2}\left(W_+\widehat YW_+^\ast-W_-\widehat YW_-^\ast\right)h^{1/2}\right)^{-1}\\
&=\left(
h^{1/2}\left(W_-(\beta-\widehat X+u\widehat Y)^{-1}W_-^\ast-uW_+\widehat YW_+^\ast\right)h^{1/2}
\right)^{-1}\\
&=h^{-1/2}W_-\left(\beta-\widehat X-u(S^\ast\widehat YS-\widehat Y)\right)^{-1}W_-^\ast\, h^{-1/2},
\end{align*}
for all $s\in]0,1[$. Since $\phi(Y)$ is trace class (actually finite rank), we conclude that
$$
g(X,Y)=\lim_{t\to\infty}\frac1t g_t(X,Y)=
-\frac1{2}\int_0^1\tr\left[\left(\beta-\widehat X-u(S^\ast\widehat YS-\widehat Y)\right)^{-1}\cal T\right]
\,\d u,
$$
where
$$
\cal T=W_-^\ast h^{-1/2}\phi(Y)h^{-1/2}W_-.
$$
To evaluate the trace, we note that the scattering matrix $S$ and the operators $\widehat X$, 
$\widehat Y$ all commute with $L_0$ while the trace class operator $\cal T$ acts non-trivially only 
on the absolutely continuous spectral subspace of $L_0$. It follows that
\begin{align}
&\tr\left[\left(\beta-\widehat X-u(S^\ast\widehat YS-\widehat Y)\right)^{-1}\cal T\right]\nonumber\\
=&\tr\left[\left(\one-u(\beta-\widehat X)^{-1}(S^\ast\widehat YS-\widehat Y)\right)^{-1}
(\beta-\widehat X)^{-1}\cal T\right]\label{tridonsat}\\
=&\sum_{\mu=\pm}\int_0^\pi\sum_{\sigma=\pm}
\langle\psi_{\sigma,k,\mu}|\left(\one-u(\beta-\widehat X)^{-1}(S^\ast\widehat YS-\widehat Y)\right)^{-1}
(\beta-\widehat X)^{-1}\cal T|\psi_{\sigma,k,\mu}\rangle\,\d k.\nonumber
\end{align}
Set 
\[
A(\eta)=\int_{-\infty}^\infty\e^{-\eta|t|}
\langle\psi_{\sigma,k,\pm}|\e^{\i tL_0}\cal T\e^{-\i tL_0}|\psi_{\sigma',k',\pm}\rangle\frac{\d t}{2\pi},
\]
\[B(\eta)=\eta\int_{0}^\infty\e^{-\eta t}
\langle\psi_{\sigma,k,\pm}|{\cal F}|\psi_{\sigma',k',\pm}\rangle\frac{\d t}{2\pi},
\]
where 
\[
{\cal F}=W_-^\ast h^{-1/2}
\left(\e^{-t\cal L^\ast}k(Y)\e^{-t\cal L}
-\e^{t\cal L^\ast}k(Y)\e^{t\cal L}\right)h^{-1/2}W_-.
\]
By the intertwining property of the wave operator, we have
\begin{align*}
\e^{\i tL_0}\cal T\e^{-\i tL_0}&=W_-^\ast\e^{\i tL}h^{-1/2}\phi(Y)h^{-1/2}\e^{-\i tL}W_-\\
&=W_-^\ast h^{-1/2}\e^{t\cal L^\ast}\phi(Y)\e^{t\cal L}h^{-1/2}W_-\\
&=-\frac{\d\ }{\d t}W_-^\ast h^{-1/2}\e^{t\cal L^\ast}k(Y)\e^{t\cal L}h^{-1/2}W_-,
\end{align*}
and an integration by parts yields that 
\begin{equation}
A(\eta)=B(\eta),
\label{Sat1}
\end{equation}
for any $\eta >0$. 
Let us now take the limit $\eta\downarrow0$ in this formula. Since $\psi_{\sigma,k,\pm}$
is a generalized eigenfunction of $L_0$ to the eigenvalue $\pm\varepsilon(k)$,
we get, on the left hand side of \eqref{Sat1},
$$
\langle\psi_{\sigma,k,\pm}|\cal T|\psi_{\sigma',k',\pm}\rangle\!\!
\int_{-\infty}^\infty\!\e^{-\eta|t|\pm\i  t(\varepsilon(k)-\varepsilon(k'))}
\frac{\d t}{2\pi} \to\langle\psi_{\sigma,k,\pm}|\cal T|\psi_{\sigma',k,\pm}\rangle
\delta(\varepsilon(k)-\varepsilon(k')).
$$
Using \eqref{eq:klimit}, the Abel limit\footnote{See footnote \ref{GenLim} on page \pageref{GenLim}}
on the right hand side of \eqref{Sat1} yields
\begin{align*}
\frac1{2\pi}\langle\psi_{\sigma,k,\pm}|&W_-^\ast h^{-1/2}\left(h^{1/2}W_-\widehat YW_-^\ast h^{1/2}
-h^{1/2}W_+\widehat YW_+^\ast h^{1/2}
\right)h^{-1/2}W_-|\psi_{\sigma',k',\pm}\rangle\\
&=\frac1{2\pi}\langle\psi_{\sigma,k,\pm}|\widehat Y-S^\ast\widehat YS|\psi_{\sigma',k',\pm}\rangle\\
&=\frac1{2\pi}\langle\psi_{\sigma,k,\pm}|\widehat Y-S_\pm(k)^\ast\widehat YS_\pm(k)|\psi_{\sigma',k,\pm}\rangle
\delta(k-k'),
\end{align*}
and we conclude that
\begin{equation}
\langle\psi_{\sigma,k,\pm}|\cal T|\psi_{\sigma',k,\pm}\rangle=\frac1{2\pi}
\langle\psi_{\sigma,k,\pm}|\widehat Y-S_\pm(k)^\ast\widehat YS_\pm(k)|\psi_{\sigma',k,\pm}\rangle
\varepsilon'(k).
\label{Tmatrixelt}
\end{equation}
Note that the operator $\widehat Y$ acts on the fiber $\mathfrak h_{k\pm}$ as the matrix
\begin{equation}
\left.\widehat Y\right|_{\mathfrak h_{k\pm}}=\left[
\begin{array}{cc}
Y_L&0\\0&Y_R
\end{array}
\right].
\label{YhatAction}
\end{equation}
Relation \eqref{Tmatrixelt} allows us to write
\begin{align*}
\sum_{\sigma=\pm}
\langle\psi_{\sigma,k,\mu}&|\left(I-u(\beta-\widehat X)^{-1}(S^\ast\widehat YS-\widehat Y)\right)^{-1}
(\beta-\widehat X)^{-1}\cal T|\psi_{\sigma,k,\mu}\rangle\\
&=\tr_{\mathfrak h_{k\mu}}\left[
\left(\one-u(\beta-\widehat X)^{-1}(S_\mu(k)^\ast\widehat YS_\mu(k)-\widehat Y)\right)^{-1}\right.\\
&\qquad\qquad\times\left.\vphantom{\left[\left(\widehat X\right)^{-1}\right]}
(\beta-\widehat X)^{-1}
(\widehat Y-S_\mu(k)^\ast\widehat YS_\mu(k))\right]\frac{\varepsilon'(k)}{2\pi}\\
&=\frac{\d\ }{\d u}\tr_{\mathfrak h_{k\mu}}\left[
\log\left(\one-u(\beta-\widehat X)^{-1}(S_\mu(k)^\ast\widehat YS_\mu(k)-\widehat Y)\right)\right]
\frac{\varepsilon'(k)}{2\pi}.
\end{align*}
Inserting the last identity into \eqref{tridonsat} and  integrating  over $u$ we derive 
\begin{align*}
g(X,Y)&=-\sum_{\mu=\pm}\int_0^\pi\tr_{\mathfrak h_{k\mu}}\left[
\log\left(\one-(\beta-\widehat X)^{-1}(S_\mu(k)^\ast\widehat YS_\mu(k)-\widehat Y)\right)\right]
\frac{\d\varepsilon(k)}{4\pi}\\
&=-\sum_{\mu=\pm}\int_0^\pi\log{\mathrm{det}}_{\mathfrak h_{k\mu}}
\left(\one-(\beta-\widehat X)^{-1}(S_\mu(k)^\ast\widehat YS_\mu(k)-\widehat Y)\right)
\frac{\d\varepsilon(k)}{4\pi}.
\end{align*}

{\noindent\bf Remark.} The last formula retains its validity in a much broader context. It holds for
an arbitrary number of infinite harmonic reservoirs coupled to a  finite harmonic system as
long as the scattering approach sketched here applies. Furthermore, the formal
analogy between our Hilbert space treatment of harmonic dynamics and quantum
mechanics suggests  that quasi-free quantum systems could be  also studied  by a similar
scattering approach. That is indeed the case, see Section \ref{sect:EBBM}.
\begin{figure}[htbp]
\begin{center}
    \includegraphics[scale=0.9]{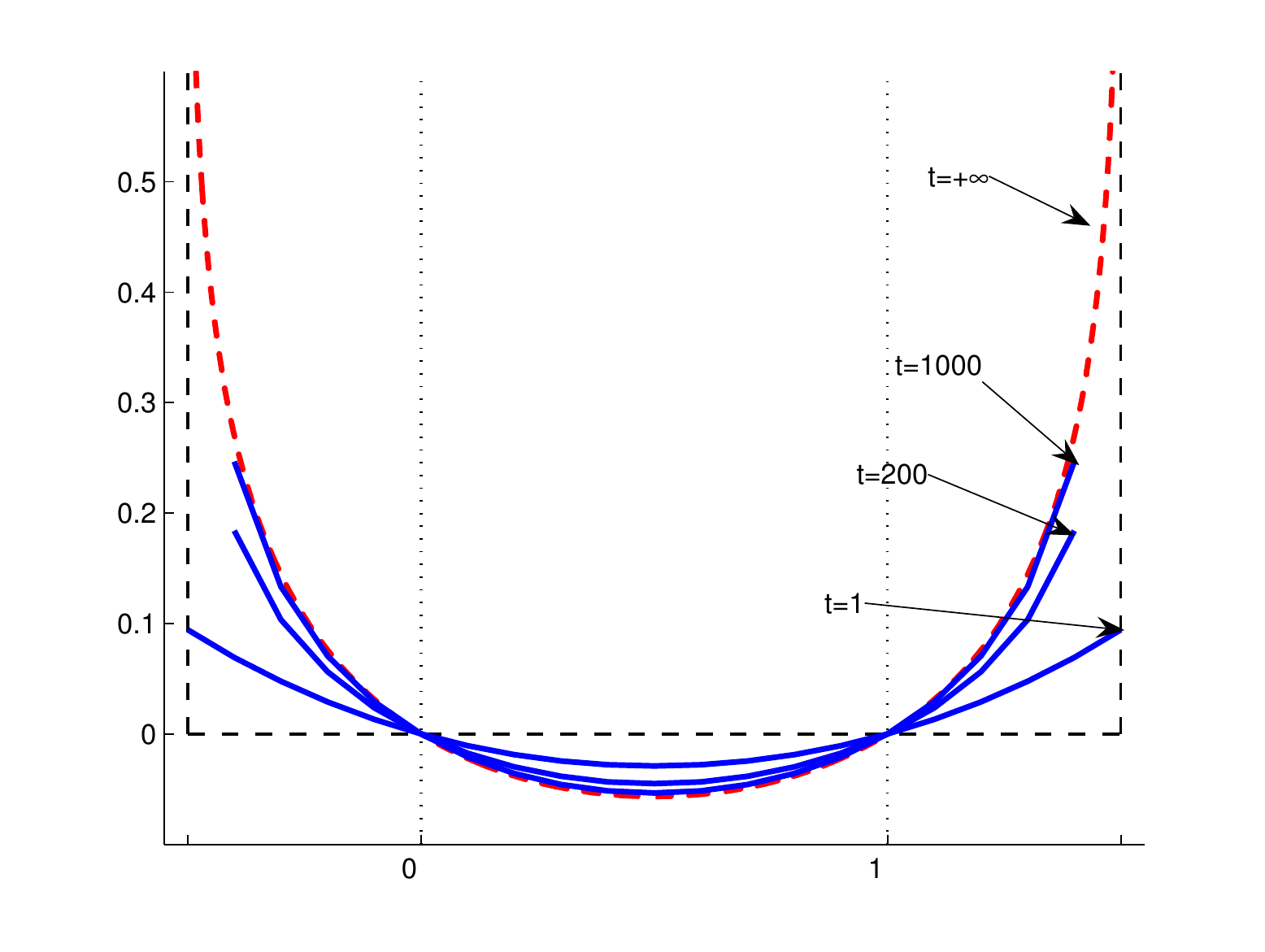}
    \caption{Solid lines: the generating function $\alpha\mapsto t^{-1}e_t(\alpha)$ for various values of
    $t$ and finite reservoirs ($N=20$, $M=300$). The slope at $\alpha=1$
    is $\omega_X(\Sigma^t)$, compare with Figure \ref{fig:sigma}.
    Dashed line: the limiting function 
    $\alpha\mapsto e(\alpha)$ for infinite reservoirs.  }
    \label{fig:eofalpha}
\end{center}
\end{figure}

\bigskip
Invoking \eqref{SAction} and \eqref{YhatAction} leads to our final result
\begin{equation}
g(X,Y)=-\kappa\log\left(1+\frac{(Y_R-Y_L)\left[(X_R-X_L)-(Y_R-Y_L)\right]}{(\beta-X_R)(\beta-X_L)}
\right).
\label{ESfuncForm}
\end{equation}
Note that $g(X,Y)$ is finite for $-T_R^{-1}<Y_R-Y_L<T_L^{-1}$ and $+\infty$ otherwise.
Since $e_t(\alpha)=g_t(X,\alpha X)$, one has
$$
e(\alpha)=\lim_{t\to\infty}\frac1te_t(\alpha)=-\kappa
\log\left(1+\frac{(T_L-T_R)^2}{T_LT_R}\alpha(1-\alpha)\right),
$$
which is finite provided $2|\alpha-1/2|<(T_L+T_R)/|T_L-T_R|$ and $+\infty$ otherwise
(see Figure \ref{fig:eofalpha}).
Note also the explicit symmetries $g(X,X-Y)=g(X,Y)$ and $e(1-\alpha)=e(\alpha)$ inherited from
the finite time Evans-Searles symmetries \eqref{FTESsym} and \eqref{FTGESsym}.
\index{symmetry!Evans-Searles}

\medskip
\begin{exo}\label{Exo:Dlog}
Let $\rr\ni x\mapsto A(x)$ be a differentiable function with values in the trace class
operators on a Hilbert space. Show that if
$\|A(x_0)\|<1$ then $x\mapsto\tr\log(\one+A(x))$ is differentiable at $x_0$ and
$$
\left.\frac{\d\ }{\d x}\tr\log(\one+A(x))\right|_{x=x_0}=\tr((\one+A(x_0))^{-1}A'(x_0)).
$$
{\sl Hint}: use the formula
$$
\log(1+a)=\int_1^\infty\left(\frac1{t}-\frac1{t+a}\right)\d t,
$$
valid for $|a|<1$.
\end{exo}

\section{The central limit theorem}
\label{sect:CLT}

\index{theorem!central limit}\myindex{CLT|see{theorem, central limit}}
As a first application of the generalized Evans-Searles functional $g(X,Y)$, we derive
a central limit theorem (CLT) for the current fluctuations. To this end, let
us decompose the mean currents into its expected value and a properly normalized
fluctuating part, writing
$$
\frac1t\int_0^t\Phi_{j,s}\,\d s=\frac1t\int_0^t\omega_X(\Phi_{j,s})\,\d s+
\frac1{\sqrt t}\delta\Phi_j^t,
$$
for $j\in\{L,R\}$. By Definition \eqref{gXYdef}, the expected mean current is given by
$$
\frac1t\int_0^t\omega_{X}(\Phi_{j,s})\,\d s
=-\left.\partial_{Y_j}\frac1tg_t(X,Y)\right|_{Y=0},
$$
while the fluctuating part is centered, $\omega_X(\delta\Phi_{j,t})=0$, with covariance
$$
\omega_X(\delta\Phi_j^t\delta\Phi_k^t)=
\left.\partial_{Y_k}\partial_{Y_j}\frac1tg_t(X,Y)\right|_{Y=0}.
$$
For large $t$, the expected mean current converges to the NESS expectation
$$
\lim_{t\to\infty}\frac1t\int_0^t\omega_{X}(\Phi_{j,s})\,\d s=\omega_{X,+}(\Phi_j).
$$
To study the large time asymptotics of the current fluctuations
$\delta{\bf\Phi}^t=(\delta\Phi_L^t,\delta\Phi_R^t)$ we consider
the characteristic function
\begin{equation}
\omega_X\left(\e^{\i Y\cdot\delta{\bf\Phi}^t}\right)=
\omega_X\left(\e^{\i\sum_jY_j\frac1{\sqrt t}\int_0^t\left(\Phi_{j,s}-\omega_X(\Phi_{j,s})\right)\,\d s}\right),
\label{CLTgf}
\end{equation}
\ie the Fourier transform of their distribution. To control the limit
$t\to\infty$, we need a technical result which is the object of the following exercise.

\medskip
\begin{exo}
\label{Ex:vitali}
Show that for a given  $\beta_L>0$ and $\beta_R>0$  there exists $\epsilon>0$ such that
the function $Y\mapsto g_t(X,Y)$ is analytic in
$D_\epsilon=\{Y=(Y_L,Y_R)\in\cc^2\,|\,|Y_L|<\epsilon,|Y_R|<\epsilon\}$ and
satisfies
\begin{equation}
\sup_{\atop{Y\in D_\epsilon}{t>0}}\left|\frac1tg_t(X,Y)\right|<\infty.
\label{VitBound}
\end{equation}
{\sl Hint}: start with  \eqref{GESFormOne} and use the identity  $\log\det(\one-T)=\tr (\log(\one-T))$ and
the factorization $\log(\one-z)=-zf(z)$ to obtain the bound
$|\log\det(\one-T)|\le\|T\|_1f(\|T\|)$ where
$\|T\|_1=\tr(\sqrt{T^\ast T})$ denotes the trace norm of $T$.
\end{exo}

\medskip
The convergence result of the preceding section and the uniform bound \eqref{VitBound}
imply that 
\[
\lim_{t\to\infty}\frac1t g_t(X,Y)=g(X,Y),
\]
uniformly for $Y$ in compact subsets of 
$D_\epsilon$, that all the derivatives w.r.t. $Y$ of $\frac1t g_t(X,Y)$ are uniformly bounded 
on such compact subsets and converge uniformly to the corresponding derivatives 
of $g(X,Y)$ (see Theorem \ref{PropVitali} in Appendix \ref{appx:Vitali}).
For $Y\in\cc^2$ and $t>0$ large enough, Equ. \eqref{CLTgf} can be written as
$$
\omega_X\left(\e^{\i Y\cdot\delta{\boldsymbol\Phi}^t}\right)
=\exp\left[t\left(\frac1tg_t\left(X,\frac{Y}{\i\sqrt t}\right)
-\frac{Y}{\i\sqrt t}\cdot\left({\boldsymbol\nabla}_Y\frac1tg_t\right)(X,0)\right)\right],
$$
and the Taylor expansion of $\frac1t g_t(X,Y)$ around $Y=0$ yields
$$
\frac1tg_t\left(X,\frac{Y}{\i\sqrt t}\right)
-\frac{Y}{\i\sqrt t}\cdot\left({\boldsymbol\nabla}_Y\frac1tg_t\right)(X,0)=
-\frac1{2t}\sum_{j,k}\left(\partial_{Y_j}\partial_{Y_k}\frac1tg_t\right)(X,0)Y_jY_k+O(t^{-3/2}),
$$
from which we conclude that
\begin{equation}
\lim_{t\to\infty}\omega_X\left(\e^{\i Y\cdot\delta{\bf\Phi}^t}\right)=
\e^{-\frac12 Y\cdot{\bf D}Y},
\label{GaussD}
\end{equation}
with a covariance matrix ${\bf  D}=[{D}_{jk}]$ given by
$$
{D}_{jk}=\lim_{t\to\infty}\left(\partial_{Y_j}\partial_{Y_k}\frac1t g_t\right)(X,0)
=\left(\partial_{Y_j}\partial_{Y_k}g\right)(X,0).
$$
Evaluating the right hand side of these identities yields
$$
D_{11}=D_{22}=-D_{12}=-D_{21}=\kappa\left(T_L^2+T_R^2\right).
$$
Since the right hand side of \eqref{GaussD} is the Fourier transform of the centered
Gaussian measure on $\rr^2$ with covariance ${\bf D}$, the L\'evy-Cram\'er continuity 
theorem (see \eg Theorem 7.6 in \cite{Bi1}) implies that the current fluctuations
$\delta{\bf\Phi}^t$ converge in law to  this Gaussian, \index{theorem!central limit}
\ie that for all bounded continuous functions $f:\rr^2\to\rr$
\begin{equation}
\lim_{t\to\infty}\omega_X(f(\delta{\bf\Phi}^t))=\int f(\phi,-\phi)\e^{-\phi^2/2\mathfrak{d}}\,
\frac{\d\phi}{\sqrt{2\pi\mathfrak{d}}},
\label{eq:CLTcurr}
\end{equation}
where $\mathfrak{d}=\kappa\left(T_L^2+T_R^2\right)$. Note in particular that the fluctuations
of the left and right mean currents are opposite to each other in this limit.

\medskip
\begin{exo}
\label{Exo:ECons}
Use the CLT \eqref{eq:CLTcurr} and the results of Exercise
\ref{Exo:SteadyFlux} to show that
$$
\frac1t\int_0^t\left(\Phi_{L,s}+\Phi_{R,s}\right)\,\d s\longrightarrow 0,
$$
in probability as $t\to\infty$, \ie that for any $\epsilon>0$ the probability
\begin{equation}
\omega_X\left(\left\{\left|\frac1t\int_0^t\left(\Phi_{L,s}+\Phi_{R,s}\right)\,\d s\right|\ge\epsilon\right\}\right),
\label{eq:offdiagproba}
\end{equation}
tends to zero as $t\to\infty$.
\end{exo}

\medskip
It is interesting to compare the equilibrium  ($T_L=T_R$) and the non-equilibrium 
($T_L\not=T_R$) case. In the first case  the expected mean currents 
vanish (recall that in this  case $\omega_{X,+}$ is the equilibrium state) while in the second they 
are non-zero. In both cases the fluctuations of the mean currents have similar qualitative
features at the CLT scale $t^{-1/2}$. In particular they are always symmetrically distributed  w.r.t. $0$.

\section{Linear response theory near equilibrium}
\label{sect:ClassicLinearResponse}

The linear response theory for our harmonic chain model follows trivially from the formula 
for steady heat fluxes derived in Exercise \ref{Exo:SteadyFlux}. Our goal in this section, however,  
is to  present a derivation of the linear response theory based on the functionals $g_t(X, Y)$ and 
$g(X, Y)$. This derivation, which  follows the ideas of Gallavotti \cite{Ga}, is applicable to 
any time-reversal invariant dynamical system for which the conclusions of Exercise \ref{Ex:vitali} 
hold. For additional information and a general axiomatic approach to derivation of linear response 
theory based on functionals $g_t(X, Y)$ and  $g(X, Y)$  we refer the reader to 
\cite{JPR}.\index{linear response}

Starting from
$$
-\lim_{t\to\infty}\left.\partial_{Y_{L/R}}\frac1tg_t(X,Y)\right|_{Y=0}=\omega_{X,+}(\Phi_{L/R}),
$$
and using the fact that the derivative and the limit can be interchanged (as we learned in the
previous section) one gets
\beq
-\left.\partial_{Y_{L/R}}g(X,Y)\right|_{Y=0}=\omega_{X,+}(\Phi_{L/R}).
\label{SFlux}
\eeq

\medskip
{\bf\noindent Remark.} The main result of Section \ref{sect:GF}, which expresses the
Evans-Searles function $g(X,Y)$ in terms of the on-shell scattering matrix, immediately implies
\begin{align*}
&\omega_{X,+}(\Phi_{L/R})\\
&=\left.\partial_{Y_{L/R}}
\sum_{\mu=\pm}\int_0^\pi{\mathrm{tr}}_{\mathfrak h_{k\mu}}\left[\log
\left(\one-(\beta-\widehat X)^{-1}(S_\mu(k)^\ast\widehat YS_\mu(k)-\widehat Y)\right)\right]
\frac{\d\varepsilon(k)}{4\pi}\right|_{Y=0}\\
&=\sum_{\mu=\pm}\int_0^\pi{\mathrm{tr}}_{\mathfrak h_{k\mu}}
\left[(\beta-\widehat X)^{-1}(P_{L/R}-S_\mu(k)^\ast P_{L/R}S_\mu(k))\right]
\frac{\d\varepsilon(k)}{4\pi},
\end{align*}
which can be interpreted as a classical version of the Landauer-B\"uttiker formula
(see Exercise \ref{Exo:LBFormula}).\index{formula!Landauer-B\"uttiker}

\bigskip
The Onsager matrix ${\bf L}=[L_{jk}]_{j,k\in\{L,R\}}$ defined by\index{Onsager matrix}
$$
L_{jk}=\left.\partial_{X_k}\omega_{X,+}(\Phi_j)\right|_{X=0},
$$
describes the response of the system to weak thermodynamic forces. Taylor's formula
$$
\omega_{X,+}(\Phi_j)=\sum_{k}L_{jk}X_k+o(X),\qquad (X\to0),
$$
expresses the steady currents to the lowest order in the driving forces. From \eqref{SFlux},
we deduce that
$$
L_{jk}=\left.-\partial_{X_k}\partial_{Y_j}g(X,Y)\right|_{X=Y=0}.
$$
The ES symmetry $g(X,X-Y)=g(X,Y)$ further leads to
\begin{align*}
\partial_{X_k}\partial_{Y_j}g(X,Y)&=
\partial_{X_k}\partial_{Y_j}g(X,X-Y)\\
&=-\partial_{X_k}(\partial_{Y_j}g)(X,X-Y)\\
&=-(\partial_{X_k}\partial_{Y_j}g)(X,X-Y)-(\partial_{Y_k}\partial_{Y_j}g)(X,X-Y),
\end{align*}
so that
\begin{equation}
\left.\partial_{X_k}\partial_{Y_j}g(X,Y)\right|_{X=Y=0}
=\left.-\frac12\partial_{Y_k}\partial_{Y_j}g(X,Y)\right|_{X=Y=0},
\label{symlemma}
\end{equation}
and hence
\begin{equation}
L_{jk}=\left.\frac12\partial_{Y_k}\partial_{Y_j}g(0,Y)\right|_{Y=0}.
\label{LFormTwo}
\end{equation}
Since the function $g(0,Y)$ is $C^2$ at $Y=0$, we conclude from \eqref{LFormTwo} that the 
Onsager reciprocity relation\index{relation!Onsager reciprocity}
$$
L_{jk}=L_{kj},
$$
hold.

\medskip
\begin{exo}
In regard to Onsager relation, open systems with  {\em two} thermal reservoirs are special.  
Show that  the Onsager relation  follow from the conservation law 
\[
\omega_{X, +}(\Phi_L) +\omega_{X, +}(\Phi_R)=0.
\]
Time-reversal invariance plays no role in this argument! What is the physical origin of this derivation? 
Needless to say, the derivation of Onsager reciprocity relation described in 
this section directly  extends to open classical systems coupled to more than $2$ thermal reservoirs 
to which this exercise does not apply. 
\end{exo}

\medskip
The positivity of entropy production implies
$$
0\le\omega_{X,+}(\sigma_X)=\sum_j\omega_{X,+}(\Phi_j)X_j=\sum_{j,k}L_{jk}X_jX_k+o(|X|^2),
$$
so that the Onsager matrix is positive semi-definite. In fact, looking back at Section 
\ref{sect:CLT}, we observe that the Onsager matrix coincide, up to a constant factor, with the 
covariance of the current fluctuations at equilibrium,
$$
{\bf L}=\left.\frac12\,{\bf D}\right|_{X=0}.
$$
This is of course the celebrated Einstein relation.\index{relation!Einstein}

For our harmonic chain model the Green-Kubo formula for the Onsager matrix can be derived by an 
explicit computation. In the  following exercises we outline a derivation that extends to general 
time-reversal invariant dynamical systems.\index{formula!Green-Kubo}

\medskip

\begin{exo}
Show that the Green-Kubo formula holds in the Ces\`aro sense
$$
L_{jk}=\lim_{t\to\infty}\frac1{t}\int_0^t
\left[\frac12\int_{-s}^s\omega_0(\Phi_j\Phi_{k,\tau})\,\d\tau\right]\,\d s.
$$
{\sl Hint}: using the results of the previous section, rewrite \eqref{LFormTwo} as
$$
L_{jk}=\lim_{t\to\infty}\left.\partial_{Y_k}\partial_{Y_j}\frac1{2t} g_t(0,Y)\right|_{Y=0},
$$
and work out the derivatives.
\end{exo}
\begin{exo}
Using the fact${}^\dagger$
that $\langle\delta_x|\e^{\i t\sqrt{I-\Delta}}|\delta_y\rangle=O(t^{-1/2})$ as $t\to\infty$ ($\delta_x$
is the Kronecker delta at $x\in\zz$), show that 
$\omega_0(\Phi_j\Phi_{k,t})=O(t^{-1})$. Invoke the Hardy-Littlewood Tauberian
theorem (see, \eg \cite{Ko}) to conclude that the Kubo formula
$$
L_{jk}=\lim_{t\to\infty}\frac12\int_{-t}^t\omega_0(\Phi_j\Phi_{k,\tau})\,\d\tau,
$$
holds.

\noindent${}^\dagger$This follows from a simple stationary phase estimate.
\end{exo}

\medskip

\section{The Evans-Searles fluctuation theorem}
\label{sect:ClassicESFT}

The central limit theorem derived in Section \ref{sect:CLT} shows that, for large $t$, typical 
fluctuations of the mean current ${\bf\Phi}^t$ with respect to its expected value
$\omega_X({\bf\Phi}^t)$
are small, of the order $t^{-1/2}$. In the same regime $t\to\infty$, the theory of large deviations 
provides information on the probability of occurrence of bigger fluctuations, of the order $1$.
More precisely, the existence of the
limit\footnote{The distribution $Q^t$ of the mean current ${\bf\Phi}^t$ was introduced in 
Exercise \ref{Exo:Qt}},
\begin{equation}
g(X,Y)=\lim_{t\to\infty}\frac{1}{t}\log\int\e^{-tY\cdot s}\,\d Q^t(s),
\label{gRap}
\end{equation}
and the regularity of the function $Y\mapsto g(X,Y)$
allow us to apply the G\"artner-Ellis theorem  (see Exercise \ref{Exo:ESLDP} below) to obtain
the Large Deviation Principle (LDP)\index{large deviation principle}\index{theorem!G\"artner-Ellis}
\myindex{LDP|see{large deviation principle}}
\begin{align*}
-\inf_{s\in{\rm int} (G)}I_X(s)
\le\liminf_{t\to\infty}\frac1t\log Q^t(G)
\le\limsup_{t\to\infty}\frac1t\log Q^t(G)
\le-\inf_{s\in{\rm cl}({G})}I_X(s),
\end{align*}
for any Borel  set $G\subset\rr^2$. Here, ${\rm int} (G)$ denotes the interior of $G$,
${\rm cl}({G})$ its closure, and the rate function $I_X:\rr^2\to[-\infty,0]$ is given by
$$
I_X(s)=-\inf_{Y\in\rr^2}\left(Y\cdot s+g(X,Y)\right).
$$
The symmetry $g(X, Y)=g(X, X-Y)$ implies   
\begin{equation}
I_X(-s)= X\cdot s + I_X(s).
\label{ES-rate}
\end{equation}
The last relation  is sometimes called the Evans-Searles symmetry for the rate function. 
\index{symmetry!Evans-Searles}

\begin{figure}[htbp]
\begin{center}
    \includegraphics[scale=0.6]{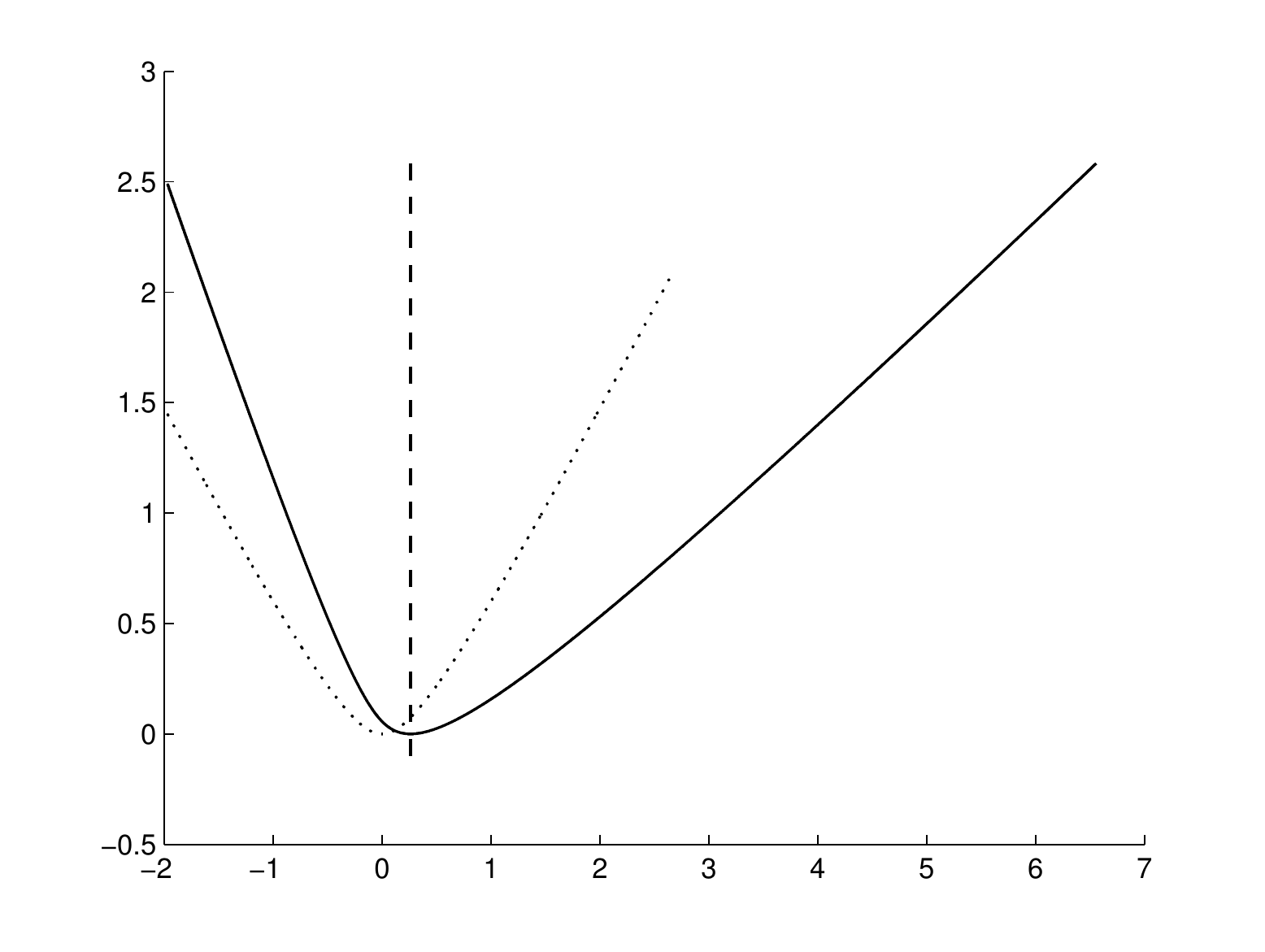}
    \caption{The rate function $I_X(s,-s)$ (solid line). Notice the asymmetry which reflects the fact that
     $X_L>X_R$. The dashed vertical line marks the position of the
     mean current $\omega_{X,+}(\Phi_L)>0$. In contrast, the rate function $I_X(s,-s)$ in the
     absence of forcing, $X_L=X_R$, (dotted line) is symmetric around zero. }
    \label{fig:IX}
\end{center}
\end{figure}

\medskip
\begin{exo} 
Show  that
$$
I_X(s_L,s_R)=\left\{
\begin{array}{ll}
+\infty&\text{if } s_L+s_R\not=0,\\[10pt]
\displaystyle F(\theta)&\text{if }s_L=-s_R=\displaystyle\frac\kappa{\beta_0}\sinh \theta,
\end{array}
\right.
$$
where
$$
F(\theta)=\kappa\left[
2\,\sinh^2\frac\theta2-\frac{\delta}{\beta_0}\sinh \theta
-\log\left(\left(1-\frac{\delta^2}{\beta_0^2}\right)\cosh^2\frac\theta2\right)
\right ],
$$
$\beta_0=\beta-(X_L+X_R)/2$ and $\delta=(X_L-X_R)/2$. Show that $I_X(s_L,s_R)$
is strictly positive (or $+\infty$) except for $s_L=-s_R=\omega_{X,+}(\Phi_L)$ where it vanishes. 
Compare with  Figure \ref{fig:IX}.
\end{exo}

\medskip
The LDP provides the most powerful formulation of the Evans-Searles or transient fluctuation theorem.
In particular, it gives fairly precise information on the rate at which the measure $Q^t$ concentrates 
on the diagonal $\{(\phi,-\phi)\,|\,\phi\in\rr\}$ (recall Exercise \ref{Exo:ECons}):
the probability \eqref{eq:offdiagproba} decays 
super-exponentially as $t\to\infty$ for any $\epsilon>0$. Taking this fact as well as the continuity
of the function $F(\theta)$ into account, we observe that for any interval $J\subset\rr$
one has
\index{theorem!Evans-Searles fluctuation}\index{theorem!transient fluctuation}
$$
\lim_{t\to\infty}\frac1t\log Q^t(J\times\rr)=-\inf_{s\in J}I_X(s,-s).
$$
A rough interpretation of this formula 
$$
\omega_X\left(\left\{\frac1t\int_0^t\Phi_{L,s}\,\d s=\phi\right\}\right)\sim\e^{-t I_X(\phi,-\phi)},
$$
identifies $I_X(-\phi,\phi)$ as the rate of exponential decay of the probability for the mean
current to deviate from its expected value $\omega_{X,+}(\Phi_L)$.
More precisely, one has
\begin{equation}
\lim_{\delta\downarrow0}
\lim_{t\to\infty}\frac1t\log Q^t([\phi-\delta,\phi+\delta]\times\rr)=-I_X(\phi,-\phi).
\label{minneapolis-1}
\end{equation}
The symmetry \eqref{ES-rate} implies 
$$
I_X(-\phi,\phi)=I_X(\phi,-\phi)+(X_L-X_R)\phi\ge(X_L-X_R)\phi,
$$
and it  follows that
\begin{equation}
\lim_{\delta\downarrow0}
\lim_{t\to\infty}\frac1t\log 
\frac{Q^t([-\phi-\delta,-\phi+\delta]\times\rr)}{Q^t([\phi-\delta,\phi+\delta]\times\rr)}=
-(X_L-X_R)\phi,
\label{minneapolis-2}
\end{equation}
or, in a more sloppy notation,
$$
\frac{\omega_X\left(\left\{\frac1t\int_0^t\Phi_{L,s}\,\d s=-\phi\right\}\right)}
{\omega_X\left(\left\{\frac1t\int_0^t\Phi_{L,s}\,\d s=\phi\right\}\right)}
\sim\e^{-t(X_L-X_R)\phi}.
$$
This shows that 
the mean current is exponentially more likely to flow from the hotter to the colder reservoir
than in the opposite direction, \ie on a large time scale, the probability of violating
the second law of thermodynamics becomes exceedingly small. Note also that \eqref{minneapolis-2} is (essentially) a considerably 
weaker statement then \eqref{minneapolis-1}. Relation \eqref{minneapolis-2}, after replacing $\lim$ with 
$\limsup/\liminf$ can be derived directly 
from the finite time symmetry $g_t(X, Y)=g_t(X, X-Y)$ and without invoking the  large deviation theory. 

\medskip
\begin{exo}
\label{Exo:ESLDP}\index{theorem!G\"artner-Ellis}
Check that the G\"artner-Ellis theorem (Theorem \ref{GEmultidim} in Appendix \ref{sect:GEmultiD}
applies to \eqref{gRap}, \ie
show that the function $Y\mapsto g(X,Y)$ given in Equ. \eqref{ESfuncForm}
is differentiable on the domain $\cal D=\{(Y_L,Y_R)\in\rr^2\,|\,-T_R^{-1}<Y_R-Y_L<T_L^{-1}\}$
where it is finite and that it is steep, {\sl i.e.,}
$$
\lim_{\cal D\ni Y\to Y_0}|{\boldsymbol\nabla}_Yg(X,Y)|=\infty,
$$
for $Y_0\in\partial\cal D$.
\end{exo}
\begin{exo}
Apply the G\"artner-Ellis theorem to the generating function $e(\alpha)$ to derive a LDP
for the mean entropy production rate $\Sigma^t$, \ie for the probability distribution $P^t$
of Section \ref{sect:univsym}.
\end{exo}

\medskip

\section{The Gallavotti-Cohen fluctuation theorem}
\label{sect:ClassicGCthm}

\index{theorem!Gallavotti-Cohen fluctuation}

In this section we briefly comment on the Gallavotti-Cohen fluctuation theorem for a thermally driven
harmonic chain. Let us consider the cumulant generating function of the currents in the NESS 
$\omega_{X,+}$,
$$
g_{+,t}(X,Y)=\omega_{X,+}\left(\e^{-tY\cdot{\bf\Phi}^t}\right).
$$
Evaluating the Gaussian integral yields
$$
g_{+,t}(X,Y)=-\frac12\log\det\left(
\one-D_{X,+}\left(\e^{t\cal L^\ast}k(Y)\e^{t\cal L}-k(Y)\right)\right).
$$
Proceeding as in Section \ref{sect:GF}, one shows that
$$
g_+(X,Y)=\lim_{t\to\infty}\frac1tg_{+,t}(X,Y)=\lim_{t\to\infty}\frac1tg_{t}(X,Y)=g(X,Y).
$$
Hence, $g_+(X, Y)$ and the corresponding rate functions $I_{X,+}(s)=I_{X}(s)$ inherit the symmetries 
\[
g_+(X, Y)=g_+(X, X-Y), \qquad I_{X,+}(-s)=X\cdot s + I_{X,+}(s).
\] 
Via G\"artner-Ellis theorem, the functional $I_{X, +}(s)$ control the fluctuations of ${\bf \Phi}^t$ as 
$t\rightarrow \infty$ w.r.t. $\omega_{X, +}$ and, after replacing $\omega_{X}$ with $\omega_{X, +}$ 
(so now $Q^t(f)=\omega_{X, +}(f({\bf \Phi}^t)$, etc) one can repeat the  discussion of the previous 
section line by line. The obtained results are called the Gallavotti-Cohen fluctuation theorem. 

Since $\omega_{X, +}$ is singular w.r.t. $\omega_{X}$ in the non-equilibrium case $X_L \not= X_R$, 
the Gallavotti-Cohen fluctuation theorem refers to configurations (points in the phase space) which are 
not seen  by  the Evans-Searles fluctuation theorem (and vice versa, of course). The identity
$g_+(X, Y)=g(X, Y)$, which was for the first time observed in \cite{JPR}, 
may seem surprising on the first sight. It turned out, however, that it holds for any {\em non-trivial} 
model for which the existence of $g_+(X,Y)$ and $g(X, Y)$ has been established. This point has 
been raised in \cite{JPR} to the {\em Principle of Regular Entropic Fluctuations}. Since we will not 
discuss quantum Gallavotti-Cohen fluctuation  theorem in these lecture notes, we refer the reader to 
\cite{JPR, JOPP} for additional discussion of these topics. 
\bindex{principle of regular entropic fluctuations}

\chapter{Algebraic quantum statistical mechanics of finite systems}
\label{chap:QSMfin}

We now turn to the main topic of these lecture notes: quantum statistical mechanics. 
This section is devoted to a detailed exposition of the mathematical structure of algebraic 
quantum statistical mechanics of finite quantum systems.

\section{Notation and basic facts}
\label{sect:QSMBasic}

Let $\cK$ be  a finite dimensional complex Hilbert space with  inner product $\langle \psi|\phi \rangle$
linear in the second argument\footnote{Many different Hilbert spaces will appear in
the lecture notes and in latter parts we will often denote inner product by 
$(\,\cdot\,|\,\cdot\,)$}. Recall the Schwarz inequality $\langle\psi|\phi\rangle\le\|\psi\|\,\|\phi\|$,
where equality holds iff $\psi$ and $\phi$ are collinear. In particular 
$\|\phi\|=\sup_{\|\psi\|=1}\langle \psi|\phi \rangle$.
We will use Dirac's notation: for $\psi\in\cK$, $ \langle\psi|$ denotes
the linear functional $\cK\ni\phi\mapsto\langle\psi|\phi\rangle\in\cc$ and $|\psi\rangle$ its adjoint
$\cc\ni\alpha\mapsto\alpha\psi\in\cK$. 
\nindex{(1}{$\langle\,\cdot\,\abso\,\cdot\,\rangle$}{inner product on $\cK$}
\nindex{(3}{$\langle\psi\abso$}{Dirac bra}
\nindex{(4}{$\abso\psi\rangle$}{Dirac ket}

We denote by $\cO$ the $\ast$-algebra\footnote{See Exercise \ref{exo:staralg} below.}
of all linear maps  $A:\cK\rightarrow\cK$. 
For $A\in\cO$, $\|A\|=\sup_{\|\psi\|=1}\|A\psi\|$ denotes its operator norm and $\sp(A)$ its spectrum, 
\ie the set of all eigenvalues of $A$. Let us recall some important properties of the operator norm.
Since $\|A\psi\|\le\|A\|\,\|\psi\|$, it follows that $\|AB\|\le\|A\|\,\|B\|$ for all $A, B\in\cO$.
\nindex{sp( )}{$\sp(\,\cdot\,)$}{spectrum}\nindex{norm}{$\norm\,\cdot\,\norm$}{operator norm} 
Since
$$
\|A^\ast\phi\|=\sup_{\|\psi\|=1}\langle\psi|A^\ast\phi \rangle
\sup_{\|\psi\|=1}\langle A\psi|\phi \rangle\le\|A\|\,\|\phi\|,
$$
and $A^{\ast\ast}=A$, one has $\|A^\ast\|=\|A\|$ for all $A\in\cO$. Finally, from the two inequalities
$\|A^\ast A\|\le\|A^\ast\|\,\|A\|=\|A\|^2$ and
$$
\|A\|^2=\sup_{\|\psi\|=1}\|A\psi\|^2
=\sup_{\|\psi\|=1}\langle A\psi|A\psi\rangle=\sup_{\|\psi\|=1}\langle\psi|A^\ast A\psi\rangle\le\|A^\ast A\|,
$$
on deduces the $C^\ast$-property $\|A^\ast A\|=\|A\|^2$.\index{Cstar property@$C^\ast$ property}

The identity operator is denoted by $\one$ and, whenever the meaning is clear within the context, 
we shall write $\alpha$ for  $\alpha \one$ and $\alpha\in\cc$. 
Occasionally, we shall indicate the dependence on the 
underlying Hilbert space $\cK$ by the subscript $_\cK$ ($\cO_\cK$, $\one_\cK$, etc).
\nindex{1}{$\one$}{operator unit}

To any orthonormal basis $\{e_1,\ldots,e_N\}$ of the Hilbert space $\cK$ one can associate the basis
$\{E_{ij}=|e_i\rangle\langle e_j|\,|\,i,j=1,\ldots,N\}$ of $\cO$ so that, for any $X\in\cO$,
\nindex{Eij}{$E_{ij}$}{basis of $\cO$} 
$$
X=\sum_{i,j=1}^NX_{ij}E_{ij},
$$
where $X_{ij}=\langle e_i|Xe_j\rangle$. Equipped with the inner product
$$
(X|Y)=\tr (X^\ast Y),
$$
$\cO$ becomes a Hilbert space and $\{E_{ij}\}$  an orthonormal basis of this space. 

The self-adjoint and positive parts of $\cO$ are the subsets
\nindex{Oself}{$\Os$}{self-adjoint part of $\cO$}\nindex{O+}{$\cO_+$}{positive part of $\cO$}
\begin{align*}
\Os&=\{A\in\cO\,|\,A^\ast=A\},\\
\cO_+&=\{A\in\cO\,|\,\langle\psi|A\psi\rangle\ge0\text{ for all }\psi\in\cK\}\subset\Os.
\end{align*}
We write $A\ge0$ if $A\in\cO_+$ and $A\ge B$ if $A-B\ge0$. 
Note that $A\in\cO_+$ iff $A\in\Os$ and $\sp(A)\subset[0,\infty[$. If
$A\ge0$ and $\Ker A=\{0\}$ we write $A>0$.

A linear bijection $\vartheta:\cO\rightarrow\cO$ is called a $\ast$-automorphism
\bindex{star automorphism@$\ast$-automorphism} of
$\cO$ if $\vartheta(AB)=\vartheta(A)\vartheta(B)$ and $\vartheta(A^\ast)=\vartheta(A)^\ast$.
${\rm Aut}(\cO)$ denotes the  group of all $\ast$-automorphisms of $\cO$ and $\mathrm{id}$
\nindex{Aut()}{${\rm Aut}(\cO)$}{group of $\ast$-automorphisms of $\cO$}
\nindex{id}{$\mathrm{id}$}{identity map on $\cO$}
denotes its identity. Any  $\vartheta\in\rm{Aut}(\cO)$  preserves $\Os$ and satisfies
$\vartheta(\one)=\one$ and $\vartheta(A^{-1})=\vartheta(A)^{-1}$ for all invertible $A\in\cO$.
In particular, $\vartheta((z-A)^{-1})=(z-\vartheta(A))^{-1}$ and  $\sp(\vartheta(A))=\sp(A)$.
It follows that $\vartheta$ preserves $\cO_+$ and is isometric, \ie $\|\vartheta(A)\|=\|A\|$ for all $A\in\cO$.

Let $\cK_1$ and $\cK_2$ be two complex Hilbert spaces of dimension $N_1$ and $N_2$.
Let $\{e_1^{(1)},\ldots,e_{N_1}^{(1)}\}$ and $\{e_1^{(2)},\ldots,e_{N_2}^{(2)}\}$ be orthonormal basis of 
$\cK_1$ and $\cK_2$. The tensor product $\cK_1\otimes\cK_2$
\nindex{KoK}{$\cK_1\otimes\cK_2$}{tensor product}
is defined, up to isomorphism, 
as the $N_1\times N_2$-dimensional complex Hilbert space with orthonormal basis
$\{e_{i_1}^{(1)}\otimes e_{i_2}^{(2)}\,|\,i_1=1,\ldots,N_1; i_2=1,\ldots,N_2\}$, \ie 
$\cK_1\otimes\cK_2$ consists of all linear combinations
$$
\psi=\sum_{i_1=1}^{N_1}\sum_{i_2=1}^{N_2}\psi_{i_1i_2}\,e_{i_1}^{(1)}\otimes e_{i_2}^{(2)},
$$
the inner product being determined by
$\langle e_{i_1}^{(1)}\otimes e_{i_2}^{(2)}|e_{j_1}^{(1)}\otimes e_{j_2}^{(2)}\rangle
=\delta_{i_1j_1}\delta_{i_2j_2}$. The {\sl tensor product} of two vectors 
$\psi=\sum_{i=1}^{N_1}\psi_i e_i^{(1)}\in\cK_1$ and $\phi=\sum_{i=1}^{N_2}\phi_i e_i^{(2)}\in\cK_2$
is the vector in $\cK_1\otimes\cK_2$ defined by 
$$
\psi\otimes\phi
=\sum_{i_1=1}^{N_1}\sum_{i_2=1}^{N_2}\psi_{i_1}\phi_{i_2}\, e_{i_1}^{(1)}\otimes e_{i_2}^{(2)}.
$$
The tensor product extends to a bilinear map  from $\cK_1\times\cK_2$ to $\cK_1\otimes\cK_2$.
We recall the characteristic property of the space $\cK_1\otimes\cK_2$: for any Hilbert space $\cK_3$,
any bilinear map $F:\cK_1\times\cK_2\to\cK_3$ uniquely extends to a linear map 
$\widehat F:\cK_1\otimes\cK_2\to\cK_3$ by setting $\widehat F \psi\otimes\phi=F(\psi,\phi)$.

The tensor product of two linear operators $X\in\cO_{\cK_1}$ and $Y\in\cO_{\cK_2}$ is the
linear operator on $\cK_1\otimes\cK_2$ defined by
$$
(X\otimes Y) \psi \otimes\phi=X\psi\otimes Y\phi,
$$
and $\cO_{\cK_1}\otimes\cO_{\cK_2}$ is the $\ast$-algebra generated by such operators.
Denoting by $\{E^{(1)}_{i_1i_2}\}$ and $\{E^{(2)}_{j_1j_2}\}$ the basis of $\cO_{\cK_1}$ and 
$\cO_{\cK_2}$ corresponding to the orthonormal basis $\{e_i^{(1)}\}$ and $\{e^{(2)}_j\}$ of 
$\cK_1$ and $\cK_2$, the $N_1^2\times N_2^2$ operators
$$
E_{i_1i_2,j_1j_2}=E^{(1)}_{i_1j_1}\otimes E^{(2)}_{i_2j_2}
=|e^{(1)}_{i_1}\otimes e^{(2)}_{i_2}\rangle\langle e^{(1)}_{j_1}\otimes e^{(2)}_{j_2}|,
$$
form a basis of $\cO_{\cK_1\otimes\cK_2}$. This leads to a natural  identification of  $\cO_{\cK_1\otimes\cK_2}$
and  $\cO_{\cK_1}\otimes\cO_{\cK_2}$.

If $\lambda\in\sp(A)$ we denote by
\nindex{Plambda}{$P_\lambda(\,\cdot\,)$}{spectral projection} 
$P_\lambda$  the associated spectral projection.  
When we wish to indicate its dependence on $A$ we shall write $P_\lambda(A)$. 
If $A\in\Os$, we shall denote by $\lambda_j(A)$
\nindex{lambdaj}{$\lambda_j(\,\cdot\,)$}{eigenvalues in decreasing order} 
the eigenvalues of $A$ listed with {\em multiplicities} and in {\em decreasing} order.

If $f(z)=\sum_{n=0}^\infty a_nz^n$ is analytic in the disk $|z|<r$ and $\|A\|<r$, then
$f(A)$ is defined by the analytic functional calculus,
\nindex{f(A)}{$f(A)$}{functional calculus}
$$
f(A)=\sum_{n=0}^\infty a_n A^n=\oint_{|w|=r'} f(w)(w-A)^{-1}\,\frac{\d w}{2\pi\i},
$$
for any $\|A\|<r'<r$. If $A\in\Os$ and $f:\rr\rightarrow\cc$, then $f(A)$ is defined by  the spectral theorem, \ie 
\[
f(A)=\sum_{\lambda\in \sp(A)}f(\lambda) P_\lambda.
\]
In particular, for $A\in\cO_+$, 
$$
\log A=\sum_{\lambda\in \sp(A)}\log(\lambda) P_\lambda,
$$
where $\log$ denotes the natural  logarithm. We shall always use the following conventions:
$\log 0=-\infty$ and $0\log 0=0$.\nindex{log()}{$\log(\,\cdot\,)$}{natural logarithm}
By the Lie product formula, for any $A, B\in\cO$, 
\bindex{formula!Lie product}
\begin{equation}
\e^{A+B}=\lim_{n\rightarrow \infty}\left(\e^{A/n}\e^{B/n}\right)^n
=\lim_{n\rightarrow \infty}\left(\e^{A/2n}\e^{B/n}\e^{A/2n}\right)^{n}.
\label{LieProdForm}
\end{equation}
For any $A\in\cO$, $A^\ast A\ge0$ and we set $|A|=\sqrt{A^\ast A}\in\cO_+$ and denote by
\nindex{muj}{$\mu_j(\,\cdot\,)$}{singular values}
\nindex{(A)}{$\abso A\abso$}{operator absolute value} 
$\mu_j(A)$ the singular values of $A$ (the eigenvalues of $|A|$)
listed with multiplicities and in decreasing order. Since $\|A\psi\|^2=\||A|\psi\|^2$ one has
$\Ker |A|=\Ker A$ and $\Ran |A|=\Ran A^\ast$. It follows that the map
$U:\Ran A^\ast\ni|A|\psi\mapsto A\psi\in\Ran A$
is well defined and isometric. It provides the polar decomposition $A=U|A|$.
\bindex{polar decomposition}

\medskip
\begin{exo}
\label{exo:staralg}
\bindex{algebra!complex}\bindex{algebra!commutative or abelian}\bindex{algebra!unital}
A {\sl complex algebra} is a complex vector space $\cal A$ with a product $\cal A\times\cal A\to\cal A$
satisfying the following axioms: for any $A,B,C\in\cal A$ and any $\alpha\in\cc$,\\[2pt]
(1)  $A(BC)=(AB)C$.\\[2pt]
(2)  $A(B+C)=AB+AC$.\\[2pt]
(3) $\alpha(AB)=(\alpha A)B=A(\alpha B)$.\\[2pt]
The algebra $\cal A$ is called {\sl abelian} or {\sl commutative} if $AB=BA$ for all $A,B\in\cal A$ and 
{\sl unital} if there exists $\one\in\cal A$ such that $A\one=\one A=A$ for all $A\in\cal A$.

\medskip\noindent
\bindex{algebra!$\ast$-}
A {\sl $\ast$-algebra} is a complex algebra with a map $\cal A\ni A\mapsto A^\ast\in\cal A$ such that,
for any $A,B\in\cal A$ and any $\alpha\in\cc$,\\[2pt]
(4) $A^{\ast\ast}=A$.\\[2pt]
(5) $(AB)^\ast=B^\ast A^\ast$.\\[2pt]
(6) $(\alpha A+B)^\ast=\overline{\alpha} A^\ast+B^\ast$.

\medskip\noindent
\bindex{algebra!$C^\ast$-}
A norm on a $\ast$-algebra $\cal A$ is a norm on the vector space $\cal A$ satisfying
$\|AB\|\le\|A\|\,\|B\|$ and $\|A^\ast\|=\|A\|$ for all $A,B\in\cal A$. A finite dimensional
normed $\ast$-algebra $\cal A$ is a {\sl $C^\ast$-algebra} if $\|A^\ast A\|=\|A\|^2$ 
for all $A\in\cal A$. (If $\cal A$ is infinite dimensional, one additionally requires $\cal A$ 
to be complete w.r.t. the norm topology).

\medskip\noindent
Show that if $\cK$ is a finite dimensional Hilbert space then the set $\cO$ of all linear maps 
$A:\cK\to\cK$ is a unital $C^\ast$-algebra.
\end{exo}
\begin{exo}
\label{Lowner-Heinz}
Prove the L\"owner-Heinz inequality:\bindex{inequality!L\"owner-Heinz}
if $A,B\in\cO_+$ are such that $A\ge B$ then $A^s\ge B^s$ for any $s\in[0,1]$.

\noindent{\sl Hint}: show that $(B+t)^{-1}\ge(A+t)^{-1}$ for all $t>0$ and use the identity
$$
A^s-B^s=\frac{\sin\pi s}{\pi}\int_0^\infty t^s\left(\frac1{B+t}-\frac1{A+t}\right)\,\d t.
$$
\end{exo}
\begin{exo}
\label{Duhamel}
\exop Let $A, B\in\cO$. Prove Duhamel's formula \bindex{formula!Duhamel}
\[
\e^{B}-\e^{A}=\int_0^1 \e^{sB}(B-A)\e^{(1-s)A}\,\d s.
\]
{\sl Hint}:  integrate the derivative of the function $f(s)=\e^{sB}\e^{(1-s)A}$.

\exop Iterating Duhamel's formula, prove the second order Duhamel expansion \bindex{expansion!Duhamel}
$$
\e^{B}-\e^{A}=
\int_0^1 \e^{s B}(B-A)\e^{(1-s)B}\,\d s
+\int_0^1\int_0^s\e^{uB}(B-A)\e^{(s-u)A}(A-B)\e^{(1-s)B}\,\d u\d s.
$$

\exop Let $P$ be a projection and set $Q=\one-P$. Apply the previous formula to the case
$B=PAP$ to show that
$$
P\e^AP=P\e^{PAP}P+\int_0^1\int_0^u\e^{(u-s)PAP}PAQ\e^{(1-u)A}QAP\e^{sPAP}\,\d s\d u.
$$
\end{exo}
\begin{exo}\label{paris-auto}
Let $\vartheta \in {\rm Aut}(\cO)$. Show that there exists  unitary $U_\vartheta \in {\cal O}$, 
unique up to a phase, such that $\vartheta(A)= U_\vartheta A U_{\vartheta}^{-1}$.

\noindent{\sl Hint}: show first that if $P$ is an orthogonal projection, then so is $\vartheta(P)$ and 
$\tr(P)=\tr(\vartheta(P))$. Pick an orthonormal basis $\{e_1, \cdots, e_N\}$ of $\cK$ and show that 
$\vartheta(|e_i\rangle \langle e_j|)=|e_i^\prime\rangle \langle e_j^\prime|$, where 
$\{e_1^\prime, \cdots, e_N^{\prime}\}$ is also an orthonormal basis of $\cK$. Set 
$U_{\vartheta}e_i=e_{i}^\prime$ and complete the proof.
\end{exo}
\begin{exo}\label{Exo:min-max}
\exop Let $A\in {\cal O}_{\rm self}$. Prove the min-max principle: for \bindex{min-max principle}
$j=1,\ldots,\dim\cK$,
\[
\lambda_j(A)=\sup_{S}\inf _{\atop{\psi\perp S}{\|\psi\|=1}}\langle \psi| A\psi\rangle,
\]
where supremum is taken over all subspaces $S\subset \cK$ such that $\dim S=\dim\cK- j$
(recall our convention $\lambda_1(A)\ge\lambda_2(A)\ge\cdots\ge\lambda_{\dim\cK}(A)$.
 
\exop Using the min-max principle, prove that for $A, B \in {\cal O}_{\rm self}$, 
\[
|\lambda_{j}(A)-\lambda_j(B)|\leq \|A-B\|.
\]
\end{exo}

\medskip

\section{Trace inequalities}
\label{sect:TraceInequal}

Let $\{\psi_j\}$ be an orthonormal basis of $\cK$. We recall that the trace of $A\in\cO$, 
denoted $\tr(A)$, is defined by 
\[
\tr(A)=\sum_j \langle \psi_j|A\psi_j\rangle.
\]
For any unitary $U\in\cO$, $\tr(A)=\tr (UAU^{-1})$ and $\tr(A)$ is independent of the choice 
of the basis. In particular, if $A$ is self-adjoint then $\tr (A)=\sum_j\lambda_j(A)\in\rr$ and
if $A\in\cO_+$ then $\tr\,(A)\ge0$. 

For  $p\in ]0, \infty[$ we set
\nindex{norm norm}{$\norm\,\cdot\,\norm_p$}{$p$-norm on $\cO$}
\begin{equation}
\|A\|_p =(\tr|A|^p)^{1/p}=\left(\sum_j \mu_j(A)^p\right)^{1/p}.
\label{pNormDef}
\end{equation}
$\|A\|_\infty=\max_j \mu_j(A)$ is the usual operator norm of $A$. The function
$]0, \infty[\,\ni  p \mapsto \|A\|_p$ is real analytic,  monotonically decreasing, and 
\begin{equation}
\lim_{p\rightarrow \infty}\|A\|_p=\|A\|_\infty. 
\label{jeep}
\end{equation}
For $p\in[1,\infty]$, the map $\cO\ni A\mapsto\|A\|_p$ is a unitary invariant norm. Since 
$\dim\cK<\infty$, all these norms are equivalent and induce the same topology on $\cO$.

Let $A=U|A|$ be the polar decomposition of $A$ and denote by $\{\psi_j\}$ an orthonormal basis
of eigenvectors of $|A|$. Then
$$
\tr(BA)=\sum_j\langle\psi_j|BU|A|\psi_j\rangle=\sum_j\mu_j(A)\langle\psi_j|BU\psi_j\rangle,
$$
from which we conclude that
\begin{equation}
|\tr(BA)|\le\sum_j\mu_j(A)|\langle\psi_j|BU\psi_j\rangle|\le\|B\|\sum_j\mu_j(A)=\|B\|\,\|A\|_1.
\label{sec-in}
\end{equation}
In particular,
\[
|\tr(A)|\leq \|A\|_1.
\]

The basic trace inequalities are:
\bet \label{theor-trace-in}
\ben
\item The Peierls-Bogoliubov inequality: for $A, B\in \Os$, \bindex{inequality!Peierls-Bogoliubov}
\[
\log\frac{\tr(\e^A\e^B)}{\tr (\e^B)}\geq \frac{\tr (A\e^B)}{\tr (\e^B)}.
\]
\item The Klein inequality: for $ A, B \in\cO_+$, \bindex{inequality!Klein}
\[
\tr(A\log A- A\log B)\geq \tr(A-B),
\]
with equality iff $A=B$. 
\item The H\"older inequality: for $A, B\in\cO$ and $p, q\in [1,\infty]$ satisfying
\bindex{inequality!H\"older} $p^{-1} + q^{-1}=1$,
\[
\|AB\|_1\leq \|A\|_p \|B\|_q.
\]
\item The Minkowski inequality: for $A, B\in\cO$ and $p\in [1,\infty]$, \bindex{inequality!Minkowski}
\[
 \|A+ B\|_p\leq \|A\|_p + \|B\|_p.
\]
\een
\eet

\bigskip
\demo (1) For $\lambda\in\sp(A)$ we set
$$
p_\lambda=\frac{\tr(P_\lambda(A)\e^B)}{\tr(\e^B)},
$$
so that $p_\lambda\in[0,1]$ and $\sum_\lambda p_\lambda=1$.
The convexity of the exponential function and Jensen's inequality imply
$$
\frac{\tr (\e^A\e^B)}{\tr(\e^B)}
=\sum_\lambda \e^\lambda p_\lambda\ge\e^{\sum_\lambda\lambda p_\lambda}
=\e^{\tr (A\e^B)/\tr(\e^B)}.
$$

\medskip\noindent(2) If $\Ker B\not\subset\Ker A$, then  the left-hand side in (2) is  $+\infty$  
and the inequality holds trivially. Assuming $\Ker B\subset\Ker A$, we set
$$
p_{\lambda,\mu}=\tr(P_\lambda(A)P_\mu(B)),
$$
for $(\lambda,\mu)\in\sp(A)\times\sp(B)$ so that $p_{\lambda,\mu}\in[0,1]$,
$\sum_{\lambda,\mu}p_{\lambda,\mu}=1$ and $p_{\lambda,0}=\delta_{\lambda,0}p_{0,0}$.
Then, we can write
\[
\tr(A\log A-A\log B)=\sum_{\atop{\lambda,\mu}{\mu\not=0}}\lambda\log(\lambda/\mu)p_{\lambda,\mu}.
\]
The inequality $x\log x \geq x-1$, which holds for $x\geq 0$, implies that for $\lambda\ge0$
and $\mu>0$,
$$
\lambda\log\frac\lambda\mu=\mu\,\frac\lambda\mu\log\frac\lambda\mu
\ge\mu\left(\frac\lambda\mu-1\right)=\lambda-\mu,
$$
and so 
\[
\tr(A\log A-A\log B)\geq\sum_{\atop{\lambda,\mu}{\mu\not=0}}(\lambda-\mu)p_{\lambda,\mu}
=\sum_{\lambda,\mu}(\lambda-\mu)p_{\lambda,\mu}
=\tr (A-B).
\]
If the equality holds, then we must have
$$
\sum_{\atop{\lambda,\mu}{\mu\not=0}}\mu\left[\frac\lambda\mu\log\frac\lambda\mu
-\left(\frac\lambda\mu-1\right)\right]p_{\lambda,\mu}=0,
$$
where all the terms in the sum are non-negative. Since $x\log x=x-1$ iff $x=1$, it follows that 
$p_{\lambda,\mu}=0$ for $\lambda\not=\mu\not=0$. We have already noticed that
$p_{\lambda,0}=0$ for $\lambda\not=0$, hence we have  $p_{\lambda,\mu}=0$ for
$\lambda\not=\mu$ and it follows that
$$
P_\lambda(A)P_\mu(B)P_\lambda(A)=0=P_\mu(B)P_\lambda(A)P_\mu(B),
$$
for $\lambda\not=\mu$. Since
$$
P_\lambda(A)=\sum_\mu P_\lambda(A)P_\mu(B)P_\lambda(A)
=P_\lambda(A)P_\lambda(B)P_\lambda(A),
$$
we must have $P_\lambda(B)\ge P_\lambda(A)$ and $\sp(A)\subset\sp(B)$. By symmetry, 
the reverse inequalities also hold and hence  $B=A$.

\medskip\noindent(3) Equ. \eqref{jeep} implies that it suffices to consider the case $1<p<\infty$.
Denote by $AB=U|AB|$, $A=V|A|$ and $B=W|B|$  the polar decompositions of $AB$, $A$ 
and $B$. Then
$$
\|AB\|_1=\tr|AB|=\tr (U^\ast AB)=\tr(U^\ast V|A|W|B|)
=\lim_{\epsilon\downarrow0}\tr( U^\ast V(|A|+\epsilon)W(|B|+\epsilon)).
$$
The function
\[
F_\epsilon(z)= \tr (U^\ast V(|A| +\epsilon)^{pz} W(|B|+\epsilon)^{q(1-z)}),
\]
is entire analytic and bounded on  the strip $0\le\Re z\le1$. 
For any $y\in \rr$, the bound \eqref{sec-in} yields 
\[
|F_\epsilon(\i y)|\leq \tr((|B| +\epsilon)^q), \qquad |F_\epsilon(1+\i y)|\leq \tr ((|A|+\epsilon)^p).
\]
Hence, by Hadamard's three lines theorem (see, \eg \cite{RS2}), for any $z$ in the strip 
$0 \leq \Re z \leq 1$, 
\[
|F_\epsilon(z)|\leq \left[\tr((|A| +\epsilon)^p)\right]^{\Re z} \left[\tr((|B| +\epsilon)^q)\right]^{1-\Re z}.
\]
Substituting   $z=1/p$ we get 
\[
|\tr (U^\ast V(|A| +\epsilon)W(|B|+\epsilon)) |\leq \||A|+\epsilon\|_p\||B|+\epsilon\|_q,
\]
and the limit $\epsilon \downarrow 0$ yields the statement.

\medskip\noindent(4) Again, it suffices to consider the case $1<p<\infty$.  Let $q$ be such that 
$p^{-1}+ q^{-1}=1$. We first observe that 
\begin{equation}
 \| A\|_p=\sup_{\|C\|_q=1}|\tr(AC)|.
\label{jeep2}
 \end{equation}
Indeed, the  H\"older inequality implies 
\[
 \sup_{\|C\|_q=1}|\tr(AC)|\le\sup_{\|C\|_q=1}\|AC\|_1
\leq  \| A\|_p.
\]
On the other hand, if $C=\|A\|_p^{-p/q}|A|^{p/q}U^\ast$ where $A=U|A|$ denotes the polar 
decomposition of $A$, then $\|C\|_q=1$ and $\tr(AC)=\|A\|_p$, and so \eqref{jeep2} holds. 
Finally, \eqref{jeep2} implies 
\[
\|A+B\|_p =\sup_{\|C\|_q=1}|\tr((A+B)C)|\leq \sup_{\|C\|_q=1}|\tr(AC)| +  \sup_{\|C\|_q=1}|\tr(BC)|=\|A\|_p+ \|B\|_p.
\]
\qed

\bigskip
We shall also need:
\bet
The Araki-Lieb-Thirring inequality: for $A, B\in\cO_+$, $p>0$ and $r\geq 1$,
\bindex{inequality!Araki-Lieb-Thirring}
\[ \tr \left((A^{1/2}B A^{1/2})^{rp}\right)\leq  \tr \left((A^{r/2}B^r A^{r/2})^p\right).\]
\label{theor-trace-in1}
\eet
\demo
By an obvious limiting argument (replacing $A$ and $B$ with $A+\epsilon$ and $B+\epsilon$)
it suffices to prove the theorem in the case $A, B>0$. We  split the proof into four steps.

\medskip{\noindent\em Step 1.} If $A, B>0$, then for $0\leq s\leq  1$, $\|A^sB^\s\|\leq  \|AB\|^s$.

{\noindent\em Proof.} Let $\phi, \psi \in \cK$ be  unit vectors
 and 
\[F(z)=\frac{(\phi| A^{z}B^z\psi)}{\|AB\|^{z}}.
\]
The function $F(z)$ is entire analytic and bounded on the strip  $0\le\Re z\le1$. 
For $y\in \rr$ one has $|F(\i y)|\leq 1$,  $|F(1+\i y)|\leq 1$, and so by the three lines theorem, 
$|F(z)|\leq 1$ for $0 \leq \Re z \leq 1$. Taking $z=s$, we deduce that 
\[|(\phi| A^{s}B^s\psi)|\leq \|AB\|^s,\]
and 
\[
\|A^sB^s\|=\sup_{\|\phi\|=\|\psi\|=1}|(\phi| A^{s}B^s\psi)|\leq \|AB\|^s.
\]

\medskip{\noindent\em Step 2.} If $A, B>0$, then for $s\geq   1$, $\|A^sB^\s\|\geq   \|AB\|^s$.

{\noindent\em Proof.} Let $\tilde A=A^s$, $\tilde B=B^s$. Then by Step 1, $\|\tilde A^{1/s}\tilde B^{1/s}\|\leq \|\tilde A\tilde B\|^{1/s}$, 
and the result follows.

{\noindent\em Step 3.} Set $X_r=B^{r/2}A^{r/2}$, $Y_r=X_r^\ast X_r=A^{r/2}B^r A^{r/2}$. 
Let $N=\dim \cK$ and denote by $\lambda_1(r)\ge\cdots\ge\lambda_N(r)$ the eigenvalues of $Y_r$
listed  with multiplicities. Then for $1\leq n\leq N$,
\begin{equation}
 \prod_{j=1}^n\lambda_j(r) \geq\prod_{j=1}^n \lambda_j(1)^r.
\label{ihes-tues-1}
 \end{equation} 
{\noindent\em Proof.} Let $\cH={\cal K}^{\wedge n}$ be the $n$-fold anti-symmetric tensor
product of $\cK$ and $\Gamma_n(Y_q)=Y_q^{\wedge n}$ (the reader not familiar with this concept
may consult Section \ref{sect:SecondQuantiz}).  Step 2 yields the inequality
\begin{align*}
\|\Gamma_n (Y_r)\|&=\|\Gamma_n (X_r)^\ast \Gamma_n(X_r)\|=\|\Gamma_n(X_r)\|^2 =
\|\Gamma_n(B)^{r/2}\Gamma_n(A)^{r/2}\|^2\\[4pt]
&\geq \|\Gamma_n(B)^{1/2}\Gamma_n(A)^{1/2}\|^{2r}=\|\Gamma_n(X_1)\|^{2r}=\|\Gamma_n(Y_1)\|^r,
\end{align*}
Since $\|\Gamma_n(Y_r)\|=\prod_{j=1}^n \lambda_j(r)$, \eqref{ihes-tues-1} follows.

\medskip{\noindent\em Step 4.} For $1\leq n\leq N$,
\begin{equation}
 \sum_{j=1}^n \lambda_j(r)^p\geq \sum_{j=1}^n\lambda_j(1)^{rp}.
\label{ihes-tues-3}
\end{equation}
{\noindent\em Proof.} Set $a_j(r)=\log\lambda_j(r)$. Then, by Step 3,  $a_j(r)$ is a decreasing 
sequence of real numbers satisfying 
\[
\sum_{j=1}^n a_j(r)\geq \sum_{j=1}^n ra_j(1),
\]
for all $n$. We have to show that for all $n$, 
\begin{equation}
\sum_{j=1}^n\e^{pa_j(r)}\geq \sum_{j=1}^n \e^{pra_j(1)}.
\label{ihes-home-0}
\end{equation}
Let $y_+=\max(y, 0)$. We claim that for all $y \in \rr$ and all $n$, 
\begin{equation}
\sum_{j=1}^n (a_j(r)-y)_+\geq \sum_{j=1}^n (ra_j(1)-y)_+.
\label{ihes-home1}
\end{equation}
This relation is obvious if $ra_1(1)-y\le0$. Otherwise, let $k\leq n$ be such that
$$
ra_1(1)-y\ge\cdots\ge ra_k(1)-y\ge0\ge ra_{k+1}(1)-y\ge\cdots\ge ra_n(1)-y.
$$
Then $\sum_{j=1}^n(ra_j(1)-y)_+=\sum_{j=1}^k(ra_j(1)-y)$ and it follows that
\begin{align*}
\sum_{j=1}^n (a_j(r)-y)_+&\geq \sum_{j=1}^k (a_j(r)-y)_+\geq \sum_{j=1}^k (a_j(r)-y)\\[3mm]
&\geq \sum_{j=1}^k(ra_j(1)-y)=\sum_{j=1}^n(ra_j(1)-y)_+.
\end{align*}
The relation \eqref{ihes-home1} and the identity 
\[
\e^{px}=p^2\int_\rr (x-y)_+\e^{p y}\d y,
\]
imply \eqref{ihes-home-0} and \eqref{ihes-tues-3} follows.
In the case $n=N$ the relation  \eqref{ihes-tues-3} reduces to the Araki-Lieb-Thirring inequality. 
\qed

\bigskip
Theorem \ref{theor-trace-in1} and the Lie product formula \eqref{LieProdForm} imply:
\bec\label{GTcor}
For $A, B \in\Os$ the function
\[
[1,\infty[\ni p \mapsto \|\e^{B/p}\e^{A/p}\|_{p}^{p}=\tr ([\e^{A/p}\e^{2B/p}\e^{A/p}]^{p/2})
\] 
is monotonically decreasing and 
\[
\lim_{p\to\infty}\|\e^{B/p}\e^{A/p}\|_{p}^{p}=\tr(\e^{A+B}).
\]
In particular, the Golden-Thompson inequality holds,\bindex{inequality!Golden-Thompson}
\[
\tr (\e^{A}\e^{B})=\|\e^{B/2}\e^{A/2}\|_{2}^{2}\geq \tr(\e^{A+B}).
\]
\eec

\medskip
\begin{exo}
\label{Exo:Holder}
\exop Prove the following generalization of H\"older's inequality: \index{inequality!H\"older}
\begin{equation}
\|AB\|_r\le\|A\|_p\,\|B\|_q,
\label{HolderTwoGen}
\end{equation}
for $p,q,r\in[1,\infty]$ such that $p^{-1}+q^{-1}=r^{-1}$. 

\noindent{\sl Hint}: use the polar decomposition
$B=U|B|$ to write $|AB|^2=|B|C^2|B|$ with $C=\sqrt{U^\ast|A|^2U}$. Invoke the Araki-Lieb-Thirring
inequality to show that $\tr (|AB|^r)\le\tr (|C^r|B|^r|)=\|C^r|B|^r\|_1$. Conclude the proof by applying 
the H\"older inequality. 

\exop Using \eqref{HolderTwoGen}, show that
$$
\|A_1\cdots A_n\|_r\le\prod_{j=1}^n\|A_j\|_{p_j},
$$
provided $\sum_jp_j^{-1}=r^{-1}$.
\end{exo}
\begin{exo}
\label{Exo:ABBA}
Show that for any $A\in\cO$ and $p\in[1,\infty]$ one has $\|A^\ast\|_p=\|A\|_p$. 
In particular, if $A,B\in\Os$ then
\begin{equation}
\|AB\|_p=\|BA\|_p.
\label{ABBAId}
\end{equation}
\end{exo}
\begin{exo}\label{late-gt} Let $A, B \in \Os$. Prove that the function 
\[
[1,\infty[\ni p \mapsto \|\e^{B/p}\e^{A/p}\|_{p}^{p}=\tr ([\e^{A/p}\e^{2B/p}\e^{A/p}]^{p/2})
\] 
is strictly decreasing unless $A$ and $B$ commute (in which case the function is constant). Deduce 
that the Golden-Thompson inequality is strict unless $A$ and $B$ commute. \index{inequality!Golden-Thompson}

\noindent{\sl Hint}: show first that the function is real analytic. Hence, if the function is not strictly decreasing, it must be constant. If the function 
is constant, then its values at $p=2$ and $p=4$ are  equal and  
\[ \tr (\e^{A}\e^B)= \tr (\e^{A/2}\e^{B/2}\e^{A/2}\e^{B/2}).\]
This identity is equivalent to $\tr ([ \e^{A/2}\e^{B/2}-\e^{B/2}\e^{A/2}] [ \e^{A/2}\e^{B/2}-\e^{B/2}\e^{A/2}]^\ast)=0$,  and 
so $\e^{A/2}\e^{B/2}=\e^{B/2}\e^{A/2}$.
\end{exo}

\medskip
\bec\label{logconvcor}
For $A,B\in\cO_+$ and $p\ge1$ the function
$$
\rr\ni\alpha\mapsto\log\|A^\alpha B^{1-\alpha}\|_p^p,
$$
is convex.
\eec

\demo As in the proof of Theorem \ref{theor-trace-in1} we can assume that $A$ and $B$
are non-singular. We first note that for any $s\in]0,1[$ the Araki-Lieb-Thirring inequality implies
\begin{align*}
\|A^sB^s\|_p^p&=\tr\left(\left[B^sA^{2s}B^s\right]^{p/2}\right)
=\tr\left(\left[\left(B^sA^{2s}B^s\right)^{1/s}\right]^{ps/2}\right)\\[3mm]
&\le\tr\left(\left[BA^2B\right]^{ps/2}\right)=\|AB\|_{ps}^{ps}.
\end{align*}
Applying the H\"older inequality \eqref{HolderTwoGen}, the identity \eqref{ABBAId} and the 
previous inequality one gets, for $\alpha,\beta\in\rr$ and 
$\lambda\in]0,1[$,
\begin{align*}
\|A^{\lambda\alpha+(1-\lambda)\beta} B^{1-(\lambda\alpha+(1-\lambda)\beta)}\|_p^p
&=\|A^{\lambda\alpha}A^{(1-\lambda)\beta}B^{(1-\lambda)(1-\beta)}B^{\lambda(1-\alpha)}\|_p^p\\
&=\|B^{\lambda(1-\alpha)}A^{\lambda\alpha}A^{(1-\lambda)\beta}B^{(1-\lambda)(1-\beta)}\|_p^p\\
&\le\|B^{\lambda(1-\alpha)}A^{\lambda\alpha}\|_{p/\lambda}^p\,
\|A^{(1-\lambda)\beta}B^{(1-\lambda)(1-\beta)}\|_{p/(1-\lambda)}^p\\
&=\|A^{\lambda\alpha}B^{\lambda(1-\alpha)}\|_{p/\lambda}^p\,
\|A^{(1-\lambda)\beta}B^{(1-\lambda)(1-\beta)}\|_{p/(1-\lambda)}^p\\
&\le\|A^{\alpha}B^{1-\alpha}\|_{p}^{\lambda p}\,
\|A^{\beta}B^{1-\beta}\|_{p}^{(1-\lambda)p}.
\end{align*}
Taking the logarithm of both sides  yields the result.
\qed

\section{Positive and completely positive maps on $\cO$}
\label{sect:CPmaps}

Denoting by $\{e_1,\ldots,e_N\}$ the standard basis of $\cc^N$, a vector $\psi\in\cK\otimes\cc^N$
has a unique representation
$$
\psi=\sum_{j=1}^N\psi_j\otimes e_j,
$$
where $\psi_j\in\cK$ is completely determined by 
$\langle\phi|\psi_j\rangle=\langle\phi\otimes e_j|\psi\rangle$ for all $\phi\in\cK$. Accordingly, an operator
$X\in\cO_{\cK\otimes\cc^N}$ can be represented as a $N\times N$ block matrix
$$
X=\left[
\begin{array}{cccc}
X_{11}&X_{12}&\cdots&X_{1N}\\
X_{21}&X_{22}&\cdots&X_{2N}\\
\vdots&\vdots&\ddots&\vdots\\
X_{N1}&X_{N2}&\cdots&X_{NN}
\end{array}
\right],
$$
where $X_{ij}\in\cO_{\cK}$ is completely determined by
$\langle\phi|X_{ij}\psi\rangle=\langle\phi\otimes e_i|X\psi\otimes e_j\rangle$ for all $\phi,\psi\in\cK$,
so that
$$
X\psi=\sum_{i,j=1}^N(X_{ij}\psi_j)\otimes e_i.
$$
In particular, 
$X$ is non-negative iff
$$
\sum_{i,j}\langle\psi_i|X_{ij}\psi_j\rangle\ge0,
$$
for all $\psi_1,\ldots,\psi_N\in\cK$.
Note that since $\cO_\cK\otimes\cO_{\cc^N}$ is isomorphic to $\cO_{\cK\otimes\cc^N}$, the same
block matrix representation holds for $X\in\cO_\cK\otimes\cO_{\cc^N}$.

Let $\Phi:\cO_\cK\to\cO_{\cK'}$ be a linear map. $\Phi$ is called {\sl positive} if \bindex{map!positive}
$\Phi(\cO_{\cK+})\subset\cO_{\cK'+}$. One easily shows that if $\Phi$ is positive, then
$\Phi(X^\ast)=\Phi(X)^\ast$ for all $X\in\cO_{\cK}$.
$\Phi$ is called {\sl $N$-positive} if the map 
$\Phi\otimes\one_N:\cO_\cK\otimes\cO_{\cc^N}\to\cO_{\cK'}\otimes\cO_{\cc^N}$ is
positive, where $\one_N$ is the identity map on $\cO_{\cc^N}$. Note that if 
$X\in\cO_\cK\otimes\cO_{\cc^N}$ has the block matrix representation $[X_{ij}]$, then
$\Phi\otimes\one_N(X)\in\cO_{\cK'}\otimes\cO_{\cc^N}$ is represented by the block matrix
$[\Phi(X_{ij})]$.
If $\Phi$ is $N$-positive for all $N$, then it is called {\sl completely positive} (CP).
\bindex{map!completely positive}\bindex{map!unital}\bindex{map!trace preserving}
$\Phi$ is called {\sl unital} if $\Phi(\one_\cK)=\one_{\cK'}$
and {\sl trace preserving} if $\tr\,(\Phi(X))=\tr\,(X)$ for all $X\in\cO_\cK$.

\begin{example}
Suppose that $\cK=\cK_1\otimes \cK_2$ and let $\Phi:\cO_\cK\rightarrow\cO_{\cK_1}$ be the 
unique map satisfying\index{partial trace}\nindex{tr}{$\tr_{\cK}(\,\cdot\,)$}{partial trace}
\[
\tr_{\cK}((B\otimes \one_{\cK_2})A)=\tr_{\cK_1}(B\Phi(A)),
\]
for all $A\in\cO_{\cK}$, $B\in\cO_{\cK_1}$. $\Phi(A)$  is called  the partial trace of $A$ over 
$\cK_2$ and we shall denote it by $\tr_{\cK_2}(A)$.  If $\{\chi_j\}$ is an orthonormal 
basis of $\cK_2$, then the matrix elements of $\tr_{\cK_2}(A)$ are 
\[
\langle \psi | \tr_{\cK_2}(A)\varphi\rangle
=\sum_k \langle\psi\otimes \chi_k|A\, \varphi\otimes \chi_k\rangle.
\]
The map $A\mapsto \tr_{\cK_2}(A)$ is obviously linear, positive (in fact $A>0$ 
implies that $\tr_{\cK_2}(A)>0$) and trace preserving. To show that it is completely positive, 
we note that if $[X_{ij}]$ is a positive block matrix then
$$
\sum_{i,j}\langle\psi_i|\tr_{\cK_2}(X_{ij})\psi_j\rangle=\sum_k\sum_{i,j}
\langle\psi_i\otimes \chi_k|X_{ij}\, \psi_j\otimes \chi_k\rangle\ge0.
$$
\label{partTrace}
\end{example}
\begin{exo}
Show that the following maps are completely positive:\newline
1. A $\ast$-automorphism $\vartheta:\cO\to\cO$.\newline
2. $\cO_\cK\ni X\mapsto\Phi(X)=X\otimes\one_{\cK'}\in\cO_{\cK\otimes\cK'}$.\newline
3. $\cO_\cK\ni X\mapsto\Phi(X)=VXV^\ast\in\cO_\cK$, where $V\in\cO_\cK$.
\label{exo:CPmap}
\end{exo}

\medskip
The following result, due to Stinespring, gives a characterization of CP maps.
\bep\label{PropKraus}
 The linear map $\Phi:\cO_\cK\to\cO_{\cK'}$ is completely positive iff there exists a finite family of
operators $V_\alpha:\cK\to\cK'$ such that
\begin{equation}
\Phi(X)=\sum_{\alpha}V_\alpha XV_\alpha^\ast,
\label{KrausForm}
\end{equation}
for all $X\in\cO_\cK$. Moreover, $\Phi$ is unital iff 
$\sum_\alpha V_\alpha V^\ast_\alpha=\one_{\cK'}$ and trace preserving iff 
$\sum_\alpha V^\ast_\alpha V_\alpha=\one_{\cK}$.
\eep

{\noindent\bf Remark.} The right hand side of \eqref{KrausForm} is called a Kraus representation 
of the completely positive map $\Phi$. Such a representation is not unique.
\bindex{representation!Kraus}

\begin{example}Let $U$ be a unitary operator on $\cK_1\otimes\cK_2$. By Example \ref{partTrace}
and Exercise \ref{exo:CPmap}, the map
$$
\Phi(X)=\frac{\tr_{\cK_2}(U(X\otimes\one_{\cK_2})U^\ast)}{\dim\cK_2},
$$
is completely positive and unital on $\cO_{\cK_1}$. A Kraus representation is given by
$$
\Phi(X)=\sum_{i,j=1}^{\dim\cK_2}V_{i,j}XV_{i,j}^\ast,
$$
where
$$
V_{i,j}=\frac1{\sqrt{\dim\cK_2}}\sum_{k,l=1}^{\dim\cK_1}
|e_k\rangle\langle e_k\otimes f_i|U e_l\otimes f_j\rangle\langle e_l|,
$$
and $\{e_j\}$, $\{f_k\}$ are orthonormal basis of $\cK_1$ and $\cK_2$.
\end{example}

{\noindent\bf Proof of Proposition \ref{PropKraus}.} The fact that a map $\Phi$ defined by
Equ. \eqref{KrausForm} is completely positive  follows from Part 2 of Exercise \ref{exo:CPmap}.
To prove the reverse implication, let $\Phi:\cO_\cK\to\cO_{\cK'}$ be completely positive and denote
by $E_{ij}=|\chi_i\rangle\langle\chi_j|$ the basis of $\cO_\cK$ associated to the orthonormal basis 
$\{\chi_i\}$ of $\cK$. Since
$$
\sum_{i,j=1}^{\dim\cK}\langle\psi_i|E_{ij}\psi_j\rangle
=\left|\sum_{i=1}^{\dim\cK}\langle\psi_i|\chi_i\rangle\right|^2\ge0,
$$
the block matrix $[E_{ij}]$ is positive and hence so is the block matrix $M=[\Phi(E_{ij})]$, an
operator on $\cK'\otimes\cc^{\dim\cK}$. Let $e_i$ be the standard basis of $\cc^{\dim\cK}$ and
define the operator $Q_i:\cK'\otimes\cc^{\dim\cK}\to\cK'$ by
$Q_i\sum_j\psi_j\otimes e_j=\psi_i$, so that $\Phi(E_{ij})=Q_iMQ_j^\ast$. If
$$
M=\sum_{k=1}^{\dim\cK\times\dim\cK'}\lambda_k|\phi_k\rangle\langle\phi_k|,
$$
is a spectral representation of $M$, then
\begin{equation}
\Phi(E_{ij})=\sum_{k=1}^{\dim\cK\times\dim\cK'}\lambda_k Q_i|\phi_k\rangle\langle\phi_k|Q_j^\ast.
\label{PhiEq}
\end{equation}
For each $k=1,\ldots,\dim\cK\times\dim\cK'$ define a linear operator
$V_k:\cK\to\cK'$ by $V_ke_i=\sqrt{\lambda_k}Q_i\phi_k$ for $i=1,\ldots,\dim\cK$.
Then, we can rewrite \eqref{PhiEq} as
$$
\Phi(E_{ij})=\sum_{k=1}^{\dim\cK\times\dim\cK'}V_kE_{ij}V_k^\ast,
$$
and since any $X\in\cO_\cK$ can be written as $X=\sum_{i,j}X_{ij}E_{ij}$ we have
$$
\Phi(X)=\sum_{k=1}^{\dim\cK\times\dim\cK'}V_kXV_k^\ast.
$$
The last statement of Proposition \ref{PropKraus} is obvious.
\hfill$\square$

\medskip
\begin{definition}A linear map $\Phi:\cO_{\cK}\to\cO_{\cK'}$ such that,
for all $X\in\cO_{\cK}$,
\begin{equation}
\Phi(X)^\ast\Phi(X)\le\Phi(X^\ast X),
\label{SchwarzIneq}
\end{equation}
is called a Schwarz map and \eqref{SchwarzIneq} is called the
Schwarz inequality.\bindex{inequality!Schwarz}\bindex{map!Schwarz}
\end{definition}

\bep\label{SchwarzLemma}
Any $2$-positive map $\Phi:\cO_{\cK}\to\cO_{\cK'}$ is a Schwarz map.
\eep

\demo For any $X\in\cO_{\cK}$, the $2\times2$ block matrix
$$
[A_{ij}]=
\left[
\begin{array}{cc}
\one&X\\
X^\ast&X^\ast X
\end{array}
\right],
$$
is non-negative. Indeed, for any $\psi_1,\psi_2\in\cK$ one has
$$
\sum_{i,j}\langle\psi_i|A_{ij}\psi_j\rangle=\|\psi_1+X\psi_2\|^2\ge0.
$$
If $\Phi$ is $2$-positive, then the block matrix $[\Phi(A_{ij})]$ is also non-negative and hence
$$
\sum_{i,j}\langle\phi_i|A_{ij}\phi_j\rangle=\|\phi_1+\Phi(X)\phi_2\|^2
+\langle\phi_2|(\Phi(X^\ast X)-\Phi(X)^\ast\Phi(X))\phi_2\rangle\ge0,
$$
for all $\phi_1,\phi_2\in\cK'$. Setting $\phi_1=-\Phi(X)\phi_2$ yields the Schwarz inequality.
\qed

\medskip
\begin{exo}
Let $\Phi:\cO_{\cK}\to\cO_{\cK'}$ be a linear map and denote by $\Phi^\ast$ its adjoint 
w.r.t. the inner product $(\,\cdot\,|\,\cdot\,)$, that is
$$
(X|\Phi(Y))=\tr_{\cK'}(X^\ast\Phi(Y))=\tr_{\cK}(\Phi^\ast(X)^\ast Y)=(\Phi^\ast(X)|Y).
$$
1. Show that $\Phi^\ast$ is positive iff $\Phi$ is positive.\newline
2. Show that  $\Phi^\ast$ is $N$-positive iff  $\Phi$ is $N$-positive.\newline
3. Show that $\Phi^\ast$ is trace preserving iff $\Phi$ is unital.
\end{exo}

\medskip

\section{States}
\label{sect:States}

An element $\rho \in\cO_+$ is called a density matrix or a {\em state} if $\tr(\rho)=1$. We denote 
by ${\mathfrak S}$ the collection of all states. We shall identify a state $\rho$ with the linear functional
\bindex{density matrix}\bindex{state}\nindex{Sa}{${\mathfrak S}$}{set of states on $\cO$}
$$
\begin{array}{llll}
\rho:&\cO&\to&\cc\\
&A&\mapsto&\tr(\rho A).
\end{array}
$$
With this identification, $\mathfrak S$ can be characterized as the set of all linear functionals
$\phi:\cO\to\cc$ which are positive ($\phi(A)\ge0$ for all $A\in\cO_+$) and normalized ($\phi(\one)=1$).
In models that arise in physics the elements of $\cO$ 
describe observables of the physical system under consideration. The physical states are described 
by elements of ${\mathfrak S}$. If $A$ is self-adjoint and $A=\sum_{\alpha\in \sp(A)} \alpha P_\alpha$ is its 
spectral decomposition, then the possible outcomes of a measurement of $A$ are the 
eigenvalues of $A$. If  the system is in a state $\rho$, the probability that $\alpha$ is observed 
is $\tr (\rho P_\alpha)=\tr(P_\alpha\rho P_\alpha)\in[0,1]$. In particular,  
\[
\rho(A)=\tr(\rho A),
\]
is the expectation value of the observable $A$ and its variance is 
\[
\Delta_{\rho}(A)=\rho((A-\rho(A))^2)=\rho(A^2)-\rho(A^2).
\]
Note that if $A\in\cO_+$, then $\rho(A)\ge0$.
For  $A, B\in\Os$ the  Heisenberg uncertainty principle takes the form  \index{uncertainty principle}
\[
\frac{1}{2}|\rho(\i [A,B])|\leq \sqrt{\Delta_\rho(A)}\sqrt{\Delta_{\rho}(B)}.
\]
If $\Phi:\cO_\cK\to\cO_{\cK'}$ is a positive, unit preserving map, then its adjoint
$\Phi^\ast$ is positive and trace preserving. In particular, it maps states
$\rho\in\mathfrak S_{\cK'}$ into states $\Phi^\ast(\rho)\in\mathfrak S_{\cK}$ in such a way that
$$
\rho(\Phi(A))=\Phi^\ast(\rho)(A),
$$
\ie $\Phi^\ast(\rho)=\rho\circ\Phi$.

\section{Entropy}
\label{sect:Entropy}

Let $\rho$ be a state. \nindex{s()}{$\s(\rho)$}{$\Ran\rho$, support of a state}\bindex{support}
The orthogonal projection on the subspace $\Ran\rho=(\Ker \rho)^{\perp}$ is called the support
of $\rho$ and is denoted ${\rm s}(\rho)$. We shall use the notation $\rho\ll\nu$ iff
\nindex{ll}{$\nu\ll\omega$}{$\Ran\nu\subset\Ran\omega$}
\nindex{perp}{$\nu\perp\omega$}{$\Ran\nu\perp\Ran\omega$}
${\rm s}(\rho)\leq {\rm s}(\nu)$, that is, iff $\Ran\rho\subset\Ran\nu$, and $\rho\perp\nu$ iff 
${\rm s}(\rho)\perp {\rm s}(\nu)$, that is, iff $\Ran\rho\subset\Ker\nu$. 
Two states $\rho$ and $\nu$ are called equivalent if $\rho\ll\nu$ and $\nu\ll\rho$.
A state $\rho$ is called faithful if ${\rm s}(\rho)=\one$, \ie
if $\rho>0$. The set ${\mathfrak S}$ and the set of all faithful states ${\mathfrak S}_{\rm f}$ are convex subsets 
of $\cO_+$. A state $\rho$ is called pure if $\rho=|\psi\rangle\langle\psi|$ for some unit
vector $\psi$. The state \bindex{state!faithful}\bindex{state!pure}\bindex{state!equivalent}
\bindex{state!chaotic}
\begin{equation}
\rho_{\rm ch}=\frac{\one}{\dim \cK},
\label{ChaoStateDef}
\end{equation}
is called chaotic. If $A$ is self-adjoint, we denote 
\[
\rho_A=\frac{\e^A}{\tr(\e^A)}.
\]
The state $\rho_A$ is faithful and $\rho_{A}=\rho_{B}$ iff $A$\nindex{rA}{$\rho_A$}{$\e^A/\tr(\e^A)$}
and $B$ differ by a constant. If $\cK=\cK_1\otimes \cK_2$ and $\rho\in {\mathfrak S}_\cK$, then 
$\rho_{\cK_1}=\tr_{\cK_2}(\rho)\in {\mathfrak S}_{\cK_1}$ and 
$\rho_{\cK_2}=\tr_{\cK_1}(\rho)\in {\mathfrak S}_{\cK_2}$. 

The von Neumann entropy of a state $\rho$, defined by
\bindex{entropy!von Neumann}\nindex{S}{$S(\rho)$}{von Neumann entropy}
\[
S(\rho)=-\tr(\rho \log \rho)=-\sum_{\lambda\in\sp(\rho)}\lambda\log \lambda.
\]
is the non-commutative extension of the Gibbs or Shannon entropy of a probability
distribution. It is characterized by  the following dual variational principles.

\bet\label{var-ent-1}\index{variational principle}
{\rm (1)} For any $\rho\in\mathfrak S$, one has
$$
S(\rho)=\min_{A\in \Os}\log \tr (\s(\rho)\e^A) - \rho(A).
$$
{\rm\noindent(2)} For any $A\in\Os$, one has
\[
\log\tr(\e^A)=\max_{\rho\in{\mathfrak S}}\rho (A) + S(\rho).
\]
\eet
{\bf Remark.} Adopting the decomposition $\cK=\Ran\rho\oplus\Ker\rho$, the minimum in (1) is
achieved at $A$ iff $A=(\log(\rho|_{\Ran\rho})\oplus B)+c$ where $B$ is an arbitrary self-adjoint
operator on $\Ker\rho$ and $c$ an arbitrary real constant. An alternative formulation of (1) is
$$
S(\rho)=\inf_{A\in \Os}\log \tr (\e^A) - \rho(A).
$$
The maximizer in (2) is unique and given by $\rho=\rho_A$. 

\bigskip
\demo
{\noindent\rm(2)} Let $G_A(\rho)=\rho(A) + S(\rho)$. Since $\log\rho_A=A-\log\tr(\e^A)$, 
Klein's inequality implies\index{inequality!Klein}
\[
\log \tr(\e^A)-G_A(\rho)=\tr (\rho(\log \rho-\log \rho_A))\ge\tr(\rho-\rho_A)=0,
\]
for any $\rho\in\mathfrak S$, with equality iff $\rho=\rho_A$. Thus,
\begin{equation}
G_A(\rho)\le\log \tr(\e^A),
\label{GAless}
\end{equation}
with equality iff $\rho=\rho_A$.

\medskip{\noindent\rm (1)} We decompose $\cK=\Ran\rho\oplus\Ker\rho$ and set $P=\s(\rho)$ and 
$Q=\one-P$. Since $G_A(\rho)=G_{PAP}(\rho)$, we can invoke \eqref{GAless} within the subspace 
$\Ran\rho$ to write
$$
G_{A}(\rho)\le\log\tr(P\e^{PAP}P),
$$
where equality holds iff $\rho=P\e^{PAP}P/\tr(P\e^{PAP}P)$, \ie 
$PAP=\log(\rho|_{\Ran\rho})+c$ for some real constant $c$. A second order Duhamel expansion 
(see Part 3 of Exercise \ref{Duhamel}) further yields\index{expansion!Duhamel}
$$
\tr(P\e^AP)=\tr(P\e^{PAP}P)+\int_0^1 u\,\tr\left(\e^{uPAP/2}PAQ\e^{(1-u)A}QAP\e^{uPAP/2}\right)\,\d u,
$$
so that $\tr(P\e^{PAP}P)\le\tr(P\e^AP)$ with equality iff $QAP=0$. We conclude that
$G_{A}(\rho)\le\log\tr(P\e^{A}P)$ and hence $S(\rho)\le\log\tr(P\e^{A}P)-\rho(A)$ where
equality holds iff $A=(\log(\rho|_{\Ran\rho})\oplus B)+c$.
\qed

\bigskip
An immediate consequence of Theorem \ref{var-ent-1} is 
\bec
{\noindent\rm(1)} The function ${\mathfrak S}\ni\rho\mapsto S(\rho)$ is concave.

{\noindent\rm(2)} The function $\Os\ni A\mapsto \log \tr(\e^A)$ is convex.
\eec

Further basic properties of the entropy functional are:
\bet\label{ent-prop-1}
{\rm (1)} The map ${\mathfrak  S}\ni \rho \mapsto S(\rho)$ is continuous.\newline
{\noindent\rm (2)} $0 \leq S(\rho)\leq \log \dim \cK$. Moreover, $S(\rho)=0$ iff $\rho$ is pure  and 
$S(\rho)=\log \dim \cK$ iff $\rho$ is chaotic.\newline
{\rm (3)} For any unitary $U$, $S(U\rho U^{-1})=S(\rho)$.\newline
{\rm (4)} $S(\rho_A)= \log \tr (\e^A)-\tr(A\rho_A)$.\newline
{\rm (5)} If $\cK=\cK_1\otimes \cK_2$, then $S(\rho)\leq S(\rho_{\cK_1}) + S(\rho_{\cK_2})$ 
where the equality holds if and only if $\rho=\rho_{\cK_1}\otimes\rho_{\cK_2}$
(recall that $\rho_{\cK_1}=\tr_{\cK_2}(\rho)$).
\eet

{\noindent\bf Remark.} To (1): the Fannes inequality\bindex{inequality!Fannes}
$$\newline
|S(\rho)-S(\nu)|\le\|\rho-\nu\|_1\log\frac{\dim\cK}{\|\rho-\nu\|_1},
$$
holds provided $\|\rho-\nu\|_1<1/3$. See, \eg \cite{OP}.

\bigskip
\demo The proofs of (1)--(4) are easy and left to the reader. To prove (5) we invoke the variational
principle to write
$$
S(\rho)\le\min_{(A,B)\in\cO_1\times\cO_2}\log \tr (\s(\rho)\e^{A\otimes\one+\one\otimes B})
-\rho(A\otimes\one+\one\otimes B).
$$
Setting $\rho_j=\rho_{\cK_j}$, the support $\s(\rho_1)$ satisfies  
$1=\tr_{\cK_1}(\s(\rho_1)\rho_1)=\tr_\cK((\s(\rho_1)\otimes\one)\rho)$
and, by the definition of the support,  we must have $\s(\rho_1)\otimes\one\ge\s(\rho)$ and a similar
inequality for $\s(\rho_2)$. It follows that $\s(\rho_1)\otimes \s(\rho_2)\ge\s(\rho)$  and
therefore 
\[\tr (\s(\rho)\e^{A\otimes\one+\one\otimes B})\le\tr(\s(\rho_1)\e^A\otimes \s(\rho_2)\e^{B}).\]
Thus, we can write
\begin{align*}
S(\rho)&\le\min_{(A,B)\in\cO_1\times\cO_2}\log\tr(\s(\rho_1)\e^A)+\log\tr(\s(\rho_2)\e^{B})
-\rho_1(A)-\rho_2(B)\\[3mm]
&=S(\rho_1)+S(\rho_2).
\end{align*}
Moreover, equality holds iff the variational principle has a minimizer of the form
$A\otimes\one+\one\otimes B$,  which, by the remark
after Theorem \ref{ent-prop-1}, is possible only if $\rho=\rho_1\otimes\rho_2$.
\qed

\section{Relative entropies}
\label{sect:Entropies}

The R\'enyi relative entropy (or $\alpha$-relative entropy) of two states $\rho,\nu$ is defined for 
$\alpha\in]0,1[$ by \bindex{entropy!R\'enyi}\nindex{Sa}{$S_\alpha(\rho\abso\nu)$}{R\'enyi entropy}
\[
S_\alpha(\rho|\nu)=\log \tr(\rho^\alpha\nu^{1-\alpha}).
\]
This quantity will play an important role in these lecture notes. According to our convention
$\log(\rho|_{\Ker\rho})=-\infty$, and
$$
\rho^\alpha=\e^{\alpha\log\rho}=\e^{\alpha\log(\rho|_{\Ran\rho})}\oplus0|_{\Ker\rho}.
$$
\index{inequality!H\"older}%
The H\"older inequality implies  that $S_\alpha(\rho|\nu)\in [-\infty, 0]$. $S_\alpha(\rho|\nu)=-\infty$ iff 
$\rho\perp \nu$ (that is, if $\rho$ and $\nu$ are mutually singular). In terms of the spectral data of 
$\rho$ and $\nu$,
\begin{equation}
S_\alpha(\rho|\nu)=\log\left[
\sum_{\atop{(\lambda,\mu)\in\sp(\rho)\times\sp(\nu)}{\lambda\not=0,\mu\not=0}}
\lambda^\alpha \mu^{1-\alpha}\tr(P_\lambda(\rho)P_\mu(\nu))\right],
\label{SalphaForm}
\end{equation}
and so if $\rho\not\perp \nu$, then  $]0,1[\,\ni \alpha \mapsto S_\alpha(\rho|\nu)\in]-\infty,0]$ extends 
to a real-analytic function on $\rr$. The basic properties of R\'enyi's relative entropy are: 

\bep \label{propSalphaprops}
Suppose that $\rho\not\perp \nu$. Then:
\ben
\item $S_0(\rho|\nu)=\log\nu({\rm s}(\rho))$ and
$S_1(\rho|\nu)=\log\rho({\rm s}(\nu))$.
\item The map $\rr \ni \alpha \mapsto S_\alpha(\rho|\nu)$ is  convex.
\item $S_\alpha(U\rho U^{-1}|U\nu U^{-1})=S_\alpha(\rho|\nu)$ for any unitary $U$.
\item Suppose that  $\s(\rho)=\s(\nu)$. Then the  map $\rr \ni \alpha \mapsto S_\alpha(\rho|\nu)$ 
is strictly convex iff $\rho\not=\nu$. 
\item $S_\alpha(\rho|\nu)=S_{1-\alpha}(\nu|\rho)$.
\een
\eep
\demo (1), (3) and (5) are obvious. 
(2) Follows from the following facts, easily derived from \eqref{SalphaForm},
\begin{align*}
\partial_\alpha S_\alpha(\rho|\nu)
&=\sum_{(\lambda,\mu)\in\sp(\rho)\times\sp(\nu)}p_{\lambda,\mu}\log(\lambda/\mu)=\theta_\alpha,\\
\partial_\alpha^2S_\alpha(\rho|\nu)
&=\sum_{(\lambda,\mu)\in\sp(\rho)\times\sp(\nu)}p_{\lambda,\mu}
\left[\log(\lambda/\mu)-\theta_\alpha\right]^2
\ge0,
\end{align*}
where
\[
p_{\lambda,\mu}=\frac{\lambda^\alpha\mu^{1-\alpha}\tr(P_\lambda(\rho)P_\mu(\nu))}
{\displaystyle\sum_{(\lambda,\mu)\in\sp(\rho)\times\sp(\nu)}
\lambda^\alpha\mu^{1-\alpha}\tr(P_\lambda(\rho)P_\mu(\nu))}\ge0,\]
\[\sum_{(\lambda,\mu)\in\sp(\rho)\times\sp(\nu)}p_{\lambda,\mu}=1.
\]

\medskip\noindent(4) 
Invoking analyticity, we further deduce that either $\partial_\alpha^2S_\alpha(\rho|\nu)=0$ for all 
$\alpha\in\rr$, or $\partial_\alpha^2S_\alpha(\rho|\nu)>0$ except possibly on a discrete subset of $\rr$.
In the former case $S_\alpha(\rho|\nu)=(1-\alpha)S_0(\rho|\nu)+\alpha S_1(\rho|\nu)$ is an affine
function of $\alpha$. In the latter case $S_\alpha(\rho|\nu)$ is strictly convex. 

Suppose now that $\s(\rho)=\s(\nu)$. Without loss of generality, we can assume that 
$\rho$ and $\nu$ are faithful. If $\rho=\nu$ then $S_\alpha(\rho|\nu)=0$ for all $\alpha\in\rr$. 
Reciprocally, if $\partial_\alpha^2S_\alpha(\rho|\nu)$
vanishes identically then $\theta=\partial_\alpha S_\alpha(\rho|\nu)$ is constant and
$\lambda=\e^\theta\mu$ whenever $\tr(P_\lambda(\rho)P_\mu(\nu))\not=0$. It follows from
$$
1=\tr(\rho)=\sum_{\lambda,\mu}\lambda\,\tr(P_\lambda(\rho)P_\mu(\nu))
=\e^\theta\sum_{\lambda,\mu}\mu\,\tr(P_\lambda(\rho)P_\mu(\nu))=\e^\theta\tr(\nu)=\e^\theta,
$$
that $\theta=0$. Repeating the argument in the proof of Part (2) of Theorem \ref{theor-trace-in}
leads to the conclusion that $\rho=\nu$.
\qed

\bigskip
The following theorem, a variant of the celebrated Kosaki's variational formula (\cite{Kos,OP}), 
is deeper. The result and its proof were communicated to us by R. Seiringer 
(unpublished). The proof will be given in Section \ref{sect:Modular} as an illustration of the power 
of the modular structure to be introduced there.\bindex{formula!Kosaki}

\bet\label{KosakiThm}
For $\alpha \in ]0,1[$,
\[
S_\alpha(\rho|\nu)=\inf_{A\in C(\rr_+,\cO)}\log\left[\frac{\sin\pi\alpha}{\pi}
\int_0^\infty t^{\alpha-1}\left(\frac{1}{t}\rho(|A(t)^\ast|^2)+\nu(|\one-A(t)|^2)\right)\d t\right],
\]
where $C(\rr_+,\cO)$ denotes the set of all continuous functions $\rr_+\ni t\mapsto A(t)\in\cO$.
\index{variational principle} Moreover, the infimum is achieved for
\[ 
A(t)=t\int_0^\infty\e^{-s\rho}\nu\e^{-st\nu}\d s, 
\]
and this is the unique minimizer if either $\rho $ or $\nu$ is faithful.
\eet

An immediate consequence of Kosaki's variational formula is {\sl Uhlmann's monotonicity theorem,} 
\cite{Uh}:\bindex{inequality!Uhlmann}\bindex{theorem!Uhlmann monotonicity}

\bet\label{Uhlmann}
If $\Phi:\cO_\cK\to\cO_{\cK'}$ is a unital Schwarz map, then\index{map!Schwarz}
$$
S_\alpha(\rho\circ\Phi|\nu\circ\Phi)\ge S_\alpha(\rho|\nu),
$$
for all $\alpha\in[0,1]$ and $\rho,\nu\in\mathfrak S_{\cK'}$.
\eet

\demo With $\hat\rho=\rho\circ\Phi$ and $\hat\nu=\nu\circ\Phi$, Kosaki's formula reads
$$
S_\alpha(\hat\rho|\hat\nu)=\inf_{A\in C(\rr_+,\cO_\cK)}\log\left[\frac{\sin\pi\alpha}{\pi}
\int_0^\infty t^{\alpha-1}\left(\frac{1}{t}\hat\rho(|A(t)^\ast|^2)+\hat\nu(|\one-A(t)|^2)\right)\d t\right].
$$
Since $\Phi$ is a unital Schwarz map, for any $A\in\cO_\cK$ one has
$$
\hat\rho(|A^\ast|^2)=\rho(\Phi(AA^\ast))\ge\rho(\Phi(A)\Phi(A)^\ast)=\rho(|\Phi(A)^\ast|^2),
$$
as well as
\begin{align*}
\hat\nu(|\one-A|^2)&=\nu(\Phi((\one-A)^\ast(\one-A)))\\
&\ge\nu(\Phi(\one-A)^\ast\Phi(\one-A))\\
&=\nu((\one-\Phi(A))^\ast(\one-\Phi(A)))=\nu(|\one-\Phi(A)|^2).
\end{align*}
It follows that
\begin{align*}
S_\alpha(\hat\rho|\hat\nu)&\ge\inf_{A\in C(\rr_+,\cO_\cK)}\log\left[\frac{\sin\pi\alpha}{\pi}
\int_0^\infty t^{\alpha-1}\left(\frac{1}{t}\rho(|\Phi(A(t))^\ast|^2)+\nu(|\one-\Phi(A(t))|^2)\right)\d t\right]\\
&=\inf_{A\in \Phi(C(\rr_+,\cO_\cK))}\log\left[\frac{\sin\pi\alpha}{\pi}
\int_0^\infty t^{\alpha-1}\left(\frac{1}{t}\rho(|A(t)^\ast|^2)+\nu(|\one-A(t)|^2)\right)\d t\right],
\end{align*}
where $\Phi(C(\rr_+,\cO_\cK))=\{\Phi\circ A\,|\,A\in C(\rr_+,\cO_\cK)\}$. Since $\Phi$ is continuous,
one has $\Phi(C(\rr_+,\cO_\cK))\subset C(\rr_+,\cO_{\cK'})$, and the result follows from Kosaki's
formula.
\qed

\bigskip
Another consequence of Theorem \ref{KosakiThm} is the celebrated Lieb's concavity theorem.
\bindex{entropy!joint concavity}\bindex{theorem!Lieb concavity}

\bet\label{LiebConcavAlpha}
For $\alpha \in[0,1]$, the map ${\mathfrak S}\times {\mathfrak S}\ni (\rho, \nu)\mapsto S_\alpha(\rho|\nu)$
is jointly concave, i.e.,
$$
S_\alpha(\lambda\rho+(1-\lambda)\rho'|\lambda\nu+(1-\lambda)\nu')\ge
\lambda S_\alpha(\rho|\nu)+(1-\lambda)S_\alpha(\rho'|\nu'),
$$
for any $\rho,\rho',\nu,\nu'\in\mathfrak S$ and any $\lambda\in[0,1]$.
\eet
\demo The result is obvious for $\lambda=0$ and for $\lambda=1$. Hence, we assume
$\lambda\in]0,1[$ in the following. For $\alpha=0$ and for $\alpha=1$, the result follows from Part (1)
of Proposition \ref{propSalphaprops}, the concavity of the logarithm, and the fact that
$\s(\lambda\rho+(1-\lambda)\rho')\ge\s(\rho)$. For $\alpha\in]0,1[$ and $A\in C(\rr_+,\cO)$, the map
$$
(\rho,\nu)\mapsto F_A(\rho,\nu)=
\frac{\sin\pi\alpha}{\pi}
\int_0^\infty t^{\alpha-1}\left(\frac{1}{t}\rho(|A(t)^\ast|^2)+\nu(|\one-A(t)|^2)\right)\d t,
$$
is affine. The concavity of the logarithm implies that the map $(\rho,\nu)\mapsto\log F_A(\rho,\nu)$
is concave. Therefore, the function $(\rho,\nu)\mapsto S_\alpha(\rho|\nu)$ being the infimum
of a family of concave functions, it is itself concave (see the following exercise).
\qed

\medskip
\begin{exo}
\label{SalphaUSC}
Let $C\subset\rr^n$ be a convex set and $\mathcal F$ a nonempty set of real valued functions on
$C$. Set $F(x)=\inf\{f(x)\,|\,f\in{\mathcal F}\}$.

\exop Show that if the elements of $\mathcal F$ are concave then $F$ is concave.

\exop Show that if the elements of $\mathcal F$ are continuous then $F$ is 
upper semi-continuous, \ie 
$$
\limsup_{x\to x_0}F(x)\le F(x_0),
$$
for all $x_0\in\rr^n$.

\exop Show that the function $(\rho,\nu)\mapsto S_\alpha(\rho|\nu)$ is upper semi-continuous
on $\mathfrak S\times\mathfrak S$.
\end{exo}

\medskip
The relative entropy of the state $\rho$ w.r.t. the state  $\nu$ is defined by\bindex{entropy!relative}
\nindex{Sa}{$S(\rho\abso\nu)$}{relative entropy}
\[
S(\rho|\nu)=\left\{
\begin{array}{ll}
\tr (\rho(\log \nu -\log \rho))&\text{if }\rho\ll\nu.\\[6pt]
-\infty&\text{otherwise}.
\end{array}
\right.
\]
Equivalently, in terms of the spectral data of $\rho$ and $\nu$, one has
$$
S(\rho|\nu)=\sum_{(\lambda,\mu)\in\sp(\rho)\times\sp(\nu)}\lambda(\log\mu-\log\lambda)
\tr(P_\lambda(\rho)P_\mu(\nu)).
$$
For $\nu\in\mathfrak S$ and $A\in\Os$, we define
$$
\e^{A+\log\nu}=\lim_{n\to\infty}\left(\e^{A/n}\nu^{1/n}\right)^n.
$$
It s not difficult to show that, according to the decomposition $\cK=\Ran\nu\oplus\Ker\nu$,
$$
\e^{A+\log\nu}=\e^{\s(\nu)A\s(\nu)|_{\Ran\nu}+\log(\nu|_{\Ran\nu})}\oplus 0_{\Ker\nu}.
$$
With this definition, the relative entropy functional has the following variational characterizations:

\bet\label{RelEntVar}\index{variational principle}
{\rm(1)} For any $\rho,\nu\in{\mathfrak S}$, one has 
\[
S(\rho|\nu)=\inf_{A\in \Os}\log \tr(\e^{A+\log \nu})-\rho(A).
\] 

{\noindent\rm(2)} For any $A\in\Os$ and $\nu\in\mathfrak S$, one has
$$
\log\tr(\e^{A+\log\nu})=\max_{\rho\in\mathfrak S}S(\rho|\nu)+\rho(A).
$$
\eet

{\noindent\bf Remark.} If $\rho$ and $\nu$ are equivalent, then the infimum in (1) is achieved at $A$
iff $\s(\nu)A\s(\nu)|_{\Ran\nu}=\log(\rho|_{\Ran\nu})-\log(\nu|_{\Ran\nu})+c$ where $c$ is an arbitrary 
real constant. The maximizer in (2) is unique and given by 
$\rho=\e^{A+\log\nu}/\tr(\e^{A+\log\nu})$.

\bigskip
\demo (2) Set $G_{\nu,A}(\rho)=S(\rho|\nu)+\rho(A)$ and 
$\widetilde\nu=\e^{A+\log\nu}/\tr(\e^{A+\log\nu})$. Note that $G_{\nu,A}(\rho)=-\infty$
if $\rho\not\ll\nu$ while $G_{\nu,A}(\nu)=\nu(A)>-\infty$. Thus, it suffices to consider $\rho\ll\nu$
in which case one has
$$
\rho\left(\log\widetilde\nu\right)=\rho(A)+\rho(\log\nu)-\log\tr(\e^{A+\log\nu}).
$$
Klein's inequality yields\index{inequality!Klein}
$$
\log\tr(\e^{A+\log\nu})-G_{\nu,A}(\rho)=\tr\left(\rho\left(\log\rho-\log\widetilde\nu\right)\right)
\ge\tr(\rho-\widetilde\nu)=0,
$$
with equality iff $\rho=\widetilde\nu$.

\medskip\noindent{\rm(1)} We first consider the case $\rho\not\ll\nu$. Then, there exists a projection
$P$ such that $\Ran P\perp\Ran\nu$ and $\rho(P)>0$. Since $\e^{\lambda P+\log\nu}=\nu$,
it follows that
$$
\log\tr(\e^{\lambda P+\log\nu})-\rho(\lambda P)=-\lambda\rho(P)\to-\infty=S(\rho|\nu),
$$
as $\lambda\to\infty$. On the other hand, for any $\rho\ll\nu$ and $A\in\Os$, (2) implies that
$$
S(\rho|\nu)\le\log\tr(\e^{A+\log \nu})-\rho(A),
$$
with equality iff $\rho=\e^{A+\log\nu}/\tr(\e^{A+\log\nu})$. If $\nu\ll\rho$, this means that
equality holds iff $\s(\nu)A\s(\nu)|_{\Ran\nu}=\log(\rho|_{\Ran\nu})-\log(\nu|_{\Ran\nu})$
up to an arbitrary additive constant. If $\nu\not\ll\rho$, \ie if $\s(\rho)<\s(\nu)$,
then 
$A_\lambda=\log(\rho|_{\Ran\rho})\oplus\lambda\one_{\Ker\rho}-\log(\nu|_{\Ran\nu})\oplus0_{\Ker\nu}$
is such that, with $d=\dim\Ran\nu-\dim\Ran\rho$,
$$
\log\tr(\e^{A_\lambda+\log \nu})-\rho(A_\lambda)=\log\left(1+\e^{\lambda}d\right)+S(\rho|\nu)
\to S(\rho|\nu),
$$
as $\lambda\to-\infty$.
\qed

\bigskip
As an immediate consequence of Theorem \ref{RelEntVar} we note, for later reference
\bec\label{PBgen}
For any state $\nu\in\mathfrak S$ and any self-adjoint observable $A\in\cO$ one has
$$
\tr(\e^{\log\nu+A})\ge\e^{\nu(A)}.
$$
\eec
The  basic properties of  the relative entropy functional are: 
\bep\label{PropSprops}
\ben
\item $S(\rho|\nu)\leq 0$ with equality iff $\rho=\nu$.
\item $S(\rho|\rho_{\rm ch})= S(\rho)-\log \dim \cK$.
\item $S(U\rho U^{-1}|U\nu U^{-1})= S(\rho|\nu)$ for any unitary $U$.
\item \[S(\rho_A|\rho_B)=\log \frac{\tr(\e^A)}{\tr(\e^B)}-\tr(\rho_A(A-B)).\]
\item For any $\rho,\nu\in\mathfrak S$ one has\index{entropy!R\'enyi}
\begin{equation}
S(\rho|\nu)=\lim_{\alpha\downarrow0} \frac{S_\alpha(\nu|\rho)}\alpha
=\lim_{\alpha\uparrow1}\frac{S_\alpha(\rho|\nu)}{1-\alpha}.
\label{SLimSalpha}
\end{equation}
In particular, if $\rho\ll\nu$ then $S_0(\nu|\rho)=S_1(\rho|\nu)=0$ and
\begin{equation}
S(\rho|\nu)=\left.\frac{\d\ }{\d\alpha} S_\alpha(\nu|\rho)\right|_{\alpha=0}
=- \left.\frac{\d\ }{\d\alpha} S_\alpha(\rho|\nu)\right|_{\alpha=1}.
\label{SDerivSalpha}
\end{equation}
\item If $\Phi:\cO_\cK\to\cO_{\cK'}$ is a unital Schwarz map then,\index{map!Schwarz}
for any $\rho,\nu\in\mathfrak S_{\cK}$,
$$
S(\rho\circ\Phi|\nu\circ\Phi)\ge S(\rho|\nu).
$$ 
\item The map $(\rho, \nu)\mapsto S(\rho|\nu)$ is continuous on 
${\mathfrak S} \times {\mathfrak S}_{\rm f}$ and upper semi-continuous on 
${\mathfrak S}\times{\mathfrak S}$.
\item If $\s(\nu)=\s(\rho)$, then $S_\alpha(\rho|\nu)\geq \alpha S(\nu|\rho)$.
\een
\eep
\demo Part (1) follows from Klein's inequality.\index{inequality!Klein}
Parts (2), (3) and (4) are obvious. Part (5) is easy
and left to the reader. Part (6) follows from \eqref{SLimSalpha} and Uhlmann's monotonicity
theorem (Theorem \ref{Uhlmann}).\index{inequality!Uhlmann}
The  upper semi-continuity of the map $(\rho, \nu)\mapsto S(\rho|\nu)$ follows from 
(5) and part 3 of Exercise \ref{SalphaUSC}. A direct proof goes as follows.  
Let us fix $(\rho_0,\nu_0)\in{\mathfrak S}\times{\mathfrak S}$. Define
 $\lambda_0=\min\{\lambda\in\sp(\nu_0)\,|\,\lambda>0\}$, and for $\nu\in\mathfrak S$ set
$$
Q_\nu=\sum_{\atop{\lambda\in\sp(\nu)}{\lambda>\lambda_0/2}}P_\lambda(\nu).
$$
Let $0<\varepsilon<\lambda_0/2$. We know from perturbation theory that
$$
\lim_{\nu\to\nu_0}Q_\nu=\s(\nu_0),\qquad
\lim_{\nu\to\nu_0}\nu Q_\nu=\nu_0,
$$
and that
$$
\nu^{(\varepsilon)}=\nu Q_\nu+(\one-Q_\nu)\varepsilon\ge\nu,
$$
provided $\nu$ is close enough to $\nu_0$. It follows that
$$
S(\rho|\nu)=\rho(\log\nu)-S(\rho)\le\rho(\log\nu^{(\varepsilon)})-S(\rho).
$$
Since $\nu^{(\varepsilon)}\ge\epsilon>0$ it follows from the analytic functional calculus that
$$
\lim_{\nu\to\nu_0}\log\nu^{(\varepsilon)}=\log\left(\lim_{\nu\to\nu_0}\nu^{(\varepsilon)}\right)
=\log(\nu_0|_{\Ran\nu_0})\oplus\log\varepsilon|_{\Ker\nu_0},
$$
and hence, using Theorem \ref{ent-prop-1} (1), we deduce
\begin{align*}
\limsup_{(\rho,\nu)\to(\rho_0,\nu_0)}S(\rho|\nu)
&\le\lim_{(\rho,\nu)\to(\rho_0,\nu_0)}\rho(\log\nu^{(\varepsilon)})-S(\rho)\\
&=\rho_0(\log\nu_0|_{\Ran\nu_0}\oplus 0|_{\Ker\nu_0})-S(\rho_0)+(1-\rho_0(\s(\nu_0)))\log\varepsilon.
\end{align*}
If $\rho_0\not\ll\nu_0$ then $1-\rho_0(\s(\nu_0))>0$ and letting $\varepsilon\downarrow0$ we conclude
that
$$
\limsup_{(\rho,\nu)\to(\rho_0,\nu_0)}S(\rho|\nu)\le-\infty=S(\rho_0|\nu_0).
$$
If $\rho_0\ll\nu_0$ then $1-\rho_0(\s(\nu_0))=0$ and $\Ker\nu_0\subset\Ker\rho_0$ so that
$$
\limsup_{(\rho,\nu)\to(\rho_0,\nu_0)}S(\rho|\nu)
\le\rho_0(\log\nu_0)-S(\rho_0)=S(\rho_0|\nu_0).
$$
Finally, we observe that if $\nu_0>0$, then $\nu\ge\lambda_0/2$ for all $\nu$ sufficiently
close to $\nu_0$. Hence $\lim_{\nu\to\nu_0}\log\nu=\log\nu_0$ and
$$
\lim_{(\rho,\nu)\to(\rho_0,\nu_0)}S(\rho|\nu)=S(\rho_0|\nu_0).
$$
Property (8) is a direct consequence of the convexity of $\alpha\mapsto S_\alpha(\rho|\nu)$ and
Equ. \eqref{SDerivSalpha}.
\qed

{\noindent\bf Remark.} The following example shows that the function
$(\rho,\nu)\mapsto S(\rho|\nu)$ is not continuous on ${\mathfrak S}\times{\mathfrak S}$.
Setting
$$
\rho_n=\left[\begin{array}{cc}
1-1/n&0\\
0&1/n
\end{array}\right],\qquad
\nu_n=\left[\begin{array}{cc}
1&0\\
0&0
\end{array}\right],
$$
one has $S(\rho_n|\nu_n)=-\infty$ for all $n\in\nn^\ast$, so
$$
\lim_{n\to\infty}S(\rho_n|\nu_n)=-\infty\not=S(\lim_{n\to\infty}\rho_n|\lim_{n\to\infty}\nu_n)=0.
$$

\medskip
As a direct consequence of Theorem \ref{LiebConcavAlpha} and Relation \eqref{SLimSalpha},
we have:\index{entropy!joint concavity}
\bet\label{thm:Sconcav}
The map ${\mathfrak S}\times {\mathfrak S}\ni (\rho, \nu)\mapsto S(\rho|\nu)$ is jointly concave, that is, 
for $\lambda\in [0,1]$ and $\rho,\rho',\nu,\nu'\in\mathfrak S$,
$$
S(\lambda\rho +(1-\lambda)\rho'|\lambda\nu+(1-\lambda)\nu')
\geq\lambda S(\rho|\nu)+(1-\lambda)S(\rho'|\nu').
$$
\eet
\begin{exo}
\label{Exo:IsoSalpha}
Use Uhlmann's monotonicity theorem to show that \index{theorem!Uhlmann monotonicity}
$$
S_\alpha(\rho\circ\vartheta|\nu\circ\vartheta)=S_\alpha(\rho|\nu),
$$
for all $\rho,\nu\in\mathfrak S$ and $\vartheta\in{\rm Aut}(\cO)$.
\end{exo}

\medskip

\section{Quantum hypothesis testing}
\label{sect:QHT}\bindex{hypothesis testing}

Since the pioneering work of Pearson \cite{Pe}, hypothesis testing has played an important 
role in theoretical and applied statistics (see, \eg \cite{Be}). In the last decade, the 
mathematical structure and basic results of classical hypothesis testing have been extended to the 
non-commutative setting. A clear exposition of the basic results of quantum hypothesis testing can 
be found in \cite{ANSV,HMO}. 

It was recently observed in \cite{JOPS} that there is a close relation between 
recent developments in the field of quantum hypothesis testing and the developments in 
non-equilibrium statistical mechanics. In this section we describe the setup of quantum hypothesis 
testing following essentially \cite{ANSV}. We will discuss the relation
to non-equilibrium statistical mechanics in Section \ref{sect:LThypo}.

Let $\nu$ and $\rho$ be two states and $p\in \,]0,1[$. Suppose that we know a  priori that  the system 
is with probability $p$ in the state $\rho$ and with probability $1-p$ in the state $\nu$.  By performing 
a measurement we wish to decide with minimal error probability what is the true state of the system. 
The following procedure is known as {\sl quantum hypothesis testing.} 
A {\em test} $P$ \bindex{test} is an orthogonal projection in $\cO$. On the basis of the outcome of the test 
(that is, a measurement of $P$) one decides whether the system is in the state $\rho$ or $\nu$. More
precisely, if the outcome of the test is $1$, one decides that  the system is in the state $\rho$ 
(Hypothesis I)  and if the outcome is $0$, one decides that the system is in the state $\nu$ 
(Hypothesis II). $\rho(\one-P)$ is the error probability of  accepting II if I is true and $\nu(P)$ is the  
error probability  of accepting  I if II is true. The average error probability is \bindex{error probability}
\[
D_p(\rho, \nu, P)=p\rho(\one -P) + (1-p)\nu(P),
\]
\nindex{DpP}{$D_p(\rho, \nu, P)$}{error probability of the test $P$}%
and we  are interested in minimizing $D_p(\rho, \nu, P)$ w.r.t. $P$.  Let  
\[
D_p(\rho, \nu)=\inf\{ D_p(\rho, \nu, P)\,|\, P\in\Os, \,P^2=P\}.
\]
\nindex{Dp}{$D_p(\rho, \nu)$}{minimal error probability}
The set of all orthogonal projections is a norm closed subset of $\cO$ and so the infimum
on the right-hand side is achieved at some projection $P$. The quantum Bayesian distinguishability 
problem is to identify the orthogonal projections $P$ such that $D_p(\rho, \nu,P)=D_p(\rho, \nu)$. 
Let $P_{\rm opt}$ be the orthogonal projection onto the range 
of \nindex{Popt}{$P_{\mathrm{opt}}$}{optimal test (Neyman-Pearson)}
\[
\left((1-p)\nu-p\rho\right)_+,
\]
where $x_+=(|x|+x)/2$ denotes the positive part of $x$. The following result was proven in
\cite{ANSV}, where the reader can find references to the previous works 
on the subject. 

\bet\label{QHTthm}
\ben
\item
\[
D_p(\rho,\nu)=D_p(\rho, \nu, P_{\rm opt})= \frac{1}{2}\left(1 -\|(1-p)\nu-p\rho\|_1\right).
\]
Moreover, $P_{\rm opt}$ is the unique minimizer of the functional $P\mapsto D_p(\rho, \nu, P)$.
\item 
\[
D_p(\rho, \nu)=\min \{D_p(\rho, \nu, T)\,|\,T\in\Os, 0\leq T\leq \one\}.
\]
\item For $\alpha \in [0,1]$, 
\[
 D_p(\rho,\nu)\leq p^{\alpha}(1-p)^{1-\alpha}\tr (\rho^{\alpha}\nu^{1-\alpha}).
\]
\een
\eet
{\noindent\bf Remark.} Part (1) is the quantum version of the Neyman-Pearson lemma.
\index{Neyman-Pearson}
Part (3) is the quantum analog of the Chernoff bound in classical hypothesis testing.
In quantum information theory the quantity
\[
\zeta_{QCB}(\rho, \nu) =-\log\min_{\alpha\in[0,1]}\tr(\rho^\alpha \nu^{1-\alpha})
=-\min_{\alpha\in[0,1]}S_\alpha(\rho|\nu),
\]
is called the Chernoff distance between the states $\rho$ and $\nu$. We shall prove a lower bound on the function $D_p(\rho,\nu)$ in 
Section \ref{sect:Modular}.\nindex{zQCB}{$\zeta_{QCB}(\rho, \nu)$}{Chernoff distance}
\bindex{Chernoff!distance}

\bigskip
\demo
(1)--(2) Set $A=(1-p)\nu-p\rho$ so that, for $T\in\Os$, $0\le T\le\one$, we can write
$$
D_p(\rho,\nu,T)=\tr\left(p\rho(\one-T)+(1-p)\nu T\right)=p+\tr(TA)\ge p+\tr(TA_+),
$$
where equality holds iff $\Ran T\subset\Ker A_-=\Ran A_+$. It follows that $P_{\rm opt}$ is the 
unique minimizer and
$$
D_p(\rho,\nu,P_{\rm opt})=p+\tr(A_+)=p+\frac12\tr\,(A+|A|)=\frac12(1+\tr\,(|A|)).
$$

\medskip\noindent(3) (Following S. Ozawa, private communication. The original proof can be found 
in \cite{ANSV}).  
Setting $B=p\rho$ and $C=(1-p)\nu$ and given (1), one has to show that
\[ 
\tr(B^\alpha C^{1-\alpha})\ge\frac{1}{2} \tr(B+C-|B-C|),
\]
for all $B,C\in\cO_+$ and $\alpha\in[0,1]$. With $A=C-B$, one clearly has
\begin{equation}
B\le B+A_+, 
\label{ozbasic1}
\end{equation}
and since $C-B\le(C-B)_+$, one also has
\begin{equation}
C\le B+A_+. 
\label{ozbasic2}
\end{equation}
We shall make repeated use of the L\"owner-Heinz inequality (Exercise \ref{Lowner-Heinz}).
\index{inequality!L\"owner-Heinz} From \eqref{ozbasic1} and the fact that $B^\alpha\ge0$ we get
\begin{equation}
\tr(B^\alpha(B^{1-\alpha}-C^{1-\alpha}))\le\tr(B^\alpha((B+A_+)^{1-\alpha}-C^{1-\alpha})). 
\label{ozstep1}
\end{equation}
From \eqref{ozbasic2} we deduce that
\[
(B+A_+)^{1-\alpha}-C^{1-\alpha}\ge 0. 
\]
Thus,   \eqref{ozbasic1} and \eqref{ozstep1} imply 
\begin{align*}
\tr(B^\alpha(B^{1-\alpha}-C^{1-\alpha}))
&\le\tr((B+A_+)^\alpha((B+A_+)^{1-\alpha}-C^{1-\alpha}))\\
&=\tr(B+A_+)-\tr((B+A_+)^\alpha C^{1-\alpha}) . 
\end{align*}
Using again Inequality \eqref{ozbasic2}, and the fact that $C\ge0$, we obtain
\[
\tr(B^\alpha(B^{1-\alpha}-C^{1-\alpha})) 
\le \tr(B+A_+)-\tr(C^\alpha C^{1-\alpha})=\tr(B-C+A_+).
\]
This inequality can be  rewritten as 
\[ 
\tr(B^\alpha C^{1-\alpha})\ge\tr(C-A_+), 
\]
and since $A_+=A+A_-$,
\[
\tr(B^\alpha C^{1-\alpha})\ge\tr(C-A-A_-)=\tr(C-(C-B)-A_-)=\tr(B-A_-). 
\]
Combining the last  two inequalities we finally get
\[
\tr(B^\alpha C^{1-\alpha})
\ge\frac{1}{2}\tr(B+C-A_+-A_-)=\frac{1}{2}\tr(B+C-|B-C|),
\]
as required.
\qed

\section{Dynamical systems}
\label{sect:DynSys}

A {\sl dynamics} on the $\ast$-algebra $\cO$ is a continuous one-parameter subgroup of
\bindex{dynamics}
\index{star automorphism@$\ast$-automorphism!group}
\nindex{tta}{$\tau^t$}{dynamics}
$\rm{Aut}(\cO)$, \ie a map $\rr\ni t\mapsto \tau^t\in\rm{Aut}(\cO)$ satisfying
$\tau^t\circ \tau^s=\tau^{t+s}$ for all $t,s\in\rr$ and $\lim_{t\to0}\|\tau^t(A)-A\|=0$
for all $A\in\cO$. Such a map automatically satisfies $\tau^0=\rm{id}$ and $(\tau^t)^{-1}=\tau^{-t}$ for
all $t\in\rr$. Moreover, since $\tau^t$ is isometric and $\cO$ is a finite dimensional vector space, 
the continuity is uniform
$$
\lim_{\epsilon\to0}\sup_{\atop{\|A\|=1}{t\in\rr}}\|\tau^{t+\epsilon}(A)-\tau^t(A)\|=0,
$$
and the map $t\mapsto\tau^t(A)$ is differentiable (in fact entire analytic). In terms of the generator
$$
\delta(A)=\left.\frac{\d\ }{\d t}\tau^t(A)\right|_{t=0},
$$
\nindex{delta}{$\delta(\,\cdot\,)$}{generator of a dynamics}%
\index{star automorphism@$\ast$-automorphism!group!generator}%
one has $\tau^t(A)=\e^{t\delta}(A)$. Clearly, $\delta(\one)=0$, $\delta(AB)=\delta(A)B + A\delta(B)$
and $\delta(A)^\ast=\delta(A^\ast)$ hold for all $A,B\in\cO$.
We call {\sl dynamical system} a pair $(\cO,\tau^t)$, where $\tau^t$ is a dynamics on $\cO$.
\bindex{dynamical system}

If $H\in\Os$, then
\begin{equation}\tau^t(A)=\e^{\i t H}A\e^{-\i t H},
\label{ihes-morn-1}
\end{equation}
is a dynamics on $\cO$. One of the special features of finite quantum systems is that the converse
is true. Given a dynamical system $(\cO,\tau^t)$, there exists $H\in\Os$ such that \eqref{ihes-morn-1}
holds for all $t\in\rr$. Moreover, $H$ is uniquely determined up to a constant. It can be explicitly
constructed  as follows. Let $\delta$ be the generator of $\tau^t$. Let $\{\psi_j\}$ be an orthonormal 
basis of $\cK$ and $E_{ij}=|\psi_i\rangle\langle\psi_j|$ the corresponding basis of $\cO$. Let
\[
H= \frac{1}{\i}\sum_{j}\delta(E_{ji})E_{ij}.
\]
The relation   $\sum_jE_{ji}E_{ij}=\sum_jE_{jj}=\one$ implies
$$
\sum_{j}\delta(E_{ji})E_{ij}+\sum_{j}E_{ji}\delta(E_{ij})=\delta(\one)=0,
$$
and  
\begin{align*}
\i[H,E_{kl}]&=\sum_j\delta(E_{ji})E_{ij}E_{kl}+E_{kl}E_{ji}\delta(E_{ij})\\
&=\delta(E_{ki})E_{il}+E_{ki}\delta(E_{il})=\delta(E_{ki}E_{il})=\delta(E_{kl}).
\end{align*}
Hence  $\i[H,X]=\delta(X)$ for all $X\in\cO$ and  \eqref{ihes-morn-1}
follows.

\bigskip
{\noindent\bf Remark.} From the above discussion, the reader familiar with the theory of Lie groups 
will recognize that ${\rm Aut}(\cO)$ is a simply connected Lie group with Lie algebra 
$$
\mathfrak{aut}(\cO)=\{\d_X=\i[X,\,\cdot\,]\,|\,X\in\Os\},
$$
and bracket $[\d_X,\d_Y]=\d_{\i[X,Y]}$. Since $\d_X=\d_Y$ iff $X-Y$ is a real multiple of the identity,
the dimension of ${\rm Aut}(\cO)$ is given by $\dim_\rr(\Os)-1=(\dim\cK)^2-1$.

\bigskip
According to the basic principles of quantum mechanics, if $H$ is the energy observable of
the system, \ie its Hamiltonian,  \bindex{Hamiltonian} then the group $\tau^t(A)=\e^{\i tH}A\e^{-\i tH}$ describes 
its time evolution in the Heisenberg picture. If the system was in the state $\rho$ at time $t=0$
then the expectation value of the observable $A$ at time $t$ is given by 
$$
\tr(\rho\tau^t(A))=\rho(\tau^t(A))=\rho\circ\tau^t(A).
$$
In the Schr\"odinger picture the state $\rho$ evolves in time as $\tau^{-t}(\rho)$ and in what 
follows we adopt the shorthands 
\[
A_t=\tau^t(A), \qquad \rho_t=\tau^{-t}(\rho)=\rho\circ\tau^t.
\]
Clearly, $\rho_t(A)=\rho(A_t)$.\index{Heisenberg picture}\index{Schr\"odinger picture}

\section{Gibbs states, KMS condition and variational principle}
\label{sect:KMS}

For the dynamical system $(\cO,\tau^t)$, with Hamiltonian $H$, the state of thermal equilibrium 
at inverse temperature $\beta$ is described by the Gibbs canonical ensemble
\bindex{Gibbs!canonical ensemble}
\[
\rho_\beta=\frac{\e^{-\beta H}}{\tr(\e^{-\beta H})}.
\] 
Note that, for any $A,B\in\cO$, one has
$$
\rho_\beta(AB)=\frac{\tr(\e^{-\beta H}AB)}{\tr(\e^{-\beta H})}=
\frac{\tr(B\e^{-\beta H}A)}{\tr(\e^{-\beta H})}
=\frac{\tr(\e^{-\beta H}\tau^{-\i\beta}(B)A)}{\tr(\e^{-\beta H})}=\rho_\beta(\tau^{-\i\beta}(B)A).
$$
We say that a state $\rho$ satisfies the Kubo-Martin-Schwinger (KMS) condition at inverse 
temperature $\beta$, or, for short, that $\rho$ is a $\beta$-KMS state if\bindex{KMS!condition}
\begin{equation}
\rho(AB)=\rho(\tau^{-\i \beta}(B)A),
\label{kms-1}
\end{equation}
holds for all $A, B\in\cO$. The $\beta$-KMS condition \eqref{kms-1} plays a central role in 
algebraic quantum statistical mechanics.  For the finite quantum system considered in this
section it is a characterization of the Gibbs state $\rho_\beta$.

\bep
$\rho$ is a $\beta$-KMS state iff $\rho=\rho_\beta$.\bindex{state!KMS}
\eep 
\demo It remains to show that if $\rho$ is $\beta$-KMS, then $\rho=\rho_\beta$. Setting
$X=\rho\e^{\beta H}$ and  $A=\e^{\beta H}C$ in the KMS condition
\begin{equation*}
\tr(\rho\e^{\beta H}B\e^{-\beta H}A)=\tr(\rho AB),
\end{equation*}
yields $\tr(XBC)=\tr(XCB)$ for all $B,C\in\cO$. Since this is equivalent to
$\tr([X,B]C)=0$, we conclude that $[X,B]=0$ for all $B\in\cO$ and hence
that $X=\alpha\one$ for some constant $\alpha$. This means that $\rho=\alpha\e^{-\beta H}$. The
constant $\alpha$ is now determined by the normalization condition $\tr(\rho)=1$.
\qed

The Gibbs canonical ensemble can be also characterized by a variational principle.  
The number $ E=\rho_\beta(H)$ is the expectation value of the energy in the state $\rho_\beta$. 
Since 
\[
\frac{\d\ }{\d\beta}\rho_\beta(H)=-\rho_\beta((H-E)^2)\leq 0,
\]
the function $\beta \mapsto \rho_\beta(H)$ is decreasing  and is strictly decreasing unless $H$ 
is constant. If $E_{\rm min}=\min\sp(H)$ and $E_{\rm max}=\max\sp(H)$, then 
\[
\lim_{\beta \rightarrow -\infty}\rho_\beta(H)=E_{\rm max}, 
\qquad 
\lim_{\beta \rightarrow \infty}\rho_\beta(H)=E_{\rm min}.
\]
Note also that 
$\lim_{\beta\rightarrow \pm \infty}\rho_\beta=\rho_{\pm \infty}$ where 
\[
\rho_{+\infty/-\infty}=\frac{P_{\min/\max}}{\tr(P_{\min/\max})},
\]
and $P_{\min/\max}$ denote the spectral projection of $H$ associated to its eigenvalue
$E_{\min/\max}$. Hence to any $ E\in [E_{\min}, E_{\max}]$ one can associate a unique 
$\beta \in [-\infty, \infty]$ such that 
\begin{equation}
\rho_{ \beta}(H)= E.
\label{ihes-poss}
\end{equation}
We adopt the shorthands 
\[
S(\beta)=S(\rho_\beta), \qquad P(\beta)=\log \tr(\e^{-\beta H}).
\]
The function $P(\beta)$ is called the {\em pressure} (or {\em free energy}). Note that
\bindex{pressure}\bindex{free energy}%
\begin{equation}
S(\beta)=\beta  E + P(\beta).
\label{ihes-last}
\end{equation}
If ${\mathfrak S}_{ E}=\{\rho\in\mathfrak S\,|\,\rho(H)=E\}$ and  $\nu\in {\mathfrak S}_{ E}$,  then 
\index{variational principle}%
\[
S(\nu)= S(\nu)- \beta \nu(H)+  \beta  E
\leq \max_{\rho\in\mathfrak S}\{ S(\rho)-\beta \rho(H)\}+  \beta E
=\log \tr (\e^{- \beta H}) + \beta E,
\]
and so 
\[ 
S(\nu)\leq S( \beta),
\]
where equality holds iff $\nu=\rho_\beta$. Hence, we have proven the Gibbs variational principle:
\index{Gibbs!variational principle}
\bet
Let $ E\in [E_{\min}, E_{\max}]$ and let $\beta$ be given by \eqref{ihes-poss}. Then
\[
\max_{\rho\in{\mathfrak S}_E}S(\rho) =S(\beta),
\]
and the unique maximizer is the Gibbs state $\rho_{ \beta}$.
\eet

Note that neither the KMS condition nor the Gibbs variational principle require $\beta$ to be positive.  
The justification of the physical restriction $\beta > 0$ typically involves some form of the second law 
of thermodynamics. Recall that $\beta=\beta(E)$ is uniquely specified by  \eqref{ihes-poss}. 
Considering $S(E)=S(\beta(E))$ as the function of $E$, the differentiation of relation 
\eqref{ihes-last} w.r.t. $E$ yields 
\[
\frac{\d S}{\d E}=\beta,
\]
and the second law  $\frac{\d S}{\d E}\geq 0$ (the increase of entropy with energy ) requires 
$\beta \geq 0$.  An alternative approach goes as follows. Let an external force act on the system
during the time interval $[0,T]$ so that its Hamiltonian becomes time dependent,
$H(t)=H+V(t)$. We assume that $V(t)$ depends continuously on $t$  and vanishes for
$t\not\in]0,T[$. Let $U(t)$ be the corresponding unitary propagator, \ie the  solution of the time-dependent Schr\"odinger
equation 
\[
\i\frac{\d\ }{\d t}U(t)=H(t)U(t),\qquad U(0)=\one.
\]
Suppose that at $t=0$ the system was in the Gibbs state $\rho_\beta$. At the later time $t>0$,
its state is given by  $\rho_{\beta,t}=U(t)\rho_\beta U(t)^\ast$ and the work performed on 
the system by the external force during the time interval $[0, T]$ is
\[
\Delta E=\rho_{\beta,T}(H)-\rho_{\beta}(H)=\int_0^T\frac{\d\ }{\d t} \rho_{\beta, t}(H)\,\d t.
\]
The change of relative entropy $S(\rho_{\beta,t}|\rho_\beta)$ over the time interval $[0,T]$ equals 
\[
\Delta S=S(\rho_{\beta, T}|\rho_{\beta})- S(\rho_{\beta}|\rho_\beta)
=\int_0^T \frac{\d\ }{\d t} S(\rho_{\beta, t}|\rho_\beta)\,\d t 
=-\beta \int_0^T\frac{\d\ }{\d t} \rho_{\beta, t}(H)\,\d t,
\]
and so
\[ 
\Delta S = -\beta \Delta E.
\]
If $V(t)$ is non-trivial in the sense that $\rho_{\beta, T}\not=\rho_\beta$, then
$\Delta S=S(\rho_{\beta, T}|\rho_\beta)<0$. The second law of thermodynamics, more precisely the
fact that one can not extract work from a system in thermal equilibrium, requires that $\Delta E\ge0$.
Hence, negative values of $\beta$ are not allowed by thermodynamics.

\bigskip
The above discussion can be generalized as follows. Let $N\in\Os$ be an observable 
such that $[H, N]=0$ ($N$ is colloquially called a {\em charge}). Let $\beta$ and $\mu$ be real 
parameters and let\index{charge}
\[
\rho_{\beta, \mu}=\frac{\e^{-\beta(H-\mu N)}}{\tr(\e^{-\beta(H-\mu N)})},
\]
be the $\beta$-KMS state for the dynamics generated by $H-\mu N$.
Denote $\rho_{\beta, \mu}(H)= E$, $\rho_{\beta, \mu}(N)=\varrho$, 
$S(\beta, \mu)=S(\rho_{\beta, \mu})$, $P(\beta, \mu)=\log \tr (\e^{-\beta(H-\mu N)})$.
Then \bindex{KMS!state}\bindex{state!KMS}
\begin{equation}
S(\beta, \mu)=\beta ( E -\mu \varrho) +P(\beta, \mu).
\label{ihes-last-2}
\end{equation}
If ${\mathfrak S}_{ E, \varrho}=\{\rho\in {\mathfrak S}\,|\,\rho(H)= E, \rho(N)=\varrho\}$, then 
\[
\max_{\rho\in {\mathfrak S}_{E, \varrho}}S(\rho)= S(\beta, \mu),
\]
with unique maximizer $\rho_{\beta, \mu}$. The parameter $\mu$ is interpreted as  
chemical potential \bindex{chemical potential}
associated to the charge $N$ and the state $\rho_{\beta, \mu}$ describes the system in  thermal 
equilibrium at inverse temperature $\beta$ and chemical potential $\mu$. Considering 
$\beta=\beta(E, \varrho)$ and $\mu=\mu(E, \varrho)$ as functions of $E$ and $\varrho$ we see 
from \eqref{ihes-last-2} that 
\[
\frac{\partial S}{\partial E}=\beta, \qquad \frac{\partial S}{\partial \varrho}=-\beta\mu.
\]

Although  in general $\rho_{\beta, \mu}$  is not a $\beta$-KMS state for the dynamics $\tau^t$, 
if $A$ and $B$ commute with $N$, then $\tau^t(A)=\e^{\i t(H-\mu N)}A\e^{-\i t(H-\mu N)}$
and the $\beta$-KMS condition
$$
\rho_{\beta, \mu}(\tau^{-\i\beta}(B)A)=\rho_{\beta, \mu}(AB),
$$
is satisfied. In other words, if $\mu\not=0$, the physical observables must be invariant under the gauge 
group \bindex{gauge group} $\gamma^\theta(A)=\e^{\i\theta N}A\e^{-\i\theta N}$. The generalization of these results to 
the  case of  several  charges is straightforward.

\section{Perturbation theory}
\label{sect:Perturbations}

Let $(\cO,\tau^t)$ be a dynamical system with Hamiltonian $H$ and let $V\in\Os$ be a perturbation. 
In this section we consider the perturbed dynamics $\tau_V^t$ generated by the Hamiltonian $H+V$,
\bindex{dynamics!perturbed}\nindex{ttb}{$\tau_V^t$}{perturbed dynamics}
\[
\tau_V^t(A)=\e^{\i t(H+V)}A\e^{-\i t (H+V)}.
\]
If $\delta$ denotes the generator of $\tau^t$, then the generator of $\tau_V^t$ is given by
$$
\delta_V=\i[H+V,\,\cdot\,]=\delta+\i[V,\,\cdot\,]=\delta+\d_V,
$$
and one easily checks that the map
$\rr\ni t\mapsto\gamma_V^t\in{\rm Aut}(\cO)$ defined by
$$
\gamma_V^t=\tau_V^t\circ\tau^{-t}=\e^{t(\delta+\d_V)}\circ\e^{-t\delta},
$$
has the following properties:
\ben
\item $\tau_V^t=\gamma_V^t\circ\tau^t$.
\item $\left(\gamma_V^t\right)^{-1}=\tau^t\circ\gamma_V^{-t}\circ\tau^{-t}$.
\item $\gamma_V^{t+s}=\gamma_V^s\circ\tau^s\circ\gamma_V^t\circ\tau^{-s}$.
\item $\gamma_V^{0}={\rm id}$ and $\partial_t\gamma_V^t=\gamma_V^t\circ\d_{\tau^t(V)}$.
\een
Integration of Relation (4) yields the integral equation
$$
\gamma_V^t={\rm id}+\int_0^t\gamma_V^s\circ\d_{\tau^s(V)}\,\d s,
$$
which can be iterated to obtain
\begin{align*}
\gamma_V^t={\rm id}+\sum_{n=1}^{N-1}&\int_{0\le s_1\le\cdots\le s_n\le t}
\d_{\tau^{s_n}(V)}\circ\cdots\circ\d_{\tau^{s_1}(V)}\,\d s_1\cdots\d s_n\\
+&\int_{0\le s_1\le\cdots\le s_{N}\le t}
\gamma_V^{s_{N}}\circ\d_{\tau^{s_{N}}(V)}\circ\cdots
\circ\d_{\tau^{s_1}(V)}\,\d s_1\cdots\d s_{N}.
\end{align*}
Since $\gamma_V^t$ is isometric and $\|d_{\tau^t(V)}\|=\|\i[\tau^t(V),\,\cdot\,]\|\le 2\|V\|$, we can
bound the norm of the last term  by
$$
\int_{0\le s_1\le\cdots\le s_{N}\le t}(2\|V\|)^{N}\,\d s_1\cdots\d s_{N}
\le\frac{(2\|V\|t)^{N}}{N!},
$$
and conclude that the Dyson expansion\bindex{expansion!Dyson}
$$
\gamma_V^t={\rm id}+\sum_{n=1}^\infty\int_{0\le s_1\le\cdots\le s_n\le t}
\d_{\tau^{s_n}(V)}\circ\cdots\circ\d_{\tau^{s_1}(V)}\,\d s_1\cdots\d s_n,
$$
converges in norm for all $t\in\rr$, and uniformly for $t$ in compact intervals.
Using Relation (1), we conclude that
$$
\tau_V^t=\tau^t+\sum_{n=1}^\infty\int_{0\le s_1\le\cdots\le s_n\le t}
\d_{\tau^{s_n}(V)}\circ\cdots\circ\d_{\tau^{s_1}(V)}\circ\tau^t\,\d s_1\cdots\d s_n,
$$
which we can rewrite as
\[
\tau_V^t(A)=\sum_{n=0}^\infty(\i t)^n\int_{0\leq s_1 \leq \cdots \leq s_n\leq 1}
[\tau^{ts_n}(V), [\cdots, [\tau^{ts_1}(V), \tau^t(A)]\cdots]]\,\d s_1\cdots\d s_n.
\]
Finally, we note that since $\tau^z(V)$, $\tau^z(A)$ and $\tau_V^z(A)$ are entire analytic 
functions of $z$ and $\|\tau^z\|\le\e^{2|\Im z|\,\|H\|}$, the above expression provides an expansion
of $\tau_V^z(A)$ which converges uniformly for $z$ in compact subsets of $\cc$.

Similar conclusions hold for the interaction picture propagator
\[
{\rm E}_V(t)=\e^{\i t (H+V)}\e^{-\i t H}.
\]
It satisfies:
\begin{enumerate}[{\ \ \rm (1')}]
\item $\e^{\i t(H+V)}=\rE_V(t)\e^{\i tH}$ and $\tau_V^t(A)=\rE_V(t)\tau^t(A)\rE_V(t)^{-1}$.
\item $\rE_V(t)^{-1}=\rE_V(t)^\ast=\tau^t(\rE_V(-t))$.
\item $\rE_V(t+s)=\rE_V(s)\tau^{s}(\rE_V(t))$.
\item $\rE_V(0)=\one$ and $\partial_t \rE_V(t)=\i \rE_V(t)\tau^t(V)$.
\end{enumerate}
\noindent Integrating relation (4') yields, after iteration,
\[
\rE_V(t)=\sum_{n=0}^\infty(\i t)^n\int_{0\leq s_1 \leq \cdots \leq s_n\leq 1}
\tau^{ts_n}(V)\cdots\tau^{ts_1}(V)\,\d s_1\cdots \d s_n.
\]
This expansion is uniformly convergent for $t$ in compact subsets of $\cc$. In  particular,
\begin{equation}
\rE_V(\i \beta)=
\sum_{n=0}^\infty(-\beta)^n\int_{0\leq s_1\leq\cdots\leq s_n\leq1}
\tau^{\i \beta s_n}(V)\cdots\tau^{\i \beta s_1}(V)\,\d s_1\cdots \d s_n.
\label{EViBetaExp}
\end{equation}
Using Relation (1') with $t=\i\beta$  we can express the perturbed KMS-state
\index{state!perturbed KMS}
\[
\rho_{\beta V}=\frac{\e^{-\beta (H + V)}}{\tr (\e^{-\beta(H+ V)})},
\]
in terms of the unperturbed one $\rho_\beta=\e^{-\beta H}/\tr (\e^{-\beta H})$ as 
\begin{equation}
\rho_{\beta V}(A)=\frac{\rho_\beta(A\,\rE_V(\i \beta))}{\rho_\beta(\rE_V(\i \beta))}.
\label{ihes-morn}
\end{equation}
Using this last  formula one can compute the perturbative expansion of $\rho_{\beta V}(A)$ 
w.r.t. $V$. To control this expansion, we need the following estimate.
\bep The bound
\begin{equation}
|\rho_{\beta}(\rE_{\alpha V}(\i \beta))-1|\leq\e^{|\alpha\beta| \|V\|}-1,
\label{ihes-esti}
\end{equation}
holds for any $\beta\in\rr$, $V\in\Os$ and $\alpha\in\cc$.
\eep
\demo
Using Duhamel formula\index{formula!Duhamel}
$$
\frac{\d\ }{\d s}\,\e^{-\beta(H+s\alpha V)}
=-\alpha\int_0^\beta\e^{-(\beta-u)(H+s\alpha V)}V\e^{-u(H+s\alpha V)}\,\d u,
$$
we can write
\begin{equation}
\rho_{\beta}(\rE_{\alpha V}(\i \beta)-\one)
=\int_0^1\frac{\d\ }{\d s}\rho_\beta(\rE_{s\alpha V}(\i\beta))\,\d s
=-\alpha\beta\int_0^1f_\beta(s) \,\d s,
\label{fbetacont}
\end{equation}
where
$$
f_\beta(s)=\frac{\tr( V\e^{-\beta(H+s\alpha V)})}{\tr(\e^{-\beta H})}.
$$
Starting with the simple bound
$$
|f_\beta(s)|\le\|V\|\frac{\|\e^{-\beta(H+s\alpha V)}\|_1}{\tr(\e^{-\beta H})},
$$
and setting $\alpha=a+\i b$ with $a,b\in\rr$, we estimate the numerator on the right hand side by 
the H\"older inequality (Part 2 of Exercise \ref{Exo:Holder}) applied to the Lie product formula,
\index{inequality!H\"older}\index{formula!Lie product}
\begin{align*}
\|\e^{-\beta(H+saV+\i sbV)}\|_1&=\lim_{n\to\infty}\|(\e^{-\beta(H+saV)/n}\e^{-\i\beta sbV/n})^n\|_1\\
&\le\limsup_{n\to\infty}\|\e^{-\beta(H+saV)/n}\|_n^n\,\|\e^{-\i\beta sbV/n}\|_\infty^n
=\tr(\e^{-\beta(H+saV)}).
\end{align*}
For $s\in[0,1]$, the Golden-Thompson inequality further leads to\index{inequality!Golden-Thompson}
$$
\frac{\tr(\e^{-\beta(H+saV)})}{\tr(\e^{-\beta H})}
\le\frac{\tr(\e^{-\beta H}\e^{-\beta saV})}{\tr(\e^{-\beta H})}
=\rho_\beta(\e^{-\beta sa V})\le\e^{s|\beta\alpha|\,\|V\|},
$$
so that, finally,
$$
|f_\beta(s)|\le\|V\|\,\e^{s|\beta\alpha|\,\|V\|}.
$$
Using Equ. \eqref{fbetacont}, we derive
\[
|\rho_{\beta}(\rE_V(\i \beta))-1|\leq  |\alpha\beta|\|V\|\int_0^1 \e^{s |\alpha\beta|\|V\|}\d s
=\e^{|\alpha\beta| \|V\|}-1.
\]
\qed

\bigskip
Replacing $V$ with $\alpha V$ and using the expansion
\eqref{EViBetaExp}, we can write 
\[
\rho_\beta(A\,\rE_{\alpha V}(\i \beta))= \sum_{n=0}^\infty \alpha^n c_n(A),
\]
where $c_0(A)=\rho_\beta(A)$ and 
\[
c_n(A)=(-\beta)^n\int_{0\leq s_1 \leq \cdots \leq s_n\leq 1}
\rho_\beta(A\tau^{\i \beta s_n}(V)\cdots\tau^{\i \beta s_1}(V))\,\d s_1\cdots \d s_n.
\]
It follows from the estimate \eqref{ihes-esti} that the entire function
$\cc\ni\alpha\mapsto\rho_{\beta}(\rE_{\alpha V}(\i\beta))$ has no zero in the disk
\[
|\alpha|<\frac{\log 2}{|\beta|\|V\|}.
\]
Hence, Equ. \eqref{ihes-morn} shows that the function $\cc\ni\alpha\mapsto\rho_{\beta (\alpha V)}(A)$
is analytic on this disk. Writing
$$
\rho_{\beta (\alpha V)}(A)=\sum_{n=0}^\infty \alpha^n b_{n}(A),
$$
Relation \eqref{ihes-morn} yields
\[
\sum_{n=0}^\infty \alpha^n c_n(A)=
\left(\sum_{n=0}^\infty \alpha ^n b_{n}(A)\right)\left(\sum_{n=0}^\infty \alpha^n c_n(\one)\right),
\]
and we conclude that for all $n$, 
\[
c_n(A)= \sum_{j=0}^n b_{j}(A)c_{n-j}(\one).
\]
Thus, with coefficients $b_{n}(A)$ given by the recursive formula
$$
b_0(A)=c_0(A)=\rho_{\beta}(A),\qquad b_n(A)=c_n(A)-\sum_{j=0}^{n-1}b_j(A)c_{n-j}(\one),
$$
we can write
\begin{equation}
\rho_{\beta V}(A)=\sum_{n=0}^\infty b_n(A),
\label{rhobetaVexpand}
\end{equation}
provided $|\beta|\,\|V\| < \log 2$. 

\medskip
\begin{exo}
\label{Exo:KuboMari}
\noindent Show that the expression
\begin{equation}
\langle A| B\rangle_\beta=\int_0^1\rho_\beta(A^\ast \tau^{\i \beta s}(B))\,\d s
=\frac1\beta\int_0^\beta\rho_\beta(A^\ast \tau^{\i s}(B))\,\d s,
 \label{kubo-mari}
\end{equation}
defines an inner product on $\cO$. It is called Kubo-Mari or Bogoliubov  scalar product,
Duhamel two point function or canonical correlation.
\bindex{Kubo-Mari inner product}
\index{Bogoliubov inner product|see{Kubo Mari inner product}}
\index{Duhamel two point function|see{Kubo Mari inner product}}
\index{canonical correlation|see{Kubo Mari inner product}}%
\nindex{(5}{$\langle\,\cdot\,\abso\,\cdot\,\rangle_\beta$}{Kubo-Mari inner product}
\end{exo}
\begin{exo}
\label{Exo:bncoeffs}
Show that the first coefficients  $b_1(A)$ and $b_2(A)$ can be written as
\begin{align*}
b_{1}(A)&=-\beta\langle V|\widehat A\rangle_\beta,\\
b_2(A)&=\beta^2\!\int_0^1\!\d s\!\int_0^s\!\d s'\!\left[
\rho_\beta(\widehat A\tau^{\i\beta s}(V)\tau^{\i\beta s'}(V))
-\rho_{\beta}(\widehat A\tau^{\i \beta s}(V))\rho_\beta(V)
-\rho_{\beta}(\widehat A\tau^{\i \beta s'}(V))\rho_\beta(V)
\right],
\end{align*}
where $\widehat A=A-\rho_\beta(A)$.
\end{exo}

\medskip

\section{The standard representations of $\cO$}
\label{sect:Representation}
\index{modular!structure}

In this and the following sections we introduce the so called modular structure associated with the
$\ast$-algebra $\cO=\cO_\cK$.  Historically, the structure was unveiled in  the work of Araki and Woods
\cite{AWo} on
the equilibrium states of a free Bose gas and  linked to the KMS condition by Haag, Hugenholtz
and Winnink \cite{HHW}. 
After the celebrated works of Tomita \cite{To} and 
Takesaki \cite{Ta}, modular theory became an essential tool in the study of operator algebras.

For us, the main purpose of modular theory is to provide a framework which will
allow us to describe a quantum system in a way that is robust enough to survive the thermodynamic 
limit. While familiar objects like Hamiltonians or density matrices will lose their meaning in this limit,
the notions that we are about to introduce: standard representation, modular groups and operators,
Connes cocycles, relative Hamiltonians, Liouvilleans, etc, will continue to make sense in the context
of extended quantum systems. As a {\sl rule of thumb,} a result that holds for 
 finite quantum systems and   can be formulated
 in terms of robust objects of modular theory will  remain valid  for extended systems. 

\bigskip
Let $\cH$ be an auxiliary Hilbert space and denote by $\cL(\cH)$ 
the $\ast$-algebra of all linear operators on $\cH$. A subset $\cal A\subset\cL(\cH)$ is called 
{\sl self-adjoint,} written $\cal A^\ast=\cal A$, if $A^\ast\in\cal A$ for all $A\in\cal A$. A self-adjoint
subset $\cal A\subset\cL(\cH)$ is a {\sl $\ast$-subalgebra} if it is a vector subspace such 
that $AB\in\cal A$ for all $A,B\in\cal A$. A {\sl representation} of $\cO$ in $\cH$ is a linear map
\bindex{representation!of a $\ast$-algebra}
$\phi:\cO\to\cL(\cH)$ such that $\phi(AB)=\phi(A)\phi(B)$ and $\phi(A^\ast)=\phi(A)^\ast$
for all $A,B\in\cO$. A representation is {\sl faithful} if the map $\phi$ is injective, \ie
if $\Ker\phi=\{0\}$. A faithful representation of $\cO$ in $\cH$ is therefore an isomorphism
\bindex{representation!faithful}
between $\cO$ and the $\ast$-subalgebra $\phi(\cO)\subset\cL(\cH)$.
A vector $\psi\in\cH$ is called {\sl cyclic} for the representation $\phi$
\bindex{vector!cyclic}\bindex{vector!separating}
if $\cH=\phi(\cO)\psi$. It is called {\sl separating} if $\phi(A)\psi=0$ implies that $A=0$.
Two representations $\phi_1:\cO\to\cL(\cH_1)$ and $\phi_2:\cO\to\cL(\cH_2)$ are called
{\sl equivalent} if there exists a unitary $U:\cH_1\to\cH_2$ such that $U\phi_1(A)=\phi_2(A)U$
for all $A\in\cO$. \bindex{representation!equivalent}

Let $\cal A$ and $\cal B$ be subsets of $\cL(\cH)$. $\cal A\vee\cal B$ denotes the smallest 
$\ast$-subalgebra of $\cL(\cH)$ containing $\cal A$ and $\cal B$.
$\cal A'$ denotes the {\sl commutant} of $\cal A$, \ie the set of all elements of
$\cL(\cH)$ which commute with all elements of $\cal A$. If $\cal A$ is self-adjoint, then
$\cal A'$ is a $\ast$-subalgebra.\nindex{A}{$\cal A'$}{commutant}\bindex{commutant}

A {\sl cone} in the Hilbert space $\cH$ is a subset $\cC\subset\cH$ such that $\lambda\psi\in\cC$
for all $\lambda\ge0$ and all $\psi\in\cC$.\bindex{cone}
If $\cal M\subset\cal H$, then
$$
\widehat{\cal M}=\{\phi\in\cH\,|\,\langle\psi|\phi\rangle\ge0\text{ for all }\psi\in\cal M\},
$$
is a cone. A cone $\cC\subset\cH$ is called {\sl self-dual} if $\widehat\cC=\cC$.
\bindex{cone!self-dual}
We have already noticed that $\cO$, viewed as a complex vector space, becomes a Hilbert space
when equipped with the inner product \nindex{(2}{$(\,\cdot\,\abso\,\cdot\,)$}{inner product on $\cH_\cO$}
$$
(\xi|\eta)=\tr\,(\xi^\ast\eta).
$$
In the sequel, in order to distinguish this Hilbert space from the $\ast$-algebra $\cO$ we shall
denote the former by $\cH_\cO$. Thus, $\cO$ and $\cH_\cO$ are the same set, but
carry distinct algebraic structures. We will use lower case greeks $\xi,\eta,\ldots$ to denote 
elements of the Hilbert space $\cH_\cO$ and upper case romans $A,B,\ldots$ to denote 
elements of the $\ast$-algebra $\cO$. \nindex{Ho}{$\cH_\cO$}{standard representation space}

\bigskip
{\noindent\bf Remark.}
Let $\psi\mapsto\overline{\psi}$ denote an arbitrary complex conjugation (\ie an anti-unitary involution)
on the Hilbert space $\cK$. One easily checks that the map \index{complex conjugation}
$|\psi\rangle\langle\varphi|\mapsto\psi\otimes\overline{\varphi}$ extends to a unitary operator from 
$\cH_{\cO_\cK}$ to $\cK\otimes{\cK}$. Thus, the Hilbert space $\cH_{\cO_\cK}$ 
is isomorphic to $\cK\otimes{\cK}$.

\bigskip
To any $A\in\cO$ we can associate two elements $L(A)$ and $R(A)$ of $\cL(\cH_\cO)$  by
\nindex{L()}{$L(\,\cdot\,)$}{(Left) standard representation}\nindex{R()}{$R(\,\cdot\,)$}{(Right) standard representation}
$$
L(A):\xi\mapsto A\xi,\qquad R(A):\xi\mapsto\xi A^\ast.
$$
The map $\cO\ni A\mapsto L(A)\in\cL(\cH_\cO)$ is clearly linear and satisfies
$L(AB)=L(A)L(B)$. Moreover, for all $\xi,\eta\in\cH_\cO$ one has
$$
(\xi|L(A)\eta)=\tr\,(\xi^\ast A\eta)=\tr\,((A^\ast\xi)^\ast\eta)=(L(A^\ast)\xi|\eta),
$$
so that $L(A^\ast)=L(A)^\ast$. In short, $L$ is a representation of the $\ast$-algebra $\cO$ on 
the Hilbert space $\cH_\cO$. In the same way one checks that $R:\cO\to\cL(\cO)$ is antilinear and 
satisfies $R(AB)=R(A)R(B)$ as well as $R(A^\ast)=R(A)^\ast$.

\bep\label{prop:StdOne}
\ben
\item The maps $L$ and $R$ are isometric and hence injective.
\item $L(\cO)=\{L(A)\,|\,A\in\cO\}$ and $R(\cO)=\{R(A)\,|\,A\in\cO\}$ are
$\ast$-subalgebras of $\cL(\cH_\cO)$ isomorphic to $\cO$.
\item $L(\cO)\cap R(\cO)=\cc\one$.
\item $L(\cO)\vee R(\cO)=\cL(\cH_\cO)$.
\item $L(\cO)'=R(\cO)$.
\item $R(\cO)'=L(\cO)$.
\een
\eep

\demo (1)--(2) For $A\in\cO$, one has
\begin{align*}
\|L(A)\|^2=\sup_{\|\xi\|=1}\|L(A)\xi\|^2&=\sup_{\tr\, (\xi^\ast\xi)=1}\tr((A\xi)^\ast(A\xi))\\
&=\sup_{\tr\,(\xi\xi^\ast)=1}\tr\,((\xi\xi^\ast)(A^\ast A))\le\|A^\ast A\|=\|A\|^2.
\end{align*}
On the other hand, if $\psi$ is a normalized eigenvector of $A^\ast A$ to its maximal eigenvalue
$\|A^\ast A\|$ and $\xi=|\psi\rangle\langle\psi|$, then $\|\xi\|=1$ and 
$$
\|L(A)\xi\|=\|A\xi\|=\langle\psi|A^\ast A\psi\rangle=\|A^\ast A\|,
$$
so that we can conclude that $\|L(A)\|=\|A\|$.
$L$ is a linear map and $\Ker L=\{0\}$. Thus, $L$ is injective 
and is an $\ast$-isomorphism between $\cO$ and its image $L(\cO)$. The same argument holds for 
$R$. 

\medskip\noindent(3) If $T\in L(\cO)\cap R(\cO)$, then there exists $A,B\in\cO$ such that $A\xi=\xi B$
for all $\xi\in\cH_\cO$. Setting $\xi=\one$ we deduce $A=B$. It follows that $[A,\xi]=0$ for all 
$\xi\in\cO$ and hence $A$ must be a multiple of the identity.

\medskip\noindent(4) Let $T\in\cL(\cH_\cO)$ and denote by $\{E_{ij}\}$ the
orthogonal basis of $\cH_\cO$ associated to some orthogonal basis $\{e_i\}$ of $\cK$. Setting
$T_{ij,kl}=(E_{ij}|TE_{kl})$, one has
$$
TE_{kl}=\sum_{i,j,k,l}T_{ij,kl}E_{ij}.
$$
Since 
$E_{ij}=|e_i\rangle\langle e_j|=|e_i\rangle\langle e_k|e_k\rangle\langle e_l|e_l\rangle\langle e_j|=
E_{ik}E_{kl}E_{lj}=L(E_{ik})R(E_{jl})E_{kl}$, we can write
$$
T=\sum_{i,j,k,l}T_{ij,kl}L(E_{ik})R(E_{jl}),
$$
which shows that the subalgebras $L(\cO)$ and $R(\cO)$ generate all of $\cL(\cH_\cO)$.

\medskip\noindent(5)--(6) For any $A,B\in\cO$ and $\xi\in\cH_\cO$ on has 
$L(A)R(B)\xi=A\xi B=R(B)L(A)\xi$ which shows that $R(\cO)\subset L(\cO)'$ and 
$L(\cO)\subset R(\cO)'$. Let $T\in L(\cO)'$ so that $[T,L(A)]=0$ for all $A\in\cO$. Set $B=T\one$,
then
$$
T\xi=TL(\xi)\one=L(\xi)T\one=L(\xi)B=\xi B=R(B^\ast)\xi,
$$
for all $\xi\in\cH_\cO$. Hence, $T=R(B^\ast)$ and we conclude that $L(\cO)'\subset R(\cO)$. A similar
argument shows that $R(\cO)'\subset L(\cO)$. 
\qed

\bep\label{prop:StdTwo}
\ben
\item The map $J:\xi\mapsto\xi^\ast$ is a anti-unitary involution of the Hilbert space $\cH_\cO$.
\item $JL(\cO)J^\ast=L(\cO)'$.
\item $\cH_\cO^+=\cO_+$ is a self-dual cone of the Hilbert space $\cH_\cO$.
\item $J\xi=\xi$ for all $\xi\in\cH_\cO^+$.
\item $JXJ=X^\ast$ for all $X\in L(\cO)\cap L(\cO)'$. 
\item $L(A)JL(A)\cH_\cO^+\subset\cH_\cO^+$ for all $A\in\cO$. 
\een
\eep
\demo
\noindent(1) $J$ is clearly antilinear and involutive. Since
$$
(\xi|J\eta)=\tr\,(\xi^\ast \eta^\ast)=\overline{\tr\,(\xi\eta})=\overline{(J\xi|\eta)},
$$
$J$ is also antiunitary. 

\medskip\noindent(2) For all $A\in\cO$ and $\xi\in\cH_\cO$ one has
$JL(A)J\xi=(A\xi^\ast)^\ast=\xi A^\ast=R(A)\xi$ which
implies $JL(A)J=R(A)$.

\medskip\noindent(3) The fact that $\cH_\cO^+=\cO_+$ is a cone is obvious. It is also clear that if
$\xi,\eta\in\cH_\cO^+$ then $(\xi|\eta)\ge0$ so that $\cH_\cO^+\subset\widehat{\cH_\cO^+}$. 
To prove the reverse 
inclusion, let $\xi\in\widehat{\cH_\cO^+}$. Then $(\eta|\xi)\ge0$ for all $\eta\in\cH_\cO^+$. 
In particular, with
$\eta=|\psi\rangle\langle\psi|$, we get $(\eta|\xi)=\langle\psi|\xi\psi\rangle\ge0$ from which we conclude
that $\xi\in\cH_\cO^+$.

\medskip\noindent(4)--(5) are obvious and (6) follows from the fact that
$$
L(A)JL(A)\xi=A\xi A^\ast\ge0,
$$
for all $\xi\ge0$.
\qed

\bigskip
The faithful representation $L:\cO\to\cL(\cH_\cO)$ is called {\sl standard representation} of $\cO$,
$J$ is called the {\sl modular conjugation} and the cone $\cH_\cO^+$ is called the {\sl natural cone.} 
\bindex{representation!standard}\bindex{cone!natural}\bindex{modular!conjugation}
\nindex{J}{$J$}{modular conjugation}\nindex{Ho}{$\cH_\cO^+$}{natural cone}
The map
$$
\mathfrak S\ni\nu\mapsto\xi_\nu=\nu^{1/2}\in\cH_\cO^+,
$$
is clearly a bijection between the set of states and the unit vectors in $\cH_\cO^+$. For all $A\in\cO$,
one has
$$
(\xi_\nu|L(A)\xi_\nu)=\tr\,(\nu^{1/2}A\nu^{1/2})=\nu(A).
$$
$\xi_\nu$ is called the{\sl  vector representative} of the state $\nu$ in the standard representation.
\bindex{vector!representative of a state}\nindex{xn}{$\xi_\nu$}{vector representative of the state $\nu$}%
Note that a unit vector $\xi\in\cH_\cO^+$ is cyclic for the standard representation iff $\xi>0$, \ie
iff the corresponding state is faithful and in this case, for any $\eta\in\cH_\cO$, one has
$\eta=L(A)\xi$ with $A=\eta\xi^{-1}$. Since $L(A)\xi=0$ iff $\Ran\xi\subset\Ker A$, $\xi$ is a 
separating vector iff $\xi>0$.\index{vector!cyclic}\index{vector!separating}\index{state!faithful}

\medskip
\begin{exo}
\label{Exo:GNS}
{\sl (The GNS representation)}
Let $\nu$ be a state and define $\cal H_\nu$ to be the vector space of all linear maps
$\xi:\Ran\nu\to\cK$, equipped with the inner product\nindex{Hn}{$\cH_\nu$}{GNS space}
$$
(\xi|\eta)_\nu=\tr_{\Ran\nu}(\nu\xi^\ast\eta)=\tr_\cK(\eta\nu\xi^\ast).
$$
\exop Show that $\cal H_\nu$ is a Hilbert space and that $\pi_\nu:\cO\to\cL(\cH_\nu)$ defined by 
$\pi_\nu(A)\xi=A\xi$ is a representation of $\cO$ in $\cH_\nu$. \nindex{pn}{$\pi_\nu$}{GNS representation}

\exop Denote by  $\eta_\nu:\Ran\nu\hookrightarrow\cK$ the canonical injection
$\eta_\nu\psi=\psi$. Show that $\eta_\nu$ is a cyclic vector for the representation $\pi_\nu$ and that
$$
\nu(A)=(\eta_\nu|\pi_\nu(A)\eta_\nu)_\nu,
$$
for all $A\in\cO$.\index{vector!cyclic}

\exop A cyclic representation of $\cO$ associated to a state $\nu$ is a representation $\pi$ 
of $\cO$ in a Hilbert space $\cH$ such that:\index{representation!cyclic}\\
(i) there exists a vector $\psi\in\cH$ which is cyclic for $\pi$.\\
(ii) $\nu(A)=(\psi|\pi(A)\psi)$ for all $A\in\cO$.\\
Show that any cyclic representation of $\cO$ associated to the  state $\nu$ is equivalent to
the above representation $\pi_\nu$.

\noindent{\sl Hint}: show that $\pi(A)\psi\mapsto\pi_\nu(A)\eta_\nu$
defines a unitary map from $\cH$ to $\cH_\nu$.

Thus, up to equivalence, there is only one cyclic representation of $\cO$ associated to a state
$\nu$. This representation is called the Gelfand-Naimark-Segal (GNS) representation of $\cO$
induced by $\nu$.\bindex{representation!GNS}

\exop Show that the map $U:\cH_\nu\ni\xi\mapsto\xi\nu^{1/2}\in\cH_\cO$ is a partial isometry
which intertwine the GNS representation and the standard representation
$$
U\pi_\nu(A)\xi=L(A)U\xi.
$$
Show that if $\nu$ is faithful, then $U$ is unitary so that these two representations 
are equivalent.\

\exop Let $\psi \mapsto \overline{ \psi}$ be a complex conjugation on $\cK$. We have already 
remarked that the map 
$U(|\psi\rangle\langle\varphi|)=\psi\otimes\overline{\varphi}$ extends to a unitary operator from 
$\cH_{\cO_\cK}$ to $\cK\otimes{\cK}$. Show that under this unitary the standard representation 
transforms as follows. \newline
(i) $UR(A)U^{-1}=A\otimes \one$ and $UL(A)U^{-1}=\one \otimes A$. \newline
(ii) $UJU^{-1}\psi \otimes \phi=\overline{\phi}\otimes \overline{\psi}$. \newline
(iii) $U\xi_\nu= \sum_j \lambda_{j}^{1/2}\psi_j \otimes \overline{\psi_j}/\tr (\nu)^{1/2}$, where 
$\lambda_j$'s are the eigenvalues of $\nu$ listed with multiplicities and $\psi_j$'s are the
corresponding eigenfuntions. 
\end{exo}

\medskip
Let $\tau^t$ be a dynamics on $\cO$ generated by the Hamiltonian $H$. Since\index{dynamics}
$$
L(\tau^t(A))=L(\e^{\i tH}A\e^{-\i tH})=L(\e^{\i tH})L(A)L(\e^{-\i tH})
=\e^{\i tL(H)}L(A)\e^{-\i tL(H)},
$$
the self-adjoint operator $L(H)$ seems to play the role of the Hamiltonian in the standard
\index{Hamiltonian}representation. If $\nu$ is a state and $\xi_\nu\in\cH_\cO^+$ its
vector representative, then \index{vector!representative of a state}
$$
\nu(\tau^t(A))=(\xi_\nu|L(\tau^t(A))\xi_\nu)=(\e^{-\i tL(H)}\xi_\nu|L(A)\e^{-\i tL(H)}\xi_\nu).
$$
The state vector thus evolves according to $\e^{-\i tL(H)}\xi_\nu=\e^{-\i tH}\xi_\nu$.
Note that this vector is generally not an element of the natural cone. Indeed, since
$\nu_t=\e^{-\i tH}\nu\e^{\i tH}$, its vector representative is given by
$$
\xi_{\nu_t}=\nu_t^{1/2}=\e^{-\i tH}\nu^{1/2}\e^{\i tH}=L(\e^{-\i tH})R(\e^{-\i tH})\xi_\nu,
$$
which is generally distinct from $\e^{-\i tH}\xi_\nu$. On the other hand, by Part (5) of Proposition
\ref{prop:StdOne}, one has
$$
L(\e^{\i tH})R(\e^{\i tH})L(A)R(\e^{-\i tH})L(\e^{-\i tH})=
L(\e^{\i tH})L(A)L(\e^{-\i tH})=L(\tau^t(A)),
$$
so that the unitary group (recall that $R$ is anti-linear)
$$
L(\e^{\i tH})R(\e^{\i tH})=\e^{\i tL(H)}\e^{-\i tR(H)}=\e^{\i t(L(H)-R(H))},
$$
also implements the dynamics $\tau^t$ in the standard representation. We call the self-adjoint
generator
$$
K=L(H)-R(H)=[H,\,\cdot\,],
$$
the {\sl standard Liouvillean} of the dynamics.\bindex{Liouvillean!standard}
\nindex{K}{$K$}{standard Liouvillean}

\medskip
\begin{exo}\label{Exo:StdLiouvilleChar}
\exop Show that if $\nu$ is a faithful state on $\cO$ then the natural cone of $\cH_\cO$ can be 
written as
$$
\cH_{\cO}^+=\{L(A)JL(A)\xi_\nu\,|\,A\in\cO\}.
$$
Conclude that the unitary group $\e^{\i tX}$ preserves the natural cone iff $JX+XJ=0$.

\exop Show that the standard Liouvillean $K$ is the only self-adjoint operator on $\cH_\cO$ such that, 
for all $A\in\cO$ and $t\in\rr$,
$$
\e^{\i tK}L(A)\e^{-\i tK}=L(\tau^t(A)),
$$
with the additional property that $\e^{-\i t K}\cH_\cO ^+\subset\cH_\cO^+$.
(See Proposition \ref{UpChar} for a generalization of this result.)

\exop Show that the spectrum of $K$ is given by
$$
\sp(K)=\{\lambda-\mu\,|\,\lambda,\mu\in\sp(H)\}.
$$
Note in particular that if $\dim\cK=n$ then $0$ is at least $n$-fold degenerate eigenvalue of $K$.
\end{exo}

\medskip

\section{The modular structure of $\cO$}
\label{sect:Modular}

\subsection{Modular group and modular operator}
In Section \ref{sect:KMS}  we have shown that, given a dynamics $\tau^t$ generated by the 
Hamiltonian $H$, $\e^{-\beta H}/\tr(\e^{-\beta H})$ is the unique $\beta$-KMS state.\index{state!KMS}
Modular theory starts with the reverse  point of view. Given a faithful state $\rho$,  the dynamics
generated by the Hamiltonian $-\beta^{-1}\log\rho$ is the unique dynamics with respect to which $\rho$
is a $\beta$-KMS state. This dynamics might not be in itself physical but it will lead to a remarkable 
mathematical structure with profound physical implications. For historical reasons 
the reference value of $\beta$ is taken to be $-1$. The dynamics 
\[
\varsigma_{\rho}^t(A)=\e^{\i t \log \rho}A\e^{-\i t \log \rho},
\]
is called the {\sl modular dynamics} or {\sl modular group} of the state $\rho$. Its generator is given by
$$
\delta_\rho(A)=\i[\log\rho, A].
$$
\bindex{modular!dynamics}\bindex{modular!group}%
\nindex{sot}{$\varsigma_{\omega}^t$}{modular group of $\omega$}%
The $(\beta=-1)$-KMS condition can be written as
\[
\rho(AB)=\rho(\varsigma_{\rho}^{\i}(B)A).
\] 
According to the previous section, the standard Liouvillean of the modular dynamics is the
self-adjoint operator on $\cH_\cO$ defined by\index{Liouvillean!standard}
$$
K_\rho=L(\log\rho)-R(\log\rho),
$$
and one has
\[
L(\varsigma_\rho^t(A))=\Delta_\rho^{\i t}L(A)\Delta_\rho^{-\i t},
\]
where the positive operator $\Delta_\rho=\e^{K_\rho}$ is called the {\sl modular operator} of the state 
$\rho$. Its action on a vector $\xi\in\cH_\cO$ is described by\bindex{modular!operator}
\nindex{Do}{$\Delta_\omega$}{modular operator of $\omega$}
\[
\Delta_\rho\xi=\e^{L(\log\rho)-R(\log\rho)}\xi=
L(\rho)R(\rho^{-1})\xi=\rho\xi\rho^{-1}.
\]
More generally, for $z\in\cc$,
$$
\Delta_\rho^z\xi=\e^{z(L(\log\rho)-R(\log\rho))}\xi=
L(\rho^z)R(\rho^{-z})\xi=\rho^z\xi\rho^{-z}, 
$$
and in particular
\begin{equation}
J\Delta_\rho^{1/2}A\xi_\rho=(\Delta_\rho^{1/2}A\xi_\rho)^\ast
=(\rho^{1/2}(A\rho^{1/2})\rho^{-1/2})^\ast
=A^\ast\xi_\rho,
\label{ModularCharact}
\end{equation}
for any $A\in\cO$. The last relation completely characterizes the modular conjugation $J$ and the
square root of the modular operator $\Delta_\rho^{1/2}$ as the anti-unitary and positive factors of 
the (unique) polar decomposition of the anti-linear map $A\xi_\rho\mapsto A^\ast\xi_\rho$.
\index{modular!conjugation}

Generalizing the Kubo-Mari inner product \eqref{kubo-mari}, we shall call\index{Kubo-Mari inner product}
$$
\langle A|B\rangle_\rho=\int_0^1 \rho(A^\ast \varsigma_\rho^{-\i u}(B))\d u,
$$
the standard correlation of $A,B\in\cO$ w.r.t. $\rho$\index{standard correlation}
\nindex{(5}{$\langle\,\cdot\,\abso\,\cdot\,\rangle_\rho$}{standard correlation w.r.t. $\rho$}

\subsection{Connes cocycle and relative modular operator}
The modular groups of two faithful states $\rho$ and $\nu$ are related by their
{\sl Connes' cocycle}, the family of unitary elements of $\cO$  defined by 
\[
[D\rho: D\nu]^t=\rho^{\i t}\nu^{-\i t}=\e^{\i t  \log \rho}\e^{-\i t \log \nu}.
\] 
Indeed, one has\bindex{cocycle!Connes}\nindex{[1}{$[D\rho: D\nu]^t$}{Connes cocycle}
\begin{equation}
[D\rho: D_\nu]^t \varsigma_{\nu}^t(A)[D\nu: D\rho]^t=\varsigma_\rho^{t}(A),
\label{ConnesConnect}
\end{equation}
for all $A\in\cO$ and any $t\in\rr$. The Connes cocycles have the following immediate properties: 
\ben
\item $[D\rho: D\nu]^t[D\nu: D\omega]^t=[D\rho: D\omega]^t$.
\item $([D\rho: D\nu]^t)^{-1} = [D\nu: D\rho]^t$.
\item $[D\rho: D\nu]^{t}\varsigma_\nu^{t}([D\rho: D_\nu]^{s})=[D\rho: D\nu]^{t+s}$.
\een
They are obviously defined for any $t\in\cc$ and \eqref{ConnesConnect} as well as
(1)--(3) remain valid. The operator 
\[
[D\rho: D\nu]^{-\i }=\rho\nu^{-1},
\]  
satisfies 
\[
\nu(A[D\rho: D\nu]^{-\i})=\rho(A),
\] 
and is the non-commutative Radon-Nikodym derivative of $\rho$ w.r.t. $\nu$. The R\'enyi
relative entropy can be  expressed in terms of the Connes cocycle as\index{entropy!R\'enyi}
\index{Radon-Nikodym derivative}
$$
S_{\alpha}(\rho|\nu)=\log \nu([D\rho: D\nu]^{-\i \alpha}).
$$
The {\sl relative modular dynamics} of two faithful states $\rho$ and $\nu$ is defined by 
\[
\varsigma_{\rho|\nu}^t(A)=\rho^{\i t }A\nu^{-\i t}=\e^{\i t \log \rho}A\e^{-\i t \log \nu}.
\]
\bindex{modular!dynamics!relative}\nindex{so}{$\varsigma_{\rho\abso \nu}^t$}{relative modular group}
It is related to the modular dynamics of $\rho$ and $\nu$ by the Connes cocycles,
\[
\varsigma_{\rho|\nu}^t(A)=
[D\rho: D\nu]^{t}\varsigma_{\nu}^t(A)=\varsigma_{\rho}^t(A)[D\rho: D\nu]^{t}.
\]
Its standard Liouvillean is given by\index{Liouvillean!standard}
$$
K_{\rho|\nu}=L(\log\rho)-R(\log\nu),
$$
and the corresponding {\sl relative modular operator} $\Delta_{\rho|\nu}=\e^{K_{\rho|\nu}}$ is
a positive operator acting in $\cH_\cO$ as\bindex{modular!operator!relative}
$$
\Delta_{\rho|\nu}\xi=L(\rho)R(\nu^{-1})\xi=\rho\xi\nu^{-1}.
$$
More generally, for $z\in\cc$,
$$
\Delta_{\rho|\nu}^z\xi=L(\rho^z)R(\nu^{-z})\xi=\rho^z\xi\nu^{-z},
$$
and in particular\nindex{Do}{$\Delta_{\rho\abso \nu}$}{relative modular operator}
$$
J\Delta_{\rho|\nu}^{1/2}A\xi_\nu=A^\ast\xi_\rho,
$$
for any $A\in\cO$. Again, this relation characterizes completely $\Delta_{\rho|\nu}^{1/2}$ as the
positive factor of the polar decomposition of the anti-linear map $A\xi_\nu\mapsto A^\ast\xi_\rho$.

In the standard representation of $\cO$ the relative modular dynamics is described by
\index{representation!standard}
\[
L(\varsigma_{\rho|\nu}^t(A))=\Delta_{\rho|\nu}^{\i t }L(A)\Delta_{\rho|\nu}^{-\i t },
\]
and the relative entropies of $\rho$ w.r.t. $\nu$ are given by\index{entropy!relative}
\begin{align*}
S_\alpha(\rho|\nu)&=\log(\xi_\nu|\Delta_{\rho|\nu}^\alpha\xi_\nu),\\[3mm]
S(\rho|\nu)&=(\xi_\rho|\log \Delta_{\nu|\rho}\xi_\rho).
\end{align*} 
The {\sl relative Hamiltonian} of $\rho$ with respect to $\nu$ is the self-adjoint element of $\cO$
defined by\bindex{Hamiltonian!relative}\nindex{lrn}{$\ell_{\rho\abso \nu}$}{relative Hamiltonian}
\begin{equation}
\ell_{\rho|\nu}=\left.\frac1\i\frac{\d\ }{\d t}[D\rho:D\nu]^t\right|_{t=0}=\log \rho - \log \nu.
\label{RelatHDef}
\end{equation}
Since $\delta_\rho=\delta_\nu+\i[\ell_{\rho|\nu},\,\cdot\,]$, $\ell_{\rho|\nu}$ is the perturbation that 
links the modular dynamics $\varsigma_{\nu}^t$ and $\varsigma_{\rho}^t$,
\ie with the notation of Section \ref{sect:Perturbations},\index{dynamics!perturbed}
$$
\varsigma_{\nu \ell_{\rho|\nu}}^t=\varsigma_{\rho}^t.
$$
Further immediate properties of the relative Hamiltonian are:
\label{RelatHProps}
\ben
\item For any $\vartheta \in {\rm Aut}(\cO)$, 
$\ell_{\rho \circ \vartheta^{-1}|\nu\circ \vartheta^{-1}}=\vartheta(\ell_{\rho|\nu})$.
\item $S(\rho|\nu)=-\rho(\ell_{\rho|\nu})$.
\item $\log \Delta_{\rho|\nu}=\log \Delta_{\nu} + L(\ell_{\rho|\nu})$. 
\item $\log \Delta_{\rho}=\log \Delta_{\nu} + L(\ell_{\rho|\nu})-R(\ell_{\rho|\nu})$. 
\item $\ell_{\rho|\nu}  +\ell_{\nu|\omega}=\ell_{\rho|\omega}$.
\een

\bigskip
At this point, the reader could ask about the need for such abstract constructions.
To answer these  concerns let us make more precise the introductory remarks
made at the beginning of Section \ref{sect:Representation}. After taking the thermodynamic limit,
the Hamiltonian $H$ generating the dynamics and the density matrices defining the states
will lose their meaning. So will any expression explicitly involving $H$ or density matrices.
What will remain is an infinite dimensional algebra $\cO$ describing the quantum observables of
the system, a group $\tau^t$ of $\ast$-automorphisms of $\cO$ describing quantum dynamics
and states, positive, normalized linear functionals on $\cO$. 
The modular group $\varsigma_\rho$ will also survive as a group of $\ast$-automorphisms 
of $\cO$ and the modular operator $\Delta_\rho$ will survive as a positive self-adjoint operator 
on the Hilbert space carrying the standard representation of $\cO$. In the same way, relative
modular groups and operators will be available after the thermodynamic limit.
These objects will become our handles to manipulate states. Modular theory allows us to recover,
in the infinite dimensional case, the algebraic structure of the set of states which is clearly
visible in the finite dimensional case. For example, the formula
$$
[0,1]\ni\alpha\mapsto S_\alpha(\rho|\nu)=\log\tr\, (\rho^\alpha\nu^{1-\alpha}),
$$
obviously makes sense if $\rho$ and $\nu$ are density matrices (even in an infinite dimensional
Hilbert space---it follows from H\"older's inequality that the product 
$\rho^\alpha\nu^{1-\alpha}$ is trace class). Thinking of $\rho$ and $\nu$ as linear functionals, 
it is not clear how to make sense of such a product. The alternative formula
$$
S_\alpha(\rho|\nu)=\log(\xi_\nu|\Delta_{\rho|\nu}^\alpha\xi_\nu),
$$
provides  a more general expression which  makes sense even if $\rho$ and $\nu$ are not
associated to density matrices.

From a purely mathematical point of view, modular theory unravels the structures hidden 
in the traditional presentations of quantum statistical mechanics. These structures  often allow 
for simpler and mathematically more natural  proofs of classical results in quantum statistical 
mechanics with an additional advantage that the proofs typically extend to the general von Neumann
algebra setting. We should illustrate this point on three examples at the end of this\index{algebra!von Neumann}
section\footnote{A perhaps most famous  application of modular theory in mathematics is Alain Connes work 
on the general classification and structure theorem of type III factors for which he was awarded the 
Fields medal in 1982.}.
 
\medskip
\begin{exo}\label{Exo:RelatDelta}
Let $\rho$ and $\nu$ be two faithful states on $\cO$.
\exop Show that $\Delta_{\rho|\nu}^{-1}=J\Delta_{\nu|\rho}J$.
\exop Let $\tau^t$ be a dynamics on $\cO$. Show that \index{modular!operator!relative}
$$
\Delta_{\rho\circ\tau^t|\nu\circ\tau^t}=\e^{-\i tK}\Delta_{\rho|\nu}\e^{\i tK}.
$$
where $K$ is the standard Liouvillean of $\tau^t$. \index{Liouvillean!standard}
\end{exo}

\medskip

\subsection{Non-commutative $L^p$-spaces} For $p\in [1,\infty]$, we denote by $L^p(\cO)$ the
Banach space $\cO$ equipped with the $p$-norm\nindex{LpO}{$L^p(\cO)$}{$\cO$ equiped with the $p$-norm $\norm\,\cdot\,\norm_p$}\index{inequality!H\"older}
\eqref{pNormDef}. It follows from H\"older's inequality (Part (3) of Theorem \ref{theor-trace-in})
that if $p^{-1}+q^{-1}=1$ then $L^q(\cO)$ is the dual Banach space to $L^p(\cO)$ with respect to 
the duality $(\xi|\eta)=\tr(\xi^\ast\eta)$. Note in particular that $L^2(\cO)=\cH_\cO$.

While the standard representation will provide a natural extension of $L^2(\cO)$ in  the 
infinite dimensional setting that arises in the thermodynamic limit, there are no such extensions for
the Banach spaces $L^p(\cO)$ for $p\not=2$. Infinite dimensional extensions of those 
spaces which depend on a reference state were introduced  by Araki and Masuda \cite{AM}.
We describe here their finite dimensional counterparts and relate them to the spaces $L^p(\cO)$.
 
Let $\omega$ be a faithful state. For $p\in [2, \infty]$ we set
\nindex{normz}{$\norm\,\cdot\,\norm_{\omega, p}$}{Araki-Masuda $p$-norm}
\[
\|\xi\|_{\omega, p}=\max_{\nu \in {\mathfrak S}}\|\Delta_{\nu|\omega}^{\frac12-\frac1p}\xi\|_2.
\]
One easily checks that this is a norm on $\cO$ and we denote by $L^p(\cO,\omega)$ the 
corresponding Banach space. Note that $\|\xi\|_{\omega, 2}=\|\xi\|_2$ so that
\nindex{LpOomega}{$L^p(\cO,\omega)$}{Araki-Masuda $L^p$-space}%
$L^2(\cO,\omega)=L^2(\cO)=\cH_\cO$ for any faithful state $\omega$. For $p\in[1,2]$, we define
$L^p(\cO,\omega)$ to be the dual Banach space of $L^q(\cO,\omega)$ for  $p^{-1}+q^{-1}=1$
w.r.t. the duality $(\xi|\eta)=\tr(\xi^\ast\eta)$.\bindex{Araki-Masuda $L^p$-space}

\bet\label{LpIso}
For $p\in[1,\infty]$ one has $\|\xi\|_{\omega, p}=\|\xi\omega^{1/p-1/2}\|_p$, \ie
the map
$$
\begin{array}{ccc}
L^p(\cO)&\to&L^p(\cO,\omega)\\
\xi&\mapsto&\xi\omega^{1/2-1/p},
\end{array}
$$
is a surjective isometry.
\eet
\demo
For $p\in[2,\infty]$ one has $r=p/(p-2)\in[1,\infty]$ and if $\nu\in{\mathfrak S}$, then 
$\nu^{(p-2)/p}\in L^r(\cO)$ with $\|\nu^{(p-2)/p}\|_r=\|\nu\|_1=1$. By definition of the relative 
modular operator, one further has\index{modular!operator!relative}
$$
\|\Delta_{\nu|\omega}^{\frac{p-2}{2p}}\xi\|_2^2
=(\nu^{\frac{p-2}{2p}}\xi\omega^{-\frac{p-2}{2p}}|\nu^{\frac{p-2}{2p}}\xi\omega^{-\frac{p-2}{2p}})
=\tr(\nu^{\frac{p-2}{p}}\xi^\ast\omega^{-\frac{p-2}{p}}\xi).
$$
Noting that $1-1/r=2/p$, we can write 
$$
\|\xi\|_{\omega, p}^2=\max_{\|\eta\|_r=1}\tr(\eta\xi^\ast\omega^{-\frac{p-2}{p}}\xi)
=\|\xi^\ast\omega^{-\frac{p-2}{p}}\xi\|_{p/2}
=\|\omega^{-\frac{p-2}{2p}}\xi\|_{p}^2.
$$
We conclude using the fact that $\|\omega^{-\frac{p-2}{2p}}\xi\|_{p}=\|\xi\omega^{-\frac{p-2}{2p}}\|_{p}$
(recall Exercise \ref{Exo:ABBA}).
For $p\in[1,2]$ we have, with $q^{-1}=1-p^{-1}\in[2,\infty]$,
$$
\|\xi\|_{\omega, p}=\sup_{\eta\not=0}\frac{|\tr(\xi^\ast\eta)|}{\|\eta\|_{\omega,q}}
=\sup_{\eta\not=0}\frac{|\tr(\xi^\ast\eta)|}{\|\omega^{-\frac{q-2}{2q}}\eta\|_q}
=\sup_{\nu\not=0}\frac{|\tr(\xi^\ast\omega^{\frac{q-2}{2q}}\nu)|}{\|\nu\|_q}
=\|\xi^\ast\omega^{\frac{q-2}{2q}}\|_p.
$$
Since $(q-2)/2q=-(p-2)/2p$, we get
$$
\|\xi\|_{\omega, p}
=\|\xi^\ast\omega^{-\frac{p-2}{2p}}\|_p
=\|\omega^{-\frac{p-2}{2p}}\xi\|_p=\|\xi\omega^{-\frac{p-2}{2p}}\|_p.
$$
\qed

\medskip
\begin{exo}\label{Exo:LpCone}
\exop Denote by $L^p_+(\cO,\omega)$ the image of the cone \nindex{LpOomega+}{$L^p_+(\cO,\omega)$}{Araki-Masuda positive cone}
$L^p_+(\cO)=\{\xi\in L^p(\cO)\,|\,\xi\ge0\}$ by the isometry of Theorem \ref{LpIso},
$$
L^p_+(\cO,\omega)=\{A\omega^{1/2-1/p}\,|\,A\in\cO_+\}.
$$
Show that, with $p^{-1}+q^{-1}=1$, the dual cone to $L^p_+(\cO,\omega)$ is $L^q_+(\cO,\omega)$,
\ie that \index{cone!dual}
$$
(\eta|\xi)\ge0
$$
for all $\xi\in L^p_+(\cO,\omega)$ iff $\eta\in L^q_+(\cO,\omega)$. (Note that
$L^2_+(\cO,\omega)=\cH_\cO^+$, the natural cone.)

\exop Show that
$$
L^p_+(\cO,\omega)=\{\lambda\Delta_{\rho|\omega}^{1/p}\xi_\omega\,|\,\rho\in\mathfrak S,\lambda>0\}.
$$
\end{exo}

\medskip
We finish this section with several examples of applications of the modular structure. The first one
is a proof of Kosaki's variational formula.\index{formula!Kosaki}

{\noindent\bf Proof of Theorem \ref{KosakiThm}.}
We extend the definition of the relative modular operator to pairs of non-faithful states. As already 
noticed (just before Exercise \ref{Exo:GNS}), if $\nu\in\mathfrak S$ is not faithful then its vector 
representative $\xi_\nu\in\cH_\cO$ is not cyclic for the standard representation. 
\index{vector!representative of a state}\index{vector!cyclic} In fact 
$\cO\xi_\nu=\{A\xi_\nu\,|\,A\in\cO\}$ is the proper subspace of $\cH_\cO$ given by
$$
\cO\xi_\nu=\{\eta\in\cH_\cO\,|\,\Ker\eta\supset\Ker\nu\}=\{\eta\in\cH_\cO\,|\,\eta(\one-\s(\nu))=0\}.
$$
Accordingly, one has the orthogonal decomposition
$$
\cH_\cO=\cO\xi_\nu\oplus[\cO\xi_\nu]^\perp,
$$
where
$$
[\cO\xi_\nu]^\perp=\{\eta\in\cH_\cO\,|\,\Ker\eta\supset\Ran\nu\}
=\{\eta\in\cH_\cO\,|\,\eta\s(\nu)=0\}.
$$
For $\rho,\nu\in\mathfrak S$, we define
the linear operator $\Delta_{\rho|\nu}$ on $\cH_\cO$ by
$$
\Delta_{\rho|\nu}: \xi\mapsto \rho\xi[(\nu|_{\Ran\nu})^{-1}\oplus0|_{\Ker\nu}].
$$
One easily checks that $\Delta_{\rho|\nu}$ is non-negative, with\index{modular!operator!relative}
$\Ker\Delta_{\rho|\nu}=\{\xi\in\cH_\cO\,|\,\s(\rho)\xi\s(\nu)=0\}$. We note in particular that
\begin{equation}
J\Delta_{\rho|\nu}^{1/2}(L(A)\xi_\nu\oplus\eta)=\s(\nu)L(A)^\ast\xi_\rho,
\label{DeltaRNGeneral}
\end{equation}
for any $A\in\cO$ and $\eta\in[\cO\xi_\nu]^\perp$. 

Starting from the identity $\tr(\rho^\alpha\nu^{1-\alpha})=(\xi_\nu|\Delta_{\rho|\nu}^\alpha\xi_\nu)$ 
and using the integral formula of Exercise \ref{Lowner-Heinz} we write, for $\alpha\in]0,1[$,
\[
\tr(\rho^{\alpha}\nu^{1-\alpha})=
\frac{\sin\pi\alpha}{\pi}\int_0^\infty t^{\alpha-1}\left(
\xi_\nu|\Delta_{\rho|\nu}(\Delta_{\rho|\nu} + t)^{-1}\xi_\nu
\right)\d t.
\]
For $A\in\cO$ one has,
$$
\rho(|A^\ast|^2)=\|L(A)^\ast\xi_\rho\|^2=\|\s(\nu)L(A)^\ast\xi_\rho\|^2+\|QL(A)^\ast\xi_\rho\|^2,
$$
where $Q=\one-\s(\nu)$ is the orthogonal projection on $\Ker\nu$. By Equ. \eqref{DeltaRNGeneral},
we obtain
\begin{align*}
\rho(|A^\ast|^2)&=\|J\Delta_{\rho|\nu}^{1/2}L(A)\xi_\nu\|^2+\|QL(A)^\ast\xi_\rho\|^2\\
&=(\xi_\nu|L(A^\ast)\Delta_{\rho|\nu}L(A)\xi_\nu)+\rho(AQA^\ast),
\end{align*}
from which we deduce
\begin{align*}
\frac1t\rho(|A^\ast|^2)+\nu(|\one-A|^2)
&=\frac{1}{t}(\xi_\nu|L(A^\ast)\Delta_{\rho|\nu}L(A)\xi_\nu)+(\xi_\nu| L(|\one-A|^2)\xi_\nu)\\
&+\frac1t\rho(AQA^\ast).
\end{align*}
With some elementary algebra, this identity leads to 
$$
\left(\xi_\nu|\Delta_{\rho|\nu}(\Delta_{\rho|\nu} + t)^{-1}\xi_\nu\right)=
\frac1t\rho(|A^\ast|^2)+\nu(|\one-A|^2)-R_A,
$$
where
\[
R_A=\frac1t\rho(AQA^\ast)+\left\|(\one+\Delta_{\rho|\nu}/t)^{1/2}(L(A)
-(\one+\Delta_{\rho|\nu}/t)^{-1})\xi_\nu\right\|^2.
\]
Since $R_A\geq 0$, we get
\[
\tr(\rho^\alpha\nu^{1-\alpha})\leq 
\frac{\sin\pi\alpha}{\pi}\int_0^\infty t^{\alpha-1}\left[
\frac{1}{t}\rho(|A(t)^\ast|^2)+\nu(|\one-A(t)|^2)\right]\d t,
\]
for all $A\in C(\rr_+,\cO)$, with equality iff $R_{A(t)}=0$ for all $t>0$. Since $\Delta_{\rho|\nu}\ge0$, 
this happens iff $(\one+\Delta_{\rho|\nu}/t)L(A(t))\xi_\nu=\xi_\nu$ and $\rho^{1/2}A(t)Q=0$ for all $t>0$.
The first condition is equivalent to
$$
(\one-A(t))\nu=\frac{1}{t}\rho A(t)\s(\nu).
$$
An integration by parts shows that the function
$$
A_{\rm opt}(t)= t\int_0^\infty \e^{-s\rho}\nu\e^{-s t \nu}\d s,
$$
satisfies this condition as well as $A_{\rm opt}(t)Q=0$ so that $R_{A_{\rm opt}(t)}=0$. 
This proves Kosaki's variational principle. 

Suppose that $B(t)\in C(\rr_+,\cO)$ is such that $A(t)=A_{\rm opt}(t)+B(t)$ is also minimizer. 
It follows that $B(t)$ satisfies the two conditions 
\begin{align}
tB(t)\nu+\rho B(t)\s(\nu)&=0,\label{BoptCondone}\\
\rho^{1/2}B(t)(\one-\s(\nu))&=0,\label{BoptCondtwo}
\end{align}
for all $t>0$. Let $\phi$ be an eigenvector of $\nu$ to the eigenvalue $p>0$. Condition 
\eqref{BoptCondone} yields $(\rho+tp)B(t)\phi=0$ which implies $B(t)\phi=0$. We conclude that 
$B(t)\s(\nu)=0$ and Condition \ref{BoptCondtwo} further yields $\rho^{1/2}B(t)=0$.
It follows that if either $\nu$ or $\rho$ is faithful then $B(t)=0$.
\qed

\bigskip
As a second application of modular theory, we give an alternative proof of Uhlmann's monotonicity 
theorem.\index{inequality!Uhlmann}\index{theorem!Uhlmann monotonicity}

{\noindent\bf Proof of Theorem \ref{Uhlmann}.}  To simplify notation, we shall set $\hat\nu=\Phi^\ast(\nu)$ and $\hat\rho=\Phi^\ast(\rho)$. 
In terms of the extended modular operator, one has
$$
S_\alpha(\rho|\nu)=\log\tr(\rho^\alpha\nu^{1-\alpha})=\log(\xi_\nu|\Delta_{\rho|\nu}^\alpha\xi_\nu),
$$
and we have to show that
\begin{equation}
(\xi_{\hat\nu}|\Delta_{\hat\rho|\hat\nu}^\alpha\xi_{\hat\nu})
\ge(\xi_\nu|\Delta_{\rho|\nu}^\alpha\xi_\nu),
\label{UTarget}
\end{equation}
for all $\alpha\in[0,1]$.

Consider the orthogonal decomposition
$\cH_{\cO_\cK}=\cO_\cK\xi_{\hat\nu}\oplus[\cO_\cK\xi_{\hat\nu}]^\perp$.
For $A\in\cO_\cK$ and $\eta\in[\cO_\cK\xi_{\hat\nu}]^\perp$, the Schwarz inequality 
\eqref{SchwarzIneq} yields
\begin{align*}
\|\Phi(A)\xi_\nu\|^2&=(\xi_\nu|\Phi(A)^\ast\Phi(A)\xi_\nu)\\
&\le(\xi_\nu|\Phi(A^\ast A)\xi_\nu)
=\nu(\Phi(A^\ast A))=\hat\nu(A^\ast A)\\
&=(\xi_{\hat\nu}|A^\ast A\xi_{\hat\nu})=\|A\xi_{\hat\nu}\|^2\\
&\le\|A\xi_{\hat\nu}\|^2+\|\eta\|^2=\|A\xi_{\hat\nu}\oplus\eta\|^2,
\end{align*}
which shows that the map $A\xi_{\hat\nu}\oplus\eta\mapsto\Phi(A)\xi_\nu$ is well defined
as a linear contraction $T_\nu:\cH_{\cO_\cK}\to\cH_{\cO_{\cK'}}$. The map $T_\rho$ is defined
in a similar way.

For $A\in\cO_{\cK}$ and $\eta\in[\cO_\cK\xi_{\hat\nu}]^\perp$, one has
\begin{align*}
J\Delta_{\rho|\nu}^{1/2}T_\nu(A\xi_{\hat\nu}\oplus\eta)
&=J\Delta_{\rho|\nu}^{1/2}T_\nu(A\s(\hat\nu)\xi_{\hat\nu}\oplus\eta)\\
&=J\Delta_{\rho|\nu}^{1/2}\Phi(A\s(\hat\nu))\xi_\nu\\
&=\s(\nu)\Phi(A\s(\hat\nu))^\ast\xi_\rho=\s(\nu)\Phi(\s(\hat\nu)A^\ast)\xi_\rho\\
&=\s(\nu)T_\rho \s(\hat\nu)A^\ast\xi_{\hat\rho}\\
&=\s(\nu)T_\rho J\Delta_{\hat\rho|\hat\nu}^{1/2}(A\xi_{\hat\nu}+\eta),
\end{align*}
from which we conclude that $\Delta_{\rho|\nu}^{1/2}T_\nu=K\Delta_{\hat\rho|\hat\nu}^{1/2}$
where $K=J\s(\nu)T_\rho J$ is a contraction. It follows that for $\varepsilon>0$
$$
\Delta_{\rho|\nu}^{1/2}T_\nu(\Delta_{\hat\rho|\hat\nu}^{1/2}+\varepsilon)^{-1}
=K\Delta_{\hat\rho|\hat\nu}^{1/2}(\Delta_{\hat\rho|\hat\nu}^{1/2}+\varepsilon)^{-1},
$$
and since $\sup_{x\ge0}x/(x+\varepsilon)=1$ one has
$\|\Delta_{\rho|\nu}^{1/2}T_\nu(\Delta_{\hat\rho|\hat\nu}^{1/2}+\varepsilon)^{-1}\|\le1$.
The entire analytic function
$$
F(z)=(\xi|(\Delta_{\hat\rho|\hat\nu}^{1/2}+\varepsilon)^{-z}T_\nu^\ast
\Delta_{\rho|\nu}^{z}T_\nu(\Delta_{\hat\rho|\hat\nu}^{1/2}+\varepsilon)^{-z}\xi),
$$
thus satisfies
$$
|F(z)|\le\frac{1}{\varepsilon^2}\|\Delta_{\rho|\nu}+\one\|\,\|\xi\|^2,
\qquad
|F(\i t)|\le\|\xi\|^2,\qquad |F(1+\i t)|\le\|\xi\|^2,
$$
on the strip $0\le\Re z\le 1$. By the three lines theorem $|F(z)|\le\|\xi\|^2$
on this strip. Setting $z=\alpha\in[0,1]$, we conclude that
$$
(T_\nu\xi|\Delta_{\rho|\nu}^{\alpha}T_\nu\xi)
\le(\xi|(\Delta_{\hat\rho|\hat\nu}^{1/2}+\varepsilon)^{2\alpha}\xi).
$$
Letting $\varepsilon\downarrow0$ we get
$$
(T_\nu\xi|\Delta_{\rho|\nu}^{\alpha}T_\nu\xi)
\le(\xi|\Delta_{\hat\rho|\hat\nu}^{\alpha}\xi),
$$
and \eqref{UTarget} follows from the fact that $T_\nu\xi_{\hat\nu}=\Phi(\one)\xi_\nu=\xi_\nu$.
\qed

\bigskip
As a last illustration of the use of modular theory, we prove a lower bound for quantum
hypothesis testing which complements Theorem \ref{QHTthm}. Our proof is an abstract version of 
similar results proven in \cite{ANSV,HMO}, 
where reader can find references for the previous works on the subject. 
The extension of our proof to the general von Neumann algebra setting can be found in 
\cite{JOPS}.\index{algebra!von Neumann}\index{hypothesis testing}

Let $D_p(\rho,\nu)=D_p(\rho,\nu,P_{\rm opt})$ be as in Section 1.3.7. Let $\Delta_{\rho|\nu}$ be 
the modular operator defined in the proof of Theorem 1.14, and let $\mu_{\rho|\nu}$ be the spectral 
measure for $\Delta_{\rho|\nu}$ and $\xi_\nu$. 

\medskip
\bep\label{QHTlower}
\[
D_p(\rho, \nu)\geq \frac12 \min(p,1-p)\mu_{\rho|\nu}([1,\infty[).
\]
\eep
\demo Let $P$ be an orthogonal projection (a test). By Equ. \eqref{DeltaRNGeneral}, one has
\begin{align*}
D_p(\rho,\nu,P)&=p\|(\one-P)\xi_\rho\|^2+(1-p)\|P\xi_\nu\|^2\\
&\ge p\|\s(\nu)(\one-P)\xi_\rho\|^2+(1-p)\|P\xi_\nu\|^2\\
&\ge p\|\Delta_{\rho|\nu}^{1/2}(\one-P)\xi_\nu\|^2+(1-p)\|P\xi_\nu\|^2\\
&\ge\min(p,1-p)\left(\|\Delta_{\rho|\nu}^{1/2}(\one-P)\xi_\nu\|^2+\|P\xi_\nu\|^2\right)\\
&\ge \min(p,1-p)(\xi_\nu|((\one-P)\Delta_{\rho|\nu}(\one-P)+P\one P)\xi_\nu).
\end{align*}
Let $F$ be the characteristic function of the interval $[1,\infty[$. Since $\one\ge F(\Delta_{\rho|\nu})$
and $\Delta_{\rho|\nu}\ge F(\Delta_{\rho|\nu})$, we further have
$$
D_p(\rho,\nu,P)\ge \min(p,1-p)(\xi_\nu|((\one-P)F(\Delta_{\rho|\nu})(\one-P)
+PF(\Delta_{\rho|\nu}) P)\xi_\nu).
$$
From the identity
$$(\one-P)F(\Delta_{\rho|\nu})(\one-P)+PF(\Delta_{\rho|\nu}) P-\frac12F(\Delta_{\rho|\nu})
=(\one-2P)F(\Delta_{\rho|\nu})(\one-2P),
$$
we deduce $(\one-P)F(\Delta_{\rho|\nu})(\one-P)+PF(\Delta_{\rho|\nu}) P\ge\frac12F(\Delta_{\rho|\nu})$
which allows us to conclude
$$
D_p(\rho,\nu,P)\ge \frac12\min(p,1-p)(\xi_\nu|F(\Delta_{\rho|\nu})\xi_\nu),
$$
for all orthogonal projections $P\in\cO$. Finally we note that
\begin{align*}
D_p(\rho,\nu)=\min_{P}D_p(\rho,\nu,P)&\ge\frac12\min(p,1-p)(\xi_\nu|F(\Delta_{\rho|\nu})\xi_\nu)\\
&=\frac12 \min(p,1-p)\mu_{\rho|\nu}([1,\infty[),
\end{align*}
which concludes the proof.
\qed

\medskip
\begin{exo}
Prove the following generalization of Kosaki's variational formula: for any $\rho,\nu\in\mathfrak S$,
$B\in\cO$ and $\alpha\in]0,1[$ one has\index{formula!Kosaki}
$$
\tr\left(B^\ast\rho^\alpha B\nu^{1-\alpha}\right)=\inf_{A\in C(\rr_+,\cO)}
\frac{\sin\pi\alpha}{\pi}\int_0^\infty t^{\alpha-1}\left[
\frac{1}{t}\rho(|A(t)^\ast|^2)+\nu(|B-A(t)|^2)\right]\d t.
$$
\end{exo}

\medskip

\chapter{Entropic functionals and fluctuation relations of finite quantum systems}
\label{chap:FINthermo}

\section{Quantum dynamical systems}
\label{sect:QDS}

Our starting point is a quantum dynamical system  $(\cO, \tau^t, \omega)$ on a finite dimensional
Hilbert space $\cK$,  where $\rr\ni t\mapsto\tau^t$ is a continuous group of $\ast$-automorphisms 
of $\cO$, and $\omega$ a faithful state. We denote by $\delta$ the generator of $\tau^t$ and by
$H$ the corresponding Hamiltonian.\index{dynamical system}

As in our discussion of the thermally driven harmonic chain in Chapter \ref{chap:HarmoChain},
time-reversal invariance (TRI) will play an important role in the sequel. An anti-linear
$\ast$-automorphism $\Theta$ of $\cO$ is called time-reversal of $(\cO,\tau^t)$ if%
\bindex{time reversal invariance}\myindex{TRI|see{time reversal invariance}}
$$
\Theta\circ \Theta={\rm id}, \qquad \tau^t \circ \Theta=\Theta \circ \tau^{-t}.
$$
A state $\omega$ is called TRI iff $\omega(\Theta(A))=\omega(A^\ast)$.
The quantum dynamical system $(\cO,\tau^t,\omega)$ is called TRI if there exists a
time-reversal $\Theta$ of $(\cO,\tau^t)$ such that $\omega$ is TRI.

\medskip
\begin{exo}\label{tri-bas}Suppose that $\Theta$ is a time-reversal of 
$(\cO,\tau^t)$. Show that there exists an anti-unitary $U_\Theta:\cK\rightarrow\cK$, unique up to a phase, such that 
$\Theta(A)=U_\Theta A U_\Theta^{-1}$  and deduce that $\tr(\Theta(A))=\tr(A^\ast)$.
Show that $\Theta(H)=H$ and that a state $\omega$ is TRI iff $\Theta(\omega)=\omega$. 

\noindent{\sl Hint:} Recall Exercise \ref{paris-auto}.
\end{exo}

\section{Entropy balance}
\label{sect:EntropyBalance}

The relative Hamiltonian  of $\omega_t$ w.r.t.\;$\omega$,\index{Hamiltonian!relative} 
$\ell_{\omega_t|\omega}=\log \omega_t -\log \omega$, is easily seen to satisfy:
\bep\label{prop:cocy0}
\ben
\item For all $t, s\in\rr$ the additive cocycle property\index{cocycle}
\begin{equation}
\ell_{\omega_{t+s}|\omega}= \ell_{\omega_t|\omega} + \tau^{-t}(\ell_{\omega_s|\omega}),
\label{cocyprop}
\end{equation}
holds.
\item If $(\cO, \tau, \omega)$ is TRI, then
\begin{equation}
 \Theta (\ell_{\omega_t|\omega}) =
-\tau^t (\ell_{\omega_t|\omega}),
\label{coc-ihes}
\end{equation}
for all $t\in\rr$.
\een
\eep

Differentiating the cocycle relation \eqref{cocyprop} we obtain
$$
\frac{\d}{\d t}\ell_{\omega_t|\omega}=\tau^{-t}(\sigma),
$$
where
$$
\sigma =\left.\frac{\d}{\d t}\ell_{\omega_t|\omega}\right|_{t=0}=-\i [H,\log \omega]=\delta_\omega(H),
$$
(recall that $\delta_\omega$ denotes the generator of the modular group of $\omega$). 
\index{modular!group} Thus, we can write
\begin{equation}
\ell_{\omega_t|\omega}=\int_0^t \sigma_{-s}\,\d s,
\label{ellsigma}
\end{equation}
and the relation $S(\omega_t|\omega)= -\omega_t(\ell_{\omega_t|\omega})$ yields the
quantum mechanical version of Equ. \eqref{ClassicEbalance},
\[
S(\omega_t|\omega)=-\int_{0}^t \omega(\sigma_s)\,\d s.
\]
We shall refer to this  identity  as the {\sl entropy balance equation} and call $\sigma$
the {\sl entropy production} observable.\bindex{entropy!balance}\bindex{entropy!production}
\bep\label{bas-ep}
$\omega(\sigma)=0$ and  if $(\cO,\tau^t,\omega)$ is TRI then 
$\Theta (\sigma)=-\sigma$.
\eep
\demo
$\omega(\sigma)=-\i\,\tr(\omega[H,\log\omega])=\i\,\tr(H[\omega,\log\omega])=0$.
Differentiating \eqref{coc-ihes} at $t=0$ one derives the second statement. 
\qed

\bigskip
An immediate consequence of the entropy balance equation is that the mean entropy production 
rate over the time interval $[0, t]$,
$$
\Sigma^t=\frac1t\int_0^t\sigma_s\,\d s,
$$
has a non-negative expectation
\begin{equation}
\omega(\Sigma^t)=\frac{1}{t}\int_0^t \omega(\sigma_s) \d s \geq 0.
\label{ent-ba-equ}
\end{equation}
Introducing the entropy observable $S=-\log\omega$ (so $S_t=\tau^t(S)=-\log\omega_{-t}$), we see 
that 
\begin{equation}
\Sigma^t=\frac{1}{t}(S_t-S),\qquad\frac{\d\ }{\d t}S_t|_{t=0}=\sigma.
\label{e-obs}
\end{equation}
The observable $S$ cannot survive the thermodynamic limit. However, 
the relative Hamiltonian and all other objects defined in this section do. All relations  
except \eqref{e-obs} remain valid after the thermodynamic limit is taken.

\medskip
\begin{exo}\label{Exo:StdLiouvillePerturb}
Assume that the quantum dynamical system $(\cO,\tau^t,\omega)$ is in a steady state,
$\omega(\tau^t(A))=\omega(A)$ for all $A\in\cO$ and $t\in\rr$. Denote by $K$ the
standard Liouvillean of $\tau^t$ and by $\delta_\omega$ the generator of the modular group 
of $\omega$: $\varsigma_\omega^t=\e^{t\delta_\omega}$. Consider the perturbed dynamical system $(\cO,\tau_V^t,\omega)$ associated to
 $V\in\cO_{\rm self}$ (see Section \ref{sect:Perturbations}).
\exop Show that its entropy production observable is
given by\index{Liouvillean!standard}
$$
\sigma=\delta_\omega(V).
$$
\exop Show that its standard Liouvillean is given by
$$
K_V \xi= K\xi+V\xi - JVJ\xi.
$$
\end{exo}

\medskip

\section{Finite time Evans-Searles symmetry}
\label{sect:FTQES}

At this point, looking back at Section \ref{sect:univsym}, one may think that, for
TRI quantum dynamical systems, the universal ES relation \eqref{ESRelation}\bindex{relation!Evans-Searles}
holds between the spectral measure $P^t$ of $\Sigma^t$ associated to $\omega$
$$
\omega(f(\Sigma^t))=P^t(f)=\int f(s)\,\d P^t(s),
$$
and its reversal $\bPt(f)=\omega(f(-\Sigma^t))$. To check this point, we first note that,
by Proposition \ref{bas-ep},
$$
\Theta(\Sigma^t)=\frac1t\int_0^t\Theta(\tau^s(\sigma))\,\d s=-\frac1t\int_0^t\tau^{-s}(\sigma)\,\d s
=-\tau^{-t}(\Sigma^t),
$$
which is the quantum counterpart of Equ. \eqref{SigmatTRI} and  \eqref{SigmatTRIagain}. Note that 
this relation implies that $s\in\sp(\Sigma^t)$ iff $-s\in\sp(\Sigma^t)$ and that the eigenvalues $\pm s$ have 
equal multiplicities. Furthermore, 
$$
\bPt(f)=\omega_t\circ\tau^{-t}(f(-\Sigma^t))=\omega_t\circ\Theta(f(\Sigma^t))
=\omega_{-t}(f(\Sigma^t))=\omega(f(\Sigma^t)\omega_{-t}\omega^{-1}),
$$
which, using \eqref{ellsigma}, can be rewritten as
$$
\bPt(f)=\omega\left(f(\Sigma^t)\e^{\log\omega-t\Sigma^t}\e^{-\log\omega}\right).
$$
If $\omega$ is not a steady state then $\log\omega$ and $\Sigma^t$ do not commute and 
hence we can not conclude, as in the classical case, that $\bPt(f)$ equals
$\omega\left(f(\Sigma^t)\e^{-t\Sigma^t}\right)$. Our naive attempt to generalize the ES relation 
\eqref{ESRelation} to quantum dynamical systems thus failed because quantum mechanical 
observables do not commute.

\medskip
\begin{exo}
Show that the ES-relation 
$$
\omega\left(\e^{-\alpha t\Sigma^t }\right)=\omega\left(\e^{-(1-\alpha) t\Sigma^t }\right).
$$
holds for all $t$ if and only if $[H, \omega]=0$.

\noindent{\sl Hint}:  the relation implies  $\omega(\e^{-t \Sigma^t})=1$. By Golden-Thompson
inequality, \index{inequality!Golden-Thompson} 
\[
\omega(\e^{-t \Sigma^t})=\tr (\e^{\log \omega}\e^{\log \omega_{-t}-\log \omega})
\geq\tr(\e^{\log \omega_{-t}})=1,
\]
and equality holds iff $\omega$ and $\omega_{-t}$ commute (recall Exercise \ref{late-gt}). 
Differentiating $\omega\omega_t=\omega_t\omega$ at $t=0$ deduce that $H\omega^2=\omega^2 H$.
\end{exo}

\medskip
As noticed in Section \ref{sect:univsym}, the ES relation \eqref{ESRelation} is equivalent to the 
ES-symmetry \eqref{FTESsym} \index{symmetry!Evans-Searles} of the Laplace transform of the 
measure $P^t$. We recall also that this Laplace transform is related to the relative entropy
\index{entropy!relative} through Equ. \eqref{SalphaFirst}. It is therefore natural to check for 
the ES-symmetry of the function
$$
\alpha\mapsto S_\alpha(\omega_t|\omega).
$$
Assuming TRI, we have 
\[
\tr (\omega_t^{\alpha}\omega^{1-\alpha})=\tr (\Theta(\omega^{1-\alpha}\omega_t^{\alpha}))=
\tr (\omega^{1-\alpha}\omega_{-t}^{\alpha})=\tr (\omega_t^{1-\alpha}\omega^\alpha),
\]
where we used that $\tr (\Theta(A))=\tr (A^\ast)$ (Exercise \ref{tri-bas}). Thus, 
$$
S_\alpha(\omega_t|\omega)=\log\tr\,(\omega_t^\alpha\omega^{1-\alpha})
=\log\tr\,(\omega^{(1-\alpha)/2}\omega_t^{\alpha}\omega^{(1-\alpha)/2}),
$$
satisfies the ES-symmetry.\index{ES-symmetry|see{symmetry, Evans-Searles}}

\bigskip
In our non-commutative framework one may also  define the entropy-like
functional\nindex{e(pt)a}{$e_{p,t}(\alpha)$}{entropic pressure functional}
$$
\rr\ni\alpha\mapsto
e_{p,t}(\alpha) =
\log\tr\left[\left(\omega^{(1-\alpha)/p} \omega_{t}^{2\alpha/p}
\omega^{(1-\alpha)/p}\right)^{p/2}\right].
$$
For reasons that will become clear later, we restrict the real parameter $p$ to $p\ge1$.
Since $\log\omega_t=\log\omega+\ell_{\omega_t|\omega}$, Corollary \ref{GTcor} yields
$$
e_{\infty,t}(\alpha)=\lim_{p\to\infty}e_{p,t}(\alpha)
=\log \tr (\e^{(1-\alpha)\log \omega +\alpha\log\omega_t})
=\log \tr (\e^{\log \omega +\alpha \ell_{\omega_t|\omega}}).
$$
We shall call the $e_{p, t}(\alpha)$  {\em entropic pressure functionals}.\bindex{entropic pressure}
Their basic properties are: 
\bep\label{epprops}
\ben
\item The function $[1,\infty]\ni p\mapsto e_{p,t}(\alpha)$ is continuous and monotonically 
decreasing.
\item The function $\rr\ni\alpha\mapsto e_{p,t}(\alpha)$ is real-analytic and convex.
It satisfies $e_{p,t}(0)=e_{p,t}(1)=0$ and
$$
e_{p,t}(\alpha)\left\{\begin{array}{ll}
\le 0&\text{for }\alpha\in[0,1],\\[4pt]
\ge 0&\text{otherwise}.
\end{array}
\right.
$$
\item $e_{p,t}(\alpha)=e_{p,-t}(1-\alpha)$.
\item $\partial_\alpha e_{p,t}(\alpha)|_{\alpha=0}=\omega(\ell_{\omega_t|\omega})=S(\omega|\omega_t)$ and 
$\partial_\alpha e_{p,t}(\alpha)|_{\alpha=1}=\omega_t(\ell_{\omega_t|\omega})=-S(\omega_t|\omega)$.
\item $\partial_\alpha^2 e_{\infty,t}(\alpha)|_{\alpha=0}
=\langle\ell_{\omega_{t}|\omega}|\ell_{\omega_{t}|\omega}\rangle_\omega
-\omega(\ell_{\omega_t|\omega})^2.$
\item $\partial^2_\alpha e_{2,t}(\alpha)|_{\alpha=0}
=\omega(\ell_{\omega_{t}|\omega}^2)-\omega(\ell_{\omega_t|\omega})^2$.
\item If $(\cO, \tau^t, \omega)$ is TRI, then the finite time quantum Evans-Searles 
{\rm (ES)} symmetry holds,\index{symmetry!Evans-Searles}
\begin{equation}
e_{p,t}(\alpha)=e_{p,t}(1-\alpha).
\label{QESsymmetry}
\end{equation}
\een
\label{ihes-sunday-1}
\eep
\demo
 (1) Continuity is obvious. Writing
\begin{equation}
e_{p,t}(\alpha)=\log\|\omega_t^{\alpha/p}\omega^{(1-\alpha)/p}\|_p^p,
\label{Altepalpha}
\end{equation}
monotonicity follows from Corollary \ref{GTcor}.

\medskip\noindent(2) Analyticity easily follows from the analytic functional calculus and convexity 
is a consequence of Corollary \ref{logconvcor}. The value taken by $e_{p,t}$ at $\alpha=0$ and 
$\alpha=1$ is evident and the remaining inequalities follow from convexity. 

\medskip\noindent(3) Unitary invariance of the trace norms and
Identity \eqref{ABBAId} give
\begin{align*}
\|\omega_t^{\alpha/p}\omega^{(1-\alpha)/p}\|_p&=
\|\e^{-\i t H}\omega^{\alpha/p}\e^{\i tH}\omega^{(1-\alpha)/p}\|_p\\
&=\|\omega^{\alpha/p}\e^{\i tH}\omega^{(1-\alpha)/p}\e^{-\i t H}\|_p\\
&=\|\omega^{\alpha/p}\omega_{-t}^{(1-\alpha)/p}\|_p
=\|\omega_{-t}^{(1-\alpha)/p}\omega^{\alpha/p}\|_p.
\end{align*}

\medskip\noindent(4) We consider only $p\in[1,\infty[$. The limiting case $p=\infty$ will be treated 
in the proof of Assertion (5). We set 
$T(\alpha)=\omega^{(1-\alpha)/p}\omega_t^{2\alpha/p}\omega^{(1-\alpha)/p}$ so that
$$
\left.\vphantom{T^{p/2}}\partial_\alpha e_{p,t}(\alpha)\right|_{\alpha=0}
=\left.\partial_\alpha\tr\,(T(\alpha)^{p/2})\right|_{\alpha=0}.
$$
Let $\Gamma$ be a closed contour on the right half-plane $\Re z>0$ encircling the strictly positive
spectrum of $T(0)=\omega^{2/p}$. Since $\alpha\mapsto T(\alpha)$ is continuous, $\Gamma$ can be
chosen in such a way that it encloses the spectrum of $T(\alpha)$ for $\alpha$ small enough. 
Hence, with $f(z)=z^{p/2}$, we can write
$$
\tr\,T(\alpha)^{p/2}=\oint_\Gamma f(z)\tr\,((z-T(\alpha))^{-1})\frac{\d z}{2\pi\i},
$$
so that
$$
\left.\partial_\alpha\tr\,(T(\alpha)^{p/2})\right|_{\alpha=0}
=\oint_\Gamma f(z)\tr\left[(z-T(0))^{-1}T'(0) (z-T(0))^{-1}\right]
\frac{\d z}{2\pi\i}.
$$
An elementary calculation gives
$$
T'(0)=\frac2p\,\omega^{1/p}\ell_{\omega_t|\omega}\,\omega^{1/p},
$$
and the cyclicity of the trace allows us to write
\begin{align*}
\left.\partial_\alpha\tr\, (T(\alpha)^{p/2})\right|_{\alpha=0}
&=\frac2p\,\oint_\Gamma f(z)\tr\left[(z-\omega^{2/p})^{-2}\omega^{2/p}\ell_{\omega_t|\omega}\right]
\frac{\d z}{2\pi\i}\\
&=\frac2p\,\tr\left[f'(\omega^{2/p})\omega^{2/p}\ell_{\omega_t|\omega}\right]=
\tr\,\omega\ell_{\omega_t|\omega}=S(\omega|\omega_t).
\end{align*}
The second statement also follows by taking (3) into account and observing that
$S(\omega|\omega_{-t})=S(\omega_t|\omega)$.
 
\medskip\noindent (5) Setting $T(\alpha)=\e^{\log\omega+\alpha\ell_{\omega_t|\omega}}$, we have
$T(0)=\omega$ and
\begin{align*}
\left.\vphantom{T^{P/2}}\partial_\alpha e_{\infty,t}(\alpha)\right|_{\alpha=0}&=\tr \,(T'(0)),\\[4pt]
\left.\vphantom{T^{P/2}}\partial_\alpha^2 e_{\infty,t}(\alpha)\right|_{\alpha=0}&=\tr\,(T''(0))-(\tr\,(T'(0)))^2.
\end{align*}
Iterating Duhamel's formula (recall Exercise \ref{Duhamel}), we can write
$$
T(\alpha)=\omega+\alpha\int_0^1\omega^{1-s}\ell_{\omega_t|\omega}\omega^{s}\,\d s
+\alpha^2\int_0^1\int_0^u\omega^{1-u}\ell_{\omega_t|\omega}\omega^{s}
\ell_{\omega_t|\omega}\omega^{u-s}\,\d s\,\d u
+O(\alpha^3),
$$
so that
$$
\tr\,(T'(0))=\int_0^1\tr\left[\omega^{1-s}\ell_{\omega_t|\omega}\omega^{s}\right]\,\d s
=\omega(\ell_{\omega_t|\omega})=S(\omega|\omega_t),
$$
which proves (4) in the special case $p=\infty$, and
\begin{align*}
\tr\,(T''(0))&=2\int_0^1\int_0^u\tr\left[\omega^{1-s}\ell_{\omega_t|\omega}\omega^{s}
\ell_{\omega_t|\omega}\right]\d s\,\d u\\
&=2\int_0^1\int_s^1\tr\left[\omega^{1-s}\ell_{\omega_t|\omega}\omega^{s}
\ell_{\omega_t|\omega}\right]\d u\,\d s\\
&=2\int_0^1(1-s)\tr\left[\omega^{1-s}\ell_{\omega_t|\omega}\omega^{s}\ell_{\omega_t|\omega}
\right]\d s\\
&=2\int_0^1s\,\tr\left[\omega^{s}\ell_{\omega_t|\omega}\omega^{1-s}\ell_{\omega_t|\omega}
\right]\d s.
\end{align*}
Taking the mean of the last two expressions, we get
\begin{align*}
\tr\,(T''(0))
&=\int_0^1\tr\left[\omega^{1-s}\ell_{\omega_t|\omega}\omega^{s}\ell_{\omega_t|\omega}
\right]\d s\\
&=\int_0^1\omega\left(\varsigma_\omega^{\i s}(\ell_{\omega_t|\omega})\ell_{\omega_t|\omega}
\right)\d s,
\end{align*}
and hence
$$
\left.\vphantom{T^{P/2}}\partial_\alpha^2 e_{\infty,t}(\alpha)\right|_{\alpha=0}=
\int_0^1\left[
\omega\left(\varsigma_\omega^{\i s}(\ell_{\omega_t|\omega})\ell_{\omega_t|\omega}
\right)-\omega(\ell_{\omega_t|\omega})^2\right]\d s.
$$

\medskip\noindent(6) Follows easily from the fact that 
$e_{2,t}(\alpha)=S_\alpha(\omega_t|\omega)=\log\tr\,(\omega_t^\alpha\omega^{1-\alpha})$.

\medskip\noindent(7) Under the TRI assumption one has $\Theta(\omega)=\omega$, 
$\Theta(\omega_t)=\omega_{-t}$, 
$$
\Theta\left(\left(\omega^{(1-\alpha)/p} \omega_{t}^{2\alpha/p}
\omega^{(1-\alpha)/p}\right)^{p/2}\right)=\left(\omega^{(1-\alpha)/p} \omega_{-t}^{2\alpha/p}
\omega^{(1-\alpha)/p}\right)^{p/2},
$$
and hence $e_{p,t}(\alpha)=e_{p,-t}(\alpha)$. The result now follows from Assertion (3).
\qed

\bigskip
According to our rule of thumb, we reformulate the definition of the functionals
$e_{p,t}(\alpha)$ in terms which are susceptible to survive the  thermodynamic limit.
We first note that
\[
e_{2,t}(\alpha)=S_\alpha(\omega_t|\omega)
=\log(\xi_\omega|\Delta_{\omega_t|\omega}^\alpha\xi_\omega),
\]
while Theorem \ref{RelEntVar} (2) yields the variational principle
$$
e_{\infty,t}(\alpha)=\max_{\rho\in\mathfrak S} S(\rho|\omega)+\alpha\rho(\ell_{\omega_t|\omega}).
$$
Moreover, Equ. \eqref{Altepalpha} and Theorem \ref{LpIso} immediately lead to 
$$
e_{p,t}(\alpha)=\log\|\Delta_{\omega_t|\omega}^{\alpha/p}\xi_\omega\|_{\omega,p}^p,
$$
for $p\in[1,\infty[$.

\medskip
\begin{exo}
Show that
$$
e_{\infty,t}(\alpha)
=\log(\xi_\omega|\e^{\log\Delta_\omega+\alpha L(\ell_{\omega_t|\omega})}\xi_\omega).
$$
\end{exo}

\begin{exo}
Show that the function $[1,\infty]\ni p\mapsto e_{p,t}(\alpha)$ is strictly decreasing unless $H$ and $\omega$ 
commute.  

\noindent{\sl Hint}: recall Exercise \ref{late-gt}.
\end{exo} 

\medskip

\section{Quantum transfer operators}
\label{sect:UpGroup}

For $p\in[1,\infty]$  we define a linear map $U_p(t): \cH_\cO\rightarrow\cH_\cO$ by 
$$
U_p(t)\xi=\e^{-\i tH}\xi\omega^{-\frac12+\frac1p}\e^{\i tH}\omega^{\frac12-\frac1p}.
$$
In terms of Connes cocycles and relative modular dynamics, one has\index{cocycle!Connes}
\begin{equation}
U_p(t)\xi
=\e^{-\i tH}\xi\e^{\i tH}[D\omega_t : D\omega]^{\i(\frac12-\frac1p)} 
=\e^{-\i tH}\xi\e^{\i t\varsigma_{\omega}^{\i(\frac12-\frac1p)}(H)}.
\label{UpForm}
\end{equation}
One easily checks that $\rr\ni t\mapsto U_p(t)$ is a group of operators on $\cH_\cO$ which
satisfies
\begin{equation}
(\xi|U_p(t)\eta)=(U_q(-t)\xi|\eta),
\label{UpAdjoint}
\end{equation}
for all $\xi,\eta\in\cH_\cO$ with $p^{-1} + q^{-1}=1$. The following result elucidates the nature
of this group: it is the unique isometric implementation of the dynamics on the Banach space
$L^p(\cO,\omega)$ which preserves the positive cone $L^p_+(\cO,\omega)$.
\index{Araki-Masuda $L^p$-space}

\bep\label{UpChar}
\ben
\item $t\mapsto U_p(t)$ is a group of isometries of $L^p(\cO,\omega)$.
\item $U_p(t)L^p_+(\cO,\omega)\subset L^p_+(\cO,\omega)$.
\item $U_p(-t)L(A)U_p(t)=L(\tau^t(A))$ for any $A\in\cO$.
\item $U_p(t)$ is uniquely characterized by Properties {\rm (1)--(3)}.
\een
\eep

The groups $U_p$ are natural  non-commutative generalizations of the classical Ruelle 
transfer operators. \index{transfer operator} We call {\sl $L^p$-Liouvillean} of the quantum dynamical
system $(\cO, \tau^t, \omega)$ the generator $L_p$ of $U_p$, \bindex{Liouvillean!$L^p$}
\nindex{Lp}{$L_p$}{$L^p$-Liouvillean}
$$
U_p(t)=\e^{-\i t L_p}.
$$
From Equ. \eqref{UpForm} we immediately get
$$
L_p\xi=H\xi-\xi\varsigma_{\omega}^{\i(\frac12-\frac1p)}(H).
$$
Interpreting \eqref{UpAdjoint} in terms of the duality between $L^p(\cO,\omega)$ and 
$L^q(\cO,\omega)$, we can write
$$
L_p^\ast=L_q.
$$
Note that, in the special case $p=2$, $L_2=L_2^\ast$ coincide with the standard Liouvillean $K$ 
of the dynamics $\tau^t$.\index{Liouvillean!standard}
\bet\label{LpSpec}
For any $p\in[1,\infty]$ one has
$$
\sp(L_p)=\sp(K)=\{\lambda-\mu\,|\,\lambda,\mu\in\sp(H)\}.
$$
\eet
\begin{exo}
This is the continuation of Exercise \ref{Exo:StdLiouvillePerturb}.
Show that the $L^p$-Liouvillean of the perturbed dynamical system
$(\cO,\tau_V^t,\omega)$ is given by
$$
L_p \xi= K\xi+V\xi - J\varsigma_\omega^{-\i(\frac12-\frac1p)}(V)J\xi.
$$
\end{exo}

\medskip
Interestingly enough, one can relate the groups $U_p$ to the entropic pressure functionals introduced 
in the previous section. The resulting formulas are particularly well suited to investigate the large 
time limit of these functionals. 

\bet\label{quantum-transfer}
For $\alpha \in [0,1]$, 
$$
e_{p,t}(\alpha)=\log\|\e^{-\i tL_{p/\alpha}}\xi_\omega\|_{\omega,p}^p,
$$
holds provided $p \in [1, \infty[$. In the special case $p=2$, this reduces to
$$
e_{2,t}(\alpha)=\log\,(\xi_\omega|\e^{-\i tL_{1/\alpha}}\xi_\omega).
$$
\eet
With the help of Theorem \ref{LpIso}, the proof of  the last theorem reduces to elementary calculations. 

\bigskip
{\noindent\bf Proof of Proposition \ref{UpChar} and Theorem \ref{LpSpec}.} 
Let $K$ be the standard Liouvillean of $(\cO,\tau^t,\omega)$. Since
 $\e^{-\i tK}\xi=\e^{-\i tH}\xi\e^{\i tH}$, it is obvious
that $\e^{-\i tK}$ is a group of isometries of $L^p(\cO)$ which preserves the positive 
cone $L^p_+(\cO)$. Denote by $V_p:L^p(\cO)\to L^p(\cO,\omega)$ the isometry 
defined in Theorem  \ref{LpIso}. Theorem \ref{LpSpec} and
Properties (1) and (2) of Proposition \ref{UpChar} follow from the facts that 
$U_p(t)=V_p\e^{-\i tK}V_p^{-1}$ and $L^p_+(\cO,\omega)=V_pL^p_+(\cO)$. To
prove Property (3) we note that $V_p\in R(\cO)=L(\cO)'$, so that
\begin{align*}
U_p(-t)L(A)U_p(t)&=V_p\e^{\i tK}V_p^{-1}L(A)V_p\e^{-\i tK}V_p^{-1}\\
&=V_p\e^{\i tK}L(A)\e^{-\i tK}V_p^{-1}\\
&=V_pL(\tau^t(A))V_p^{-1}=L(\tau^t(A)).
\end{align*}

\medskip\noindent (4) Let $\rr\ni t\mapsto U^t$ be a group of linear operators on $\cH_\cO$ satisfying
Properties (1)--(3) and set $V^t=L(\e^{\i tH})U^t$. The group property implies that
$$
(V^t)^{-1}=(U^t)^{-1} L(\e^{\i tH})^{-1}=U^{-t}L(\e^{-\i tH}),
$$
so that, by Property (3),
\begin{align*}
L(\tau^t(A))=U^{-t}L(A)U^t
&=(V^t)^{-1}L(\e^{\i tH})L(A)L(\e^{-\i tH})V^t\\
&=(V^t)^{-1}L(\e^{\i tH}A\e^{-\i tH})V^t=(V^t)^{-1}L(\tau^t(A))V^t,
\end{align*}
for all $A\in\cO$. Setting $A=\tau^{-t}(B)$ we conclude that
$$
V^tL(B)=L(B)V^t,
$$
for all $B\in\cO$, \ie $V^t\in L(\cO)'=R(\cO)$. Using the group property of $U^t$ one
easily shows that $t\mapsto V^t$ is also a group. It follows that $V^t=R(\e^{\i t \widetilde H})$ for some 
$\widetilde H\in\cO$. Thus, for any $A\in\cO$, one has
$$
U^tA\omega^{\frac12-\frac1p}=\e^{\i tH}A\omega^{\frac12-\frac1p}\e^{-\i t\widetilde H^\ast}
=\e^{\i tH}A\e^{-\i t H^\#}\omega^{\frac12-\frac1p},
$$
where $H^\#=\varsigma_\omega^{-\i(\frac12-\frac1p)}(\widetilde H^\ast)$. Exercise \ref{Exo:LpCone} (1) 
and Property (2) imply that $\e^{\i tH}A\e^{-\i tH^\#}\in\cO_+$  for any $A\in\cO_+$.
Since any self-adjoint element of $\cO$ is a real linear combination of elements of $\cO_+$,
it follows that
$$
\e^{\i tH}A\e^{-\i tH^\#}=\left(\e^{\i tH}A\e^{-\i tH^\#}\right)^\ast=
\e^{\i tH^{\#\ast}}A\e^{-\i tH},
$$
for any $A\in\cO_{\rm self}$.
This identity extends by linearity to arbitrary $A\in\cO$. Differentiation at $t=0$ yields
\begin{equation}
(H-H^{\#\ast})A=A(H^{\#}-H).
\label{HAAH}
\end{equation}
Setting  $A=\one$, we deduce that $H^\#+H^{\#\ast}=2H$, and hence that
$H^\#=H+\i T$ with $T\in\cO_{\rm self}$.  Relation \eqref{HAAH} now implies
$TA=AT$ for all $A\in\cO$ so that $T=\lambda\one$ for some $\lambda\in\rr$. It follows that
$H^\#=\varsigma_\omega^{\i(1/2-1/p)}(H)^\ast-\i\lambda$ and hence
$U^t=\e^{\lambda t}U_p(t)$. Property (1) finally imposes $\lambda=0$.
\qed

\bigskip

\section{Full counting statistics}
\label{sect:FCS}
\bindex{full counting statistics}

The functional
\[
e_{2,t}(\alpha)=S_\alpha(\omega_t|\omega)
=\log(\xi_\omega|\Delta_{\omega_t|\omega}^\alpha\xi_\omega)=
\log(\xi_\omega|\e^{\alpha\log \Delta_{\omega_t|\omega}}\xi_\omega),
\]
can be interpreted in  spectral terms. If we denote by $Q^t$ the spectral measure of 
the self-adjoint operator
$$
-\frac1t\log \Delta_{\omega_t|\omega}
=-\frac1t\log\Delta_\omega-\frac1tL(\ell_{\omega_t|\omega})
=-\frac1t\log\Delta_\omega-L(\Sigma^{-t}),
$$
for the vector $\xi_\omega$ then
\begin{equation}
e_{2,t}(\alpha)=\log \left[ \int_\rr \e^{-\alpha ts}\d Q^t(s)\right].
\label{PtTMdef}
\end{equation}
As explained at the end of Section \ref{sect:univsym}, the ES symmetry \eqref{QESsymmetry}
\index{symmetry!Evans-Searles} can be expressed in terms of the measure $Q^t$ in the 
following familiar form (see \cite{TM}).
Let ${\mathfrak r}: \rr\rightarrow \rr$  be the reflection ${\mathfrak r}(s)=-s$, and let 
$\bQt=Q^t \circ {\mathfrak r}$ be  the reflected spectral measure.
\bep\label{prop-mat-tas}
Suppose that $(\cO, \tau^t, \omega)$ is TRI. Then the  measures $Q^t$ and 
$\bQt$ are mutually absolutely continuous and 
$$
\frac{\d\bQt}{\d Q^t}(s)=\e^{-ts}.
$$
\eep
The measure $Q^t$ is not the spectral measure of any observable in $\cO$ and on the first
sight one may question its physical relevance. Its interpretation is somewhat striking and is 
linked to concept of {\em Full Counting Statistics} (FCS) of repeated quantum measurement 
of the entropy observable $S=-\log \omega$. To our knowledge, this interpretation
goes back to Kurchan \cite{Ku} (see also \cite{DRM}).\index{FCS|see{full counting statistics}}

At time $t=0$, with  the system  in the state $\omega$, we perform a measurement of $S$.
The possible outcomes of the measurement are eigenvalues of $S$  and $s\in\sp(S)$ 
is observed with  probability $\omega(P_s)$, where 
$P_s$ is the spectral projection of $S$ onto its eigenvalue $s$. After the measurement, the state 
of the system reduces to
$$
\frac{\omega P_s}{\omega(P_s)},
$$
and this state now evolves according to
$$
\frac{\e^{-\i tH}\omega P_s\e^{\i tH}}{\omega(P_s)}.
$$
A second measurement of $S$ at time $t$ yields the result $s'\in\sp(S)$ with
probability
$$
\frac{\tr\left(\e^{-\i tH}\omega P_s\e^{\i tH}P_{s'}\right)}{\omega(P_s)}.
$$
Thus, the joint probability distribution of the two measurement is given by
$$
\tr\left(\e^{-\i tH}\omega P_s\e^{\i t H}P_{s'}\right),
$$
and the probability distribution of the mean rate of change of entropy, $\phi=(s'-s)/t$, is given by 
$$
{\mathbb P}_t(\phi)=\sum_{s'-s=t\phi} \tr\left(\e^{-\i tH}\omega P_s\e^{\i t H}P_{s'}\right).
$$
It follows that
\[
{\rm tr}(\omega_t^{1-\alpha}\omega^{\alpha})= 
\sum_{s,s'}\e^{-\alpha(s'-s)} 
{\rm tr}\left(\e^{-\i tH}\omega P_s\e^{\i t H}P_{s'}\right)
=\sum_{\phi}{\mathbb P}_t(\phi)\e^{-t\alpha\phi}.
\]
and we conclude that
$$
e_{2,-t}(\alpha)=e_{2,t}(1-\alpha)=\log\left[\sum_\phi {\mathbb P}_t(\phi)\e^{-t\alpha\phi}\right].
$$
Comparison with Equ. \eqref{PtTMdef} allows us to conclude that the spectral measure
$Q^{-t}$ coincide with the distribution ${\mathbb P}_t(\phi)$. Consequently,
applying Proposition \ref{ihes-sunday-1}, the expectation and variance of $\phi$ w.r.t. ${\mathbb P}_t$ 
are given by
\[
{\mathbb E}_t(\phi)=\left.-\frac1t\partial_\alpha e_{2,-t}(\alpha)\right|_{\alpha=0}
=-\frac1t\omega(\ell_{\omega_{-t}|\omega})
=\omega(\Sigma^t),
\]
\[
{\mathbb E}_t(\phi^2)-{\mathbb E}_t(\phi)^2
=\left.\frac1{t^2}\partial_\alpha^2 e_{2,-t}(\alpha)\right|_{\alpha=0}
=\frac1{t^2}\left(\omega(\ell_{\omega_{-t}|\omega}^2)-\omega(\ell_{\omega_{-t}|\omega})^2\right)
=\omega({\Sigma^t}^2)-\omega(\Sigma^t)^2.
\]
They coincide with the expectation and variance of $\Sigma^t$ w.r.t. $\omega$. However,
we warn the reader that such a relation does not hold true for higher order cumulants.

Note that time-reversal invariance played no role in the identification of 
$\bar Q^{\kern1pt -t}$ with
${\mathbb P}_t(\phi)$. However if $(\cO, \tau, \omega)$ is TRI, then 
$\bar Q^{\kern1pt -t}=Q^t$ and
Proposition \ref{prop-mat-tas} translates into the fluctuation relation 
$$
\frac{{\mathbb P}_t(-\phi)}{{\mathbb P}_t(\phi)}=\e^{-t\phi},
$$
where $\phi\in(\sp(S)-\sp(S))/t$.

\section{On the choice of reference state}
\label{sect:ChoiceOne}

Starting with entropy production, all the objects that we have introduced so far depend on the choice 
of the reference state $\omega$. In this subsection we shall indicate by a subscript this dependence 
on $\omega$ (hence, $\sigma_\omega$ is the entropy production of \index{entropy!production}
$(\cO, \tau^t, \omega)$, etc.).

If $\omega$ and $\rho$ are two faithful states on $\cO$, then
\[
\sigma_\omega-\sigma_\rho=\i[\ell_{\omega|\rho},H]
=\left.-\frac{\d\ }{\d t}\tau^t(\ell_{\omega|\rho})\right|_{t=0},
\]
and hence
\[
\Sigma^t_\omega-\Sigma^t_\rho=\frac1t \int_0^t\frac{\d\ }{\d s}\tau^s(\ell_{\omega|\rho})\,\d s
=\frac{\tau^{t}(\ell_{\omega|\rho})-\ell_{\omega|\rho}}t.
\]
Consequently,
$$
\|\Sigma^t_\omega-\Sigma^t_\rho\|=\|\ell_{\omega|\rho}\|O(t^{-1}).
$$
Thus, $\Sigma^t_\omega$ and $\Sigma^t_\rho$ become indistinguishable for large $t$.
A similar result holds for the properly normalized entropic functionals. For example:
 
\bep\label{Prop:EquivState}
For all $\alpha\in\rr$ and $t\in\rr$ one has the estimate
\[
\left|\frac1te_{\infty,t,\omega}(\alpha)-\frac1te_{\infty,t,\rho}(\alpha)\right|
\leq (|1-\alpha| +|\alpha|) \frac{\|\ell_{\omega|\rho}\|}t.
\]
\eep
\demo 
We have
\begin{align*}
\tr (\e^{\log \omega +\alpha \ell_{\omega_t|\omega}})&= 
\tr (\e^{\log \rho +\alpha\ell_{\rho_t|\rho} +(1-\alpha)\ell_{\omega|\rho} +\alpha \ell_{\omega_t|\rho_t}})\\
&\leq \tr (\e^{\log \rho +\alpha\ell_{\rho_t|\rho}}\e^{(1-\alpha)\ell_{\omega|\rho} +\alpha \ell_{\omega_t|\rho_t}})\\
&\leq \e^{(|1-\alpha| +|\alpha|)\|\ell_{\omega|\rho}\|}\tr (\e^{\log \rho +\alpha\ell_{\rho_t|\rho}}),
\end{align*}
where we have used the  Golden-Thompson inequality (Corollary \ref{GTcor}).
\index{inequality!Golden-Thompson} Taking logarithms, we get 
\[
e_{\infty,t,\omega}(\alpha)-e_{\infty,t,\rho}(\alpha)
\leq (|1-\alpha| +|\alpha|)\|\ell_{\omega|\rho}\|.
\]
Reversing the roles of $\omega$ and $\rho$ and using that 
$\|\ell_{\omega|\rho}\|=\|\ell_{\rho|\omega}\|$ we deduce the statement.
\qed

\bigskip

\section{Compound systems}
\label{sect:MultiSys}

Consider the quantum dynamical system $(\cO, \tau^t, \omega)$ describing a compound
system made of $n$ subsystems. The underlying Hilbert space is given by a tensor product
\[
\cK=\bigotimes_{j=1}^n\cK_j,
\]
and
\begin{equation}
\cO=\bigotimes_{j=1}^n\cO_j,
\label{Oprod}
\end{equation}
where $\cO_j=\cO_{\cK_j}$ is the algebra of observables of the $j$-th subsystem. We
identify $A_j\in\cO_{j}$ with 
$\one_{\otimes_{i=1}^{j-1}\cK_i}\otimes A_j \otimes \one_{\otimes_{i=j+1}^{n}\cK_i}\in\cO$.

We assume that the  reference state $\omega$ has the product structure
\begin{equation}
\omega(A_1\otimes\cdots\otimes A_n)=\prod_{j=1}^n\omega_j(A_j),
\label{omegaprod}
\end{equation}
where $\omega_j$ is a faithful state on $\cO_j$. According to the above convention, 
$\omega_j$ is identified with the positive operator 
$\one_{\otimes_{i=1}^{j-1}\cK_i}\otimes\omega_j \otimes \one_{\otimes_{i=j+1}^{n}\cK_i}$, so that
$\log\omega_j$ is a self-adjoint element of $\cO$ and
\[
\log \omega =\sum_{j=1}^n \log\omega_j.
\]
Accordingly, the entropy production observable of the system can be written 	as \index{entropy!production}
\[
\sigma =\i [\log \omega, H]=\sum_{j}\sigma_j,
\]
where $\sigma_j=\i[\log \omega_j,H]$. Similarly, the relative Hamiltonian $\ell_{\omega_t|\omega}$
decomposes as\index{Hamiltonian!relative}
\[
\ell_{\omega_t|\omega}=\sum_{j=1}^n\ell_{\omega_{jt}|\omega_j},
\]
where
\[
\ell_{\omega_{jt}|\omega_j}=\tau^{-t}(\log \omega_j)-\log \omega_j=\int_0^t \tau^{-s}(\sigma_j)\d s.
\]
If the system $(\cO, \tau^t, \omega)$ is TRI with time-reversal $\Theta$, we shall always assume 
that 
\[
\Theta(\omega_j)=\omega_j.
\]
This  implies
\[
\Theta(\sigma_j)=-\sigma_j.
\]
For $\balpha=(\alpha_1, \cdots, \alpha_n)\in\rr^n$ we denote 
$\omega^\balpha =\omega_1^{\alpha_1}\cdots  \omega_n^{\alpha_n}$.
Similarly, 
\[
\omega_t^\balpha=\e^{-\i t H}\omega^\balpha\e^{\i t H}=\prod_{j=1}^n \omega_{jt}^{\alpha_j}.
\]
We also denote ${\bf 1}=(1,\ldots, 1)$ and ${\bf 0}=(0, \ldots, 0)$. 
The multi-parameter entropic pressure functionals are defined for $t\in\rr$ and \index{entropic pressure}
\nindex{eptA}{$e_{p,t}(\balpha)$}{multi-parameter entropic pressure functional} $\balpha\in\rr^n$ by 
\[
e_{p,t}(\balpha)=
\begin{cases}
\log\tr\left[\left(
\omega^{\frac{{\bf 1}-\balpha}p} \omega^{\frac{2\balpha}p}_t\omega^{\frac{{\bf 1}-\balpha}p}
\right)^{\frac{p}{2}}\right] & \text{for }1\le p<\infty,\\[5mm]
\log\tr\left(\e^{\log\omega+\sum_j\alpha_j\ell_{\omega_{jt}|\omega_j}}\right) &\text{for }p=\infty.
\end{cases}
\]
These functionals are natural generalizations of the functionals introduced in Section \ref{sect:FTQES}
and  have very similar properties: 
\bep\label{ihes-baby-1}
\ben
\item The function $[1,\infty]\ni p\mapsto e_{p,t}(\balpha)$ is continuous and monotonically 
increasing.
\item The function $\rr^n\ni\balpha\mapsto e_{p,t}(\balpha)$ is real-analytic, convex, and
$e_{p,t}({\bf 0})=e_{p,t}({\bf 1})=0$.
\item $e_{p,t}(\balpha)=e_{p,-t}({\bf 1}-\balpha)$.
\item $\partial_{\alpha_j}e_{p,t}(\balpha)|_{\balpha={\bf 0}}=\omega(\ell_{\omega_{jt}|\omega_j})$.
\item 
$$
\partial_{\alpha_k}\partial_{\alpha_j} e_{\infty,t}(\balpha)|_{\balpha={\bf 0}}
=\langle\ell_{\omega_{kt}|\omega_k}|\ell_{\omega_{jt}|\omega_j}\rangle_\omega
-\omega(\ell_{\omega_{kt}|\omega_k})\omega(\ell_{\omega_{jt}|\omega_j}).
$$
\item 
$$\partial_{\alpha_k}\partial_{\alpha_j} e_{2,t}(\balpha)|_{\balpha={\bf 0}}
=\frac1{2}\int_{0}^t\int_0^t
\omega\left((\sigma_{ks}-\omega(\sigma_{ks}))(\sigma_{ju}-\omega(\sigma_{ju}))\right)\d s \d u.
$$
\item If $(\cO,\tau^t,\omega)$ is TRI, then the finite time Evans-Searles {\rm (ES)} 
symmetry holds:  \index{symmetry!Evans-Searles}
\[
e_{p,t}(\balpha)=e_{p,t}({\bf 1}-\balpha).
\]
\een
\eep

The proof, which is similar to the proof of Proposition \ref{ihes-sunday-1}, is left as an exercise.

\bigskip
In order to express the multi-parameter entropic pressure functionals in terms of the
modular structure of $(\cO,\omega)$, we have to extend the definition of relative modular
operator. Let us briefly indicate how to proceed. The main problem is that $\omega_j$
is not a state on $\cO$ (it is not properly normalized, and cannot be normalized in the
thermodynamic limit since the dimensions of the Hilbert spaces $\cK_i$ diverge in this limit).
However, as a state on $\cO_j$, $\omega_j$  has a modular group $\varsigma_{\omega_j}$
\index{modular!group} and a modular operator $\Delta_{\omega_j}$ \index{modular!operator}
such that
$$
\varsigma_{\omega_j}^s(A)=\Delta_{\omega_j}^{\i s}A\Delta_{\omega_j}^{-\i s}.
$$
The formula \nindex{soS}{$\varsigma_\omega^{\bf s}$}{multi-parameter modular group}
$$
\rr^n\ni{\bf s}=(s_1,\ldots,s_n)\mapsto\varsigma_\omega^{\bf s}
=\bigotimes_{j=1}^n\varsigma_{\omega_j}^{s_j},
$$
defines an abelian group of $\ast$-automorphisms of $\cO$. With a slight abuse 
of language, we shall refer to the multi-parameter group $\varsigma_{\omega}^{\bf s}$ as 
the modular group of $\omega$. We denote by
$$
\Delta_{\omega}^{\i\bf s}=\bigotimes_{j=1}^n\Delta_{\omega_j}^{\i s_j}.
$$
the corresponding abelian unitary group. Setting
$$
\varsigma_{\omega_{t}}^{\bf s}=\tau^{-t}\circ\varsigma_{\omega}^{\bf s}\circ\tau^t,
$$
we clearly have $\varsigma_{\omega}^{\bf s}(A)=\omega^{\i\bf s}A\omega^{-\i\bf s}$ and
$\varsigma_{\omega_{t}}^{\bf s}(A)=\omega_{t}^{\i\bf s}A\omega_{t}^{-\i\bf s}$. 

The two modular groups $\varsigma_{\omega}^{\bf s}$ and $\varsigma_{\omega_t}^{\bf s}$ are 
related by 
$$
\varsigma_{\omega_t}^{\bf s}(A)=[D\omega_{t}:D\omega]^{\bf s}\varsigma_{\omega}^{\bf s}(A)
[D\omega:D\omega_t]^{\bf s},
$$
where the unitary Connes cocycle\index{cocycle!Connes}
\nindex{[}{$[D\omega_{t}:D\omega]^{\bf s}$}{multi-parameter Connes cocycle}
\begin{equation}
[D\omega_{t}:D\omega]^{\bf s}=\omega_{t}^{\i\bf s}\omega^{-\i\bf s}
=\e^{\i\sum_j s_j\tau^{-t}(\log\omega_j)}\e^{-\i\sum_j s_j\log\omega_j}
=\e^{-\i tH}\e^{\i t\varsigma_{\omega}^{\bf s}(H)},
\label{DDdefdef}
\end{equation}
satisfies the two multiplicative cocycle relations
\begin{align}
[D\omega_{t}:D\omega]^{\bf s}\varsigma_{\omega}^{\bf s}
([D\omega_{t}:D\omega]^{\bf s'})&=[D\omega_{t}:D\omega]^{\bf s+s'},\nonumber\\
\tau^{-t}([D\omega_{t'}:D\omega]^{\bf s})[D\omega_{t}:D\omega]^{\bf s}
&=[D\omega_{t+t'}:D\omega]^{\bf s}.\label{CocyTwo}
\end{align}
Thanks to the first relation, 
$$
\rr^n\ni{\bf s}\mapsto
\Delta_{\omega_{t}|\omega}^{\i\bf s}=L([D\omega_{t}:D\omega]^{\bf s})\Delta_{\omega}^{\i\bf s},
$$
defines an abelian group of unitaries on $\cH_\cO$. One easily checks that\index{Hamiltonian!relative}
$\Delta_{\omega_{t}|\omega}^{\i\bf s}\xi=\omega_{t}^{\i\bf s}\xi\omega^{-\i\bf s}$.
The relative Hamiltonian $\ell_{\omega_{jt}|\omega_j}=\tau^{-t}(\log\omega_j)-\log\omega_j$ is given by
$$
\ell_{\omega_{jt}|\omega_j}=\left.\frac1\i\frac{\d\ \ }{\d s_j}[D\omega_{t}:D\omega]^{\bf s}\right|_{\bf s=0}.
$$

Using Theorem \ref{LpIso} and the fact that $\Delta_\omega^{\balpha/p}\xi_\omega=\xi_\omega$ 
it is now easy to show that, for $p\in[1,\infty[$,
\begin{equation}
e_{p,t}(\balpha)=\log\|\Delta_{\omega_t|\omega}^{\balpha/p}\xi_\omega\|_{\omega,p}^p
=\log\|[D\omega_t:D\omega]^{-\i\balpha/p}\xi_\omega\|_{\omega,p}^p,
\label{eptmultiAM}
\end{equation}
while Theorem \ref{RelEntVar} leads to
\[
e_{\infty,t}(\balpha)=
\max_{\rho\in\mathfrak S} \left(S(\rho|\omega)
+\sum_{j=1}^n\alpha_j\rho(\ell_{\omega_{jt}|\omega_j})\right).
\]
In particular, one has
\[
e_{2,t}(\balpha)=\log(\xi_\omega|\Delta_{\omega_{t}|\omega}^\balpha\xi_\omega)
=\log\omega([D\omega_{t}:D\omega]^{-\i\balpha}).
\]

One can also generalize Theorem \ref{quantum-transfer} to the present setup. To this end, 
let $K$ be the standard Liouvillean of the dynamics $\tau^t$. With ${\bf s}\in\rr^n$, the second cocycle 
relation \eqref{CocyTwo} allows us to construct the unitary group
$$
\e^{-\i tK_{\bf s}}=R([D\omega_t:D\omega]^{{\bf s}})^\ast\e^{-\i tK},
$$
on $\cH_\cO$. By \eqref{DDdefdef}, one has
$$
\e^{-\i tK_{\bf s}}\xi=\e^{-\i tH}\xi\e^{\i tH}[D\omega_t:D\omega]^{\bf s}=
\e^{-\i tH}\xi\e^{\i t\varsigma_\omega^{\bf s}(H)},
$$
so that $K_{\bf s}=L(H)-R(\varsigma_\omega^{\bf s}(H))$. Analytic continuation
of $\e^{-\i tK_{\bf s}}$ to ${\bf s}=\i(1/2-1/p){\bf 1}$ with $p\in[1,\infty]$ yields the group 
$U_p(t)$ of isometric implementation of the dynamics on the Araki-Masuda space $L^p(\cO,\omega)$ 
introduced in Section \ref{sect:UpGroup}. \index{Araki-Masuda $L^p$-space}

For $\balpha\in{[0,1]}^n$ and $p\in[1,\infty]$, let us define
$$
L_{\frac p\balpha}=K_{\bf s},
\qquad {\bf s}=\i\left(\frac{\bf1}2-\frac{\balpha}{p}\right).
$$
From the identity
$$
\e^{-\i tL_{\frac p\balpha}}\xi_\omega=\omega_t^{\balpha/p}\omega^{1/2-\balpha/p}
=[D\omega_t:D\omega]^{-\i\balpha/p}\xi_\omega,
$$
and Equ. \eqref{eptmultiAM} we deduce
$$
e_{p,t}(\balpha)=\log\|\e^{-\i tL_{\frac p\balpha}}\xi_\omega\|_{\omega,p}^p.
$$
In the special case $p=2$, this can be rewritten as
$$
e_{2,t}(\balpha)=\log(\xi_\omega|\e^{-\i tL_{\frac1\balpha}}\xi_\omega).
$$

\begin{exo}
Show that the Connes cocycle $\Gamma({\bf s},t)=[D\omega_{t}:D\omega]^{\bf s}$ satisfies
the following differential equations,
\begin{align*}
-\i\frac{\d\ }{\d t}\Gamma({\bf s},t)&=\tau^{-t}(\varsigma_{\omega}^{\bf s}(H)-H)\Gamma({\bf s},t),
&\Gamma({\bf s},0)&=\one,\\
-\i\frac{\d\ \ }{\d s_j}\Gamma({\bf s},t)
&=\Gamma({\bf s},t)\varsigma_{\omega}^{\bf s}(\tau^{-t}(\log\omega_j)-\log\omega_j),
&\Gamma({\bf 0},t)&=\one.
\end{align*}
\end{exo}
\begin{exo}\label{Exo:LalphaLiouvPerturb}
Assume that $H=H_0+V$ with $[H_0,\omega]=0$, \ie $\omega$
is a steady state for the dynamics $\tau_0^t$ generated by $H_0$. Show that
$$
L_{\frac1\balpha}=K_0+L(V)-R(\varsigma_\omega^{\i(\balpha-{\bf1}/2)}(V)),
$$
where $K_0$ is the standard Liouvillean of $\tau_0^t$.
\end{exo}

\medskip

\section{Multi-parameter full counting statistics}
\label{sect:MultiFCS}

We  continue with  the framework of the last section and extend to compound
systems our discussion of full counting statistics  started in Section \ref{sect:FCS}.\index{full counting statistics}

With ${\bf 1}_j=(0,\ldots,1,\ldots,0)$ (a single $1$ at the $j$-th entry) we set
$$
\Delta_{\omega_{jt}|\omega_j}=\Delta_{\omega_{t}|\omega}^{{\bf1}_j}.
$$
In terms on the joint spectral measure $Q^t$ of the commuting family of self-adjoint operators
$$
-\frac1t\log\Delta_{\omega_{1t}|\omega_1},\ldots,-\frac1t\log\Delta_{\omega_{nt}|\omega_n},
$$
associated to the vector $\xi_\omega$ one has, for $\balpha\in\rr^n$,
$$
(\xi_\omega|\Delta_{\omega_t|\omega}^\balpha\xi_\omega)
=(\xi_\omega|\e^{\sum_j\alpha_j\log\Delta_{\omega_{jt}|\omega_j}}\xi_\omega)
=\int\e^{-t\balpha\cdot{\bf s}}\,\d Q^t({\bf s}).
$$
Let ${\mathfrak r}$ denote the reflection ${\mathfrak r}({\bf s})=-{\bf s}$ on  $\rr^n$, and let 
$\bQt=Q^t \circ {\mathfrak r}$ be  the reflected spectral measure. The ES symmetry \index{symmetry!Evans-Searles}
$e_{2,t}({\bf 1}-\balpha)=e_{2,t}(\balpha)$ translates into
\bep\label{prop-mat-tas-multi}
Suppose that $(\cO, \tau^t, \omega)$ is TRI. Then the  measures $Q^t$ and 
$\bQt$ are mutually absolutely continuous and 
$$
\frac{\d\bQt}{\d Q^t}({\bf s})=\e^{-t{\bf 1\cdot s}}.
$$
\eep

To interpret this result, considered the vector observable 
\[
{\bf S}=(-\log \omega_1, \cdots, -\log \omega_n).
\]
Since the $\omega_j$'s commute, the components of $\bf S$ can be simultaneously measured.
Let $P_{\bf s}$ denote the joint spectral projection of $\bf S$ to  the eigenvalue ${\bf s}\in\sp({\bf S})$.
The joint probability distribution of two measurements is 
$$
\tr\left(\e^{-\i tH}\omega P_{\bf s}\e^{\i t H}P_{\bf s'}\right).
$$
Denote by ${\mathbb P}_t({\boldsymbol \phi})$ the
induced probability distribution of the vector ${\boldsymbol\phi}=({\bf s'-s})/t$ which describes
the mean rate of change of $\bf S$ between the two measurements. For $\balpha\in \rr^n$ one has, 
by Proposition \ref{ihes-baby-1} (3),
\begin{align*}
(\xi_\omega|\Delta_{\omega_{-t}|\omega}^{\balpha}\xi_\omega)=
(\xi_\omega|\Delta_{\omega_t|\omega}^{{\bf1}-\balpha}\xi_\omega)&=
\tr(\omega_t^{{\bf 1}-\balpha}\omega^\balpha)\\
&=\sum_{\bf s,s'}\e^{-\sum_j \alpha_j(s_j'-s_j) }\tr(\e^{-\i t H}\omega P_{\bf s}\e^{\i t H}P_{\bf s'})\\
&=\sum_{\boldsymbol\phi}\e^{-\sum_j t\alpha_j\phi_j }{\mathbb P}_t({\boldsymbol\phi}).
\end{align*}
As in Section \ref{sect:FCS}, we can conclude that the spectral measure $\bar Q^{\kern1pt -t}$ coincide with 
the probability distribution ${\mathbb P}_t$. Assertion (4) and (6) of Proposition \ref{ihes-baby-1} 
yield the expectation and covariance of $\boldsymbol\phi$ w.r.t. ${\mathbb P}_t$,
\[
{\mathbb E}_t(\phi_j)=\left.-\frac1t\partial_{\alpha_j}e_{2,-t}(\balpha)\right|_{\balpha=0}
=-\frac1t\omega(\ell_{\omega_{j(-t)}|\omega_j})
=\frac1t\int_0^t \omega(\sigma_{js})\d s,
\]
\begin{align*}
{\mathbb E_t}(\phi_j\phi_k)-{\mathbb E}_t(\phi_j){\mathbb E}_t(\phi_k)
&=\left.\frac1{t^2}\partial_{\alpha_j}\partial_{\alpha_k} e_{2,-t}(\balpha)\right|_{\balpha=0}\\
&=\frac1{2t^2}\int_{0}^t\int_0^t
\omega\left((\sigma_{js}-\omega(\sigma_{js}))(\sigma_{ku}-\omega(\sigma_{ku}))\right)\d s \d u.
\end{align*}
If the system is TRI then $\bar Q^{\kern1pt -t}=Q^t$ and Theorem \ref{prop-mat-tas-multi} yields the ES 
fluctuation relation
$$
\frac{{\mathbb P}_t(-{\boldsymbol\phi})}{{\mathbb P}_t({\boldsymbol\phi})}
=\e^{-t{\bf 1}\cdot{\boldsymbol\phi}}.
$$

\medskip
\begin{exo}The above formula for the covariance of the full counting statistics implies that
$$
A_{jk}=
\int_{0}^t\int_0^t
\omega\left((\sigma_{js}-\omega(\sigma_{js}))(\sigma_{ku}-\omega(\sigma_{ku}))\right)\d s \d u,
$$
is symmetric, $A_{jk}=A_{kj}$. Prove this directly, starting from the definition
$\sigma_j=-\i[H,\log\omega_j]$.

\noindent{\sl Hint}: show that
\begin{align*}
\int_{0}^t\int_0^t[\sigma_{js},\sigma_{ku}]\d s \d u
&=[\log\omega_j,\log\omega_k]+\tau^t([\log\omega_j,\log\omega_k])\\
&-[\tau^t(\log\omega_j),\log\omega_k]-[\log\omega_j,\tau^t(\log\omega_k)].
\end{align*}
\end{exo}
\begin{exo}\label{Exo:AbelianOmega}
Check that the tensor product structure \eqref{Oprod} was never used in
the last two  sections. More precisely, replacing Assumption \eqref{omegaprod} 
with
$$
\log\omega=\sum_{j=1}^n Q_j,
$$
where $(Q_1,\ldots,Q_n)$ is a commuting family of self-adjoint elements of $\cO$,
and defining $\omega_j=\e^{Q_j}$ so that
$$
\omega^\balpha=\e^{\sum_{j=1}^n\alpha_jQ_j},
$$
show that all the results of the two sections hold without modification.
\end{exo}

\medskip

\section{Control parameters and fluxes}
\label{sect:ControlDynSys}

Suppose that our quantum dynamical system $(\cO_X, \tau_X, \omega_X)$ depends on some 
control parameters $X=(X_1, \cdots, X_n)\in {\mathbb R}^n$. One can think of  $X_j$'s as 
mechanical or thermodynamical forces acting on the system. We denote by $H_X$ the Hamiltonian 
generating the dynamics $\tau_X^t$, by $\sigma_X$ the entropy production observable, etc. 
We assume that $\omega_0$ is $\tau_0^t$ invariant and refer to the value $X=0$ as {\em equilibrium}.
Note that this implies $\sigma_0=0$. We adopt the shorthands $\tau^t=\tau_0^t$, 
$\omega=\omega_0$. \index{flux}

\begin{definition} A vector-valued observable ${\bf \Phi}_X= (\Phi_X^{(1)}, \cdots, \Phi_X^{(n)})
\in\cO^n_{\rm self}$, is called a flux relation if, for all $X$,\bindex{relation!flux}
\nindex{FX}{${\bf \Phi}_X$}{flux relation}
$$
\sigma_X=\sum_{j=1}^n X_j \Phi_X^{(j)}.
$$
\end{definition}
In what follows we will consider a family of quadruples
$(\cO, \tau_X^t, \omega_X, {\bf \Phi}_X)_{X\in \rr^n}$, where ${\bf \Phi}_X$ is  a given flux relation.
Somewhat colloquially, we will refer to $\Phi_X^{(j)}$ as the flux (or current) observable
associated to the force $X_j$. In concrete models, physical requirements typically select a unique 
flux relation ${\bf \Phi}_X$.

If $(\cO_X, \tau_X^t,\omega_X)_{X\in \rr^N}$ are time-reversal invariant (TRI),
\index{time reversal invariance} we shall always assume that 
\begin{equation}
\Theta_X({\bf \Phi}_X)= -{\bf \Phi}_X.
\label{ThetaXPhiX}
\end{equation}
This assumption implies that  $\omega_X({\bf \Phi}_X)=0$ for all $X$. 

\bigskip
{\noindent\bf Notation.} For $\nu\in{\mathfrak S}$,  $\vartheta\in{\rm Aut}(\cO)$, 
${\bf A}=(A_1, \ldots, A_n)\in\cO^n$,  and $Y=(Y_1,\ldots,Y_n)\in\cc^n$
we shall use the shorthands
\begin{align*}
\nu({\bf A})&=(\nu(A_1), \cdots, \nu(A_n))\in\cc^n,\\[3mm]
\vartheta({\bf A})&=(\vartheta(A_1), \cdots \vartheta(A_n))\in\cO^n, \\[3mm]
\tau^t({\bf A})&={\bf A}_t=(\tau^t(A_1), \cdots, \tau^t(A_n))\in\cO^n,\\[1mm]
Y\cdot{\bf A}&=\sum_{j=1}^n Y_j A_j\in\cO.
\end{align*}

\bigskip
The relative Hamiltonian of $\omega_{Xt}$ w.r.t. $\omega_X$ is given by\index{Hamiltonian!relative}
\[
\ell_{\omega_{Xt}|\omega_X}=\int_0^t\tau_X^{-s}(\sigma_X)\,\d s
=X \cdot\int_0^t {\bf \Phi}_{X(-s)}\d s
=\sum_{j=1}^n X_j\int_0^t \tau_X^{-s}(\Phi_X^{(j)})\d s.
\]
We generalize the $p=\infty$ entropic pressure functional\bindex{entropic pressure!generalized}
$$
e_{\infty,t}(\alpha)=\log\tr\left(\e^{\log\omega_X+\alpha\ell_{\omega_{Xt}|\omega_X}}\right),
$$ 
by introducing\nindex{etXY}{$e_t(X,Y)$}{generalized entropic pressure functional}
\begin{equation}
e_t(X,Y)=\log\tr\left(\e^{\log \omega_X+Y\cdot\int_0^t {\bf \Phi}_{X(-s)}\d s}\right),
\label{def-gen-var}
\end{equation}
where $Y\in \rr^n$. The basic properties of $e_t(X,Y)$ are summarized in the next proposition. 

\bep\label{gen-var-sym}
\ben
\item
$$
e_t(X,Y)=\sup_{\nu\in\mathfrak S}\left[S(\nu|\omega_X)+Y\cdot\int_0^t\nu({\bf \Phi}_{X{(-s)}})\,\d s\right].
$$
\item The function $\rr^n\ni Y\mapsto e_t(X,Y)$ is convex and real analytic.
\item $e_{-t}(X,Y)=e_t(X,X-Y)$. 
\item
\begin{equation}
\partial_{Y_j}e_t(X,Y)\big|_{Y=0}=\int_0^t \omega_X(\Phi_{X(-s)}^{(j)})\d s,
\label{gk-3a}
\end{equation}
\begin{align}
\partial_{Y_k}\partial_{Y_j}e_t(X,Y)\big|_{Y=0}=\int_0^t\int_0^t&
\left(\langle\Phi^{(k)}_{X(-s_1)}|\Phi^{(j)}_{X(-s_2)}\rangle_{\omega_X}\right.
\nonumber\\
-&\left.\omega_X(\Phi^{(k)}_{X(-s_1)})\omega_X(\Phi^{(j)}_{X(-s_2)})\right)\d s_2\d s_1.
\label{gk-3}
\end{align}
\item If $(\cO_X,\tau_X^t,\omega_X)_{X\in \rr^n}$ is TRI, then $e_{-t}(X,Y)=e_t(X,Y)$ and
\begin{equation}
e_t(X,Y)=e_t(X,X-Y). 
\label{ges-symmetry}
\end{equation}
\een
\eep
We shall refer to Relation \eqref{ges-symmetry} as the finite time Generalized 
Evans-Searles (GES) symmetry. Notice that\bindex{symmetry!Evans-Searles!generalized}
\[
e_t(X,\alpha X)=\log \tr (\e^{\log \omega_X + \alpha \ell_{\omega_{Xt}|\omega_X}})
=e_{\infty,t}(\alpha),
\]
which shows that the ES-symmetry of $e_{\infty,t}(\alpha)=e_{\infty,t}(1-\alpha)$
is a special case of the GES-symmetry.

\medskip
\demo
(1) follows from Theorem \ref{RelEntVar}. (2) Convexity follows from (1) and
analyticity is obvious. (3) is a consequence of the following elementary calculation:
\begin{align*}
\log \omega_X+(X-Y)\cdot\int_0^t {\bf \Phi}_{X(-s)}\d s
&=\log \omega_X+\ell_{\omega_{Xt}|\omega_X}-Y\cdot\int_0^t {\bf \Phi}_{X(-s)}\d s\\
&=\log \omega_{Xt}-Y\cdot\int_0^t {\bf \Phi}_{X(-s)}\d s\\
&=\e^{-\i tH_X}\left(\log \omega_{X}-Y\cdot\int_0^t {\bf \Phi}_{X(t-s)}\d s\right)\e^{\i tH_X}\\
&=\e^{-\i tH_X}\left(\log \omega_{X}-Y\cdot\int_0^t {\bf \Phi}_{Xs}\d s\right)\e^{\i tH_X}\\
&=\e^{-\i tH_X}\left(\log \omega_{X}+Y\cdot\int_0^{-t} {\bf \Phi}_{X(-s)}\d s\right)\e^{\i tH_X}.
\end{align*}
To prove (4) invoke Duhamel formula to differentiate \eqref{def-gen-var} \index{formula!Duhamel}
(see the proof of Assertion (5) of Proposition \ref{ihes-sunday-1}). (5)  follows from (2) and 
Assumption \eqref{ThetaXPhiX} which implies that $\Theta_X({\bf\Phi}_{X(-s)})=-{\bf\Phi}_{Xs}$,
so that
\begin{align*}
\Theta_X\left(\log \omega_X+Y\cdot\int_0^t {\bf \Phi}_{X(-s)}\d s\right)&=
\log \omega_X-Y\cdot\int_0^t {\bf \Phi}_{Xs}\d s\\
&=\log \omega_X+Y\cdot\int_0^{-t} {\bf \Phi}_{X(-s)}\d s.
\end{align*}
\qed

\bigskip

\section{Finite time linear response theory}
\label{sect:FinLinearResponse}

Finite time linear response theory is concerned with the first order perturbation theory 
(w.r.t.  $X$) of the expectation values\index{linear response}
\[
\langle {\bf\Phi}_X\rangle_t=\frac{1}{t}\int_0^t \omega_X({\bf\Phi}_{Xs}) \d s.
\]
In the discussion of linear response theory we shall always assume that functions 
$$
X\mapsto H_X, \qquad X\mapsto \omega_X, \qquad X\mapsto {\bf\Phi}_X,
$$
are continuously differentiable. This implies that the function $X\mapsto\langle{\bf\Phi}_X\rangle_t$
is continuously  differentiable for all $t$.

The finite time kinetic transport coefficients are defined by\bindex{transport coefficients}
\[ L_{jkt} =\partial_{X_k}\langle \Phi_X^{(j)}\rangle_t\big|_{X=0}.\]
Since 
\begin{equation} \langle \sigma_X\rangle_t =\sum_j X_j\langle \Phi_X^{(j)}\rangle_t =
\sum_{j,k}L_{jkt} X_jX_k+ o(|X|^2)\geq 0,
\label{ent-kin}
\end{equation}
the {\em real} quadratic form determined by the finite time Onsager matrix $[L_{jkt}]$ 
is positive definite. This fact does not depend on the TRI assumption  and  {\em does not} imply 
that $L_{jkt}=L_{kjt}$.
We shall call the relations \index{relation!Onsager reciprocity}\bindex{Onsager matrix}
\[
L_{jkt}=L_{kjt},
\]
the finite time Onsager reciprocity relations (ORR). 
As a general structural relations, they can hold only for TRI systems.

Another direct consequence of \eqref{ent-kin} is:
\bep 
Let ${\bf\Phi}_X, {\bf\widetilde\Phi}_X$ be two flux relations. Then the corresponding 
finite time transport coefficients satisfy
\[ L_{jkt} + L_{kjt} = \widetilde L_{jkt} +\widetilde L_{kjt}.
\]
If the finite time ORR hold, then $L_{jkt}=\widetilde L_{jkt}$.
\eep

The next proposition shows that the finite time ORR and Green-Kubo formula follow from the finite 
time GES symmetry. Recall our notational convention $\tau^t_{X=0}=\tau^t$, $\omega_{X=0}=\omega$, 
$\Phi_{X=0}^{(j)}=\Phi^{(j)}$, etc. \bindex{formula!Green-Kubo}

\bep\label{fin-gk}
If $(\cO_X, \tau_X^t, \omega_X)_{X\in \rr^n}$ is TRI, then
\ben
\item 
\[
L_{jkt}= \frac{1}{2}\int_{-t}^t\langle \Phi^{(k)}|\tau^s(\Phi^{(j)})\rangle_\omega\left(1 -\frac{|s|}{t}\right)\d s,
\]
\item $L_{jkt}=L_{kjt}$.
\een
\eep

\demo By Relation \eqref{gk-3a} and the TRI property one has
$$
\langle{\Phi}_X^{(j)}\rangle_t=-\partial_{Y_j}\frac1t e_t(X,Y)\big|_{Y=0},
$$
so that
\[
L_{jkt}=-\partial_{X_k}\partial_{Y_j}\frac1te_t(X,Y)\big|_{X=Y=0}.
\]
The GES-symmetry implies that
$$
-\partial_{X_k}\partial_{Y_j}\frac1te_t(X,Y)\big|_{X=Y=0}
=\frac{1}{2t}\partial_{Y_kY_j}e_t(0,Y)\big|_{Y=0},
$$
(recall the derivation of \eqref{symlemma}). Since $\omega({\bf\Phi})=0$ and 
$\omega$ is $\tau^t$ invariant, Relation \eqref{gk-3} yields
$$
L_{jkt}=\frac1{2t}\int_0^t\int_0^t
\langle \Phi^{(k)}_{-s_1}|\Phi^{(j)}_{-s_2}\rangle_\omega\d s_1\d s_2
=\frac1{2t}\int_0^t\int_0^t
\langle \Phi^{(k)}|\Phi^{(j)}_{s_1-s_2}\rangle_\omega\d s_1\d s_2.
$$
A simple change of integration variable leads to (1).
(2) follows from the equality of the mixed partial derivatives 
$\partial_{Y_k}\partial_{Y_j}e_t(0,Y)=\partial_{Y_j}\partial_{Y_k}e_t(0,Y)$. 
\qed

\bigskip

\chapter{Open quantum systems}
\label{chap:OQS}

\section{Coupling to reservoirs}
\label{sect:Coupling}

Let  $\cR_j$, $j=1, \cdots, n$, be finite quantum systems with Hilbert spaces $\cK_j$. Each $\cR_j$ is 
described by a quantum dynamical system $(\cO_j,\tau_j^t,\omega_j)$. Besides the Hamiltonian
$H_j$ which generates $\tau_j$, we assume the existence of a ``conserved charge'' $N_j$,
a self-adjoint element of $\cO_j$ such  that $[H_j,N_j]=0$. \index{charge}
It follows that $N_j$ is invariant under the dynamics $\tau_j^t$ and that the gauge group\index{gauge group}
$\vartheta_j^t(A)=\e^{\i tN_j}A\e^{-\i tN_j}$ commutes with $\tau_j^t$.
We suppose that $\cR_j$ is in thermal equilibrium at inverse temperature $\beta_j$ and chemical 
potential $\mu_j$, \ie that\index{chemical potential}
\[
\omega_j =\frac{\e^{-\beta_j(H_j-\mu_jN_j)}}{\tr (\e^{-\beta_j(H_j-\mu_jN_j)})}.
\]
The modular group of this state is given by\index{modular!group}
$$
\varsigma_{\omega_j}^t=\tau_j^{-\beta_jt}\circ\vartheta_j^{\beta_j\mu_jt}.
$$
Thus, denoting by $\delta_j=\i[H_j,\,\cdot\,]$ the generator of $\tau_j^t$ and by 
$\xi_j=\i[N_j,\,\cdot\,]$ the generator of  $\vartheta_j^t$, one has 
$$
\delta_{\omega_j}=-\beta_j(\delta_j-\mu_j\xi_j).
$$
Note that in cases where there is no conserved charge, one may simply set $N_j=\one_{\cK_j}$ so that
the gauge group becomes trivial, $\xi_j=0$, and the states $\omega_j$ independent of the chemical
potential $\mu_j$. In such cases, one can simply set $\mu_j=0$.

The joint system $\cR=\cR_1+\cdots +\cR_n$ is described by 
$$
(\cO_\cR,\tau_\cR^t,\omega_\cR)=\bigotimes_{j=1}^n(\cO_j,\tau_j^t,\omega_j).
$$
The generators of the dynamics $\tau_\cR^t$, the gauge group 
$\vartheta_\cR^t=\otimes_{j=1}^n\vartheta_j^t$ and the modular group 
$\varsigma_{\omega_\cR}^t=\otimes_{j=1}^n\varsigma_{\omega_j}^t$ are given by
$$
\begin{array}{rccclcrcl}
\delta_\cR&=&\displaystyle\sum_{j=1}^n\delta_j&=&\i[H_\cR,\,\cdot\,],&
\qquad&H_\cR&=&\displaystyle\sum_{j=1}^n H_j,\\[14pt]
\xi_\cR&=&\displaystyle\sum_{j=1}^n\xi_j&=&\i[N_\cR,\,\cdot\,],&
\qquad&N_\cR&=&\displaystyle\sum_{j=1}^n N_j,\\[14pt]
\delta_{\omega_\cR}&=&\displaystyle\sum_{j=1}^n\delta_{\omega_j}&=&\i[\log\omega_\cR,\,\cdot\,],&
\qquad& \log\omega_\cR&=&-\displaystyle\sum_{j=1}^n\beta_j(H_j-\mu_jN_j),
\end{array}
$$
with the notational convention of Section \ref{sect:MultiSys}.

Let $\cS$ be a finite quantum system described by $(\cO_\cS,\tau_\cS^t,\omega_\cS)$,
the dynamics $\tau_\cS^t$ being generated by the Hamiltonian $H_\cS$. We assume the existence
of a conserved charge $N_\cS$ such that $\i[H_\cS,N_\cS]=0$ and denote $\vartheta_\cS^t$ the
corresponding gauge group on $\cO_\cS$.

A gauge invariant coupling of $\cS$ to the system of reservoirs $\cR$ is a collection of self-adjoint 
elements $V_j\in\cO_\cS\otimes\cO_j$ such that $[N_j+N_\cS,V_j]=0$.
Denoting $V=\sum_j V_j$, the Hamiltonian
\[
H_V=H_\cR+H_\cS+V,
\]
generates a perturbation $\tau_V^t$ of the dynamics\index{dynamics!perturbed}
$\tau^t=\tau_\cS^t\otimes\tau_\cR^t$ on $\cO=\cO_\cS\otimes\cO_\cR$. 
Moreover, $\tau_V^t$ preserves the total charge $N=N_\cR+N_\cS$ and hence commutes
with the gauge group $\vartheta^t=\vartheta_\cS^t\otimes\vartheta_\cR^t$.

The quantum dynamical system $(\cO,\tau_V^t,\omega)$, where $\omega=\omega_\cS\otimes\omega_\cR$,  is called {\em open quantum system}. Open quantum systems are examples of  compound systems considered in Sections 
\ref{sect:MultiSys}--\ref{sect:MultiFCS}. \bindex{open system}

The definition of open quantum system requires some minor modifications if the particle statistics 
(bosons/fermions) is taken into account. These modifications are straightforward 
(see  Section \ref{sect:EBBM} for an example) and for simplicity of exposition we shall not 
discuss them in abstract form.

The entropy production observable  of $(\cO, \tau_V^t, \omega)$ is \index{entropy!production}
\[
\sigma=\delta_\omega(H_V).
\]
Since
$$
\delta_\omega=\delta_{\omega_\cR}+\delta_{\omega_\cS}
=-\sum_j\beta_j(\delta_j-\mu_j\xi_j)-\i[Q,\,\cdot\,],
$$
where $Q=-\log\omega_\cS$, we have
\begin{equation}
\sigma=-\sum_{j}\beta_j(\Phi_j-\mu_j{\cal J}_j)+\sigma_\cS,
\label{SigmaExpro}
\end{equation}
where
$$
\Phi_j=\delta_j(V),\qquad{\cal J}_j=\xi_j(V),\qquad\sigma_\cS=\i[H_V,Q].
$$
Observing that
\begin{equation}
\Phi_j=-\i[H_V,H_j],\qquad{\cal J}_j=-\i[H_V,N_j],
\label{FluxFormuLa}
\end{equation}
we derive
\begin{equation}
H_{jt}-H_j=-\int_0^t \Phi_{js}\d s,\qquad
N_{jt}-N_j=-\int_0^t {\cal J}_{js}\d s.
\label{PhiTegral}
\end{equation}
The observables $\Phi_j$ and ${\cal J}_j$ describe the energy and charge fluxes
out of the $j$-th reservoir $\cR_j$. The observable $\beta_j(\Phi_j-\mu_j{\cal J}_j)$ describes
entropy flux out of $\cR_j$.\index{flux}

The entropy balance equation (more precisely Inequality \eqref{ent-ba-equ}) implies \index{entropy!balance}
\begin{equation}
\begin{split}
\rho_t(Q)-\rho(Q)
&\ge\sum_{j}\beta_j \int_0^t \rho_s(\Phi_{j}-\mu_j{\cal J}_{j})\d s\\
&=\sum_j\beta_j\left[(\rho(H_j)-\rho_t(H_j))-\mu_j(\rho(N_j)-\rho_t(N_j))\right],
\end{split}
\label{snow}
\end{equation}
for any state $\rho$ on $\cO$. We note in particular that if $\rho$ is a steady state for the
dynamics $\tau_V^t$ then both sides of this inequality vanish as long as the joint system remains
finite. However, if the reservoirs become infinitely extended while the system $\cS$ remains
confined then the observable $Q$ remains well defined while $H_j$ and $N_j$
loose their meaning. A very important   feature of the   proper  mathematical formulation of \eqref{snow} in the  
thermodynamic limit is that the  left hand side still vanishes while the right hand side is typically non-zero.

Note also that 
\begin{equation}
\omega_t =Z^{-1}\e^{-Q_{-t}-\sum_j\beta_j[(H_j-\mu_j N_j)
+\int_0^t(\Phi_{j{(-s)}}-\mu_j{\cal J}_{j(-s)})\d s]},
\label{ihes-mac-zub}
\end{equation}
where 
\[
Z=\tr(\e^{-\sum_j\beta_j(H_j-\mu_jN_j)}).
\]
The density matrix $\omega_t$ expressed in the form  \eqref{ihes-mac-zub} is known as
McLennan-Zubarev dynamical ensemble.\index{McLennan-Zubarev ensemble}

\section{Full counting statistics}
\label{sect:FCSOpen}

We continue with the framework of the previous subsection and adapt our discussion of full counting 
statistics from Section \ref{sect:MultiFCS} to the open quantum system $(\cO,\tau_V^t,\omega)$.
We note that the reference state $\omega$ factorizes into a product of commuting self-adjoint
operators\index{full counting statistics}
$$
\omega=Z^{-1}\e^{-Q-\sum_{j=1}^n\beta_jH_j+\sum_{j=1}^n\beta_j\mu_jN_j}=
Z^{-1}\e^{-Q}\left(\prod_{j=1}^n\e^{-\beta_jH_j}\right)\left(\prod_{j=1}^n\e^{\beta_j\mu_jN_j}\right).
$$
Defining, according to Exercise \ref{Exo:AbelianOmega},
$$
\omega^\balpha=Z^{-\gamma_0}\e^{-\gamma_0 Q}
\left(\prod_{j=1}^n\e^{-\gamma_j\beta_jH_j}\right)\left(\prod_{j=1}^n\e^{\gamma'_j\beta_j\mu_jN_j}\right),
$$
for $\balpha=(\gamma_0,{\boldsymbol\gamma},{\boldsymbol\gamma'})\in\rr\times\rr^n\times\rr^n$
we have, 
\begin{equation}
\tr(\omega_t^{{\bf1}-\balpha}\omega^{\balpha})
=\sum_{q,{\boldsymbol\varepsilon},{\boldsymbol\nu}} 
\e^{-t(\gamma_0 q+{\boldsymbol\gamma}\cdot{\boldsymbol\varepsilon}+{\boldsymbol\gamma'}\cdot{\boldsymbol\nu})}
{\mathbb P}_t(q,{\boldsymbol\varepsilon},{\boldsymbol\nu}),
\label{OpenGenFct}
\end{equation}
where ${\mathbb P}_t(q,{\boldsymbol\varepsilon},{\boldsymbol\nu})$ is the joint probability
distribution for the mean rates of change of the commuting set of observables
$$
{\bf S}=(Q, \beta_1H_1,\ldots,\beta_nH_n,-\beta_1\mu_1N_1,\ldots,-\beta_n\mu_nN_n),
$$
between two successive joint measurements at time $0$ and $t$. The sum in \eqref{OpenGenFct}
extends over all $(q,{\boldsymbol\varepsilon},{\boldsymbol\nu})\in(\sp({\bf S})-\sp({\bf S}))/t$.
As shown in Section \ref{sect:MultiFCS}, the distribution ${\mathbb P}_t$ coincide with the joint 
spectral measure of a family of commuting relative modular operators.

Expectation and covariance of (${\boldsymbol\varepsilon},{\boldsymbol\nu})$ w.r.t. ${\mathbb P}_t$
are given by
\begin{align}
{\mathbb E}_t(\varepsilon_j)&=-\frac{\beta_j}t\int_0^t\omega_s(\Phi_{j})\d s,\nonumber\\[-6pt]
&\label{FCSexp}\\[-6pt]
{\mathbb E}_t(\nu_j)&=\frac{\beta_j\mu_j}t\int_0^t\omega_s({\cal J}_{j})\d s,\nonumber
\end{align}
and,
\begin{align}
{\mathbb E}_t(\varepsilon_j\varepsilon_k)-{\mathbb E}_t(\varepsilon_j){\mathbb E}_t(\varepsilon_k)
&=\frac{\beta_j\beta_k}{t^2}\int_{0}^t\int_0^t
\omega\left((\Phi_{js}-\omega(\Phi_{js}))(\Phi_{ku}-\omega(\Phi_{ku}))\right)\d s \d u,\nonumber\\[12pt]
{\mathbb E}_t(\nu_j\nu_k)-{\mathbb E}_t(\nu_j){\mathbb E}_t(\nu_k)
&=\frac{\beta_j\mu_j\beta_k\mu_k}{t^2}\int_{0}^t\int_0^t
\omega\left(({\cal J}_{js}-\omega({\cal J}_{js}))({\cal J}_{ku}-\omega({\cal J}_{ku}))\right)\d s \d u,
\label{FCSvar}\\[12pt]
{\mathbb E}_t(\varepsilon_j\nu_k)-{\mathbb E}_t(\varepsilon_j){\mathbb E}_t(\nu_k)
&=-\frac{\beta_j\beta_k\mu_k}{t^2}\int_{0}^t\int_0^t
\omega\left((\Phi_{js}-\omega(\Phi_{js}))({\cal J}_{ku}-\omega({\cal J}_{ku}))\right)\d s \d u.\nonumber
\end{align}
In terms of Liouvillean, the moment generating function \eqref{OpenGenFct} reads\index{Liouvillean!$L^p$}
\begin{equation}
\tr(\omega_t^{{\bf1}-\balpha}\omega^{\balpha})
=(\xi_\omega|\e^{\i t L_{\frac{1}\balpha}}\xi_\omega), 
\label{ihes-liou0}
\end{equation}
with (as derived in Exercise \ref{Exo:LalphaLiouvPerturb})
\begin{equation}
L_{\frac{1}{\balpha}}
=K_0 +L(V)-R(W_\balpha),
\label{Loneoveralphadef}
\end{equation}
where $K_0$ denotes the standard Liouvillean of the decoupled dynamics $\tau^t$,\index{Liouvillean!standard}
$$
W_\balpha=\varsigma_\omega^{\i({\boldsymbol\alpha}-{\bf1}/2)}(V)
=\sum_{j=1}^n T_j(\balpha)V_jT_j(\balpha)^{-1},
$$
and
$$
T_j(\balpha)=
\e^{-(1/2-\gamma_0)Q-\beta_j[(1/2-\gamma_j)H_j-\mu_j(1/2-\gamma'_j)N_j]}.
$$
If $(\cO,\tau^t_V,\omega)$ is TRI, then  the fluctuation relation
$$
\frac{{\mathbb P}_t(-q,-{\boldsymbol\varepsilon},-{\boldsymbol\nu})}
{{\mathbb P}_t(q,{\boldsymbol\varepsilon},{\boldsymbol\nu})}=
\e^{-t(q+{\bf 1}\cdot{\boldsymbol\varepsilon}+{\bf 1}\cdot{\boldsymbol\nu})},
$$
holds.

\section{Linear response theory}
\label{sect:LinearResponse}

We continue our discussion of open quantum systems. We now adopt the point of view of
Section \ref{sect:ControlDynSys} and describe finite time linear response theory. \index{linear response}
Let $\beta_\eq$  and $\mu_\eq$ be given equilibrium values of the inverse temperature and 
chemical potential. The thermodynamical forces $X=(X_1, \cdots, X_{2n})$ are
\[
X_j =\beta_{\eq} -\beta_j, \quad X_{n+j}=-\beta_\eq\mu_\eq+\beta_j\mu_j,\quad(j=1,\ldots,n).
\]
The reference state of the system is taken to be
\[
\omega_X=Z_X^{-1}\e^{-\beta_\eq(H_V-\mu_\eq N) +\sum_{j=1}^n (X_j H_j + X_{n+j}N_j)},
\]
where $N=N_\cR+N_\cS$ and 
$Z_X=\tr (\e^{-\beta_\eq(H_V-\mu_\eq N) +\sum_{j=1}^n (X_j H_j + X_{n+j}N_j)})$. Clearly,
\[
\omega_{0}=Z_0^{-1}\e^{-\beta_\eq(H_V-\mu_\eq N)},
\]
is the thermal equilibrium state of $(\cO,\tau_V^t)$ at inverse temperature $\beta_\eq$ and chemical 
potential $\mu_\eq$. Hence, we shall use the notation $\omega_0=\omega_\eq$. The dynamical
system $(\cO,\tau_V^t,\omega_X)$ fits into the framework of Section \ref{sect:ControlDynSys}
(with $\tau_X^t=\tau_V^t$ independent of $X$).

Note that the family of states $\omega_X$ is distinct from the one used in the previous section: it 
contains the coupling $V$. In particular, $\omega_X$  is not a product state. This is however in complete parallel 
with our  discussion of linear response theory in classical harmonic chain. If the perturbation $V$ 
remains local in the thermodynamic limit, the product state $\omega$ and the state $\omega_X$ 
describe the same thermodynamics. We shall discuss this issue  in more details in 
Section \ref{Sect:ChoiceofOmega}.

The entropy production observable of the dynamical system $(\cO,\tau_V^t,\omega_X)$ is
\begin{equation}
\sigma_X=\i[\log\omega_X,H_V]=\sum_{j=1}^n X_j\Phi_j+X_{n+j}{\mathcal J}_j,
\label{openfluxrel}
\end{equation}
where the observables \index{entropy!production}\index{flux}
$$
\Phi_j=-\i [H_V,H_j],\qquad{\mathcal J}_j=-\i [H_V,N_j],
$$ 
describe the energy  and charge flux out of the $j$-th reservoir. Clearly, \eqref{openfluxrel}
is a natural (and $X$-independent) flux relation. $\Phi_j$ is the flux associated to the thermodynamical
force $\beta_\eq-\beta_j$ and ${\mathcal J}_j$ is the flux associated to the thermodynamical force 
$-\beta_\eq\mu_\eq+\beta_j\mu_j$.  

The generalized entropic pressure is given by \index{entropic pressure!generalized}
\[
e_t(X,Y)=\log\tr\left(\e^{\log\omega_X+\sum_{j=1}^n(Y_j\int_0^t\Phi_{j(-s)}\d s
+Y_{n+k}\int_0^t{\mathcal J}_{j(-s)}\d s)}\right).
\]
Recall that the equilibrium canonical correlation is
\[
\langle A|B\rangle_{\eq}=\int_0^1 \omega_{\eq}(A^\ast\tau_V^{\i \beta s}(B))\d s.
\]
Proposition \ref{fin-gk} implies the finite Green-Kubo formulas and finite time Onsager reciprocity relations for 
energy and charge fluxes. \index{formula!Green-Kubo}\index{relation!Onsager reciprocity}
\bep\label{ihes-gk-fin}
Suppose that $(\cO,\tau_V^t,\omega_\eq)$ is TRI with time reversal $\Theta$ satisfying
$\Theta(V_j)=V_j$, $\Theta(H_j)=H_j$ and $\Theta(N_j)=N_j$ for all $j$. Then 
\begin{equation}
\begin{split}
L_{jkt}^{\rh\rh}&=\partial_{X_{k\phantom{+k}}}\left.\left(\frac{1}{t}\int_0^t\omega_X(\Phi_{js})\d s\right)\right|_{X=0}
=\frac{1}{2}\int_{-t}^t\langle \Phi_k|\Phi_{js}\rangle_\eq\left(1-\frac{|s|}{t}\right) \d s,   \\[3mm]
L_{jkt}^{\rh\rc}&=\partial_{X_{n+k}}\left.\left(\frac{1}{t}\int_0^t\omega_X(\Phi_{js})\d s\right)\right|_{X=0}
=\frac{1}{2}\int_{-t}^t\langle {\mathcal J}_k|\Phi_{js}\rangle_\eq\left(1-\frac{|s|}{t}\right) \d s, \\[3mm]
L_{jkt}^{\rc\rh}&=\partial_{X_{k\phantom{+k}}}\left.\left(\frac{1}{t}\int_0^t\omega_X({\mathcal J}_{js})\d s\right)\right|_{X=0}
=\frac{1}{2}\int_{-t}^t\langle \Phi_k|{\mathcal J}_{js}\rangle_\eq \left(1-\frac{|s|}{t}\right)\d s, \\[3mm]
L_{jkt}^{\rc\rc}
&=\partial_{X_{n+k}}\left.\left(\frac{1}{t}\int_0^t\omega_X({\mathcal J}_{js})\d s\right)\right|_{X=0}
=\frac{1}{2}\int_{-t}^t\langle{\mathcal J}_k|{\mathcal J}_{js}\rangle_\eq \left(1-\frac{|s|}{t}\right)\d s,
\end{split}
\label{ihes-gk-tr}
\end{equation}
(the indices $\rh/\rc$ stand for energy/charge) and 
\begin{align*}
L_{jkt}^{\rh\rh}&=L_{kjt}^{\rh\rh},\\[3mm]
L_{jkt}^{\rc\rc}&=L_{kjt}^{\rc\rc},\\[3mm]
L_{jkt}^{\rh\rc}&=L_{kjt}^{\rc\rh}.
\end{align*}
\eep
The special structure of open quantum systems allows for a further insight into linear response theory. 
Consider the auxiliary Hamiltonian 
\[
H_X =H_V- \mu_\eq N-\frac{1}{\beta_\eq}\sum_{j=1}^n (X_j H_j + X_{n+j}N_j),
\]
and note that 
\[
\omega_X=\frac{1}{Z_X}\e^{-\beta_{\eq}H_X},
\]
where $Z_X=\tr(\e^{-\beta_\eq H_X})$. Hence, $\omega_X$ is the $\beta_\eq$-KMS state of the dynamics 
$\tau_X^t$ generated by the Hamiltonian $H_X$.
By Equ. \eqref{PhiTegral} one has
$$
\omega_{Xt}=\e^{-\i tH_V}\omega_X\e^{\i tH_V}
=\frac{1}{Z_X}\e^{-\beta_\eq(H_X+P_t)},
$$
where
$$
P_t=-\frac1{\beta_\eq}\sum_j\left(X_j\int_0^t \Phi_{j(-s)}\d s + X_{n+j}\int_0^t 
{\mathcal J}_{j(-s)}\d s\right).
$$
We conclude that $\omega_{Xt}$ is the KMS state at inverse temperature $\beta_\eq$ of the
perturbed dynamics generated by $H_X+P_t$. Moreover, the perturbation satisfies
$P_t=O(X)$ as $X\to0$. Applying the perturbation expansion \eqref{rhobetaVexpand}
and the formula for the coefficient $b_1(A)$ derived in Exercise \ref{Exo:bncoeffs}, we obtain
$$
\omega_{Xt}(A)=\omega_X(A)
-\beta_\eq\int_0^1\omega_X\left(P_t(\tau_X^{\i s\beta_\eq}(A)-\omega_X(A))\right)\,\d s+O(|X|^2).
$$
Since $\omega_X=\omega_\eq+O(X)$ and $P_t=O(X)$, one has
\begin{align*}
\omega_X\left(P_t(\tau_X^{\i s\beta_\eq}(A)-\omega_X(A))\right)&=
\omega_\eq\left(P_t(\tau_X^{\i s\beta_\eq}(A)-\omega_\eq(A))\right)+O(|X|^2)\\
&=\omega_\eq\left(P_t\tau_X^{\i s\beta_\eq}(A)\right)-\omega_\eq(P_t)\omega_\eq(A)+O(|X|^2).
\end{align*}
From the fact that $\omega_\eq(\Phi_{js})=\omega_\eq(\Phi_j)=0$ and
$\omega_\eq({\cal J}_{js})=\omega_\eq({\cal J}_j)=0$ we deduce $\omega_\eq(P_t)=0$.
Since
$$
\tau_X^{\i s\beta_\eq}(A)=\e^{-s\beta_\eq(H_V-\mu_\eq N)}A\e^{s\beta_\eq(H_V-\mu_\eq N)}+O(X),
$$
and $[P_t,N]=0$, we can further write,
\begin{equation}
\omega_{Xt}(A)=\omega_X(A)
-\beta_\eq\int_0^1\omega_\eq\left(P_t\tau_V^{\i s\beta_\eq}(A)\right)\,\d s+O(|X|^2).
\label{LinRespInter}
\end{equation}
By Duhamel's formula one has
$$
\partial_{X_k}\e^{-\beta_\eq H_X}|_{X=0}=\int_0^{\beta_\eq}\e^{-s(H_V-\mu_\eq N)}
\left.\frac{\partial H_X}{\partial X_k}\right|_{X=0}\e^{-(\beta_\eq-s)(H_V-\mu_\eq N)}\,\d s,
$$
from which one easily derives
$$
\partial_{X_k}\omega_X(A)|_{X=0}=\left\{
\begin{array}{ll}
\langle H_k|A-\omega_\eq(A)\rangle_\eq &\text{for }1\le k\le n,\\[4pt]
\langle N_k|A-\omega_\eq(A)\rangle_\eq &\text{for }n+1\le k\le 2n.
\end{array}
\right.
$$
Finally, \eqref{LinRespInter} yields that for $1\leq k \leq n$, 
\begin{equation}
\begin{split}
\partial_{X_k}\omega_X(A_t)|_{X=0}&=\langle H_k|A-\omega_\eq(A)\rangle_\eq 
+ \int_0^t\langle { \Phi}_{k}|A_s\rangle_\eq\d s, \\[3mm]
\partial_{X_{n+k}}\omega_X(A_t)|_{X=0}&=\langle N_k|A-\omega_\eq(A)\rangle_\eq 
+ \int_0^t\langle {\mathcal J}_{k}|A_s\rangle_\eq\d s.
\end{split}
\label{ihes-gk-gener}
\end{equation}
These linear response formulas hold without time reversal assumption and for any observable $A\in\cO$.
Under the assumptions of Proposition \ref{ihes-gk-fin}, $\omega_X$ is TRI. If $A=\Phi_j$ or 
$A={\mathcal J}_j$ then $\omega_X(A)=0$. This implies $\partial_{X_k}\omega_X(A)|_{X=0}=0$ 
for $k=1,\ldots,2n$, and \eqref{ihes-gk-gener} reduces to the Green-Kubo formulas \eqref{ihes-gk-tr}. 
Using \eqref{ihes-gk-gener} it is easy to exhibit examples of open quantum systems for which finite 
time Onsager reciprocity relations fail in the absence of time reversal.

\chapter{The thermodynamic limit\\ and the large time limit}
\label{chap:TD-limit}

Apart from Section \ref{sect:TDoverview} and the first part of Section 
\ref{sect:LThypo} which should be accessible to all readers, 
this section is intended for more advanced readers
and may be skipped on first reading.

We shall describe, typically without proofs, the thermodynamic limit procedure and how one extends 
the results of the last two sections to general quantum systems. We shall also discuss the large time 
limit for infinitely extended quantum system.

\section{Overview}
\label{sect:TDoverview}

From a mathematical point of view, the dynamics of a finite quantum system $(\cO,\tau^t,\omega)$
and that of the finite classical harmonic chain of Chapter \ref{chap:HarmoChain} are very similar: both 
are described by a linear quasi-periodic propagator. In particular, the limit
$$
\lim_{t\to\infty}\omega(\tau^t(A)),
$$
does not exist, except in trivial cases. However, the Ces\`aro limit
\begin{equation}
\omega_+(A)=\lim_{T\to\infty}\frac1T\int_0^T\omega(\tau^t(A))\,\d t,
\label{BeforeNESS}
\end{equation}
exists for all $A\in\cO$ and defines a steady state $\omega_+$ of the system.

\medskip
\begin{exo}
\label{Exo:FiniteNESS}
\exop Show that for a finite quantum system $(\cO,\tau^t,\omega)$ with Hamiltonian $H$, the limit
\eqref{BeforeNESS} exists and that the limiting state $\omega_+$ is described by the density matrix
$$
\omega_+=\sum_{\lambda\in\sp(H)}P_\lambda(H)\omega P_\lambda(H).
$$

\exop For $A\in\Os$, set  $\Phi_A=\i[H,A]$. Show that
$$
\omega_+(\Phi_A)=0,
$$
for any $A$.  Conclude that, in particular, the mean entropy production rate vanishes,
$$
\omega_+(\sigma)=\lim_{t\to\infty}\omega(\Sigma^t)=0.
$$

\exop Show that the same conclusions hold if the system is infinite (\ie the Hilbert
space $\cK$ is infinite dimensional) but confined in the sense that its Hamiltonian $H$ has
purely discrete spectrum.
\end{exo}

\medskip
Thus, in order to obtain a thermodynamically 
non-trivial steady state -- with non-vanishing currents and strictly positive entropy production rate --
we need to perform a thermodynamic (TD) limit before taking the large time limit \eqref{BeforeNESS}.
In other words, some parts of the system, \eg the reservoirs of an open system, have to be 
infinitely extended.\bindex{thermodynamic limit}\myindex{TD limit|see{thermodynamic limit}}

There are two difficulties associated with the TD limit: the first one is to describe the reference state
of the extended system, the second one is to define its dynamics. These problems have been
extensively studied in the 70' and have led to the algebraic approach to quantum statistical mechanics 
and quantum field theory. Algebraic quantum statistical mechanics provides a very attractive 
mathematical framework for the description of infinitely extended quantum systems.

In algebraic quantum statistical mechanics an  extended system is described by a triple  
$(\cO, \tau^t, \omega)$,
where $\cO$ is a $C^\ast$-algebra  \index{algebra!$C^\ast$-} with identity $\one$ (recall 
Exercise \ref{exo:staralg}), $\omega$ is a state (\ie positive normalized linear functional on 
$\cO$) and $\tau^t$ is a $C^\ast$-dynamics, that is, a norm continuous  group 
of $\ast$-automorphisms of $\cO$. The triple $(\cO, \tau^t, \omega)$ is often called 
{\sl quantum dynamical system}\footnote{Such quantum dynamical systems are suitable for 
the description of spin systems or fermionic systems. In the case of bosonic system,  $\cO$ is a 
$W^\ast$-algebra, $\omega$ is a normal state, and $\tau^t$ is weakly continuous. We shall not 
discuss such  systems in these lecture notes (see, \eg \cite{Pi}).}. The observables
are elements of $\cO$,  $\omega$ describes the initial thermodynamical state 
of our system and the  group $\tau^t$ describes its time evolution. The observables evolve in time as 
$A_t=\tau^t(A)$ and the states as $\omega_t=\omega\circ\tau^t$.

Infinitely extended systems of physical interest arise as TD limit of finite 
dimensional systems. There is a number of different ways the TD limit can be realized 
in practice. In the next section we describe one of them that is suitable for spin systems and
quasi-free or locally interacting fermionic systems.

\section{Thermodynamic limit: Setup}
\label{sect:TDsetup}

One starts with a family $\{\cQ_M\}_{M\in\nn}$ of  finite quantum systems described 
by a sequence of finite dimensional Hilbert spaces $\cK_M$, algebras $\cO_{\cK_M}$, Hamiltonians 
$H_M$ and faithful states $\omega_M$. $\sigma_M$ is the entropy production observable of $\cQ_M$.
In the presence of control parameters $X\in \rr^n$ ($H_{M,X}$ and $\omega_{M,X}$ depend on $X$), 
${\boldsymbol\Phi}_{M,X}$ denotes a chosen flux relation. The number $M$ typically corresponds 
to the ``size" of $\cQ_M$. For example, $\cQ_M$ could be a spin system or Fermi gas confined to 
a box $[-M, M]^d$ of the lattice $\zz^d$.\footnote{For continuous models one may need to slightly 
modify this setup. For example, in the case of a free Fermi gas on $\rr$, $M=(L, {\cal E})$, where 
$L$ is the spatial cut-off, ${\cal E}$ is the energy cut-off, and $M\rightarrow \infty$ stands for 
the ordered limit $\lim_{L\rightarrow \infty}\lim_{{\cal E}\rightarrow\infty}$, see Exercise 
\ref{Exo:FreeGas}. The extension of our axiomatic scheme to this more general setup is 
straightforward.}  The limiting infinitely extended system is described by a
quantum dynamical system $(\cO, \tau^t, \omega)$ satisfying  the following:

\begin{enumerate}[{\bf ({A}1)}]
\item For all $M$ there is a faithful representation $\pi_M:\cO_{\cK_M}\rightarrow\cO$ such
that \index{representation!faithful}
\[
\pi_M(\cO_{\cK_M})\subset \pi_{M+1}(\cO_{\cK_{M+1}}).
\]
\item $\cO_{\rm loc}=\cup_{M}\pi_M(\cO_{\cK_M})$ is dense in $\cO$. The elements of 
$\cO_{\rm loc}$ are sometimes called {\sl local observables} of $\cO$.\bindex{local observables}
\item For $ A\in\cO_{\rm loc}$, 
$\lim_{M\rightarrow \infty}\omega_M \circ\pi_M^{-1}(A)=\omega(A)$ and 
\[
\lim_{M\rightarrow \infty}\pi_M \circ \tau_{M}^t\circ \pi_{M}^{-1}(A)=\tau^t(A),
\]
where the convergence is uniform for $t$ in compact intervals of $\rr$.

\item $\lim_{M\rightarrow \infty}\pi _M(\sigma_M)=\sigma$, exists in the norm of $\cO$. 
$\sigma$ is the entropy production observable of $(\cO, \tau^t, \omega)$.\index{entropy!production}
 
\item In the presence of control parameters $X$,
$\lim_{M\rightarrow \infty}\pi_{M}({\boldsymbol\Phi}_{M,X})={\boldsymbol\Phi}_X$ exists in the 
norm of $\cO$. ${\boldsymbol\Phi}_X$ is a flux relation of $(\cO, \tau_X^t, \omega_X)$,
\[
\sigma_X=\sum_{j=1}^n X_j\Phi_X^{(j)}.
\]
\item For $p \in [1, \infty]$ and  $\alpha,t\in\rr$ the limit \index{entropic pressure} 
\[
e_{t, p}(\alpha)=\lim_{M\rightarrow \infty} e_{M, t, p}(\alpha), 
\]
exists and is finite. In the presence of control parameters, the limit \index{entropic pressure!generalized} 
\[ 
e_{t}(X, Y)=\lim_{M\rightarrow \infty} e_{M,t}(X, Y),
\]
exists and is finitefor all $t\in\rr$ and $X,Y\in\rr^n$.
\end{enumerate}

The verification of (A1)--(A5) in the context of spin systems and Fermi gases is discussed in virtually 
any mathematically oriented monograph on statistical mechanics (see, \eg \cite{BR2}).
For such systems, the proof of (A6) is typically an easy exercise in the techniques developed 
in 70's  (see Exercise \ref{Exo:EBBTDL} below). In some models $e_{t, p}(\alpha)/e_t(X,Y)$  
may be defined/finite only for a restricted range of the parameter $\alpha/(X,Y)$ and in this case the 
fluctuation theorems need to be suitable modified (this was the case in our introductory example 
of a thermally driven harmonic chain!). 

In what follows we assume that (A1)--(A6) hold.  For reasons of space and notational simplicity  
{\sl we shall assume from the onset that all quantum systems ${\cal Q}_M$ are} TRI. Also, we 
shall discuss only  the TD/large time limit of the functionals $e_{M, 2, t}(\alpha)$ and $e_{M, t}(X, Y)$.

\section{Thermodynamic limit: Full counting statistics}
\label{sect:TDLFCS}
\index{full counting statistics}

The reader should recall the notation and results of  Section \ref{sect:FCS} where we 
introduced  full counting statistics. We have 
\[
e_{M, 2, t}(\alpha)=e_{M, 2, t}(1-\alpha)=\log \int_\rr \e^{-t \alpha \phi}\d{\mathbb P}_{M,t}(\phi),
\] 
where ${\mathbb P}_{M,t}$ is the probability distribution of the mean rate of entropy change 
associated to the repeated measurement process described in Section \ref{sect:FCS}.  

By (A6), 
$$
e_{2, t}(\alpha)=\lim_{M\rightarrow \infty}e_{M, 2, t}(\alpha), 
$$
exists for all $t$ and $\alpha$. The implications are: 
\bep\label{paris-implications}
\ben 
\item The sequence of Borel probability measures $\{{\mathbb P}_{M,t}\}$ converges 
weakly to a Borel probability measure 
${\mathbb P}_t$, {\sl i.e.}, for any bounded continuous function $f:\rr\rightarrow \rr$, 
\[
\lim_{M\rightarrow \infty}\int_\rr f \d{\mathbb P}_{M,t}=\int_\rr f \d{\mathbb P}_t.
\]
\item For all $\alpha\in \rr$, 
\[
e_{2, t}(\alpha)=\log \int_{\rr}\e^{-t \alpha \phi}\d{\mathbb P}_t(\phi). 
\]
\item $e_{2, t}(\alpha)$ is real-analytic and
\begin{equation}
e_{2, t}(\alpha)=e_{2, t}(1-\alpha).
\label{fri-es}
\end{equation}
\item All the cumulants of ${\mathbb P}_{M,t}$ converge to corresponding cumulants of 
${ \mathbb P}_t$. 
In particular, 
\[
\partial_\alpha e_{2,t}(\alpha)|_{\alpha=0}=-\int_0^t\omega(\sigma_s)\d s\leq 0.
\]
\item Let ${\mathfrak r}:\rr\rightarrow  \rr$ be the reflection ${\mathfrak r}(\phi)=-\phi$ and 
$\bar {\mathbb P}_t={\mathbb P}_t \circ {\mathfrak r}$ the reflected measure. The measures 
$\bar{\mathbb P}_t$  and ${\mathbb P}_t$ are equivalent and 
\begin{equation}
\frac{\d \bar{\mathbb P}_t(\phi)}{\d {\mathbb P}_t(\phi)}=\e^{-t \phi}.
\label{fri-es1}
\end{equation}
\een
\eep

The limiting probability measure ${\mathbb P}_t$ is called full counting statistics of the infinitely 
extended system $(\cO,\tau^t,\omega)$. Relations \eqref{fri-es} and \eqref{fri-es1} are finite time  
Evans-Searles symmetries.\index{symmetry!Evans-Searles}

Recall that ${\mathbb P}_{M,t}$ is related to the modular structure of $\cQ_M$:
${\mathbb P}_{M,t}=Q_M^{t}$, where $Q_{M}^t$ is   the spectral measure for 
\[
-\frac{1}{t}\log \Delta_{\omega_{M,t}|\omega_{M}},
\]
and the vector $\xi_{\omega_M}$.
Our next goal is to relate ${\mathbb P}_t$ to the modular structure of the infinitely extended systems 
$(\cO, \tau^t, \omega)$. We start with a brief  description of this  structure assuming that the 
reader is  familiar with the topic.\index{representation!GNS}

\medskip\noindent(1) Let $(\cH_\omega,\pi_{\omega},\xi_\omega)$ be the GNS-representation of 
$\cO$ associated  to $\omega$.  ${\mathfrak M}_\omega =\pi_{\omega}(\cO)^{\prime\prime}$ denotes 
the {\sl enveloping von Neumann algebra.} A vector $\xi\in\cH_\omega$ is called cyclic if 
${\mathfrak M}_\omega\xi$ is dense in $\cH_\omega$ and separating if $A\xi=0$ for 
$A\in {\mathfrak M}_\omega$ implies $A=0$. $\xi_\omega$ is automatically cyclic. The state $\omega$
is called {\sl modular} if $\xi_\omega$ is also separating. We assume $\omega$ to be modular.
\bindex{algebra!enveloping von Neumann}\index{vector!cyclic}\index{vector!separating}
\index{state!modular}
\bindex{modular!state}\index{modular!operator}\index{modular!group}\index{modular!conjugation}
\nindex{Mo}{${\mathfrak M}_\omega$}{enveloping von Neumann algebra of $\omega$}

\medskip\noindent(2) The  anti-linear operator $S_\omega:A \xi_\omega\mapsto A^\ast\xi_\omega$ 
with domain ${\mathfrak M}_\omega\xi_\omega$ is closable. We denote by the same letter its 
closure. Let $S_\omega =J\Delta_{\omega}^{1/2}$ be the polar decomposition of $S_\omega$.
$J$ is  the {\sl modular conjugation,} an anti-unitary involution on $\cH_\omega$, and 
$\Delta_{\omega}$ is the {\sl modular operator} of $\omega$. $\Delta_\omega$ has a trivial kernel
and $\varsigma_\omega^t(A)=\Delta_{\omega}^{\i t}A\Delta_\omega^{-\i t}$ 
is a group of $\ast$-automorphism of ${\mathfrak M}_\omega$, the {\sl modular group} of $\omega$.
\index{cone!natural}\bindex{state!normal}\nindex{No}{${\cal N}_\omega$}{set of $\omega$-normal states}

\medskip\noindent(3) The set $\cH_+=\{ AJA\xi_\omega\,|\,A\in {\mathfrak M}_\omega\}^{\rm cl}$
(cl denotes the closure in $\cH_\omega$) is the {\sl natural cone.} It is a self-dual cone in
$\cH_\omega$. A state $\nu$ on $\cO$ is called normal (or, more precisely, $\omega$-normal) if 
there exists a density matrix $\rho$ on $\cH_\omega$ such that 
$\nu(A)=\tr (\rho \pi_{\omega}(A))$. ${\cal N}_\omega$ denotes the 
collection of all $\omega$-normal states. ${\cal N}_\omega$ is a norm closed subset of the dual 
$\cO^\ast$.  Any state $\nu\in {\cal N}_\omega$ has a unique {\sl vector representative}
$\xi_\nu\in\cH_+$ such that $\nu(A)=(\xi_\nu| \pi_\omega(A)\xi_\nu)$. 
$\xi_\nu$ is cyclic iff it is separating, \ie iff $\nu$ is modular.\index{vector!representative of a state}
\index{vector!cyclic}\index{vector!separating}

\medskip\noindent(4) Let $\nu\in{\cal N}_\omega$ be a modular state. The anti-linear operator 
$S_{\nu|\omega}:A\xi_\omega\mapsto A^\ast \xi_{\nu}$ is closable 
on ${\mathfrak M}_\omega\xi_\omega$ and we denote by the same letter its closure. 
This operator has the polar decomposition $S_{\nu|\omega}=J\Delta_{\nu|\omega}^{1/2}$, where 
$J$ is the modular conjugation introduced in (2) and $\Delta_{\nu|\omega}>0$
is the relative modular operator of $\nu$ w.r.t. $\omega$.\index{modular!operator!relative}

\medskip\noindent(5) The R\'enyi relative entropy of order $\alpha \in \rr$ of a state $\nu$ w.r.t. 
$\omega$ is defined by \index{entropy!R\'enyi}
\[
S_\alpha(\nu|\omega)=\left\{
\begin{array}{ll}
\log(\xi_\omega|\Delta_{\nu|\omega}^\alpha\xi_\omega)&\text{ if  }\nu\in{\cal N}_\omega,\\[5pt]
-\infty&\text{ otherwise}.
\end{array}
\right.
\]
Its relative entropy w.r.t. $\omega$ is defined by\index{entropy!relative}
\[ 
S(\nu|\omega)=\left\{
\begin{array}{ll}
(\xi_\nu|\log \Delta_{\nu|\omega}\xi_\nu)&\text{ if  }\nu\in{\cal N}_\omega,\\[5pt]
-\infty&\text{ otherwise}.
\end{array}
\right.
\]

\medskip
To link the modular structure of the finite quantum systems $\cQ_M$ to that of 
$(\cO, \tau^t, \omega)$, in addition to (A1)--(A6) we assume:

\medskip
\begin{enumerate}[{\bf ({A}7)}]
\item Let $\varsigma_{\omega_M}^t$ be the modular group of $\omega_M$. Then for all 
$A\in\cO_{\rm loc}$, 
\[
\lim_{M \rightarrow \infty}\pi_\omega\circ\pi_M\circ \varsigma_{\omega_M}^t\circ\pi_M^{-1}(A)
=\varsigma_\omega^t\circ\pi_\omega(A),
\]
and the convergence is uniform for $t$ in compact intervals of $\rr$.
\end{enumerate}

\medskip
Again, the verification of (A7) for spin/fermionic systems is typically an easy exercise. 
Given (A1)--(A7), we have:
\bep\label{paris-implications2}
\ben 
\item Let $Q^t$ be the spectral measure for 
$-\frac{1}{t}\log \Delta_{\omega_t|\omega}$ and $\xi_\omega$. Then $Q^t={\mathbb P}_t$.
\item $\lim_{M\rightarrow \infty}S_\alpha(\omega_{M,t}|\omega_M)= S_\alpha(\omega_t|\omega)$ and 
$\lim_{M\rightarrow \infty}S(\omega_{M,t}|\omega_M)= S(\omega_t|\omega)$. In particular, 
\[
S(\omega_t|\omega)=-\int_0^t \omega(\sigma_s)\d s.
\]
\een
\eep
The proof of the last proposition is somewhat  technical and can be found in \cite{JOPP}.

Finally, we link $e_{2, t}(\alpha)$ and the full counting statistics ${\mathbb P}_t$ to quantum transfer
operators. To avoid introduction of the full machinery of the Araki-Masuda $L^p$-spaces we shall focus
here on the special case described in Exercise \ref{Exo:LalphaLiouvPerturb} (this special case 
covers open quantum systems). Suppose that the finite quantum systems $\cQ_M$ have the following 
additional structure:

\medskip
\begin{enumerate}[{\bf ({A}8)}]
\item $H_M= H_{M,0} + V_M$, where $[H_{M,0},\omega_M]=0$ and 
\[
\lim_{M\rightarrow \infty}\pi_M(V_M)=V,
\]
in the norm of $\cO$. Moreover, for any $a>0$, 
\begin{equation}
\sup_{|\alpha|<a, M}\|\varsigma_{\omega_M}^{\i \alpha}(V_M)\|<\infty.
\label{mont-st-anne1}
\end{equation}
\end{enumerate}

\medskip
(A8) is essentially an assumption on the structure of the model and is easily verifiable in practice.
(A3), (A8)  and perturbation theory imply that  the  dynamics $\tau_{M,0}^t$ generated by $H_{M,0}$ 
converges to the $C^\ast$-dynamics $\tau_0^t$, \ie that for $A\in\cO_{\rm loc}$ and uniformly for 
$t$ in compact intervals, 
\[
\lim_{M\rightarrow \infty}\pi_M \circ \tau_{M,0}^t \circ \pi_M^{-1}(A)=\tau_0^t(A).
\]
Clearly, $\omega \circ \tau_0^t=\omega$. The assumption \eqref{mont-st-anne1} and Vitali's theorem
ensure that the map 
\[
\rr\ni t \mapsto \varsigma_{\omega}^{\i t }(\pi_\omega(V))\in {\mathfrak M}_\omega,
\]
has an analytic continuation to the entire complex plane and that for $z\in \cc$, 
\[
\lim_{M\rightarrow \infty}\pi_\omega\circ\pi_M\circ\varsigma_{\omega_M}^{z}(V_M)
=\varsigma_\omega^z\circ\pi_\omega(V).
\]
Let $K_0$ be the standard Liouvillean of $(\cO, \tau_0, \omega)$. $K_0$ is the unique self-adjoint 
operator on $\cH_\omega$ satisfying \index{Liouvillean!standard}
\[
\pi_\omega(\tau_0^t(A))=\e^{\i tK_0}\pi_\omega(A)\e^{-\i tK_0}, \qquad \e^{\i tK_0}\cH_+=\cH_+,
\]
for all $t \in \rr$ and $A\in\cO$. For $\alpha \in \rr$  we set 
\[ 
L_{\frac{1}{\alpha}}=K_0 +\pi_\omega(V)- J\varsigma_\omega^{\i (\alpha-\frac{1}{2})}(\pi_\omega(V))J.
\]
$L_{\frac{1}{\alpha}}$ is a closed operator with the same domain as $K_0$. Except in trivial cases, 
$L_{\frac{1}{\alpha}}$ is not self-adjoint  unless $\alpha =1/2$. $L_2=K$ is the standard Liouvillean of 
$(\cO, \tau^t, \omega)$, {\sl i.e.}, the unique self-adjoint operator on $\cH_\omega$ such that 
\[
\pi_\omega(\tau^t(A))=\e^{\i tK}(A)\e^{-\i tK}, \qquad \e^{\i tK}\cH_+=\cH_+,
\]
for all $t\in \rr$ and $A\in\cO$.

The following result, which is of considerable conceptual and computational importance, 
 is the extension of Exercise  \ref{Exo:LalphaLiouvPerturb} to the setting of infinitely extended systems.
\bep  
For all $t$ and $\alpha$, 
\[e_{2,t}(\alpha)=(\xi_\omega|\e^{-\i t L_{\frac{1}{\alpha}}}\xi_\omega).\]
\label{mont-st-anne2}
\eep
The extension of the results of this section  to the multi-parameter/open quantum system full counting 
statistics is straightforward.

\section{Thermodynamic limit: Control parameters}
\label{sect:TDcontrol}
By (A6), the limit 
\[
e_t(X, Y)=\lim_{M\rightarrow \infty}e_{M,t}(X,Y),
\]
exists for all $t$ and $X, Y\in \rr^n$. The basic properties of $e_t(X, Y)$ are summarized in:
\bep\label{gen-var-back}
\ben
\item
\begin{equation*}
e_t(X,Y)=\sup_{\nu\in{\cal N}_{\omega_X}}\left[S(\nu|\omega_X)
+Y\cdot\int_0^t\nu({\bf \Phi}_{Xs})\,\d s\right].
\end{equation*}
\item The function $\rr^n\ni Y\mapsto e_t(X,Y)$ is convex and real analytic.
\item $e_{t}(X,Y)=e_t(X,X-Y)$. 
\item
$$
\partial_{Y_j}e_t(X,Y)\big|_{Y=0}=\int_0^t \omega_X(\Phi_{Xs}^{(j)})\d s,
$$
$$
\partial_{Y_k}\partial_{Y_j}e_t(X,Y)\big|_{Y=0}=\int_0^t\int_0^t
\left(\langle\Phi^{(k)}_{Xs_1}|\Phi^{(j)}_{Xs_2}\rangle_{\omega_X}-
\omega_X(\Phi^{(k)}_{Xs_1})\omega_X(\Phi^{(j)}_{Xs_2})\right)\d s_2\d s_1.
$$
\een
\eep
These results are the  extension of Proposition \ref{gen-var-sym} to the setting of infinitely 
extended systems. The only difference is that, for simplicity of the exposition, we have exploited 
the time reversal in the formulation of the results. \index{symmetry!Evans-Searles!generalized}

The proof of Proposition \ref{gen-var-back} can be found in \cite{JOPP} 
and we restrict ourselves to several comments. Part (3), the generalized finite time Evans-Searles 
symmetry, is of course an immediate consequences of the same property of the functionals 
$e_{M,t}(X, Y)$. The convexity of $Y\mapsto e_t(X,Y)$ follows in the same way (note  that 
convexity also follows  from (1)). The most natural way to prove the remaining parts is to use 
Araki's perturbation theory of the KMS/modular  structure (this theory is, in part, an extension of 
the results of Section \ref{sect:Perturbations} to general von Neumann algebras). 
The Kubo-Mari inner product $\langle\Phi^{(k)}_{Xs_1}|\Phi^{(j)}_{Xs_2}\rangle_{\omega_X}$ 
in Part (4) is formally similar to its finite-dimensional counterpart. It is a part of the modular 
structure that for all $A, B \in {\mathfrak M}_{\omega_X}$, the function \index{Kubo-Mari inner product}
$t\mapsto (\xi_{\omega_X}|A^\ast \varsigma_{\omega_X}^t(B)\xi_{\omega_X})$
has an analytic continuation to the strip $-1 <\Im z <0$ which is bounded on continuous on its closure. 
Then
\[
\langle\Phi^{(k)}_{Xs_1}|\Phi^{(j)}_{Xs_2}\rangle_{\omega_X}=\int_0^1
(\xi_{\omega_X}|\pi_{\omega_X}(\Phi^{(k)}_{Xs_1})
\varsigma_{\omega_X}^{-\i u}(\pi_{\omega_X}(\Phi^{(j)}_{Xs_2}))\xi_{\omega_X})\d u.
\]
 
The finite time linear response theory for family of infinitely extended systems 
$(\cO, \tau_X^t, \omega_X)$ can be developed along two complementary routes. We shall use the 
same notational conventions as in Section \ref{sect:FinLinearResponse}:
$\omega_0=\omega$, $\tau_0=\tau$, ${\boldsymbol\Phi}_0={\boldsymbol\Phi}$. Since 
\[
\langle {\boldsymbol\Phi}_{X}\rangle_t=\frac{1}{t}\int_0^t \omega_X({\boldsymbol\Phi}_{Xs})\d s
= \frac{1}{t}{\boldsymbol\nabla}_Y e_t(X, Y)|_{Y=0},
\]
we have the following: \index{linear response}
\bep\label{sun-tir}
Suppose that the map $(X, Y)\mapsto e_t(X, Y)$ is $C^2$ in an open set containing $(0,0)$. 
Then the finite time kinetic transport coefficients 
\[
L_{jkt}=\partial_{X_k}\langle \Phi_{X}^{(j)}\rangle_t|_{X=0}
=\partial_{X_k}\partial_{Y_j}e_{t}(X, Y)_{X=Y=0},
\]
satisfy :
\ben 
\item \index{formula!Green-Kubo}
\[ 
L_{jkt}=\frac{1}{2}\int_{-t}^t \langle \Phi^{(k)}|\Phi_s^{(j)}\rangle_\omega\left(1-\frac{|s|}{t}\right)\d s.
\]
\item $L_{jkt}=L_{kjt}$ \index{relation!Onsager reciprocity} and  the  quadratic form 
determined by $[L_{jkt}]$ is positive definite. 
\een
\eep
Given Proposition \ref{gen-var-back}, the proof of Proposition \ref{sun-tir} is exactly the same as the
proof of its finite dimensional counterpart (Proposition \ref{fin-gk} in Section  \ref{sect:FinLinearResponse}). 

A complementary route is based on the thermodynamical limit of the finite time finite volume linear 
response theory. This route is both technically and conceptually less satisfactory and we shall not 
discuss it here.

\section{Large time limit: Full counting statistics}
\label{sect:LTFCS}
\index{full counting statistics}

To describe fluctuations of ${\mathbb P}_t$ as $t\rightarrow \infty$ we need to assume:

\medskip
\begin{enumerate}[{\bf ({A}9)}]
\item The limit
$$
e_{2, +}(\alpha)=\lim_{t\rightarrow \infty}\frac{1}{t}e_{2, t}(\alpha),
$$
exists for $\alpha$ in some open interval ${\cal I}$ containing $[0,1]$. Moreover, the  limiting entropic
functional $e_{2,+}(\alpha)$ is differentiable on ${\cal I}$.
\end{enumerate}

\medskip
The verification of (A9) (and (A10) below) is the central step of the program. Unlike (A1)--(A8),
which are typically easily verifiable structural/thermodynamical limit properties of a given model,  
the verification of (A9) is usually a difficult analytical problem. 

The quantum Evans-Searles fluctuation theorem for the full counting statistics\index{theorem!Evans-Searles fluctuation}  
follows  from (A9)  and the G\"artner-Ellis theorem. \index{theorem!G\"artner-Ellis}We describe its 
conclusions. Without loss of generality we may assume that ${\cal I}$ is centered at
$\alpha=1/2$ (recall that we assume the system to be TRI). 
\bep\label{back-sick}
\ben
\item $e_{2, +}(\alpha)$ is convex  on ${\cal I}$, the Evans-Searles symmetry\index{symmetry!Evans-Searles}
\[
e_{2, +}(\alpha)=e_{2, +}(1-\alpha),
\]
holds, and 
\[ 
e_{2, +}^\prime(0)=-\lim_{t\rightarrow \infty}{\mathbb E}_t(\phi)=-\lim_{t\rightarrow \infty}
\frac{1}{t}S(\omega_t|\omega)=-\lim_{t\rightarrow \infty}\frac{1}{t}\int_0^t \omega(\sigma_s)\d s.
\]
The non-negative number $\langle \sigma\rangle_+ =-e_{2, +}^\prime(0)$ is called the entropy 
production of $(\cO, \tau^t, \omega)$. Notice that $\langle \sigma\rangle_+=0$ iff the function 
$e_{2, +}(\alpha)=0$ for  $\alpha \in [0,1]$.\index{entropy!production} 
\item  Let 
\[
\theta =\sup_{\alpha \in {\cal I}}e_{2,+}^\prime(\alpha)=-\inf_{\alpha \in {\cal I}}e_{2, +}^\prime(\alpha).
\]
The function
\[
 I(s)= -\inf_{\alpha \in {\cal I}}\left(\alpha s + e_{2, +}(\alpha)\right),
\]
is non-negative, convex and differentiable on $]-\theta, \theta[$.
\footnote{If $\theta<\infty$, then  $I(s)$ is linear on $]-\infty, -\theta]$ and $[\theta, \infty[$.}
 $I(s)=0$ iff $s=-\langle \sigma \rangle_+$ and the Evans-Searles symmetry implies 
\[ 
I(-s)= s + I(s).
\]
The last relation is sometimes called the Evans-Searles symmetry for the rate function. 
\item For  any open set $J\subset]-\theta,\theta[$, 
\[
\lim_{t \rightarrow \infty}\frac{1}{t}\log {\mathbb P}_t(J)=-\inf_{s\in J} I(s).
\]
\een
\eep
The interpretation of the quantum ES theorem for the full counting statistics is similar to the classical 
case. The full counting statistics concerns  the operationally defined ``mean entropy flow" across the 
system. Its expectation value converges, as $t\rightarrow \infty$, to  the entropy production 
$\langle \sigma \rangle_+$ of the model. Its fluctuations of order 1 are described by the theory of
large deviations. The specific aspect of the ES theorem is that the time reversal invariance implies the 
universal symmetry of the rate function which in turn implies that the ``mean entropy flow" is 
exponentially more likely to be positive then negative, \ie the probability of violating the second law 
of thermodynamics is exceedingly small for large $t$.  

We now describe schematically how Proposition \ref{mont-st-anne2} can be used to verify the key 
Assumption (A9).

\begin{enumerate}[(i)] 
\item In typical situations where spectral techniques are applicable the standard Liouvillean $K_0$ has 
purely absolutely continuous spectrum filling the real line except for finitely many 
embedded eigenvalues of finite multiplicity. This is precisely what happens in the study of open quantum 
systems describing a finite quantum system $\cS$ coupled to an infinitely extended  
reservoir $\cR$. Typically, $\cR$ will consists of several independent 
sub-reservoirs $\cR_j$  which are in thermal  equilibrium at inverse temperatures $\beta_j$ and 
chemical potentials $\mu_j$, but we do not need at this point to specify further the structure of 
$\cR$. The reservoir system is described by   $C^\ast$-dynamical system 
$(\cO_\cR, \tau_\cR^t, \omega_\cR)$ where  $\omega_\cR$ is stationary for the dynamics 
$\tau_\cR^t$ and assumed  to be modular. Let $(\cH_\cR,\pi_\cR,\xi_\cR)$ be the  corresponding 
GNS representation and let $K_\cR$ be the corresponding standard Liouvillean. Since 
$\omega_\cR$ is steady, $K_\cR\xi_\cR=0$.  We assume that apart from a simple 
eigenvalue at $0$, $K_\cR$ has purely absolutely continuous spectrum filling the entire real line. 
This assumption ensures that $\cR$ has strong ergodic properties and in particular that 
$(\cO_\cR,\tau_\cR^t,\omega_\cR)$ is mixing, \ie that 
$$
\lim_{|t|\rightarrow\infty}\omega_\cR(A\tau_\cR^t(B))=\omega_\cR(A)\omega_\cR(B),
$$
for $A, B \in\cO_\cR$. In the simplest nontrivial case, $\cS$ is a $2$-level system, described by the 
Hilbert space $\cc^2$ and the Hamiltonian $\sigma^{(3)}$ (the third Pauli matrix). Then the standard 
Liouvillean of the joint but decoupled system $\cS+\cR$ acts on the Hilbert space 
$\cH=\cc^2\otimes\cc^2\otimes\cH_\cR$ and has the form
\[
K_0 =(\sigma^{(3)}\otimes\one-\one\otimes\sigma^{(3)})\otimes\one+\one\otimes K_\cR.
\]
This will be precisely the case in the Spin-Fermion model which we will discuss in Section
\ref{sect:spin-fermion}. For simplicity of exposition, we assume in the following that the point spectrum 
of $K_0$ is $\{-2, 0, 2\}$, where the eigenvalues $\pm 2$ are simple and $0$ is doubly degenerate.
The rest of the  spectrum of $K_0$ is purely absolutely continuous and  fills the real line, 
see Fig. \ref{fig:Kspec}.

\begin{figure}[htbp]
\begin{center}
    \includegraphics[width=0.8\textwidth]{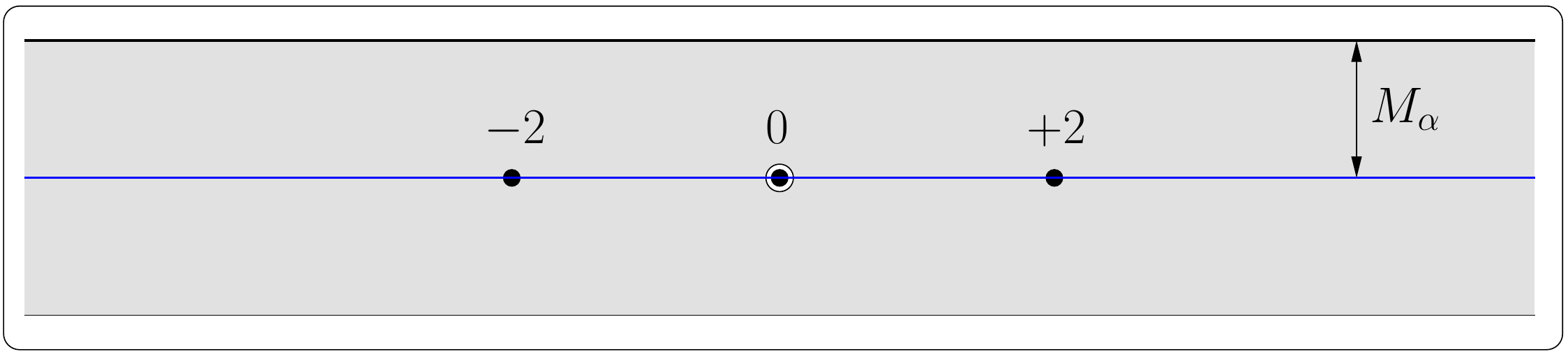}
    \caption{The point spectrum of the uncoupled standard Liouvillean $K_0$.
     The spectrum of the transfer operator $L_{\frac1\alpha}$ is contained in the grey strip.}
    \label{fig:Kspec}
\end{center}
\end{figure}

\item An application of the numerical range theorem yields that the spectrum of $L_{\frac{1}{\alpha}}$
is contained in the strip $\{ z\,|\, |\Im z|\leq M_\alpha\}$, where 
$$
M_\alpha=\|\varsigma_{\omega}^{\i (\alpha-\frac{1}{2})}(\pi_\omega(V))\|+\|\pi_\omega(V)\|.
$$
Thus, the resolvent $(z- L_{\frac{1}{\alpha}})^{-1}$ is an analytic function of $z$ on the half-plane
$\Im z>M_\alpha$.

\item By using complex deformation techniques one proves that for some $\mu>0$ and all vectors
$\xi,\eta$ in some dense subspace of $\cH$ the functions\index{complex deformation}
\[ 
z\mapsto (\xi|(z-L_{\frac{1}{\alpha}})^{-1}\eta),
\]
have a meromorphic continuation from the half-plane $\Im z> M_{\alpha}$ to the half-plane 
$\Im z>-\mu$. This extension has four simple poles located at the points $e_{\pm}(\alpha)$, 
$e(\alpha)$, $e_1(\alpha)$, where $e(\alpha)$ is the pole closest to the real axis, see Fig. 
\ref{fig:Kreso}. For symmetry reasons $e(\alpha)$ is purely imaginary. These poles are 
resonances of $L_{\frac{1}{\alpha}}$, \index{resonances} or in other words, eigenvalues of a complex deformation of 
$L_{\frac{1}{\alpha}}$. They can be computed by an application of analytic perturbation theory. 
For this purpose it is convenient to introduce a control parameter $\lambda\in\rr$ and replace the 
interaction term $V$ with $\lambda V$. The parameter $\lambda$  controls the strength of 
the coupling and analytic perturbation theory applies for small values of $\lambda$. 
One proves that given $\alpha_0>1/2$ one can find  $\Lambda >0$ such that for 
$|\alpha-\frac{1}{2}|<\alpha_0$  and $|\lambda|<\Lambda$, $\mu$ can be chosen 
independently of $\alpha$  and $\lambda$ and that the poles are analytic functions of $\alpha$. 
In particular, for $\alpha$ small enough,
\[
e(\alpha)=\i\sum_{n=1}^\infty E_n(\lambda)\alpha^n,
\]
where  each coefficient $E_n(\lambda)$ is real-analytic function of $\lambda$.  

\begin{figure}[htbp]
\begin{center}
    \includegraphics[width=0.8\textwidth]{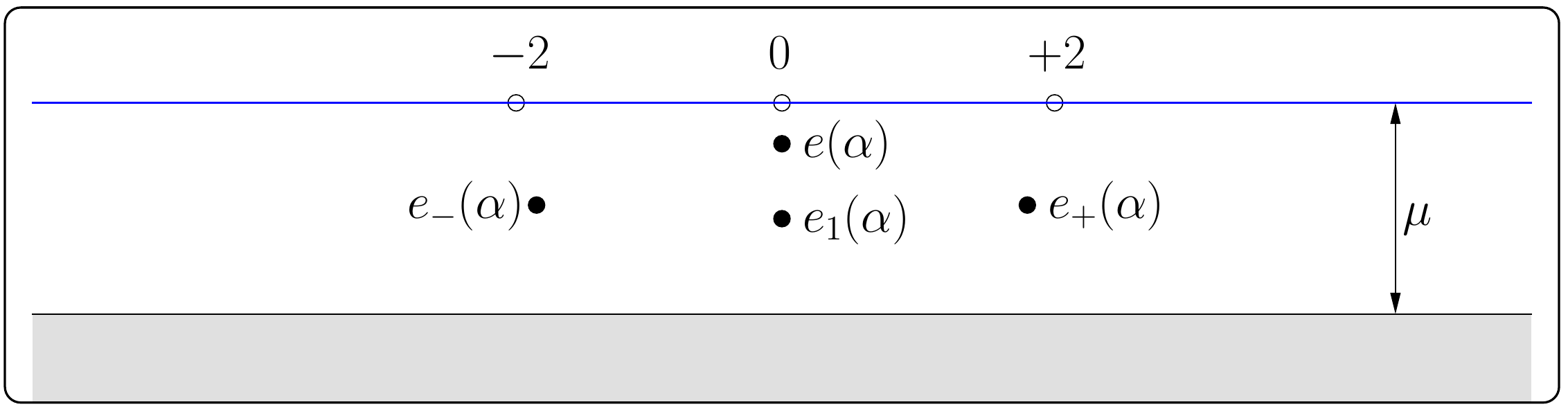}
    \caption{The resonances of the transfer operator $L_{\frac1\alpha}$.}
    \label{fig:Kreso}
\end{center}
\end{figure}

\item One now  starts with the expression 
\begin{equation}
(\xi_\omega|\e^{-\i t L_{\frac{1}{\alpha}}}\xi_\omega)=
\int_{\Re z =a}\e^{-\i t z}(\xi_\omega|(z- L_{\frac{1}{\alpha}})^{-1}\xi_\omega)\frac{\d z}{2\pi \i}, 
\label{spring}
\end{equation}
where $a >M_{\alpha}$. Moving the line of integration to $\Re z=-\mu^\prime$, where
$\mu^\prime\in]0,\mu[$ is such that the poles of the integrand are contained in 
$\{ z\,|\,\Im z>-\mu^\prime\}$ for $|\lambda| <\Lambda$ and $|\alpha-\frac{1}{2}|<\alpha_0$, and 
picking the contribution from theses poles one derives 
\begin{equation}
(\xi_\omega|\e^{-\i t L_{\frac{1}{\alpha}}}\xi_\omega)
=\e^{-\i te(\alpha)}(1 + R(t, \alpha)),
\label{spring1}
\end{equation}
where $R(t, \alpha)$ decays exponentially in $t$ as $t\rightarrow\infty$. It then follows that
\[
e_{2, +}(\alpha)=\lim_{t\rightarrow \infty}\frac{1}{t}e_{2,t}(\alpha)=-\i e(\alpha).
\]
A proper mathematical justification of \eqref{spring} and \eqref{spring1} is typically the
technically most demanding part of the argument. 

\item Recall that 
\[
\partial_\alpha e_{2, +}(\alpha)|_{\alpha=0}=E_1
=-\langle \sigma\rangle_+=-\lim_{t\rightarrow \infty}{\mathbb E}_t(\phi).
\]
Given (iv), an application of Vitali's theorem yields 
\begin{align*}
\partial_{\alpha}^2e_{2,+}(\alpha)_{\alpha=0}&= E_2 =\lim_{t \rightarrow \infty} \frac{1}{t} 
\int_0^t\int_0^t (\omega(\sigma_s\sigma_u)- \omega(\sigma_s)\omega(\sigma_u))\d s \d u\\[3mm]
&=\lim_{t\rightarrow \infty}t ({\mathbb E}_t(\phi^2) - ({\mathbb E}_t(\phi))^2).
\end{align*}

\item The arguments/estimates in (iv) extend to complex $\alpha$'s satisfying 
$|\alpha-\frac{1}{2}|<\alpha_0$ and one shows that for $\alpha$ real, 
\[
\lim_{t\rightarrow \infty} \int_\rr\e^{-\i \alpha\sqrt t(\phi- \langle \sigma\rangle_+)}\d {\mathbb P}_t(\phi) 
=\lim_{t \rightarrow \infty}\e^{\i\alpha\sqrt t\langle \sigma \rangle_+}
(\xi_\omega|\e^{-\i t L_{\frac{\sqrt t}{\i \alpha}}}\xi_\omega)=\e^{-E_2\alpha^2}.
\]
Hence, the central limit theorem holds for the full counting statistics ${\mathbb P}_t$, that is, for any 
interval $[a, b]$, \index{theorem!central limit}
\[ 
\lim_{t\rightarrow \infty}{\mathbb P}_t\left( \langle \sigma\rangle_+ 
+ \frac{1}{\sqrt{tE_2}}[a, b]\right)=\frac{1}{\sqrt{2\pi}}\int_a^b \e^{-\frac{x^2}{2}}\d x.
\]
\end{enumerate} 
The above spectral scheme is technically delicate and its implementation requires 
a number of regularity assumptions on the structure of reservoirs and the interaction $V$. On the positive side, when applicable the spectral scheme provides a wealth of information 
and a very satisfactory conceptual picture. In the classical case, the quantum transfer operators  
reduce to Ruelle-Perron-Frobenius operators and the above spectral scheme is a well-known chapter
in the theory of classical dynamical systems, see Section 5.4 in \cite{JPR} and \cite{Ba}.

\section{Hypothesis testing of the arrow of time}
\label{sect:LThypo}
\index{hypothesis testing}

Theorem \ref{QHTthm} clearly links the $p=2$ entropic functional to quantum hypothesis testing.
This link, somewhat surprisingly, can be interpreted as quantum  hypothesis testing of the second 
law of thermodynamics and arrow of time: how well can we distinguish the state 
$\omega_t=\omega\circ\tau^t$, from the same initial state evolved backward in time 
$\omega_{-t}=\omega\circ\tau^{-t}$~? More precisely, we shall investigate the asymptotic behavior of
the minimal error probability for the hypothesis testing associated to
the pair $(\omega_{-t}, \omega_t)$ as $t\to\infty$.

We start with the family of pairs $\{(\omega_{M,-t}, \omega_{M,t})\,|\,t>0\}$.
Again, the thermodynamic limit $M\to\infty$ has to be taken prior to the limit $t\to\infty$. 

Given their a priori probabilities, $1-p$ and $p$, the minimal error probability in distinguishing
the states $\omega_{M,-t/2}$ and $\omega_{M,t/2}$ is given by Theorem \ref{QHTthm},
\[
D_{M,p}(t)=\frac12\left(1-\tr\,|(1-p)\,\omega_{M,-t/2}-p\,\omega_{M,t/2}|\right).
\]
We set
\[ 
\ubar D_p(t)=\liminf_{M\rightarrow \infty}D_{M,p}(t), \qquad 
\bar D_p(t)=\limsup_{M\rightarrow\infty}D_{M,p}(t),
\]
and define the Chernoff error exponents by\bindex{Chernoff!exponents}
\[
\ubar d_p=\liminf_{t\rightarrow \infty}\frac{1}{t}\log \ubar D_p(t), \qquad 
\bar d_p=\limsup_{t\rightarrow \infty}\frac{1}{t}\log \bar D_p(t).
\]
\bet\label{EBBQHTthm}
For any $p\in ]0,1[$, 
\[ 
\ubar d_p=\bar d_p=\inf_{\alpha\in[0,1]}e_{2,+}(\alpha).
\]
Moreover, since the system is TRI the infimum is achieved at $\alpha=1/2$.
\eet

\demo We first notice that 
\[
D_{M,p}(t)=\frac12\left(1-\tr\,|(1-p)\,\omega_{M,0}-p\,\omega_{M,t}|\right).
\]
Theorem \ref{QHTthm} (3) and the existence of the limiting functional $e_{2,t}(\alpha)$ 
(for $M\to\infty$) yield the inequality
$$
\log\bar D_p(t)\leq e_{2,2t}(\alpha)+(1-\alpha)\log(1-p)+\alpha\log p,
$$
for all $\alpha\in[0,1]$. Dividing by $t$ and letting $t\to\infty$ we obtain the upper bound 
$$
\bar d_p\le\inf_{\alpha\in[0,1]}e_{2,+}(\alpha).
$$

For finite $M$, a lower bound is provided by Proposition \ref{QHTlower},
\[
D_{M,p}(t)\geq \frac12\min (p,1-p)\,{\mathbb P}_{M,t}(]0, \infty[),
\]
where ${\mathbb P}_{M,t}$ is the full counting statistics of ${\cal Q}_M$. As we have already 
discussed, the convergence of $e_{M, 2, t}(\alpha)$ to $ e_{2,t}(\alpha)$ as 
$M\rightarrow\infty$ implies that ${\mathbb P}_{M, t}$ converges weakly to the full counting statistics 
${\mathbb P}_t$ of the extended system. The Portmanteau theorem (\cite{Bi1}, Theorem 2.1) implies
\[ 
\liminf_{M\rightarrow \infty}{\mathbb  P}_{M,t}(]0,\infty[)\geq {\mathbb P}_{t}(]0,\infty[),
\]
and hence
$$
\ubar D_p(t)\ge\frac12\min (p,1-p)\,{\mathbb P}_{t}(]0, \infty[)
\ge\frac12\min (p,1-p)\,{\mathbb P}_{t}(]0,1[).
$$
Assumption (A9) and the G\"artner-Ellis theorem (or more specifically Proposition \ref{GE-THM} 
in Appendix \ref{sect:GEoneD}) imply \index{theorem!G\"artner-Ellis}
\[
\liminf_{t\rightarrow\infty}\frac{1}{t}\log{\mathbb P}_t(]0,1[)
\geq -\varphi(0),
\]
where
\[
\varphi(s)=\sup_{\alpha\in\rr}(s\alpha-e_{2,+}(\alpha)).
\]
Since
$$
\varphi(0)=-\inf_{\alpha\in\rr}e_{2,+}(\alpha)=-\inf_{\alpha\in[0,1]}e_{2,+}(\alpha),
$$
(recall that, by Proposition \ref{epprops}, $e_{2,+}(\alpha)\le0$ for $\alpha\in[0,1]$ and 
$e_{2,+}(\alpha)\ge0$ otherwise) we have
\[
\ubar d_p\ge\liminf_{t\to\infty}\frac1t\left(-\log2+\min(\log p,\log(1-p))+\log({\rm P}_t(]0,1[))\right)
\geq\inf_{\alpha\in[0,1]}e_{2,+}(\alpha).
\]
The convexity and the symmetry $e_{2,+}(1-\alpha)=e_{2,+}(\alpha)$
imply that the infimum is achieved at $\alpha=1/2$.
\qed

\bigskip

Note that the above result and its proof link the fluctuations of the full counting statistics 
${\mathbb P}_t$ as $t\rightarrow \infty$ to Chernoff error exponents in quantum hypothesis testing 
of the arrow of time. 
The TD limit plays an important role in the discussion of full counting statistics since its 
physical interpretation in terms of repeated quantum measurement is possible only for finite quantum 
systems. However, apart from the above mentioned connection with full counting statistics,
quantum hypothesis testing can be formulated in the framework of extended quantum systems
without reference to the TD limit. In fact, by considering directly an infinitely extended 
system, one can considerably refine the quantum hypothesis testing of the arrow of time. 
In the remaining part of this section we indicate how this can be done, referring the reader to 
\cite{JOPS} for proofs and additional information. 

\begin{enumerate}[(i)] 
\item We start with an infinitely extended system $\cQ$ described by the $C^\ast$-dynamical system  
$(\cO,\tau^t,\omega)$. The GNS-representation of $\cO$ \index{representation!GNS} associated to the state $\omega$ is 
denoted $(\cH_\omega,\pi_{\omega},\xi_\omega)$, and the enveloping von Neumann algebra is \index{algebra!enveloping von Neumann}
${\mathfrak M}_\omega =\pi_{\omega}(\cO)^{\prime\prime}$. We assume that $\omega$ is 
modular. The group $\pi_\omega \circ \tau^t$ extends to a weakly continuous group
$\tau_\omega^t$ of $\ast$-automorphisms of ${\mathfrak M}_\omega$. With a slight abuse of 
notation we denote the vector state $(\xi_\omega|\,\cdot\,\xi_\omega)$ on 
${\mathfrak M}_\omega$ again by $\omega$. The triple 
$({\mathfrak M}_\omega, \tau_\omega^t, \omega)$ is the $W^\ast$-quantum 
dynamical system induced by $(\cO, \tau^t, \omega)$. We denote 
$\omega_t=\omega\circ \tau_{\omega}^t$. The quantum hypothesis testing of the arrow of time 
concerns the family of pairs $\{(\omega_{-t}, \omega_{t})\,|\,t>0\}$.
 
\item Consider the following competing hypothesis: 
\begin{quote} Hypothesis I :  $\cQ$ is in the state $\omega_{t/2}$; 
\end{quote} 
\begin{quote} Hypothesis II  :  $\cQ$ is in the state $\omega_{-t/2}$; 
\end{quote} 
We know  {\em a priori} that Hypothesis I is realized  with probability $p$ and II with probability $1-p$. 
A {\em test}  is a self-adjoint projection $P\in{\mathfrak M}_\omega$ and a result of a measurement
of the corresponding  observable is a number in $\sp(P)=\{0,1\}$. If the outcome is $1$,
one accepts I, otherwise one  accepts II. The error probability of the test $P$  is
$$
D_p(\omega_{t/2}, \omega_{-t/2}, P)=p\,\omega_{t/2}(\one -P) + (1-p)\,\omega_{-t/2}(P),
$$
and 
\[ 
D_p(\omega_{t/2}, \omega_{-t/2})=\inf _{P} D_p(\omega_{t/2}, \omega_{-t/2}, P),
\]
is the minimal error probability.
\item The quantum Neyman-Pearson lemma holds:\index{Neyman-Pearson}
\begin{align*}
D_p(\omega_{t/2}, \omega_{-t/2})&=D_p(\omega_{t/2}, \omega_{-t/2},P_{\rm opt})
=\frac{1}{2}(1- \|(1-p)\omega_{-t/2}-p\omega_{t/2}\|)\\[5pt]
&=\frac{1}{2}(1- \|(1-p)\omega-p\omega_{t}\|),
\end{align*}
where $P_{\rm opt}$ is the support projection of the linear functional $((1-p)\omega_{-t/2}- p\omega_{t/2})_+$ (the positive part of 
$(1-p)\omega_{-t/2}- p\omega_{t/2}$). Just like in the classical case, the proof of the
quantum Neyman-Pearson lemma is straightforward. \index{Neyman-Pearson}

\item Let $\mu_{\omega_t|\omega}$ be the spectral measure for $\Delta_{\omega_t|\omega}$ and 
$\xi_\omega$. Then 
\[
\frac{1}{2}\min(p, 1-p)\mu_{\omega_{t}|\omega}([1, \infty[)\leq D_p(\omega_{t/2}, \omega_{-t/2})\leq 
p^\alpha(1-p)^{1-\alpha}(\xi_\omega|\Delta_{\omega_{t}|\omega}^\alpha\xi_\omega).
\]
The proof of the lower bound in exactly the same as in finite case (recall Proposition \ref{QHTlower}).
The proof of the upper bound is based on an  extension of Ozawa's argument (see the proof of 
Part (3) of Theorem \ref{QHTthm}) to the modular setting and is more subtle, 
see \cite{Og}.

\item Assuming (A9), \ie that 
\[
e_{2, +}(\alpha)=\lim_{t\to\infty}\frac1t\log(\xi_\omega|\Delta_{\omega_t|\omega}^\alpha\xi_\omega),
\]
exist and is differentiable for $\alpha$ is some interval containing $[0,1]$, then a straightforward 
application of the G\"artner-Ellis theorem yields 
\begin{equation}
\lim_{t\rightarrow \infty}\frac1t\log D_p(\omega_{t}, \omega_{-t})
=\inf_{\alpha \in [0,1]} e_{2,+}(\alpha).
\label{ChBound}
\end{equation}
Results of this type are often called quantum Chernoff bounds. Our TRI assumption implies that 
the infimum is achieved for $\alpha=1/2$.

The Chernoff bound \eqref{ChBound} quantifies the separation between the past and the future 
as time $t\uparrow \infty$. Taking $p=1/2$ and noticing that
\[ 
\frac{1}{2}(2-\|\omega_{t/2}-\omega_{-t/2}\|)=
\omega_{t/2}(\s_-(t/2))+\omega_{-t/2}(\s_+(t/2)),
\]
where $\s_\pm(t)$ is the support projection of the positive linear functional 
$(\omega_{t}-\omega_{-t})_\pm$ on ${\mathfrak M}_\omega$, we see that the Chernoff bound 
implies  
\begin{align*}
\limsup_{t \rightarrow \infty}\frac{1}{t}\log \omega_t(\s_-(t))
&\leq 2\inf_{s\in [0,1]}e_{2,+}(s),\\
\limsup_{t \rightarrow \infty}\frac{1}{t}\log \omega_{-t}(\s_+(t))
&\leq 2\inf_{s\in [0,1]}e_{2,+}(s).
\end{align*}
Therefore, as $t \uparrow \infty$, the state $\omega_t$  concentrates exponentially fast on
$\s_+(t){\mathfrak M}_\omega$ while the state $\omega_{-t}$ concentrates exponentially fast 
on $\s_-(t){\mathfrak M}_\omega$.

\item In the infinite dimensional setting one can introduce other error exponents. 
For $r\in\rr$  the Hoeffding exponents are defined  by \index{Hoefding exponents}
\begin{align*}
\bar B(r)&=\inf_{\{P_t\}}\left\{
\limsup_{t \rightarrow \infty}\frac{1}{t}\log \omega_{t/2}(\one-P_t)\,\, \bigg|\,\,
\limsup_{t\rightarrow \infty}\frac{1}{t}\log\omega_{-t/2}(P_t)<-r\right\},\\[3mm]
\ubar B(r)&=\inf_{\{P_t\}}\left\{
\liminf_{t \rightarrow \infty}\frac{1}{t}\log \omega_{t/2}(\one-P_t)\,\, \bigg|\,\,
\limsup_{t\rightarrow \infty}\frac{1}{t}\log\omega_{-t/2}(P_t)<-r\right\},\\[3mm]
B(r)&=\inf_{\{P_t\}}\left\{
\lim_{t \rightarrow \infty}\frac{1}{t}\log \omega_{t/2}(\one -P_t)\,\, \bigg|\,\, 
\limsup_{t\rightarrow \infty}\frac{1}{t}\log\omega_{-t/2}(P_t)<-r\right\},
\end{align*}
where the infimum are taken over families $\{P_t\}_{t>0}$ of orthogonal projections in 
${\mathfrak M}_\omega$ subject, in the last case, to the constraint that
$\lim_{t \rightarrow \infty}t^{-1}\log \omega_{t/2}(\one -P_t)$ exists. 

The Hoeffding exponents are increasing functions of $r$,  $\ubar B(r)\leq \bar B(r)\leq B(r)\leq 0$, and 
$\ubar B(r)=\bar B(r)= B(r)=-\infty$ if $r < 0$. The functions $\ubar B(r), \bar B(r), B(r)$ are left continuous  
and upper semi-continuous. If (A9) holds and $\langle \sigma \rangle_+>0$, then for all $r\in \rr$, 
\[
\ubar B(r)=\bar B(r)=B(r)=b(r)=-\sup_{0\leq s<1}\frac{-sr -e_{2,+}(s)}{1-s},
\]
see \cite{JOPS}. Results of this type are called quantum Hoeffding bounds. 

Let  $r>0$ and let   $P_t$ be projections in ${\mathfrak M}_\omega$   such that 
$$
\limsup_{t\rightarrow \infty}\frac{1}{t}\log \omega_{-t/2}(P_t) <-r.
$$
The Hoeffding bound  asserts 
\[
\liminf_{t\rightarrow \infty}\frac{1}{t}\log\omega_{t/2}(\one-P_t)\geq  b(r).
\]
Moreover, one can show that for a suitable choice of  $P_t$,
\[
\lim_{t\rightarrow \infty}\frac{1}{t}\log\omega_{t/2}(\one-P_t)= b(r).
\]
Hence, if $\omega_{-t/2}$ is concentrating exponentially fast on $(\one-P_t){\mathfrak M}_\omega$ 
with an exponential rate $<-r$, then $\omega_{t/2}$ is concentrating on $P_t{\mathfrak M}_\omega$ 
with the optimal exponential rate $b(r)$.
\item 
For $\epsilon \in ]0,1[$ set 
\begin{align}
&\bar B_\epsilon=\inf_{\{P_t\}}\left\{\limsup_{t \rightarrow \infty}\frac{1}{t}\log \omega_{t/2}(\one-P_t)
\,\,\bigg|\,\,\omega_{-t/2}(P_t)\leq \epsilon\right\},\nonumber\\[3mm]
&\ubar B_{\kern1.5pt\epsilon}=\inf_{\{P_t\}}\left\{\liminf_{t \rightarrow \infty}\frac{1}{t}\log \omega_{t/2}(\one-P_t)
\,\,\bigg|\,\,\omega_{-t/2}(P_t)\leq \epsilon\right\},\label{stein-1}\\[3mm]
&B_\epsilon=\inf_{\{P_t\}}\left\{\lim_{t \rightarrow \infty}\frac{1}{t}\log \omega_{t/2}(\one-P_t)
\,\,\bigg|\,\,\omega_{-t/2}(P_t)\leq \epsilon\right\},\nonumber
\end{align}
where the infimum is taken over families of tests $\{P_t\}_{t>0}$ subject, in the last case,
to the constraint that $\lim_{t \rightarrow \infty}t^{-1}\log \omega_{t/2}(\one-P_t)$ exists.  
Note that if 
\[
\beta_t(\epsilon)=\inf_{P} \{\omega_{t/2}(\one -P)\,|\, \omega_{-t/2}(P)\leq \epsilon\},
\]
then 
\[
\liminf_{t\rightarrow \infty} \frac{1}{t}\log \beta_t(\epsilon)=\ubar B_{\kern1.5pt\epsilon}, 
\qquad \limsup_{t\rightarrow \infty}\frac{1}{t}\log \beta_t(\epsilon)=\bar B_\epsilon.
\]

We also define 
\begin{align}
\bar B&=\inf_{\{P_t\}}\left\{\limsup_{t \rightarrow \infty}\frac{1}{t}\log \omega_{t/2}(\one-P_t)
\,\,\bigg|\,\,\lim_{t\rightarrow \infty}\omega_{-t/2}(P_t)=0\right\},\nonumber\\[3mm]
\ubar B&=\inf_{\{P_t\}}\left\{\liminf_{t \rightarrow \infty}\frac{1}{t}\log\omega_{t/2}(\one-P_t)
\,\,\bigg|\,\,\lim_{t\rightarrow \infty}\omega_{-t/2}(P_t)=0\right\},\label{stein-2}\\[3mm]
B&=\inf_{\{P_t\}}\left\{\lim_{t \rightarrow \infty}\frac{1}{t}\log \omega_{t/2}(\one-P_t)
\,\,\bigg|\,\,\lim_{t\rightarrow \infty}\omega_{-t/2}(P_t)=0\right\},\nonumber
\end{align}
where again in  the last case the infimum is taken over all families of tests $\{P_t\}_{t >0}$ for which
the limit $\lim_{t \rightarrow \infty}t^{-1}\log \omega_t(\one-P_t)$ exists. 

We shall call the numbers defined in \eqref{stein-1} and \eqref{stein-2} the  Stein exponents. 
Clearly, 
$\ubar B_{\kern1.5pt\epsilon} \leq \bar B_\epsilon\leq B_\epsilon$, 
$\ubar B\leq \bar B\leq B$, $\ubar B_{\kern1.5pt\epsilon}\leq \ubar B$, 
$\bar B_\epsilon \leq \bar B$, $B_\epsilon\leq B$.  If (A9) holds, then for  any $\epsilon \in ]0,1[$, 
\[\ubar B=\bar B=B=\ubar B_{\kern1.5pt\epsilon}=\bar B_\epsilon =B_\epsilon=-\langle \sigma \rangle_+,
\]
see \cite{JOPS}. Results of this type are called quantum Stein Lemma. 
\index{Stein exponent}

Stein's Lemma asserts that for any family of projections $P_t$ such that 
\begin{equation}
\sup_{t>0}\,\omega_{-t}(P_t)<1,
\label{lv-horse}
\end{equation}
one has 
\[
\liminf_{t\rightarrow \infty}\frac{1}{t}\log \omega_t(\one-P_t)\geq -2\langle \sigma\rangle_+,
\]
and that for any $\delta >0$ one can find a sequence of projections $P_t^{(\delta)}$ satisfying
\eqref{lv-horse} and 
\[
\lim_{t\rightarrow \infty}\frac{1}{t}\log \omega_t(\one-P_t^{(\delta)})\leq  
-2\langle \sigma\rangle_+ +\delta.
\]
Hence, if no restrictions are made on $P_t$ w.r.t. $\omega_{-t}$ except  \eqref{lv-horse} (which is 
needed to avoid trivial result), the optimal  exponential rate of concentration of $\omega_t$ as
$t\uparrow \infty$ is precisely twice the negative entropy production.  
\end{enumerate}

\section{Large time limit: Control parameters} 
\label{sect:LTcontrol}

We continue with the framework of Section \ref{sect:TDcontrol}. The infinitely extended systems 
$(\cO, \tau_X, \omega_X)$ are parameterized by control parameters $X\in \rr^n$. Recall the 
shorthands  $\omega=\omega_0$, $\tau=\tau_0$, ${\bf \Phi}={\bf \Phi}_0$, etc.  We assume 

\medskip
\begin{enumerate}[{\bf ({A}10)}]
\item For all $t>0$ the functional $(X, Y)\mapsto e_t(X, Y)$ has an analytic continuation to the
polydisk $D_{\delta,\epsilon}=\{(X,Y)\in\cc^n\times\cc^n\,|\,\max_j|X_j|<\delta,\max_j|Y_j|<\epsilon\}$
satisfying
$$
\sup_{\atop{(X,Y)\in D_{\delta,\epsilon}}{t>0}}\left|\frac1te_t(X,Y)\right|<\infty.
$$
In addition, the limit 
\[
e_+(X, Y)=\lim_{t\rightarrow \infty}\frac{1}{t}e_t(X, Y),
\]
exists for all $(X, Y)\in D_{\delta,\epsilon}\cap(\rr^n\times\rr^n)$.
\end{enumerate}

\medskip
As in the case of (A9), establishing (A10) for physically interesting models is typically a very difficult 
analytical problem. Although (A10) is certainly not a minimal assumption under which the results of 
this section hold (for the minimal axiomatic scheme see \cite{JOPP}), it can be verified in interesting 
examples and allows for a transparent exposition of the material of this section.  

A consequence of the first part of  (A10) is that finite time linear response theory holds for 
$(\cO,\tau_{X}^t,\omega_X)$. By Vitali's theorem, $e_+(X, Y)$ is analytic on $D_{\delta,\epsilon}$ 
and we have:
\bep\label{prop-exam}
\ben
\item For any $X\in\rr^n$ such that $\max_j|X_j|<\delta$, 
\[
\langle{\boldsymbol\Phi}_X\rangle_+=\lim_{t\to\infty}\frac1t\int_0^t
\omega_{X}\left({\boldsymbol\Phi}_{Xs}\right)\,\d s
={\boldsymbol\nabla}_Y e_{+}(X,Y)|_{Y=0}.
\]
\item The kinetic transport coefficients defined by \index{transport coefficients}
\[
L_{jk}=\partial_{X_k}\langle \Phi_X^{(j)}\rangle_+|_{X=0},
\]
satisfy
\[
L_{jk}=\lim_{t\rightarrow \infty}L_{jkt}
=\lim_{t\rightarrow \infty}\frac{1}{2}\int_{-t}^t \langle \Phi^{(k)}|\Phi_s^{(j)}\rangle_\omega
\left(1-\frac{|s|}{t}\right)\d s.
\]
\item The Onsager matrix $[L_{jk}]$ is symmetric and positive semi-definite.\index{Onsager matrix}
\item Suppose that $\omega$ is a $(\tau, \beta)$-KMS state for some $\beta >0$ and that 
$(\cO, \tau, \omega)$ is mixing, i.e., that 
$$
\lim_{t\rightarrow \infty}\omega(A\tau^t(B))=\omega(A)\omega(B),
$$
for all $A, B \in\cO$. Then 
\[
L_{jk}=\lim_{t\rightarrow \infty}\frac{1}{2}\int_{-t}^t \omega(\Phi^{(j)}\Phi^{(k)}_s) \d s.
\]
\een
\eep
Parts (1)--(3)  are an immediate consequence of Vitali's theorem (see Proposition \ref{PropVitali} in 
Appendix \ref{appx:Vitali}). Part (4) recovers the familiar form of the  Green-Kubo formula under the assumption 
that  for vanishing control parameters the infinitely extended system is in thermal equilibrium 
(and is strongly ergodic). For the proof of (4) see \cite{JOPP} or the proof of Theorem 2.3 in 
\cite{JOP2}.\index{formula!Green-Kubo}

\section{Large time limit: Non-equilibrium steady states (NESS)}
\label{sect:NESS} 

Consider our infinitely extended system $(\cO, \tau^t, \omega)$ and suppose 

\medskip
\begin{enumerate}[{\bf ({A}11)}]
\item The limit 
\[
\lim_{t\rightarrow \infty}\omega_t(A)=\omega_+(A),
\]
exists for all $A\in\cO$. $\omega_+$ is a stationary state  called the NESS of $(\cO, \tau^t, \omega)$.
\index{state!non-equilibrium steady}
\end{enumerate}

\medskip
Albeit a hard ergodic-type problem, the verification of (A11) is typically easier then the proof 
of (A9) or (A10). In fact, in all known non-trivial models satisfying  (A9)/(A10), the proof of (A11) is 
a consequence of the proof of (A9)/(A10).

The structural theory of NESS was one of the  central topics of the lecture notes 
\cite{AJPP1} and we will not discuss it here. In relation with entropic 
fluctuations, the NESS plays a central role in the Gallavotti-Cohen fluctuation theorem. We will not
enter into this subject in these lecture notes. \index{theorem!Gallavotti-Cohen fluctuation}

\section{Stability with respect to the reference state}
\label{Sect:ChoiceofOmega}

In addition to (A11), one expects that under normal conditions any normal state 
$\nu\in{\cal N}_\omega$ is in the basin of attraction of the NESS $\omega_+$, \ie that the following 
holds:

\medskip
\begin{enumerate}[{\bf ({A}12)}]
\item
\[
\lim_{t\rightarrow \infty}\nu(\tau^t(A))=\omega_+(A),
\]
for all $\nu\in{\cal N}_\omega$ and $A\in\cO$.
\end{enumerate}

\medskip
As for (A11), in all known  non-trivial models, (A12) follows from the proofs of (A9)/(A10).

(A12) is a mathematical formulation of the fact that under normal conditions the NESS and more 
generally the large time thermodynamics do not depend on local perturbations of the initial state 
$\omega$. More specifically, in the  context of open quantum systems, if  the coupling $V$  is well 
localized in the reservoirs, then in the TD limit (the $\cR_j$'s becoming infinitely extended and the 
system $\cS$ remaining finite), the effect of including $V$ in the reference state becomes negligible 
for large times. In other words, the product state $\omega$ used in Sections 
\ref{sect:Coupling}--\ref{sect:FCSOpen} and the state $\omega_X$ of Section \ref{sect:LinearResponse} 
become equivalent for large times. More generally, the system loses memory of any localized 
perturbation of its initial state.

In a similar vein one expects that, under normal conditions, the limiting entropic functionals do not 
depend on local perturbations of the initial state. To illustrate this point, we consider the functional
$e_{\infty,+}(\alpha)$ (and assume that the reader is familiar with Araki's perturbation theory
of the KMS structure).  $\omega$ has a modular group $\varsigma_\omega^t$
and if  $\omega_W$ is the KMS state (at temperature $-1$) of the perturbed group
$\varsigma_{\omega W}^t$ for some $W\in\Os$ (which, for finite systems, amounts to set
$\omega_W=\e^{\log\omega+W}/\tr(\e^{\log\omega+W})$), then \index{modular!group}
$$
\omega_W(A)=\frac{\omega(A \rE_W(-\i))}{\omega(\rE_W(-\i))},
$$
where the cocycle $\rE_W$ is given by \eqref{EViBetaExp}.\index{cocycle}
The set of states $\{\omega_W\,|\, W\in\Os\}$ is norm dense in the (norm closed) set 
${\cal N}_\omega$ of all normal state on $\cO$. Since  $\ell_{\omega_W|\omega}=W$, 
one has $\ell_{\omega_{Wt}|\omega_W}=\ell_{\omega_t|\omega}+\tau^{-t}(W)-W$ and hence
$$
\omega_+(\sigma_\omega)=\omega_+(\sigma_{\omega_W}).
$$
Similarly, for $\alpha\in]0,1[$, Proposition \ref{Prop:EquivState} holds for infinitely extended systems 
(this can be proven either via a TD limit argument or by direct application of modular theory), and so 
$$
\lim_{t\to\infty}\frac1t(e_{\infty,t,\omega}(\alpha)-e_{\infty,t,\omega_W}(\alpha))=0.
$$
Hence, 
$$
e_{\infty,+,\omega}(\alpha)=\lim_{t\to\infty}\frac1te_{\infty,t,\omega}(\alpha),
$$
exists iff 
$$
e_{\infty,+,\omega_W}(\alpha)=\lim_{t\to\infty}\frac1te_{\infty,t,\omega_W}(\alpha),
$$
exists and the limiting entropic functionals are equal. Similar stability results for other entropic 
functionals can be established under additional regularity assumptions \cite{JOPP}.

\section{Full counting statistics and quantum fluxes: a comparison}
\label{sect:Comparison}

In this section we shall focus on open quantum systems described in Chapter \ref{chap:OQS}.
For simplicity of notation we set the chemical potentials $\mu_j$ of the reservoirs $\cR_j$
to zero and deal only with energy fluxes $\Phi_j$.\index{full counting statistics}\index{flux}

Full counting statistics deals with the mean entropy/energy flow operationally defined by a repeated 
quantum measurement. It does not refer to the measurement of a single quantum observable. In 
fact, surprisingly, it gives a physical interpretation to quantities which are considered unobservable 
from the traditional point of view:  the spectral projections of a relative modular operator.
Full counting statistics is of purely quantum origin and has no counterpart in classical statistical 
mechanics. In contrast, the energy flux observables $\Phi_j$ introduced in Chapter \ref{chap:OQS}
arise by direct operator quantization of the corresponding classical observables. In this section,
we take a closer look at the relation between full counting statistics and energy flux observables.

For open quantum systems, \index{open system} the TD limit concerns only the reservoirs $\cR_j$, the  finite quantum 
system $\cS$ remaining fixed. As discussed in the previous section, if we are not interested in
transient properties then we may assume, without loss of generality, that $\omega_\cS$ is the 
chaotic state \eqref{ChaoStateDef}.  
After the TD limit is taken, the  infinitely extended reservoir $\cR_j$ is described by the quantum 
dynamical system $(\cO_j, \tau_j^t, \omega_j)$, where $\omega_j$ is a $(\tau_j,\beta_j)$-KMS state 
on $\cO_j$. The joint system $\cR=\cR_1+\cdots+\cR_n$ is described 
by 
\[
(\cO_\cR, \tau_\cR^t, \omega_\cR)=\bigotimes_{j=1}^n (\cO_j, \tau_j^t, \omega_j).
\]
The joint but decoupled system $\cS+\cR$ is described by  $(\cO, \tau^t, \omega)$ where 
\[
\cO=\cO_\cS\otimes\cO_\cR, \qquad \tau^t=\tau_\cS^t\otimes\tau_\cR^t, \qquad 
\omega=\omega_\cS\otimes\omega_\cR.
\]
The interaction of $\cS$ with $\cR_j$ is described by a self-adjoint element
$V_j\in\cO_\cS\otimes\cO_j$. The full interaction $V=\sum_j V_j$ and the corresponding
perturbed $C^\ast$-dynamics $\tau_V^t$ finally yield the quantum dynamical system
$(\cO, \tau_V^t, \omega)$ which describes the infinitely extended open 
quantum system. Without further saying, we shall always assume that all relevant quantities are 
realized as TD limit of the corresponding quantities of a sequence $\{\cQ_M\}$ of 
finite, TRI open quantum systems. In particular, that is so for the energy flux observables 
\[
\Phi_j =\delta_j(V_j),
\]
where $\delta_j$ is the generator of $\tau_j$ ($\tau_j^t=\e^{t \delta_j}$), and the 
entropy production observable \index{entropy!production}
\[
\sigma=-\sum_j \beta_j \Phi_j,
\]
of the infinitely extended open quantum system $(\cO, \tau_V^t, \omega)$.

Recall Section \ref{sect:FCSOpen}. Let ${\mathbb P}_t$ be the full counting statistics of the 
infinitely extended open systems $(\cO, \tau_V^t, \omega)$. The probability measure 
${\mathbb P}_t$ arises as the weak limit of the full counting statistics ${\mathbb P}_{M,t}$ of 
$\cQ_M$ (this realization is essential for the physical interpretation of ${\mathbb P}_t$).
Thus, it follows from Relations \eqref{FCSexp}, \eqref{FCSvar}, that
\begin{align}
\langle \varepsilon_j \rangle_+
&=\lim_{t\to\infty}{\mathbb E}_t(\varepsilon_j)=-\beta_j\omega_+(\Phi_j),\label{exam1}\\[4mm]
D_{{\rm fcs}, jk}
&=\lim_{t\to\infty}t\left({\mathbb E}_t(\varepsilon_j\varepsilon_k)
-{\mathbb E}_t(\varepsilon_j){\mathbb E}_t(\varepsilon_k)\right)\nonumber\\[4mm]
&=\beta_j\beta_k\int_{-\infty}^\infty\omega_+\left((\Phi_j-\omega_+(\Phi_j))
(\Phi_{kt}-\omega_+(\Phi_k))\right)\d t.\label{exam2}
\end{align}
Here, $\omega_+$ is the NESS of $(\cO, \tau_V^t, \omega)$ and  we have assumed that the 
correlation function
\[t\mapsto\omega_+\left((\Phi_j-\omega_+(\Phi_j))(\Phi_{kt}-\omega_+(\Phi_k))\right),\]
is integrable on $\rr$. 

The fluctuations of ${\mathbb P}_t$ as $t\rightarrow \infty$  are described by a central limit theorem 
and a large deviation principle. The  central limit theorem holds if for all $\balpha \in \rr^n$, 
\index{theorem!central limit}\index{large deviation principle}
\[
\lim_{t\rightarrow \infty}\int_{\rr^n} \e^{\i \sqrt t \balpha \cdot ({\boldsymbol \varepsilon}-\langle 
{\boldsymbol \varepsilon}\rangle_+)}\d {\mathbb P}_t({\boldsymbol\varepsilon})=\int_{\rr^n}
\e^{\i \balpha \cdot {\boldsymbol \varepsilon}} \d \mu_{{\bf D}_{\rm fcs}}({\boldsymbol \varepsilon}),
\] 
where  $\mu_{{\bf D}_{\rm fcs}}$ is  the centered Gaussian measure on $\rr^n$ with covariance 
${\bf D}_{\rm fcs}=[D_{{\rm fcs}, jk}]$. To discuss the large deviation principle, recall that 
\[
e_{2, t}({\balpha})=\log\int_{\rr^n}\e^{-t {\balpha}\cdot {\boldsymbol\varepsilon}}
\d{\mathbb P}_t({\boldsymbol\varepsilon}).
\]
Suppose that 
\[
e_{2, +}({\balpha}) =\lim_{t\rightarrow \infty} \frac{1}{t}e_t({\balpha}),
\]
exists for ${\balpha}\in \rr^n$ and satisfies the conditions of G\"artner-Ellis theorem
\index{theorem!G\"artner-Ellis} (Theorem \ref{GEmultidim} in Appendix \ref{sect:GEmultiD}). 
Then for any Borel set $G\subset \rr^d$, 
\begin{align*}
-\inf_{{\bf s}\in\interior(G)} I({\bf s}) 
\leq\liminf_{t \rightarrow \infty} 
\frac{1}{t} \log \mathbb P_t\left(G\right) 
\leq\limsup_{t \rightarrow \infty} 
\frac{1}{t} \log \mathbb P_t\left(G\right) \leq 
-\inf_{{\bf s}\in\closure(G)} I({\bf s}),
\end{align*}
where
\[
I({\bf s})=-\inf_{\balpha\in \rr^n}\left({\bf s}\cdot\balpha+e_{2, +}(\balpha)\right).
\]
Note that $I({\bf s})$ satisfies the Evans-Searles symmetry \index{symmetry!Evans-Searles}
\[
I(-{\bf s})={\bf 1}\cdot{\bf s}+I({\bf s}).
\]
For some models the central limit theorem and the large deviation principle can be \index{theorem!central limit}
proven following the spectral scheme outlined in Section \ref{sect:LTFCS} (for example, this is 
the case for Spin-Fermion model, see Section \ref{sect:spin-fermion}). For other models, scattering 
techniques are effective (see Section \ref{sect:EBBM}). In general, however, verifications of the 
central limit theorem and the  large deviation principle are difficult problems.

Let now 
\[ 
X_j =\beta_\eq-\beta_j,
\]
be the thermodynamic forces. The new reference state $\omega_X$ is the TD limit
of the states $\omega_{M, X}$ of  the finite open quantum systems ${\cal Q}_M$. Alternatively, 
$\omega_X$ can be described directly in terms of the modular structure, see 
\cite{JOP1}. $\omega_X$ is modular and normal w.r.t. $\omega$. The entropy 
production observables of $(\cO, \tau_V^t, \omega_X)$ is 
\[
\sigma_X=\sum_{j=1}^n X_j \Phi_j.
\]
The NESS $\omega_{X+}$ also depends on $X$ and, for $X=0$, reduces to a 
$(\tau_V,\beta_\eq)$-KMS state $\omega_{\beta_\eq}$.
Let $e_t(X, Y)$ be the entropic functional of the infinitely extended system $(\cO,\tau_V^t,\omega_X)$
and suppose that (A10) holds. Then Proposition \ref{prop-exam} implies that the transport coefficients
\[
L_{jk}=\partial_{X_k}\omega_{X+}(\Phi_j)|_{X=0},
\]
are defined, satisfy the Onsager reciprocity relations \index{relation!Onsager reciprocity}
\[
L_{jk}=L_{kj},
\]
and the Green-Kubo formulas \index{formula!Green-Kubo}
\[
L_{jk}=\frac{1}{2}\int_{-\infty}^\infty\omega_{\beta_{\eq}}(\Phi_j\Phi_{kt}) \d t.
\]
Here  we have assumed that the quantum dynamical system $(\cO, \tau_V^t, \omega_{\beta_\eq})$
is mixing and that the correlation function $t \mapsto \omega_{\beta_{\eq}}(\Phi_j\Phi_{kt})$ is
integrable.

The linear response theory derived for quantum fluxes $\Phi_j$ immediately yields the linear response 
theory for the full counting statistics. Indeed, it follows from the formulas \eqref{exam1} and 
\eqref{exam2} that 
\[
L_{{\rm fcs}, kj}=\partial_{X_k}\langle \varepsilon_j\rangle_+|_{X=0}=-\beta_{\eq}L_{kj}
=-\frac1{\beta_\eq}D_{{\rm fcs}, kj}|_{X=0}.
\]
The last relation also  yields the Fluctuation-Dissipation Theorem for the full counting statistics. The
Einstein relation takes the form  \index{relation!Einstein}
\[
L_{{\rm fcs}, kj}=-\frac1{2\beta_\eq}D_{{\rm fcs}, kj}|_{X=0}.
\]
and relates the kinetic transport coefficients of the full counting statistics to its fluctuations in thermal 
equilibrium. The factor $-\beta_\eq^{-1}$ is due to our choice to keep the entropic form of  the full 
counting statistics in the discussion of energy transport. In the energy form of the full counting 
statistics one considers ${\mathbb E}_t(-\varepsilon_j/\beta_j)$ and then the Einstein relation
hold in the usual form $ L_{{\rm fcs}, kj}=\frac{1}{2}D_{{\rm fcs}, kj}|_{X=0}$. The disadvantage of 
the energy form is that the Evans-Searles symmetry has to be scaled. The choice between scaling 
Einstein relations or scaling symmetries is of course of  no substance. 

At this point let us introduce a ``naive" cumulant generating function \nindex{ent}{$e_{{\rm naive}, t}({\balpha})$}{naive cumulant generating function}
\begin{equation}
e_{{\rm naive}, t}({\balpha})=
\log\omega\left(\e^{-\sum_{j=1}^n\alpha_j\beta_j\int_0^t\Phi_{js}\d s}\right),
\label{naiveedef}
\end{equation}
and the ``naive" cumulants
\[
\chi_t(k_1, \ldots, k_n)=\partial_{\alpha_1}^{k_1}\cdots \partial_{\alpha_n}^{k_n}
e_{{\rm naive}, t}({\balpha})|_{{\balpha}=0}.
\]
The function $e_{{\rm naive}, t}({\balpha})$ is just the direct quantization of the classical
cumulant generating function for the entropy transfer
$$
{\bf S^t}=(S_1^t,\ldots,S_n^t)=\int_0^t(-\beta_1\Phi_{1s},\ldots,-\beta_n\Phi_{ns})\d s,
$$
in the state $\omega$. Except in the special case $\balpha =\alpha{\bf 1}$, 
$e_{{\rm naive}, t}({\balpha})$ cannot be described in terms of classical probability, \ie 
$e_{{\rm naive}, t}({\balpha})$ {\sl is not} the cumulant generating function of a probability measure 
on $\rr^n$. If $\balpha =\alpha{\bf 1}$, then 
\[
e_{{\rm naive}, t}(\alpha{\bf 1})=\log\omega\left(\e^{\alpha \int_0^t \sigma_s\d s}\right)
=\log \int_\rr\e^{t\alpha s}\d\mu_{\omega, t}(s),
\]
where, in the GNS-representation of $\cO$ associated to $\omega$, $\mu_{\omega, t}$ is the 
spectral measure for $t^{-1}\int_0^t \pi_\omega(\sigma_s)\d s$ and $\xi_\omega$.

In general the functional $e_{{\rm naive}, t}({\balpha})$ will not satisfy the Evans-Searles symmetry,
\ie $e_{{\rm naive}, t}({\bf 1}-{\balpha})\not =e_{{\rm naive}, t}({\balpha})$,
and the same remark applies to the limiting functional
\[
e_{{\rm naive}, +}({\balpha})=\lim_{t\to\infty}\frac{1}{t}e_{{\rm naive}, t}({\balpha}),
\]
which, we assume, exists and is differentiable on some open set containing $\bf 0$. One easily checks
that the first and second order cumulants satisfy 
\begin{align*}
\partial_{\alpha_j} e_{{\rm naive}, t}({\balpha})|_{{\balpha}={\bf 0}}
&=\partial_{\alpha_j} e_{2, t}({\balpha})|_{{\balpha}={\bf 0}},\\[3mm]
\partial_{\alpha_k}\partial_{\alpha_j} e_{{\rm naive}, t}({\balpha})|_{{\balpha}={\bf 0}}
&=\partial_{\alpha_k}\partial_{\alpha_j} e_{2, t}({\balpha})|_{{\balpha}={\bf 0}},
\end{align*}
and if the limits and derivatives could be exchanged, 
\begin{align*}
\partial_{\alpha_j} e_{{\rm naive}, +}({\balpha})|_{{\balpha}={\bf 0}}
&=\partial_{\alpha_j} e_{2, +}({\balpha})|_{{\balpha}={\bf 0}},\\[3mm]
\partial_{\alpha_k}\partial_{\alpha_j} e_{{\rm naive}, +}({\balpha})|_{{\balpha}={\bf 0}}
&=\partial_{\alpha_k}\partial_{\alpha_j} e_{2, +}({\balpha})|_{{\balpha}={\bf 0}}.
\end{align*}

We summarize our observations:
\begin{enumerate}[(i)]

\item The first and second order cumulants of the full counting statistics are  the same as the 
corresponding ``naive" quantum energy flux cumulants, \ie the direct quantization of the classical 
energy flux cumulants. In general, higher order ``naive'' cumulants do not coincide with
the corresponding  cumulants of the full counting statistics.

\item The limiting expectation $\langle {\boldsymbol\varepsilon}\rangle_+$ and covariance 
${\bf D}_{{\rm fcs}}$ of the full counting statistics are expressed in terms of the NESS $\omega_+$ and 
quantized fluxes $\Phi_j$. They are direct quantization of  the  
corresponding classical expressions. The same remark applies to the central limit theorem, linear 
response theory and fluctuation-dissipation theorem. If the full counting statistics is restricted to the 
entropy production observable, then its  limiting expectation, covariance and central limit theorem 
coincide with those of the spectral measure for  $t^{-1}\int_0^t \sigma_s\d s$ and  $\omega$.

\item We emphasize: to detect the difference  between full counting statistics  and  the ``naive" cumulant generating 
function one needs to consider cumulants of at least third order. In  Chapter \ref{chap:FERMION} 
we shall illustrate this point on  some examples  of physical interest.
\end{enumerate}

\chapter{Fermionic systems}
\label{chap:FERMION}

In this section we discuss non-equilibrium statistical mechanics of fermionic systems and describe 
several  physically relevant  models to which the structural theory 
developed  in these lecture notes applies.

\section{Second quantization}
\label{sect:SecondQuantiz}

We start with some notation. Let $\cQ$ be a finite set. $\ell^2(\cQ)$ denotes the Hilbert space of all 
function $f:\cQ\rightarrow\cc$ equipped with the inner product 
\[
\langle f|g\rangle=\sum_{q\in\cQ}\bar{f(q)}g(q).
\]
The functions $\{\delta_{q}\,|\,q\in\cQ\}$, where $\delta_q(x)=1$ if $x=q$ and $0$ otherwise, form 
an orthonormal basis for $\ell^2(\cQ)$. Any Hilbert space of dimension $|\cQ|$ is isomorphic to 
$\ell^2(\cQ)$. 

Let the configuration space of a single particle be the finite set $\cQ$. Typically, $\cQ$  will be a subset 
of some lattice, but at this point we do not need to specify its structure further. The Hilbert space of 
a single particle is ${\cal K}=\ell^2(\cQ)$. If $\psi\in{\cal K}$ is a normalized wave function, then 
$|\psi(q)|^2$ is probability that the particle is located at $q\in\cQ$. The configuration space of a
system of $n$ distinguishable particles is $\cQ^n$ and $\ell^2(\cQ^n)$ is its Hilbert space.
For $q=(q_1,\ldots,q_n)\in\cQ^n$ we set 
$\delta_q(x_1,\ldots,x_n)=\delta_{q_1}(x_1)\cdots\delta_{q_n}(x_n)$. 
$\{\delta_q\,|\,q\in\cQ^n\}$ is an orthonormal basis of $\ell^2(\cQ^n)$. Let $\cK^{\otimes n}$
\nindex{Kon}{$\cK^{\otimes n}$}{$n$-fold tensor product} be the $n$-fold tensor 
product of $\cK$ with itself. Identifying $\delta_q$ with
$\delta_{q_1}\otimes\cdots\otimes\delta_{q_n}$ we obtain an isomorphism between
$\ell^2(\cQ^n)$ and $\cK^{\otimes n}$. In the following we shall identify these two spaces.

If $\psi\in\cK^{\otimes n}$ is the normalized wave function of the system of $n$ particles
and $\psi_1,\ldots,\psi_n\in\cK$ are normalized one-particle wave functions, then 
$|\langle\psi|\psi_1\otimes\cdots\otimes\psi_n\rangle|^2$ 
is the probability for the $j$-th particle to be in the state $\psi_j$, $j=1,\ldots,n$.
According to Pauli's principle, if the  particles are identical fermions, then 
this probability must vanish if at least two of the $\psi_j$'s are equal. It follows that the
multilinear functional $F(\psi_1,\ldots,\psi_n)=\langle\psi|\psi_1\otimes\cdots\otimes\psi_n\rangle$
has to vanish if at least two of its arguments coincide. Hence, for $j\not=k$,\index{Pauli principle}
$$
F(\psi_1,\ldots,\psi_j+\psi_k,\ldots,\psi_k+\psi_j,\ldots,\psi_n)=0,
$$
for any $\psi_1,\ldots,\psi_n\in\cK$. By multilinearity, this is equivalent to
\begin{align*}
0&=F(\psi_1,\ldots,\psi_j,\ldots,\psi_k,\ldots,\psi_n)+
F(\psi_1,\ldots,\psi_j,\ldots,\psi_j,\ldots,\psi_n)\\
&+F(\psi_1,\ldots,\psi_k,\ldots,\psi_k,\ldots,\psi_n)+
F(\psi_1,\ldots,\psi_k,\ldots,\psi_j,\ldots,\psi_n)\\
&=F(\psi_1,\ldots,\psi_j,\ldots,\psi_k,\ldots,\psi_n)+F(\psi_1,\ldots,\psi_k,\ldots,\psi_j,\ldots,\psi_n),
\end{align*}
and we conclude that  $F$ must be alternating, \ie, changing sign under transposition of two of its arguments,
\begin{equation}
F(\psi_1,\ldots,\psi_j,\ldots,\psi_k,\ldots,\psi_n)=-F(\psi_1,\ldots,\psi_k,\ldots,\psi_j,\ldots,\psi_n).
\label{AlternAting}
\end{equation}
Let $S_n$ be the group of permutations of the set $\{1,\ldots, n\}$. For $\pi\in S_n$ we set
$$
\pi \psi_1\otimes\cdots\otimes\psi_n=\psi_{\pi(1)}\otimes\cdots\otimes\psi_{\pi(n)},
$$
and extend this definition to $\cK^{\otimes n}$ by linearity. One easily checks that this action
of $S_n$ on $\cK^{\otimes n}$ is unitary. If $\pi=(jk)=\pi^{-1}$ is the transposition whose only effect is
to interchange $j$ and $k$,  then \eqref{AlternAting} is equivalent to
$$
\langle\pi\psi|\psi_1\otimes\cdots\otimes\psi_n\rangle
=\langle\psi|\pi\psi_1\otimes\cdots\otimes\psi_n\rangle
=-\langle\psi|\psi_1\otimes\cdots\otimes\psi_n\rangle,
$$
and so  $\pi\psi=-\psi$. More generally, if $\pi$ is the composition of $m$ 
transpositions, $\pi=(j_1k_1)\cdots(j_mk_m)$, then we must have $\pi\psi=(-1)^m\psi$. Any
permutation $\pi\in S_n$ can be decomposed into a product of transpositions and the corresponding
number $(-1)^m$, the signature of $\pi$, is denoted by ${\rm sign}(\pi)$ 
\nindex{sign}{${\rm sign}(\,\cdot\,)$}{signature of a permutation} (one can show that
$ {\rm sign}(\pi)=(-1)^t$ where $t$ is  the number of pairs $(j,k)\in \{1,\ldots n\}$ such that $j<k$ 
and $\pi(j)>\pi(k)$). We conclude that the wave function $\psi$  of a system of $n$ identical fermions
must satisfy
$$
\pi\psi={\rm sign}(\pi)\psi,
$$
for all $\pi\in S_n$. More explicitly, for $\pi \in S_n$ the wave function $\psi$ satisfies 
\begin{equation}
\psi(x_{\pi(1)},\ldots,x_{\pi(n)})={\rm sign}(\pi)\psi(x_1,\ldots,x_n).
\label{minn-snow}
\end{equation}
Functions satisfying \eqref{minn-snow} are  called completely antisymmetric. The set of all completely antisymmetric functions
on $\cQ^n$ is a subspace of $\ell^2(\cQ^n)$ which we denote by $\ell_-^2(\cQ^n)$.

\medskip
\begin{exo}\label{Exo:ExteriorAlg}
\exop Show that the orthogonal projection $P_-$ on $\ell_-^2(\cQ^n)$ is given by
$$
P_-\psi=\frac1{n!}\sum_{\pi\in S_n}{\rm sign}(\pi)\pi\psi.
$$
{\sl Hint}: use the morphism property of the signature, ${\rm sign}(\pi\circ\pi')={\rm sign}(\pi){\rm sign}(\pi')$,
to show that $\pi P_-={\rm sign}(\pi)P_-$.

\exop Define the wedge product of $\psi_1,\ldots,\psi_n\in\cK$ by
\[ 
\psi_1\wedge \cdots \wedge \psi_n=\sqrt{n!}\, P_-\psi_1\otimes \cdots \otimes\psi_n,
\]
and show that
\begin{equation}
\langle \psi_1\wedge\cdots \wedge \psi_n|\phi_1\wedge\cdots\wedge\phi_n\rangle 
=\det [\langle \psi_i|\phi_j\rangle]_{1\leq i, j\leq n}.
\label{LeibnitzForm}
\end{equation}
{\sl Hint}: use Leibnitz formula\bindex{formula!Leibnitz}
$$
\det A=\sum_{\pi\in S_n}{\rm sign}(\pi)A_{1\pi(1)}\cdots A_{n\pi(n)},
$$
for the determinant of the $n\times n$ matrix $A=[A_{jk}]$.\newline

\exop Denote by ${\cal K}^{\wedge n}$ 
\nindex{Kvn}{${\cal K}^{\wedge n}$}{antisymmetric $n$-fold tensor product} the linear span of 
the set $\{\psi_1\wedge \cdots \wedge \psi_n\,|\,\psi_1,\ldots,\psi_n\in\cK\}$.
Suppose that $n\leq d=|\cQ|=\dim\cK$ and let $\{\phi_1,\ldots,\phi_d\}$ be an orthonormal basis of 
$\cK$. Prove that
\[
\{\phi_{j_1}\wedge \cdots\wedge \phi_{j_n}\,|\, 1\leq j_1 <\cdots <j_n\leq d\},
\]
is an orthonormal basis of ${\cal K}^{\wedge n}$ and deduce that
\[
\dim {\cal K}^{\wedge n}=\binom{\dim\cK}{n}.
\]
In particular, the vector space ${\cal K}^{\wedge\dim\cK}$ is one dimensional. For $n >\dim\cK$ the 
vector spaces  $\cK^{\wedge n}$ are trivial, that is, consist only of the zero vector. 
\end{exo}

\medskip
According to our identification of $\cK^{\otimes n}$ with $\ell^2(\cQ^n)$, the subspaces
$\ell_-(\cQ^n)$ and $\cK^{\wedge n}$ coincide (they are both the range of the projection $P_-$).
We denote by
\[
\Gamma_n(\cK)= \cK^{\wedge n},
\]
the Hilbert space of a system of $n$ fermions with the single particle Hilbert space ${\cal K}$. 
By definition, $\Gamma_{0}(\cK)=\cc$ is the vacuum sector.

For $A\in\cO_\cK$ and $n\ge1$, let $\Gamma_n(A)$ and $\d\Gamma_n(A)$ be the elements of 
$\cO_{\Gamma_{n}(\cK)}$ defined by 
\[\Gamma_n(A)(\psi_1\wedge\cdots\wedge\psi_n) = A\psi_1\wedge\cdots\wedge A\psi_n,\]
\[\d\Gamma_n(A)(\psi_1\wedge\cdots\wedge\psi_n)= 
A\psi_1\wedge\cdots\wedge\psi_n + \cdots +\psi_1\wedge \cdots \wedge A\psi_n.\]
For $n=0$, we define $\Gamma_0(A)$ to be the identity map on $\Gamma_0(\cK)$ and set
$\d\Gamma_0(A)=0$. One easily checks the relations
\begin{align}
\Gamma_n(A^\ast)&=\Gamma_n(A)^\ast,
&\d\Gamma_n(A^\ast)&=\d\Gamma_n(A)^\ast,\nonumber\\
\Gamma_n(AB)&=\Gamma_n(A)\Gamma_n(B),
&\d\Gamma_n(A+\lambda B)&=\d\Gamma_n(A)+\lambda\d\Gamma_n(B),\label{GammanProps}\\
\d\Gamma_n(A)&=\left.\frac{\d\ }{\d t}\Gamma_n(\e^{tA})\right|_{t=0},
&\Gamma_n(\e^{A})&=\e^{ \d \Gamma_n(A)},\nonumber
\end{align}
for $A,B\in\cO_\cK$ and $\lambda\in\cc$. The Fermionic Fock space over $\cK$ is defined by
\bindex{Fock space}\nindex{G(K)}{$\Gamma(\cK)$}{fermionic Fock space}
\[
\Gamma (\cK)= \bigoplus_{n=0}^{\dim\cK} \Gamma_n(\cK),
\]
\ie as the set of vectors $\Psi=(\psi_0,\psi_1,\ldots)$ with $\psi_n\in\Gamma_n(\cK)$ and the 
inner product
$$
\langle\Psi|\Phi\rangle=\sum_{n=0}^{\dim\cK}\langle \psi_n|\phi_n\rangle.
$$
Clearly, 
$$
\dim\Gamma(\cK)=\sum_{n=0}^{\dim\cK}\dim\Gamma_n(\cK)=
\sum_{n=0}^{\dim\cK}\binom{\dim\cK}{n}=2^{\dim\cK}.
$$ 
A normalized vector $\Psi=(\psi_0,\psi_1,\ldots)\in\Gamma(\cK)$ is interpreted as a state of a
gas of identical fermions with one particle Hilbert space $\cK$ in the following way. Setting 
$p_n=\|\psi_n\|^2$, $\phi_n=\psi_n/\|\psi_n\|$ and $\Phi^{(n)}=(0,\ldots,\phi_n,\ldots,0)$ one can write
$\Psi$ as
$$
\Psi=\sum_{n=0}^{\dim\cK}\sqrt{p_n}\,\Phi^{(n)},
$$
a coherent superposition of:
\begin{itemize}
\item a state $\Phi^{(0)}$ with no particle. Up to a phase factor, $\Phi^{(0)}$ is the
so called vacuum vector \index{vector!vacuum}\nindex{Omega}{$\Omega$}{Fock vacuum vector}
$$
\Omega=(1,0,\ldots,0),
$$
\item a state $\Phi^{(1)}$ with $1$ particle in the state $\phi_1\in\cK$; 
\item a state $\Phi^{(2)}$ with $2$ particles in the state $\phi_2\in\Gamma_2(\cK)$, etc.
\end{itemize}
Since the vectors $\Phi^{(n)}$ are mutually orthogonal, $p_n$ is the probability for $n$ particles 
to be present in the system. Pauli's principle forbid more than $\dim\cK$ particles. With a slight
abuse of notation, we shall identify the $n$-particle wave function $\phi\in\Gamma_n(\cK)$ with
the vector $\Phi=(0,\ldots,\phi,\ldots,0)\in\Gamma(\cK)$.

For $A\in\cO_\cK$ one defines $\Gamma(A)$ and $\d\Gamma(A)$ in $\cO_{\Gamma(\cK)}$ by
\[
\Gamma(A)=\bigoplus_{n=0}^{\dim\cK} \Gamma_n(A), \qquad 
\d\Gamma(A)=\bigoplus_{n=0}^{\dim\cK} \d\Gamma_n(A).
\]
Relations \eqref{GammanProps} yield\nindex{G(A)}{$\Gamma(A)$}{second quantization of $A$}
\nindex{dG(A)}{$\d\Gamma(A)$}{differential second quantization of $A$}
\begin{align}
\Gamma(A^\ast)&=\Gamma(A)^\ast,
&\d\Gamma(A^\ast)&=\d\Gamma(A)^\ast,\nonumber\\
\Gamma(AB)&=\Gamma(A)\Gamma(B),
&\d\Gamma(A+\lambda B)&=\d\Gamma(A)+\lambda\d\Gamma(B),\label{GammaProps}\\
\d\Gamma(A)&=\left.\frac{\d\ }{\d t}\Gamma(\e^{tA})\right|_{t=0},
&\Gamma(\e^{A})&=\e^{ \d \Gamma(A)}.\nonumber
\end{align}
Note that $\Gamma(A)$ is invertible iff $A$ is invertible and 
in this case $\Gamma(A)^{-1} =\Gamma(A^{-1})$. Moreover, one easily checks that
\begin{equation}
\Gamma(A)\d\Gamma(B)\Gamma(A^{-1})=\d\Gamma(ABA^{-1}).
\label{GammadGammaGamma}
\end{equation}
In particular, one has
$$
\e^{t\d\Gamma(A)}\d\Gamma(B)\e^{-t\d\Gamma(A)}
=\Gamma(\e^{tA})\d\Gamma(B)\Gamma(\e^{-tA})=\d\Gamma(\e^{tA}B\e^{-tA}).
$$
which, upon differentiation at $t=0$, yields
\begin{equation}
[\d\Gamma(A),\d\Gamma(B)]=\d\Gamma([A,B]).
\label{GammaLie}
\end{equation}
The reader familiar with Lie groups will recognize $A\mapsto\Gamma(A)$ as a representation of the
linear group ${\rm GL}(\cK)$ in $\Gamma(\cK)$ and $B\mapsto\d\Gamma(B)$ as the induced
representation of its Lie algebra $\cO_\cK$.
\begin{example} $N=\d\Gamma(\one)$ is called the number operator. \bindex{number operator}
\nindex{N}{$N$}{number operator} Since
$$
\left.N\right|_{\Gamma_n(\cK)}=n\one_{\Gamma_n(\cK)},
$$
$N$ is the observable describing the number of particles in the system.
\end{example}

We finish this section with a result which will be important in Section \ref{sect:QFree}.
\bel\label{DetLemma}
For any $A\in\cO_\cK$, one has
$$
\tr(\Gamma(A))=\det(\one+A).
$$
\eel
\demo We first prove the result for self-adjoint $A$. Let $\{\psi_1,\ldots,\psi_d\}$ be an eigenbasis
of $A$ such that $A\psi_j=\lambda_j\psi_j$. Since 
\begin{align*}
\det(\one+A)&=\prod_{j=1}^d(1+\lambda_j)=\sum_{J\subset\{1,\ldots,d\}}\prod_{k\in J}\lambda_k\\
&=\sum_{n=0}^d\sum_{\atop{J\subset\{1,\ldots,d\}}{|J|=n}}\prod_{k\in J}\lambda_k
=\sum_{n=0}^d\,\sum_{1\le j_1<\cdots<j_n\le d}\lambda_{j_1}\cdots\lambda_{j_n},
\end{align*}
and $\lambda_{j_1}\cdots\lambda_{j_n}=\langle\psi_{j_1}\wedge\cdots\wedge\psi_{j_n}|
\Gamma_n(A)\psi_{j_1}\wedge\cdots\wedge\psi_{j_n}\rangle$, it follows from 
Part 3 of Exercise \ref{Exo:ExteriorAlg} that
$$
\sum_{1\le j_1<\cdots<j_n\le d}\lambda_{j_1}\cdots\lambda_{j_n}
=\tr_{\Gamma_n(\cK)}(\Gamma_n(A)).
$$
Hence, 
$$
\det(\one+A)=\sum_{n=0}^d\tr_{\Gamma_n(\cK)}(\Gamma_n(A))=\tr(\Gamma(A)),
$$
holds for self-adjoint $A$. 
If $A$ is not self-adjoint, we set
$$
A(\lambda)=\frac{A+A^\ast}2+\lambda\frac{A-A^\ast}{2\i}.
$$
Clearly, $A(\lambda)$ is self-adjoint for $\lambda\in\rr$  and so 
$\det(\one+A(\lambda))=\tr(\Gamma(A(\lambda)))$. Since both sides of this identity are
analytic functions of $\lambda$ (in fact, polynomials), the identity extends  to the value $\lambda=\i$
for which $A(\i)=A$.
\qed

\bigskip

\section{The canonical anticommutation relations (CAR)}
\label{sect:CAR}

For $\psi,\psi_1,\ldots,\psi_n\in\cK$ we set \nindex{a}{$a^\#$}{fermionic creation/annihilation operators}
\begin{align*}
a^\ast(\psi)\Omega&=\psi,\\
a^\ast(\psi)(\psi_1\wedge \cdots \wedge \psi_n)
&=\psi\wedge \psi_1\wedge \cdots \wedge \psi_n.
\end{align*}
By linearity, $a^\ast (\psi)$ extends to an element of $\cO_{\Gamma(\cK)}$ which
maps $\Gamma_{n}(\cK)$ into $\Gamma_{n+1}(\cK)$ and in particular $\Gamma_{\dim\cK}(\cK)$ 
to $\{0\}$. Since $a^\ast(\psi)$ acts on a state $\Psi$ by adding to it a particle in the state $\psi$,
it is called creation operator. We note that 
$$
\psi_1\wedge \cdots \wedge \psi_n=a^\ast(\psi_1)\cdots a^\ast(\psi_n)\Omega.
$$

Similarly, one defines an element $a(\psi)$ of $\cO_{\Gamma(\cK)}$ by
\begin{align*}
a(\psi)\Omega&=0,\\
a(\psi)\psi_1&=\langle\psi|\psi_1\rangle\Omega,\\
a(\psi)(\psi_1\wedge\cdots \wedge \psi_n)&=\sum_{j=1}^n (-1)^{1+j}\langle\psi|\psi_j\rangle\,
\psi_1\wedge\cdots  \wedge\bcancel{\psi_{j}}\wedge \cdots \wedge \psi_n.
\end{align*}
$a(\psi)$  maps $\Gamma_{n}(\cK)$ into $\Gamma_{n-1}(\cK)$ and in particular $\Gamma_0(\cK)$ to $\{0\}$.
Since it acts on a state $\Psi$ by removing from it a particle in the state $\psi$, it is called
annihilation operator. In the sequel, $a^\#(\psi)$ denotes either $a^\ast(\psi)$ or $a(\psi)$. 
The basic properties of creation and annihilation operators are summarized in 

\bep\label{midterm1}
\ben
\item The map $\psi\mapsto a^\ast(\psi)$ is linear and the map $\psi\mapsto a(\psi)$ is 
anti-linear.
\item $a(\psi)^\ast = a^\ast(\psi)$.
\item The Canonical Anticommutation Relations {\rm (CAR)} hold: 
\index{CAR|see{canonical anticommutation relations}}\bindex{canonical anticommutation relations}%
\[
\{ a(\psi), a(\phi)\}=\{a^\ast(\psi), a^\ast(\phi)\}=0,\qquad
\{ a(\psi), a^\ast(\phi)\}=\langle \psi|\phi\rangle\one,
\]
where $\{A, B\}=AB+BA$ denotes the anticommutator of $A$ and $B$.
\item The family of operators ${\mathfrak A}=\{a^\#(\psi)\,|\,\psi\in\cK\}$ is irreducible in 
$\cO_{\Gamma(\cK)}$, that is, 
$$
{\mathfrak A}'=\{B\in\cO_{\Gamma(\cK)}\,|\,[A,B]=0 \text{ for all } A\in{\mathfrak A}\}=\cc\one_{\Gamma(\cK)}.
$$
\item $\|a^\ast(\psi)\|= \|a(\psi)\|=\|\psi\|$.
\item For any $A\in\cO_\cK$, 
\[ 
\Gamma(A) a^\ast(\psi)=a^\ast(A\psi)\Gamma(A),\qquad
\Gamma(A^\ast)a(A\psi)=a(\psi)\Gamma(A^\ast).
\]
In particular, if $U$ is unitary, 
\[ \Gamma(U)a^\#(\psi)\Gamma(U^\ast)=a^\#(U\psi).
\]
\item  For any $A\in\cO_\cK$, 
$$[\d\Gamma(A),a^\ast(\psi)]=a^\ast(A\psi),\qquad [\d\Gamma(A),a(\psi)]=-a(A^\ast\psi).
$$
In particular, if $A$ is self-adjoint, 
$$
\i[\d\Gamma(A),a^\#(\psi)]=a^\#(\i A\psi).
$$
\item  $a^\ast(\phi)a(\psi)=\d \Gamma(|\phi\rangle\langle\psi|)$.
\item For any $A\in\cO_\cK$ and any orthonormal basis $\{\psi_1,\ldots,\psi_d\}$ of
$\cK$ one has
$$
\d\Gamma(A)=\sum_{j,k=1}^d\langle\psi_j|A\psi_k\rangle a^\ast(\psi_j)a(\psi_k).
$$ 
\een
\eep
\demo (1) is obvious from the definitions of the creation/annihilation operators.

\medskip\noindent(2) follows from Laplace formula for developing the determinant of a $n\times n$ 
matrix $A$ along one of its row,\bindex{formula!Laplace}
\begin{equation}
\det A=\sum_{j=1}^n(-1)^{i+j}A_{ij}\det A_{(ij)},
\label{LaplaceDet}
\end{equation}
where $A_{(ij)}$ denotes the matrix obtained from $A$ be removing its $i$-th row and $j$-th column.
Indeed, by \eqref{LeibnitzForm}
\begin{align*}
\langle\phi_1\wedge\cdots\wedge\phi_{n-1}|a^\ast(\psi)^\ast\psi_1\wedge\cdots\wedge\psi_n\rangle
&=\langle a^\ast(\psi)\phi_1\wedge\cdots\wedge\phi_{n-1}|\psi_1\wedge\cdots\wedge\psi_n\rangle\\
&=\langle\psi\wedge\phi_1\wedge\cdots\wedge\phi_{n-1}|\psi_1\wedge\cdots\wedge\psi_n\rangle\\
&=\det A,
\end{align*}
where
$$
A=\left[
\begin{array}{cccc}
\langle\psi|\psi_1\rangle&\langle\psi|\psi_2\rangle&\cdots&\langle\psi|\psi_n\rangle\\
\langle\phi_1|\psi_1\rangle&\langle\phi_1|\psi_2\rangle&\cdots&\langle\phi_1|\psi_n\rangle\\
\vdots&\vdots&\ddots&\vdots\\
\langle\phi_{n-1}|\psi_1\rangle&\langle\phi_{n-1}|\psi_2\rangle&\cdots&\langle\phi_{n-1}|\psi_n\rangle
\end{array}
\right].
$$
Developing the determinant of $A$ along its first row and using the fact that
$$
\det A_{(1j)}=\langle\phi_1\wedge\cdots\wedge\phi_{n-1}|\psi_1\wedge\cdots\wedge
\bcancel{\psi_j}\wedge\cdots\wedge\psi_n\rangle,
$$ 
we obtain
$$
\det A=\sum_{j=1}^{n}(-1)^{1+j}\langle\psi|\psi_j\rangle\langle\phi_1\wedge\cdots\wedge\phi_{n-1}|
\psi_1\wedge\cdots\wedge\bcancel{\psi_j}\wedge\cdots\wedge\psi_n\rangle.
$$
Hence, 
$$
a^\ast(\psi)^\ast\psi_1\wedge\cdots\wedge\psi_n=\sum_{j=1}^{n}(-1)^{1+j}\langle\psi|\psi_j\rangle
\psi_1\wedge\cdots\wedge\bcancel{\psi_j}\wedge\cdots\wedge\psi_n,
$$
and we conclude that $a(\psi)^\ast=a^\ast(\psi)$.

\medskip\noindent(3) The relation $\{a^\ast(\psi),a^\ast(\phi)\}=0$ follows from the fact that 
$\psi\wedge\phi\wedge\psi_1\cdots\wedge\psi_n$ changes sign when $\psi$ and $\phi$
are exchanged. The relation $\{a(\psi),a(\phi)\}=0$ is obtained by conjugating the previous
relation. Finally, adding the two formulas
\begin{align*}
a^\ast( \phi)a(\psi)\psi_1\wedge\cdots\wedge\psi_n&=\sum_{j=1}^n(-1)^{j+1}\langle\psi|\psi_j\rangle
\phi\wedge\psi_1\wedge\cdots\wedge\bcancel{\psi_j}\wedge\cdots\wedge\psi_n,\\
a( \psi)a^\ast(\phi)\psi_1\wedge\cdots\wedge\psi_n&=(-1)^{1+1}\langle\psi|\phi\rangle
\psi_1\wedge\cdots\wedge\psi_n\\
&+\sum_{j=1}^n(-1)^{j+2}\langle\psi|\psi_j\rangle
\phi\wedge\psi_1\wedge\cdots\wedge\bcancel{\psi_j}\wedge\cdots\wedge\psi_n,
\end{align*}
yields the last relation $\{a^\ast(\phi),a(\psi)\}=\langle\psi|\psi_j\rangle\one$.

\medskip\noindent(4) We first notice that if $\Psi\in\Gamma(\cK)$ is such that $a(\psi)\Psi=0$ for all 
$\psi\in\cK$, then
$$
\langle \psi_n\wedge\cdots\wedge\psi_1|\Psi\rangle
=\langle a^\ast(\psi_n)\psi_{n-1}\wedge\cdots\wedge\psi_1|\Psi\rangle
=\langle \psi_{n-1}\wedge\cdots\wedge\psi_1|a(\psi_n)\Psi\rangle=0,
$$
from which we conclude that $\Psi\perp\Gamma_n(\cK)$ for $n\ge1$. Hence, $\Psi\in\Gamma_0(\cK)$,
\ie $\Psi=\lambda\Omega$ for some $\lambda\in\cc$. Let $B\in\cO_{\Gamma(\cK)}$ commute 
with all creation/annihilation operators. It follows that $a(\psi)B\Omega=Ba(\psi)\Omega=0$ for all
$\psi\in\cK$. From the previous remark, we conclude that $B\Omega=\lambda\Omega$ for some
$\lambda\in\cc$. Then, we can write
\begin{align*}
B\psi_1\wedge\cdots\wedge\psi_n&=Ba^\ast(\psi_1)\cdots a^\ast(\psi_n)\Omega\\
&=a^\ast(\psi_1)\cdots a^\ast(\psi_n)B\Omega\\
&=\lambda a^\ast(\psi_1)\cdots a^\ast(\psi_n)\Omega
=\lambda\psi_1\wedge\cdots\wedge\psi_n,
\end{align*}
which shows that $B|_{\Gamma_n(\cK)}=\lambda\one_{\Gamma_n(\cK)}$ and 
that $B=\lambda\one_{\Gamma(\cK)}$.

\medskip\noindent(5) is obvious if $\psi=0$. The CAR imply
\begin{align*}
(a^\ast(\psi)a(\psi))^2&=a^\ast(\psi)(\{a(\psi),a^\ast(\psi)\}-a^\ast(\psi)a(\psi))a(\psi)\\
&=\langle\psi|\psi\rangle a^\ast(\psi)a(\psi)-a^\ast(\psi)^2a(\psi)^2\\
&=\|\psi\|^2a^\ast(\psi)a(\psi),
\end{align*}
from which we deduce $\|a^\ast(\psi)a(\psi)\|^2=\|(a^\ast(\psi)a(\psi))^2\|=\|\psi\|^2\|a^\ast(\psi)a(\psi)\|$.
If $\psi\not=0$ then $a(\psi)\not=0$ and hence $\|a^\ast(\psi)a(\psi)\|\not=0$ so that we can conclude
$$
\|a(\psi)\|^2=\|a^\ast(\psi)\|^2=\|a^\ast(\psi)a(\psi)\|=\|\psi\|^2.
$$

\medskip\noindent(6) It follows from the definitions that
$\Gamma(A)a^\ast(\psi)\Omega=\Gamma(A)\psi=A\psi=a^\ast(A\psi)\Gamma(A)\Omega$ and
\begin{align*}
\Gamma(A)a^\ast(\psi)\psi_1\wedge\cdots\wedge\psi_n
&=\Gamma(A)\psi\wedge\psi_1\wedge\cdots\wedge\psi_n\\
&=A\psi\wedge A\psi_1\wedge\cdots\wedge A\psi_n\\
&=a^\ast(A\psi)\Gamma(A)\psi_1\wedge\cdots\wedge\psi_n.
\end{align*}
Thus, one has $\Gamma(A)a^\ast(\psi)=a^\ast(A\psi)\Gamma(A)$. By conjugation, we also get
$\Gamma(A^\ast)a(A\psi)=a(\psi)\Gamma(A^\ast)$.

\medskip\noindent(7) It follows from (6) that
$$
\e^{t\d\Gamma(A)}a^\ast(\psi)=a^\ast(\e^{tA}\psi)\e^{t\d\Gamma(A)}.
$$
Differentiation at $t=0$ yields the first relation in (7). The second is obtained by conjugation.

\medskip\noindent(8) The CAR imply 
\begin{align*}
[a^\ast(\phi)a(\psi),a^\ast(\chi)]&=a^\ast(\phi)a(\psi)a^\ast(\chi)-a^\ast(\chi)a^\ast(\phi)a(\psi)\\
&=a^\ast(\phi)a(\psi)a^\ast(\chi)+a^\ast(\phi)a^\ast(\chi)a(\psi)\\
&=a^\ast(\phi)\{a(\psi),a^\ast(\chi)\}
=\langle\psi|\chi\rangle a^\ast(\phi).
\end{align*}
On the other hand, (7) implies that 
$[\d\Gamma(|\phi\rangle\langle\psi|),a^\ast(\chi)]=\langle\psi|\chi\rangle a^\ast(\phi)$. Thus, setting
$B=a^\ast(\phi)a(\psi)-\d\Gamma(|\phi\rangle\langle\psi|)$ we get $[B,a^\ast(\chi)]=0$
for all $\chi\in\cK$. Interchanging $\phi$ and $\psi$, we obtain in the same way
$[B,a(\chi)]^\ast=-[B^\ast,a^\ast(\chi)]=0$, and so $[B, a(\chi)]=0$. Hence $B\in {\mathfrak A}^\prime$  and   (4) implies 
   that $B=\lambda\one$ for some
$\lambda\in\cc$. Since $B\Omega=0$ we conclude that $B=0$.

\medskip\noindent(9) Follows from (8) and the representation 
$A=\sum_{j,k=1}^d\langle\psi_j|A\psi_k\rangle |\psi_j\rangle\langle\psi_k|$.
\qed

\bigskip
Given a Hilbert space $\cK$, a representation of the CAR \bindex{representation!of CAR}
over $\cK$ on a Hilbert space $\cH$ is a 
pair of maps 
$$
\psi \mapsto b(\psi),\qquad\psi \mapsto b^\ast(\psi),
$$
from $\cK$ to $\cO_\cH$ 
satisfying Properties (1)--(3) of Proposition \eqref{midterm1}. Such a representation is called
irreducible if it also satisfies Property (4) with $\cO_{\Gamma(\cK)}$ replaced by $\cO_\cH$.
The particular irreducible representation $\psi\mapsto a^\#(\psi)$ on $\Gamma(\cK)$ is called the Fock 
representation.\bindex{representation!Fock} We will construct another important representation 
of the CAR in Sections \ref{sect:AWrepresentation} and \ref{Sect:JWrepres}.

\bep\label{UniqueIredCAR}
 Let $\cK$ be a finite dimensional Hilbert space and $\psi\mapsto b^\#(\psi)$ an irreducible 
representation of the CAR over $\cK$ on $\cH$. Then, there exists a unitary operator 
$U:\Gamma(\cK)\rightarrow\cH$ such that  $Ua^\#(\psi)U^\ast= b^\#(\psi)$ for all $\psi\in\cK$.  
Moreover, $U$ is unique up to a phase factor. 
\eep
In other words, any two irreducible representations of the CAR over a finite dimensional Hilbert space 
are unitarily equivalent.  A proof of Proposition \ref{UniqueIredCAR} is sketched in the
next exercise.

\medskip
\begin{exo}
Let $\cK\ni\psi\mapsto b(\psi)\in\cO_\cH$ be an irreducible representation of CAR
over the $d$-dimensional Hilbert 
space $\cK$ in the Hilbert space $\cH$. 
Denote by $\{\chi_1,\ldots,\chi_d\}$ an orthonormal basis of $\cK$ an set
$$
\widetilde N=\sum_{n=1}^db^\ast(\chi_n)b(\chi_n).
$$
\exop Show that $0\le\widetilde N\le d\one$ and $\widetilde Nb(\psi)=b(\psi)(\widetilde N-\one)$ for any 
$\psi\in\cK$.

\exop Let $\phi\in\cH$ be a normalized eigenvector to the smallest eigenvalue of $\widetilde N$. Show that
$b(\psi)\phi=0$ for all $\psi\in\cK$.

\exop Set $\cH_0=\cc\phi$ and denote by $\cH_n$ the linear span of
$\{b^\ast(\psi_1)\cdots b^\ast(\psi_n)\phi\,|\,\psi_1,\ldots, \psi_n\in\cK\}$.
Show that $\cH_n\perp\cH_m$ for $n\not=m$ and $\cH_n=\{0\}$ for $n>d$. 

\noindent{\sl Hint}: show that
$\widetilde N|_{\cH_n}=n\one_{\cH_n}$.

\exop Show that
$$
\langle b^\ast(\psi_1)\cdots b^\ast(\psi_n)\phi|b^\ast(\psi_1')\cdots b^\ast(\psi_n')\phi\rangle
=\det[\langle\psi_i|\psi_j'\rangle]_{1\le i,j\le n},
$$
and conclude that the map
$\psi_1\wedge\cdots\wedge\psi_n\mapsto b^\ast(\psi_1)\cdots b^\ast(\psi_n)\phi$ extends to an
isometry $U:\Gamma(\cK)\to\cH$.

\exop Show that $Ua^\#(\psi)U^\ast=b^\#(\psi)$.

\exop Show that $[UU^\ast,b(\psi)]=0$ for all $\psi\in\cK$ and conclude that $U$ is unitary.
\end{exo}

\medskip
One can hardly overestimate the importance of the CAR. Indeed, as we shall see, they
characterize completely the algebra of observables of a Fermi gas with a given finite-dimensional
one-particle Hilbert space $\cK$.

\bep\label{IrredCriterion}
A representation $\psi\mapsto b^\#(\psi)$ of the CAR over the finite dimensional Hilbert space
$\cK$ in $\cH$ is irreducible iff the smallest $\ast$-subalgebra of $\cO_\cH$ containing the
set ${\mathfrak B}=\{b^\#(\psi)\,|\,\psi\in\cK\}$ is $\cO_\cH$.
\eep
Note that the smallest $\ast$-subalgebra of $\cO_\cH$ containing ${\mathfrak B}$ must contain
all polynomials in the operators $b^\#(\psi)$, \ie all linear combinations of monomials
of the form $b^\#(\psi_1)\cdots b^\#(\psi_k)$. But the set of all these polynomials is obviously
a $\ast$-algebra. Hence, a representation $\psi\mapsto b^\#(\psi)$ is irreducible iff any operator 
on $\cH$ can be written as a polynomial in the operators $b^\#$. We can draw important
conclusions from this fact:

\begin{enumerate}
\item Since the Fock representation $\psi\mapsto a^\#(\psi)$ is irreducible, any 
operator on the Fock space $\Gamma(\cK)$ is a polynomial in the creation/annihilation operators
$a^\#$.

\item Any representation of the CAR over $\cK$ on a Hilbert space $\cH$ 
extends to a representation of the $\ast$-algebra $\cO_{\Gamma(\cK)}$ on $\cH$, \ie 
to a $\ast$-morphism $\pi:\cO_{\Gamma(\cK)}\to\cO_\cH$.

\item If the representation is irreducible, this morphism is an isomorphism.
\end{enumerate}

To prove Proposition \ref{IrredCriterion}, we shall need the following result, von Neumann's 
bicommutant theorem.  A subset ${\mathfrak A}\subset\cO_\cK$
is called self-adjoint if $A\in{\mathfrak A}$ implies $A^\ast\in{\mathfrak A}$ and unital if
$\one\in{\mathfrak A}$.

\bet\label{BiCommThm}
Let $\cK$ be a finite dimensional Hilbert space and $\mathfrak A$ a unital self-adjoint subset of $\cO_\cK$. 
Then its bicommutant ${\mathfrak A}''$ is the smallest $\ast$-subalgebra of $\cO_\cK$ containing 
${\mathfrak A}$.\bindex{theorem!von Neumann bicommutant}
\eet

\demo
Denote by $\cal A$ the smallest $\ast$-subalgebra of $\cO_\cK$ containing $\mathfrak A$, \ie
the set of polynomials in elements of $\mathfrak A$. One clearly has ${\cal A}'={\mathfrak A}'$ and hence
${\cal A}''={\mathfrak A}''$. Thus, it suffices to show that $\cal A={\cal A}''$ (a $\ast$-algebra satisfying
this condition is a von Neumann algebra, and we are about to show that any finite dimensional
unital $\ast$-algebra is a von Neumann algebra).\bindex{algebra!von Neumann}

Since any element of $\cal A$ commutes with all elements of $\cal A'$ one obviously have
${\cal A}\subset{\cal A}''$. We must prove the reverse inclusion.
Let $\{\psi_1,\ldots,\psi_n\}$ be a basis of  $\cK$, $\{e_1,\ldots,e_n\}$ a basis of $\cc^n$ and set
$$
\Psi=\sum_{j=1}^n\psi_j\otimes e_j\in\cH=\cK\otimes\cc^n.
$$
To any $A\in\cO_\cK$ we associate the linear operator $\widehat A=A\otimes\one\in\cO_\cH$.
It follows that $\widehat{\cal A}=\{\widehat A\,|\,A\in{\cal A}\}$ is a $\ast$-subalgebra 
of $\cO_\cH$ and  $\widehat{\cal A}\Psi=\{\widehat A\Psi\,|\,A\in{\cal A}\}$ a subspace of $\cH$.
Denote by $P$ the orthogonal projection of $\cH$ onto this subspace. We claim that 
$P\in\widehat{\cal A}'$. Indeed, for any $\widehat A\in\widehat{\cal A}$ and $\Phi\in\cH$, one has
$\widehat AP\Phi\in\widehat{\cal A}\Psi$, and hence
$$
\widehat AP\Phi=P\widehat AP\Phi.
$$
We deduce that $\widehat AP=P\widehat AP$ for all $\widehat A\in\widehat{\cal A}$ and since
$\widehat{\cal A}$ is self-adjoint, one also has
$$
P\widehat{A}=(\widehat{A}^\ast P)^\ast=(P\widehat{A}^\ast P)^\ast=P\widehat AP=\widehat AP.
$$
Since $\cal A$ is unital, so is $\widehat{\cal A}$. It follows that 
$\Psi\in\widehat{\cal A}\Psi$ and hence $P\Psi=\Psi$. Recall that $X\in\cO_\cH$ is described by a 
$n\times n$ matrix $[X_{jk}]$ of elements of $\cO_\cK$ (see Section \ref{sect:CPmaps}) via  the 
formula
$$
X(\psi\otimes e_k)=\sum_{j=1}^n(X_{jk}\psi)\otimes e_j.
$$
Consequently, one has ${\widehat{\cal A}}'=\{X=[X_{jk}]\,|\,X_{jk}\in{\cal A}'\}$.
Let $B\in{\cal A}''$.  By the previous formula,  $\widehat B\in\widehat{\cal A}''$, and so 
$\widehat B$ commutes with $P$. We conclude that
$$
\widehat B\Psi=\widehat B P\Psi=P\widehat B\Psi\in\widehat{\cal A}\Psi,
$$
and so there exists $A\in{\cal A}$ such that $\widehat B\Psi=\widehat A\Psi$, \ie
$$
B\psi_j=A\psi_j,
$$
for $j=1,\ldots,n$. We conclude that $B=A\in{\cal A}$. 
\qed

\bigskip
{\bf\noindent Proof of Proposition \ref{IrredCriterion}.} Note that
$\{b^\ast(\psi),b(\psi)\}=\|\psi\|^2\one$, so that any $\ast$-subalgebra of $\cO_\cH$ containing
$$
{\mathfrak B}=\{b^\#(\psi)\,|\,\psi\in\cK\},
$$
also contains the unital self-adjoint subset $\widetilde{\mathfrak B}={\mathfrak B}\cup\{\one\}$. 
It follows that the smallest $\ast$-subalgebra of $\cO_\cH$ containing ${\mathfrak B}$ coincide
with the smallest $\ast$-subalgebra of $\cO_\cH$ containing $\widetilde{\mathfrak B}$. Moreover, 
one clearly has $\widetilde{\mathfrak B}'={\mathfrak B}'$ and hence 
$\widetilde{\mathfrak B}''={\mathfrak B}''$.
By the von Neumann bicommutant theorem,
$\mathfrak B''$ is the smallest $\ast$-subalgebra of $\cO_\cH$ containing $\mathfrak B$.
Now the representation $\psi\mapsto b^\#(\psi)$ is irreducible iff ${\mathfrak B}'=\cc\one$, \ie
iff ${\mathfrak B}''=\cO_\cH$.
\qed

\medskip
\begin{exo}\label{Exo:ExpoMap}
Let $\cK_1$ and $\cK_2$ be two finite dimensional Hilbert spaces.
Show that there exists a unitary map 
$U:\Gamma(\cK_1\oplus\cK_2)\to\Gamma(\cK_1)\otimes\Gamma(\cK_2)$ such that
$U\Omega=\Omega\otimes\Omega$ and
$$
Ua(\psi\oplus\phi)U^\ast=a(\psi)\otimes\one+\e^{\i\pi N}\otimes a(\phi).
$$
{\sl Hint}: try to apply Proposition \ref{UniqueIredCAR}.
\end{exo}

\medskip
{\noindent\bf Remark.} Apart from a few important  exceptions, 
 the material of this and the previous section
extends with minor changes to the case where $\cK$ is an  infinite dimensional Hilbert space. For example:
\begin{enumerate}
\item The definition of the Fock space $\Gamma(\cK)$ has to be complemented with the obvious
topological condition that $\Psi=(\psi_0,\psi_1,\ldots)\in\Gamma(\cK)$ iff 
$\|\Psi\|^2=\sum_{n\in\nn}\|\psi_n\|^2<\infty$. 
\item The definition of $\Gamma_n(A)$ carries over to bounded operators $A$ on $\cK$ and 
$\|\Gamma_n(A)\|\le\|A\|^n$. Thus, $\Gamma(A)=\oplus_{n\ge0}\Gamma_n(A)$ is well defined if:
\begin{itemize} 
\item $\|A\|\le1$, and then $\|\Gamma(A)\|=\sup_{n\ge0}\|\Gamma_n(A)\|=1$. 
In particular, if $U$ is unitary, so is $\Gamma(U)$.
\item $A$ has finite rank $m$ so that $\Gamma_n(A)=0$ for $n>m$ and then
$\|\Gamma(A)\|=\sup_{n\ge0}\|\Gamma_n(A)\|\le\max(1,\|A\|^m)$. In fact, using the polar
decomposition $A=U|A|$ together with Lemma \ref{DetLemma}, one sees that $\Gamma(A)$
is trace class with $\|\Gamma(A)\|_1=\tr\,\Gamma(|A|)=\det(\one+|A|)$. By a simple approximation
argument, one can then show that $\Gamma(A)$ is well defined and trace class provided
$A$ is trace class, and Lemma \ref{DetLemma} carries over.
\end{itemize}
\item If $A$ generates a strongly continuous contraction semi-group $\e^{t A}$ on $\cK$, then 
$\d\Gamma(A)$ is defined as the generator of the strongly continuous contraction semi-group 
$\Gamma(\e^{tA})$ on $\Gamma(\cK)$. In particular, if $A$ is self-adjoint, so is $\d\Gamma(A)$.
However, some care is required since $\d\Gamma(A)$ is unbounded unless  $A=0$. If $A$ is bounded,
the dense subspace $\Gamma_{\rm fin}(\cK)=\cup_{n\ge0}(\oplus_{k\le n}\Gamma_k(\cK))$ of 
$\Gamma(\cK)$ is a core of $\d\Gamma(A)$ and on this subspace, $\d\Gamma(A)$ acts as in the
finite dimensional case. 
\item The definition of the creation/annihilation operators carries over without change.
Parts (1)--(5) of Proposition \ref{midterm1} hold with the same proofs while Parts (6)--(8) are
easily adapted. Part (9) still holds if $A$ is trace class and it follows that
$\|\d\Gamma(A)\|\le\|A\|_1$.
\item The unitary equivalence described in Exercise \ref{Exo:ExpoMap} still holds for infinite
dimensional $\cK_1$ and $\cK_2$ (prove it!).
\end{enumerate}
Proposition \ref{UniqueIredCAR} does not hold for infinite dimensional
$\cK$. In fact, there are many unitarily inequivalent irreducible representations of the CAR over $\cK$.
Also Proposition \ref{IrredCriterion} and Theorem \ref{BiCommThm} do not hold for infinite
dimensional $\cK$. In the latter, one has to replace ``smallest $\ast$-subalgebra of $\cO_\cK$''
by ``smallest weakly closed $\ast$-subalgebra of $\cO_\cK$'' (see, \eg Theorem 2.4.11 in
\cite{BR1}). Proposition \ref{IrredCriterion} has to be modified accordingly:
The representation $\psi\mapsto b^\#(\psi)$ in $\cH$ is irreducible iff any bounded operator on 
$\cH$ is a weak limit of a net of polynomials in the elements of $\mathfrak B$.

\section{Quasi-free states of the CAR algebra}
\label{sect:QFree}

We now turn to states of a free Fermi gas. Let $T\in\cO_\cK$ be a non-zero operator satisfying 
$0 \leq T <\one$. In our context, we shall refer to $T$ as {\em density operator} or just {\em density}. 
To such $T$ we associate  density matrix on $\Gamma(\cK)$ by \bindex{density}
\[
\omega_T =\frac{1}{Z_T}\Gamma\left(\frac{T}{\one -T}\right),
\]
where 
\[
Z_T=\tr \left(\Gamma\left(\frac{T}{\one -T}\right)\right).
\]
As usual, we denote by the same letter the corresponding state on 
$\cO_{\Gamma(\cK)}$. $\omega_T$ is called quasi-free state associated to the density $T$. 
Its properties are summarized in \index{state!quasi-free}
\bep\label{QuasiProps}
\ben
\item If $\phi_1, \ldots, \phi_n, \psi_1, \ldots, \psi_m\in \cK$, then 
\[
\omega_T(a^\ast(\phi_n)\cdots a^\ast(\phi_1)a(\psi_1)\cdots a(\psi_m))=
\delta_{nm}\det[\langle \psi_i| T \phi_j\rangle].
\]
In particular, $\omega_T(a^\ast(\phi)a(\psi))=\langle\psi|T\phi\rangle$.
\item $\log Z_T=-\log \det (\one-T)=-\tr (\log(\one -T))$.
\item $\omega_T(\Gamma(A))= \det (\one + T(A-\one))$.
\item $\omega_T(\d\Gamma(A))=\tr (TA)$.
\item $S(\omega_T)=-\tr(T\log T+(\one-T)\log(\one -T))$.
\item $\omega_{T_1}\ll\omega_{T_2}$ iff $\Ker T_1\subset \Ker T_2$, and then 
\[
S(\omega_{T_1}|\omega_{T_2})
=\tr\left(T_1(\log(T_2)-\log(T_1))+(\one-T_1)(\log(\one-T_2)-\log(\one-T_1))\right).
\]
\een
\eep
\demo (1) We set $Q=T(\one-T)^{-1}$, 
$A=a^\ast(\phi_n)\cdots a^\ast(\phi_1)a(\psi_1)\cdots a(\psi_m)$ and note that
$$
\e^{-\i tN}\omega_T\,\e^{\i tN}=\frac1{Z_T}\Gamma(\e^{-\i t})\Gamma(Q)\Gamma(\e^{\i t})
=\frac1{Z_T}\Gamma(\e^{-\i t}Q\e^{\i t})=\frac1{Z_T}\Gamma(Q)=\omega_T,
$$
so that
$$
\omega_T(\e^{\i tN}A\e^{-\i tN})=\omega_T(A).
$$
By Proposition \ref{midterm1} (6), we have
$$
\e^{\i tN}a^\ast(\phi_j)\,\e^{-\i tN}=a^\ast(\e^{\i t}\phi_j)=\e^{\i t}a^\ast(\phi_j),\qquad
\e^{\i tN}a(\psi_k)\,\e^{-\i tN}=a(\e^{\i t}\psi_k)=\e^{-\i t}a(\psi_k),
$$
from which we deduce that $\e^{\i tN}A\,\e^{-\i tN}=\e^{\i t(n-m)}A$, and hence that
$\omega_T(A)=0$ if $n\not=m$. We shall handle the case $n=m$ by induction on $n$.
For $n=1$, one has
\begin{align*}
\omega_T(a^\ast(\phi)a(\psi))&=Z_T^{-1}\tr(\Gamma(Q)a^\ast(\phi)a(\psi))\\
&=Z_T^{-1}\tr(a^\ast(Q\phi)\Gamma(Q)a(\psi))\\
&=Z_T^{-1}\tr(\Gamma(Q)a(\psi)a^\ast(Q\phi))\\
&=Z_T^{-1}\tr(\Gamma(Q)(\{a(\psi),a^\ast(Q\phi)\}-a^\ast(Q\phi)a(\psi)))\\
&=\langle\psi|Q\phi\rangle-\omega_T(a^\ast(Q\phi)a(\psi)),
\end{align*}
from which we deduce that 
$\omega_T(a^\ast((\one+Q)\phi)a(\psi))=\langle\psi|Q\phi\rangle$.
Since $(\one+Q)=(\one-T)^{-1}$, we finally get
$$
\omega_T(a^\ast(\phi)a(\psi))=\langle\psi|Q(\one-T)\phi\rangle=\langle\psi|T\phi\rangle.
$$
Assuming now that the result holds for $n-1$, we write
\begin{align*}
\omega_T&(a^\ast(\phi_n)\cdots a^\ast(\phi_1)a(\psi_1)\cdots a(\psi_n))\\
&=Z_T^{-1}\tr(\Gamma(Q)a^\ast(\phi_n)\cdots a^\ast(\phi_1)a(\psi_1)\cdots a(\psi_n))\\
&=Z_T^{-1}\tr(a^\ast(Q\phi_n)\Gamma(Q)a^\ast(\phi_{n-1})\cdots a^\ast(\phi_1)a(\psi_1)\cdots a(\psi_n))\\
&=\omega_T(a^\ast(\phi_{n-1})\cdots a^\ast(\phi_1)a(\psi_1)\cdots a(\psi_n)a^\ast(Q\phi_n)).
\end{align*}
Making repeated use of the CAR, 
$$
a(\psi_j)a^\ast(Q\phi_n)=\langle\psi_j|Q\phi_n\rangle-a^\ast(Q\phi_n)a(\psi_j),
\qquad a^\ast(\phi_j)a^\ast(Q\phi_n)=-a^\ast(Q\phi_n)a^\ast(\phi_j),
$$
we move the last factor $a^\ast(Q\phi_n)$ back to its original position to get
\begin{align*}
\omega_T(&a^\ast(\phi_n)\cdots a^\ast(\phi_1)a(\psi_1)\cdots a(\psi_n))
=-\omega_T(a^\ast(Q\phi_n)\cdots a^\ast(\phi_1)a(\psi_1)\cdots a(\psi_n))\\
&+\sum_{j=1}^n(-1)^{n+j}\langle\psi_j|Q\phi_n\rangle
\omega_T(a^\ast(\phi_{n-1})\cdots a^\ast(\phi_1)a(\psi_1)\cdots\bcancel{a(\psi_j)}\cdots a(\psi_n)).
\end{align*}
By the same argument as in the $n=1$ case, we deduce
\begin{align*}
\omega_T&(a^\ast(\phi_n)\cdots a^\ast(\phi_1)a(\psi_1)\cdots a(\psi_n))\\
&=\sum_{j=1}^n(-1)^{n+j}\langle\psi_j|T\phi_n\rangle
\omega_T(a^\ast(\phi_{n-1})\cdots a^\ast(\phi_1)a(\psi_1)\cdots\bcancel{a(\psi_j)}\cdots a(\psi_n)),
\end{align*}
and the induction step is achieved by Laplace formula \eqref{LaplaceDet}.\index{formula!Laplace}

\noindent(2) and (3) are immediate consequences of Lemma \ref{DetLemma},
(4) follows from (1) and Proposition \ref{midterm1} (9).

\noindent(5) We again set $Q=T(\one-T)^{-1}$ and notice that
$$
\log\Gamma(Q)=\d\Gamma(\log Q),
$$
so that, by (4),
$$
S(\omega_T)=-\omega_T\left(\log\left(Z_T^{-1}\Gamma(Q)\right)\right)
=-\omega_T(\d\Gamma(\log Q)-\log Z_T)=\log Z_T-\tr\left(T\log Q\right).
$$
Using (2), we conclude that
$$
S(\omega_T)=-\tr(\log(\one-T))-\tr(T(\log(T)-\log(\one-T))),
$$
from which the desired formula immediately follows.

\noindent(6) We set $Q_j=T_j(\one-T_j)^{-1}$ and notice that $\Ker Q_j=\Ker T_j$. It easily
follows from $\Ker T_1\subset\Ker T_2$ that $\Ker\Gamma(Q_1)\subset\Ker\Gamma(Q_2)$
and hence $\omega_{T_1}\ll\omega_{T_2}$. The remaining statement is proved in a similar
way as (5).
\qed

\bigskip
Let $h =h^\ast\in\cO_\cK$ be the one-particle Hamiltonian -- the total energy observable of a 
single fermion. The Hamiltonian of the free Fermi gas is\bindex{Hamiltonian!one-particle}
\[
H=\d\Gamma(h).
\]
Indeed, if $\{\psi_1,\ldots,\psi_d\}$ denotes an eigenbasis of $h$ such that $h\psi_j=\varepsilon_j\psi_j$,
then the state 
$$
\Psi=a^\ast(\psi_{j_1})\cdots a^\ast(\psi_{j_n})\Omega,
$$
describes $n$ fermions with energies $\varepsilon_{j_1},\ldots,\varepsilon_{j_n}$, and one has
$$
H\Psi
=\d\Gamma_n(h)\psi_{j_1}\wedge\cdots\wedge\psi_{j_n}
=\left(\sum_{i=1}^n\varepsilon_{j_i}\right)\Psi.
$$
The thermal equilibrium state at inverse temperature $\beta\in\rr$ and chemical potential 
$\mu \in \rr$ is described by the Gibbs grand canonical ensemble\index{Gibbs!grand canonical ensemble}
\[
\rho_{\beta, \mu}=\frac{\e^{-\beta(H-\mu N)}}{\tr(\e^{-\beta(H-\mu N)})}.
\]
Since 
\[
\e^{-\beta (H-\mu N)}=\e^{-\d\Gamma(\beta(h-\mu\one))}=\Gamma(\e^{-\beta (h-\mu\one)}),
\]
solving the equation
\[
\e^{-\beta(h-\mu \one)}= \frac{T}{\one -T},
\]
for $T$ we see that  the density operator of a free Fermi gas in thermal equilibrium at inverse temperature 
$\beta$ and chemical potential $\mu$ is given by 
\[
T_{\beta, \mu}=(\one + \e^{\beta(h-\mu\one)})^{-1}.
\]
$T_{\beta, \mu}$ is commonly called the Fermi-Dirac distribution. Following the notation introduced in 
Section \ref{sect:KMS}, one has\bindex{distribution!Fermi-Dirac}\index{pressure}\index{entropy}
\begin{equation}
\begin{split}
E&=\rho_{\beta,\mu}(H)=\tr(h T_{\beta,\mu}),\\[3mm]
\varrho&=\rho_{\beta,\mu}(N)=\tr(T_{\beta,\mu}),\\[3mm]
P(\beta,\mu)&= \log\tr(\e^{-\beta (H-\mu N)})
=\tr\left(\log(\one+\e^{-\beta(h-\mu\one)})\right),\\[3mm]
S(\beta,\mu)&=S(\rho_{\beta,\mu})=\beta(E-\mu\varrho)+P(\beta, \mu).
\end{split}
\label{olympia}
\end{equation}

\medskip
\begin{exo}\label{Exo:FreeGas}\index{thermodynamic limit}
The purpose of this exercise is to provide a complete discussion of the thermodynamic limit of a 
1D free Fermi gas starting from the description of a finite Fermi gas. The target system is the
ideal Fermi gas with one particle Hamiltonian $h=k^2/2$ on the one-particle Hilbert space 
$\cK=L^2(\rr,\d k/2\pi)$ in the thermal equilibrium state at inverse temperature $\beta$ and chemical 
potential $\mu$.

To describe the finite approximation, consider the operator
\[ 
(h_{L}\psi)(x)=-\frac12\psi''(x), 
\]
on $L^2([-L/2,L/2],\d x)$ with periodic boundary conditions $\psi(x+L)=\psi(x)$. $h_{L}$ is self-adjoint 
with a purely discrete spectrum consisting of simple eigenvalues $\varepsilon(k)=k^2/2$ with 
eigenfunctions $\psi_k(x)=L^{-1/2}\e^{\i k x}$,  $k\in\cQ_L=\{2\pi j/L\,|\,j\in\zz\}$. The Fourier transform 
$$
\hat\psi(k)=\langle\psi_k|\psi\rangle=\frac1{\sqrt L}\int_{-L/2}^{L/2}\psi(x)\e^{-\i kx}\,\d x,
$$ 
provides a unitary map from the position representation $L^2([-L/2,L/2])$ to the ``momentum" 
representation $\ell^2(\cQ_L)$ such that $\widehat{h_L\psi}(k)=\varepsilon(k)\hat\psi(k)$. In what 
follows, we work in the momentum representation and set $\cK_L=\ell^2(\cQ_L)$ and 
$(h_L\psi)(k)=\varepsilon(k)\psi(k)$. Let $\cal E>0$ be an energy cutoff, set 
$\cQ_{L,\cal E}=\{k\in\cQ_L\,|\,\varepsilon(k)\le\cal E\}$ and consider the free Fermi gas with single 
particle Hilbert space $\cK_{L,\cal E}=\ell^2(\cQ_{L,\cal E})$, and one-particle Hamiltonian 
$(h_{L,\cal E}\psi)(k)=\varepsilon(k)\psi(k)$. Let $E_{L,\cal E}$, $\varrho_{L,\cal E}$,  
$P_{L,\cal E}(\beta,\mu)$ be defined by \eqref{olympia}. 

\exop Prove that 
\begin{align*}
\lim_{L\to\infty}\lim_{\cal E\to\infty}\frac{E_{L,\cal E}}{L}
&=\int_{-\infty}^\infty\frac{\varepsilon(k)}{1+\e^{\beta(\varepsilon(k)-\mu)}}\frac{\d k}{2\pi},\\[3mm]
\lim_{L\to\infty}\lim_{{\cal E}\to\infty}\frac{\varrho_{L,\cal E}}{L}
&=\int_{-\infty}^\infty\frac{1}{1+\e^{\beta(\varepsilon(k)-\mu)}}\frac{\d k}{2\pi},\\[3mm]
\lim_{L\to\infty}\lim_{{\cal E}\to\infty}\frac{P_{L,\cal E}(\beta, \mu)}{L}
&=\int_{-\infty}^\infty\log(1+\e^{-\beta(\varepsilon(k)-\mu)})\frac{\d k}{2\pi}.
\end{align*}

\exop A wave function $\psi\in\cK_{L,\cal E}$ can be isometrically extended to an element of
$\cK$ by setting
$$
\widetilde\psi(k)=\sqrt L\sum_{\xi\in\cQ_{L,\cal E}}\psi(\xi)\chi_{[\xi-\pi/L,\xi+\pi/L[}(k),
$$
where $\chi_I$ denotes the indicator function of the interval $I$. Thus,  we can identify 
$\cK_{L,\cal E}$ with a finite dimensional subspace of the Hilbert space $\cK$.
Denote by $\one_{L,\cal E}$ the orthogonal projection on this subspace. Then 
$\Gamma(\one_{L,\cal E})$ is an orthogonal projection in $\Gamma(\cK)$ whose range
can be identified with $\Gamma(\cK_{L,\cal E})$. Show that we can identify the equilibrium 
density matrix
$$
\rho_{\beta,\mu,L,\cal E}=\frac{\Gamma(\e^{-\beta(h_{L,\cal E}-\mu\one)})}
{\tr(\Gamma(\e^{-\beta(h_{L,\cal E}-\mu\one)}))};
$$
of the finite Fermi gas on $\Gamma(\cK_{L,\cal E})$ with the density matrix
$$
\widetilde\rho_{\beta,\mu,L,\cal E}
=\frac{\Gamma(\e^{-\beta(h-\mu\one)}\one_{L,\cal E})}
{\tr(\Gamma(\e^{-\beta(h-\mu\one)}\one_{L,\cal E}))},
$$
on $\Gamma(\cK)$ in the sense that
$$
\tr\left(\vphantom{\widetilde\psi_1}\rho_{\beta,\mu,L,\cal E}a^\ast(\psi_1)\cdots a^\ast(\psi_n)a(\phi_m)\cdots a(\phi_1)\right)
=\tr\left(\widetilde\rho_{\beta,\mu,L,\cal E}a^\ast(\widetilde\psi_1)\cdots a^\ast(\widetilde\psi_n)
a(\widetilde\phi_m)\cdots a(\widetilde\phi_1)\right),
$$
for all $\psi_1,\ldots,\psi_n,\phi_1,\ldots,\phi_m\in\cK_{L,\cal E}$.

\exop Show that, in $\Gamma(\cK)$, the limit
$$
\widetilde\rho_{\beta,\mu,L}=\lim_{\cal E\to\infty}\widetilde\rho_{\beta,\mu,L,\cal E},
$$
exists in the trace norm and that $\widetilde\rho_{\beta,\mu,L}$ is a density matrix that can be 
identified with
$$
\rho_{\beta,\mu,L}=\frac{\Gamma(\e^{-\beta(h_L-\mu\one)})}
{\tr(\Gamma(\e^{-\beta(h_L-\mu\one)}))},
$$
on $\Gamma(\cK_L)$. Show that
$$
\slim_{L\to\infty}\widetilde\rho_{\beta,\mu,L}=0,
$$
\ie the equilibrium density matrix disappears in the thermodynamic limit $L\to\infty$.

\exop Show that,
$$
\cal D=\bigcup_{L>0,\cal E>0}\cK_{L,\cal E},
$$
is a dense subspace of $\cK$ and that for $\phi,\psi\in\cal D$ one has
$$
\lim_{L\rightarrow \infty}\lim_{{\cal E}\to\infty}\tr\left(
\widetilde\rho_{\beta,\mu,L,\cal E}a^\ast(\phi)a(\psi)\right)
=\int_{-\infty}^\infty\frac{\overline{\psi(k)}\phi(k)}{1+\e^{\beta(\varepsilon(k)-\mu)}}\frac{\d k}{2\pi}
=\langle\psi|T\phi\rangle,
$$
where $T=(\one+\e^{\beta(h-\mu)})^{-1}$.

\exop Since we have identified $\cK_{L,\cal E}$ with a subspace of $\cK$, we can also identify 
the $\ast$-algebra $\cO_{\cK_{L,\cal E}}$ with a subalgebra of the $\ast$-algebra $\cO_\cK$ of 
all bounded linear operators on $\cK$. This identification is isometric and
$$
\cO_\infty=\bigcup_{L>0,{\cal E}>0}\cO_{\cK_{L,\cal E}},
$$
is the $\ast$-algebra of all polynomials in the creation/annihilation operators $a^\#(\psi)$,
$\psi\in\cal D$. Show that the limit
$$
\rho_{\beta,\mu}(A)=\lim_{L\rightarrow \infty}\lim_{{\cal E}\to\infty}
\tr\left(\widetilde\rho_{\beta,\mu,L,\cal E}A\right),
$$
exists for all $A\in\cO_\infty$. 

\noindent{\sl Hint}: show that
$$
\lim_{L\rightarrow \infty}\lim_{{\cal E}\to\infty}\tr\left(\widetilde\rho_{\beta,\mu,L,\cal E}
a^\ast(\psi_1)\cdots a^\ast(\psi_n)a(\phi_m)\cdots a(\phi_1)\right)
=\delta_{n,m}\det[\langle\phi_j|T\psi_k\rangle],
$$
for all $\psi_1,\ldots,\psi_n,\phi_1,\ldots,\phi_m\in\cal D$.

\exop Denote by $\cO_\infty^{\mathrm{cl}}$ the norm closure of $\cO_\infty$ in $\cO_\cK$
($\cO_\infty^{\mathrm{cl}}$ is the $C^\ast$-algebra generated by $\cO_\infty$). \index{algebra!$C^\ast$-}
Show that for any
$A\in\cO_\infty^{\mathrm{cl}}$ and any sequence $A_n\in\cO_\infty$ which converges to $A$ the
limit
$$
\rho_{\beta,\mu}(A)=\lim_{n\to\infty}\rho_{\beta,\mu}(A_n),
$$
exists and is independent of the approximating sequence $A_n$. The $C^\ast$-algebra 
$\cO_\infty^{\mathrm{cl}}$ is the algebra of observables of the infinitely extended ideal Fermi gas
and $\rho_{\beta,\mu}$ is its thermal equilibrium state.
\end{exo}

\section{The Araki-Wyss representation}
\label{sect:AWrepresentation}

Araki and Wyss \cite{AWy} have discovered a specific cyclic representation of $\cO_{\Gamma(\cK)}$
associated to the quasi-free state $\omega_T$ which is of considerable conceptual and computational 
importance. Although any  two cyclic representations of $\cO_{\Gamma(\cK)}$ associated to the state 
$\omega_T$ are unitarily equivalent, the specific structure inherent to the Araki-Wyss (AW) 
representation has played a central role in many developments in non-equilibrium quantum statistical 
mechanics over the last decade.

For the purpose of this section we may assume that $T>0$ (otherwise, replace $\cK$ with $\Ran T$).
Then the quasi-free state $\omega_T$ on $\cO_{\Gamma(\cK)}$ is faithful. Set 
\begin{align*}
\cH_{\rm AW}&=\Gamma(\cK)\otimes \Gamma(\cK),\\[3mm]
\Omega_{\rm AW}&=\Omega\otimes \Omega,\\[3mm]
b_{\rm AW}^\ast(\psi)&=a^\ast((\one -T)^{1/2}\psi)\otimes\one
+\e^{\i\pi N}\otimes a(\overline{T^{1/2}\psi}),\\[3mm]
b_{\rm AW}(\psi)&=a((\one -T)^{1/2}\psi)\otimes\one
+\e^{\i\pi N}\otimes a^\ast(\overline{T^{1/2}\psi}),
\end{align*}
where $\bar\psi$ denotes the complex conjugate of $\psi\in\cK=\ell^2({\cal Q})$. For
$\Psi\in\Gamma(\cK)$,  $\bar \Psi$ denotes the complex conjugate of $\Psi$ (defined in the 
obvious way). If $A$ is a linear operator, we define the linear operator $\bar A$ by 
$\bar A\,\bar\psi=\overline{A\psi}$.

\bep\label{AWrepProp}
\ben
\item The maps $\psi\mapsto b_{\rm AW}^\#(\psi)$ define a representation of the CAR
over $\cK$ on the Hilbert space $\cH_{\rm AW}$.
\item Let $\pi_{\rm AW}$ be the induced representation of $\cO_{\Gamma(\cK)}$ on 
$\cH_{\rm AW}$. $\Omega_{\rm AW}$ is a cyclic vector for this representation \index{vector!cyclic}
and 
\begin{equation}
\omega_T(A)=(\Omega_{AW}|\pi_{\rm AW}(A)\Omega_{\rm AW}),
\label{AWomegaT}
\end{equation}
for all $A\in\cO_{\Gamma(\cK)}$. In other words, $\pi_{\rm AW}$ is a cyclic representation of 
$\cO_{\Gamma(\cK)}$ associated to the faithful state $\omega_T$.
\een
\eep
\demo
The verification of (1) is simple and we leave it  as an exercise for the reader. To check that $\Omega_{\rm AW}$ is cyclic, 
we  shall show by induction on $n+m$ that each subspace 
$D_{n,m}=\Gamma_n(\cK)\otimes\Gamma_m(\cK)$
belongs to $\pi_{\rm AW}(\cO_{\Gamma(\cK)})\Omega_{\rm AW}$. For $n+m=1$, we deduce from
$\Ran (\one-T)^{1/2}=\Ran\bar T^{1/2}=\cK$ that
$$
D_{1,0}=\{b_{\rm AW}^\ast(\psi)\Omega_{\rm AW}\,|\,\psi\in\cK\},\qquad
D_{0,1}=\{b_{\rm AW}(\psi)\Omega_{\rm AW}\,|\,\psi\in\cK\}.
$$
Assuming $D_{n,m}\subset\pi_{\rm AW}(\cO_{\Gamma(\cK)})\Omega_{\rm AW}$ for $n+m\le k$, we
observe that $\Psi\in D_{n+1,m}$ can be written as
$$
\Psi=a^\ast((\one-T)^{1/2}\psi)\otimes\one\Phi,
$$
for some $\psi\in\cK$ and $\Phi\in D_{n,m}$. Equivalently, we can write
$$
\Psi=b_{\rm AW}^\ast((\one-T)^{1/2}\psi)\Phi-\Phi'
$$
where $\Phi'=(-\one)^N\otimes a(\overline{T^{1/2}\psi})\Phi\in D_{n,m-1}$. It follows that
$\Psi\in\pi_{\rm AW}(\cO_{\Gamma(\cK)})\Omega_{\rm AW}$. A similar argument shows that
$D_{n,m+1}\subset\pi_{\rm AW}(\cO_{\Gamma(\cK)})\Omega_{\rm AW}$. Hence, the induction
property is verified for $n+m\le k+1$. Finally, \eqref{AWomegaT} follows from an elementary
calculation based on Equ. \eqref{LeibnitzForm}.
\qed

\bigskip
The triple $(\cH_{\rm AW},\pi_{\rm AW},\Omega_{\rm AW})$ is called the Araki-Wyss representation
\index{representation!Araki-Wyss} of
the CAR over $\cK$ associated to the quasi-free state $\omega_T$. Since $\omega_T$ is faithful, 
it follows from Part (2) of Proposition \ref{AWrepProp} and Part 4 of Exercise \ref{Exo:GNS} that
this representation is unitarily equivalent to the standard representation and hence 
carries the entire modular structure. The modular structure in the Araki-Wyss representation takes 
the following form.

\bep\label{AWModular}
\ben
\item The modular conjugation is given by\index{modular!conjugation}
\[
J(\Psi_1\otimes \Psi_2)= u \bar \Psi_2\otimes u \bar \Psi_1,
\]
where $u=\e^{\i\pi N(N-1)/2}$.
\item The modular operator of $\omega_T$ is \index{modular!operator}
\[
\Delta_{\omega_T}=\Gamma (\e^{k_T})\otimes\Gamma(\e^{-\overline{k_T}}),
\]
where $k_T=\log(T(\one -T)^{-1})$. In particular
$$
\log\Delta_{\omega_T}=\d\Gamma(k_T)\otimes\one-\one\otimes\d\Gamma(\overline{k_T}).
$$
\item If $\omega_{T_1}$ is the  quasi-free state of density $T_1>0$, then its relative Hamiltonian 
w.r.t. $\omega_T$ is\index{Hamiltonian!relative}
$$
\ell_{\omega_{T_1}|\omega_T}=\log\det\left((\one-T_1)(\one-T)^{-1}\right)+
\d\Gamma(k_{T_1}- k_{T}),
$$
and its relative modular operator is determined by\index{modular!operator!relative}
\[
\log\Delta_{\omega_{T_1}|\omega_{T}}=\log\Delta_{\omega_T}
+\pi_{\rm AW}(\ell_{\omega_{T_1}|\omega_T}).
\]
\item Suppose that the self-adjoint operator $h$ commutes with $T$. Then the quasi-free
state $\omega_T$ is invariant under the dynamics $\tau^t$ generated by $H=\d\Gamma(h)$.
Moreover, the operator
\[
K=\d\Gamma(h)\otimes \one -\one \otimes \d\Gamma(\bar h),
\]
is the standard Liouvillean of this dynamics.\index{Liouvillean!standard}
\een
\eep
{\noindent\bf Remark.}
Since the $\ast$-subalgebra $\cO_{\rm AW}=\pi_{\rm AW}(\cO_{\Gamma(\cK)})$ is the set of
polynomials in the $b_{\rm AW}^\#$, the dual $\ast$-subalgebra 
$\cO_{\rm AW}^\prime=J\cO_{\rm AW}J$ is the set of polynomials in the 
$b_{\rm AW}^{\prime\#}=Jb_{\rm AW}^\#J$. By Propositions \ref{prop:StdOne} and \ref{prop:StdTwo},
one has
\begin{align*}
\cO_{\rm AW}\cap\cO_{\rm AW}'&=\cc\one,\\
\cO_{\rm AW}\vee\cO_{\rm AW}'&=\cO_{\cH_{\rm AW}}.
\end{align*}

\demo We set $\Delta=\Gamma(\e^{k_T})\otimes\Gamma(\e^{-\overline{k_T}})$ and $s=\e^{\i\pi N}$.
Since $J$ is clearly an anti-unitary involution and $\Delta>0$, we deduce from the observation following 
Equ. \eqref{ModularCharact} that in order to prove (1) and (2) it suffices to show that 
$J\Delta^{1/2}A\Omega_{\rm AW}=A^\ast\Omega_{\rm AW}$  for any monomial 
$A=b_{\rm AW}^\#(\psi_n)\cdots b_{\rm AW}^\#(\psi_1)$. We shall do that by induction on the
degree $n$. 

We first compute
\begin{align*}
b_{\rm AW}^{\prime}(\psi)&=Jb_{\rm AW}(\psi)J=a^\ast(T^{1/2}\psi)s\otimes s
+\one\otimes sa(\overline{(\one-T)^{1/2}\psi}),\\[3mm]
b_{\rm AW}^{\prime\ast}(\psi)&=Jb_{\rm AW}^{\ast}(\psi)J=sa(T^{1/2}\psi)\otimes s
+\one\otimes a^\ast(\overline{(\one-T)^{1/2}\psi})s,
\end{align*}
and check that $[b_{\rm AW}^{\prime}(\psi),b_{\rm AW}^{\#}(\phi)]
=[b_{\rm AW}^{\prime^\ast}(\psi),b_{\rm AW}^{\#}(\phi)]=0$
for all $\psi,\phi\in\cK$. We thus conclude that $b_{\rm AW}^{\prime\#}(\psi)\in\cO_{\rm AW}'$.
Next, we observe that
$$
\Delta^{1/2}b_{\rm AW}(\psi)\Delta^{-1/2}=b_{\rm AW}(\e^{-k_T/2}\psi),\qquad
\Delta^{1/2}b_{\rm AW}^\ast(\psi)\Delta^{-1/2}=b_{\rm AW}^{\ast} (\e^{k_T/2}\psi).
$$
For $n=1$, the claim follows from the fact that
\begin{align*}
J\Delta^{1/2}b_{\rm AW}(\psi)\Omega_{\rm AW}
&=J\Delta^{1/2}b_{\rm AW}(\psi)\Delta^{-1/2}J\Omega_{\rm AW}\\
&=b_{\rm AW}^{\prime}(\e^{-k_T/2}\psi)\Omega_{\rm AW}\\
&=a^\ast(\e^{-k_T}T^{1/2}\psi)\otimes\one\Omega_{\rm AW}\\
&=b_{\rm AW}^{\ast}(\psi)\Omega_{\rm AW}.
\end{align*}
To deal with the induction step, let $A$ be a monomial of degree less than $n$ in the 
$b_{\rm AW}^\#$ and assume that
$J\Delta^{1/2}A\Omega_{\rm AW}=A^\ast\Omega_{\rm AW}$ for all such monomials.
Then, we can write
\begin{align*}
J\Delta^{1/2}b_{\rm AW}^\#(\psi)A\Omega_{\rm AW}
&=(J\Delta^{1/2}b_{\rm AW}^\#(\psi)\Delta^{-1/2}J)J\Delta^{1/2}A\Omega_{\rm AW}\\
&=(Jb_{\rm AW}^\#(\e^{\mp k_T/2}\psi)J)A^\ast\Omega_{\rm AW}\\
&=b_{\rm AW}^{\prime\#}(\e^{\mp k_T/2}\psi)A^\ast\Omega_{\rm AW}\\
&=A^\ast b_{\rm AW}^{\prime\#}(\e^{\mp k_T/2}\psi)\Omega_{\rm AW}\\
&=A^\ast J\Delta^{1/2}b_{\rm AW}^{\#}(\psi)\Delta^{-1/2}J\Omega_{\rm AW}\\
&=A^\ast J\Delta^{1/2}b_{\rm AW}^{\#}(\psi)\Omega_{\rm AW}\\
&=A^\ast b_{\rm AW}^{\#}(\psi)^\ast\Omega_{\rm AW},
\end{align*}
which shows that the induction property holds for all monomials of degree less than $n+1$.

\medskip\noindent(3) The first claim is an immediate consequence of the definition \eqref{RelatHDef}
of the relative Hamiltonian. Since, by Part (4) of Exercise \ref{Exo:GNS}, the Araki-Wyss 
representation is unitarily equivalent to the standard representation on $\cH_\cO$, the second claim
follows from Property (3) of the relative Hamiltonian given on page \pageref{RelatHProps}.

\medskip\noindent(4) The fact that $\omega_T$ is invariant under the dynamics $\tau^t$ is evident.
Recall from Exercise \ref{Exo:StdLiouvilleChar} that the standard Liouvillean is the unique 
self-adjoint operator $K$ on $\cH_{\rm AW}$ (the Hilbert space carrying the standard representation) 
such that the unitary group $\e^{\i tK}$ implements the dynamics and preserves the natural cone.
These two conditions can be formulated as
$$
\e^{\i tK}b_{\rm AW}(\psi)\e^{-\i tK}=b_{\rm AW}(\e^{\i th}\psi),
\qquad
JK+KJ=0,
$$
and are easily verified by $K=\d\Gamma(h)\otimes\one-\one\otimes\d\Gamma(\bar h)$.
\qed

\bigskip
{\noindent\bf Remark.} The Araki-Wyss representation of the CAR over $\cK$ immediately extends
to infinite dimensional $\cK$ and the proof of Proposition \ref{AWrepProp} carries over without 
modification. The same is true for Proposition \ref{AWModular} provided one assumes, in Part (3),
that $\log(T_1)-\log(T)$ and $\log(\one-T_1)-\log(\one-T)$ are both trace class.

\section{Spin-Fermion model} 
\label{sect:spin-fermion}
\bindex{model!spin-fermion}

The Spin-Fermion (SF) model describes a two level atom (or a spin $1/2$),  denoted $\cS$,
interacting with $n \geq 2$ independent free Fermi gas reservoirs $\cR_j$. The Hilbert space 
of $\cS$ is $\cH_\cS=\cc^2$ and its Hamiltonian is the third Pauli matrix 
$$
H_\cS=\sigma^{(3)}=\left[\begin{array}{cc}1&0\\0&-1\end{array}\right].
$$
Its initial state is $\omega_\cS=\one/2$.  
The  reservoir $\cR_j$ is described by the single particle 
Hilbert space $\cK_j=\ell^2(\cQ_j)$ and  single particle Hamiltonian $h_j$. The Hamiltonian and the
number operator of $\cR_j$ are denoted by $H_j=\d\Gamma(h_j)$ and $N_j$. The creation/annihilation
operators on the Fock space $\Gamma(\cK_j)$ are denoted by $a_j^\#$. We assume that 
$\cR_j$ is in the state
\[
\omega_{\beta_j, \mu_j}=\frac{\e^{-\beta_j(H_j-\mu_jN_j)}}{\tr (\e^{-\beta_j(H_j-\mu_jN_j)})}, 
\] 
that is, that  $\cR_j$ is in  thermal  equilibrium at inverse temperature $\beta_j$ and chemical 
potential $\mu_j$. The complete reservoir system $\cR=\sum_j\cR_j$ is described by the 
Hilbert space 
\[
\cH_\cR=\bigotimes_{j=1}^n\Gamma(\cK_j),
\]
its Hamiltonian is 
\[
H_\cR=\sum_{j=1}^n H_j,
\]
and its initial state is 
\[
\omega_\cR
=\otimes_{j=1}^n \omega_{\beta_j, \mu_j}=\frac{1}{Z}\e^{-\sum_{j=1}^n \beta_j(H_j-\mu_j N_j)},
\]
where $Z=\tr (\e^{-\sum_j \beta_j(H_j-\mu_j N_j)})$.
The Hilbert space of  the joint system $\cS+\cR$ is 
\[
{\cal H}=\cH_\cS\otimes\cH_\cR,
\] 
its initial state is $\omega=\omega_\cS\otimes\omega_\cR$, and in the absence of interaction its 
Hamiltonian is
\[
H_0= H_\cS+H_\cR.
\]
The interaction of $\cS$ with $\cR_j$ is described by 
\[
V_j=\sigma^{(1)}\otimes P_j,
\]
where $\sigma^{(1)}$ is the first Pauli matrix and $P_j$ is a self-adjoint polynomial in the field operators
\[
\varphi_j(\psi)=\frac{1}{\sqrt 2}(a_j(\psi)+a_j^\ast(\psi))\in\cO_{\Gamma(\cK_j)}.
\]
For example $P_j=\varphi_j(\psi_j)$ or $P_j=\i\varphi_j(\psi_j)\varphi_j(\phi_j)$ with
$\psi_j\perp\phi_j$. The complete interaction is 
$V=\sum_{j=1}^n V_j$ and the full (interacting) Hamiltonian of the SF model is 
\[
H_\lambda = H_0 +\lambda V,
\]
where $\lambda \in \rr$ is a coupling constant. 

\begin{exo}
Check that the SF model is an example of open quantum system as defined in Section 
\ref{sect:Coupling}. Warning: gauge invariance is broken in the SF model.
\end{exo}

\begin{exo}\label{Exo:SF}
\exop Denote by $\{e_1, e_2\}$ the standard basis of $\cH_\cS=\cc^2$. Show that the triple 
$(\cH_\cS\otimes\cH_\cS,\pi_\cS,\Omega_\cS)$, where $\pi_\cS(A)=A\otimes\one$ and 
\[
\Omega_\cS=\frac {1}{\sqrt 2}(e_1\otimes e_1 + e_2 \otimes e_2),
\]
is a GNS representation of $\cO_{\cH_\cS}$ associated to $\rho_\cS$. \index{representation!GNS}
Since $\rho_\cS$ is
faithful, this representation carries the modular structure of $\cO_\cS$. Show that the
modular conjugation and the modular operator are given by 
$J_\cS:f\otimes g\mapsto\bar g\otimes\bar f$ and $\Delta_{\omega_\cS}=\one$.

\noindent Note that this part of the exercise is the simplest non-trivial example 
of Exercise \ref{Exo:GNS} (5).

\exop Let $(\cH_{{\rm AW},j},\pi _{{\rm AW}, j}, \Omega_{{\rm AW},j})$ be the 
Araki-Wyss representation of the $j$-th reservoir associated to the quasi-free state 
$\omega_{\beta_j, \mu_j}$. \index{representation!Araki-Wyss} Show that
$\pi_{\rm SF}=\pi_\cS\otimes\pi_{{\rm AW},1}\otimes\cdots\otimes\pi_{{\rm AW}, n}$ is 
the standard representation of $\cO_\cH$ on the Hilbert space
\[
\cH_{\rm SF}=
(\cH_\cS\otimes\cH_\cS)\otimes\cH_{{\rm AW},1}\otimes\cdots\cH_{{\rm AW},n},
\]
with the cyclic vector  \index{vector!cyclic}
\[
\Omega_{\rm SF}=\Omega_\cS\otimes\Omega_{{\rm AW},1}
\otimes\cdots\otimes\Omega_{{\rm AW},n}.
\]

\exop Consider the SF model with interaction $P_j=\varphi_j(\psi_j)$. Show that in the above 
standard representation the operator $L_{\frac1\balpha}$, defined by \eqref{Loneoveralphadef},
takes the form
\begin{align}
L_{\frac1\balpha}&=(H_\cS\otimes\one_{\cH_\cS}-\one_{\cH_\cS}\otimes H_\cS)\otimes
\one_{\cH_{{\rm AW},1}}\otimes\cdots\otimes\one_{\cH_{{\rm AW},n}}\label{SFLalpha}\\
&+\sum_{j=1}^n(\one_{\cH_\cS}\otimes\one_{\cH_\cS})\otimes\one_{\cH_{{\rm AW},1}}\otimes\cdots
\otimes(\d\Gamma(h_j)\otimes\one-\one\otimes\d\Gamma(\bar h_j))
\otimes\cdots\otimes\one_{\cH_{{\rm AW},n}}\nonumber\\[3mm]
&+\lambda\sum_{j=1}^n (\sigma^{(1)}\otimes\one_{\cH_\cS})
\otimes\one_{\cH_{{\rm AW},1}}\otimes\cdots\otimes
\frac1{\sqrt2}\left(b_{{\rm AW},j}(\psi_j)+b_{{\rm AW},j}^\ast(\psi_j)\right)
\otimes\cdots\otimes\one_{\cH_{{\rm AW},n}}\nonumber\\[3mm]
&-\lambda\sum_{j=1}^M (\one_{\cH_\cS}\otimes\sigma^{(1)})
\otimes\one_{\cH_{{\rm AW},1}}\otimes\cdots\otimes
\frac1{\sqrt2}\left(b_{{\rm AW},j}'(\psi_j^{\scriptscriptstyle+})
+b_{{\rm AW},j}^{\prime\ast}(\psi_j^{\scriptscriptstyle-})\right)
\otimes\cdots\otimes\one_{\cH_{{\rm AW},n}},\nonumber
\end{align}
where $\balpha=(0,{\boldsymbol\gamma},{\boldsymbol\gamma}')$ and
$$
\psi_j^{\pm}=\e^{\pm\beta_j[(1/2-\gamma_j)h_j-\mu_j(1/2-\gamma_j')]}\psi_j.
$$
\end{exo}

Starting with the seminal papers of Davies \cite{Dav}, Lebowitz-Spohn \cite{LS2} 
and Davies-Spohn \cite{DS},
the SF model (together with the closely related Spin-Boson model) became a paradigm in 
mathematically rigorous studies of non-equilibrium quantum statistical mechanics. Despite the
number of new results obtained in the last decade many basic questions about this model are still open.

The study of  the SF model requires sophisticated analytical tools and for reasons of space
we shall not make a detailed exposition of specific results in these lecture notes.
Instead, we will  restrict ourselves to a brief description on the main new conceptual ideas that 
made the proofs of these results possible. We refer the reader to the original articles for
more details.

The key new idea, which goes back to Jak\v si\'c-Pillet \cite{JP3}, is to use modular theory and 
quantum transfer operators to study large time limits. As we have repeatedly emphasized, before
taking the limit $t\rightarrow\infty$ one must take reservoirs to their thermodynamic limit. The 
advantage of the modular structure is that it remains intact in the thermodynamic limit. In other 
words,  the basic objects and relations of the theory remain valid for infinitely extended systems. 

In the  thermodynamic limit the Hilbert spaces ${\cal K}_j$ become infinite dimensional. Under very 
general conditions the operator $L_{\frac1\balpha}$ converges to a limiting operator. In the example
of Exercise \ref{Exo:SF}, this limit has exactly the same form \eqref{SFLalpha}
on the limiting Hilbert space $\cH_{\rm SF}$ which carries representations $\psi\mapsto b_j^\#(\psi)$ of the CAR over
the infinite dimensional $\cK_j$. Moreover, the limiting moment generating function
for the full  counting statistics \eqref{OpenGenFct} is related to this operator as in \eqref{ihes-liou0}. 
Under suitable technical assumptions on the $\psi_j$'s one then can prove a large deviation principle 
for full counting statistics by a careful study of the spectral resonances of
$L_{\frac1\balpha}$. It is precisely this last  step that is technically most demanding
and requires a number of additional assumptions. The existing proofs are based on perturbation 
arguments that require small $\lambda$ and, for technical reasons, vanishing chemical potentials
$\mu_j$.   We remark that the proofs follow line by line the  spectral scheme outlined in Section \ref{sect:LTFCS} and 
we refer the interested reader to \cite{JOPP} for details and additional 
information.

For $\balpha=(0, {\bf 1}/2, {\bf 1}/2)$, the operators  $L_{\frac1\balpha}$ is 
the standard Liouvillean $K$. The spectral analysis of this operator is a key ingredient in the
proof of return to equilibrium when all reservoirs are at the same temperature. For related
results, see \cite{JP1,BFS,DJ,FM}. More generally, the spectrum of $K$ provides information 
about the normal invariant states of the system, \ie the density matrices on the space 
$\cH_{\rm SF}$ which correspond to steady states. In particular, if $K$ has no point spectrum
then the system has no normal invariant state and hence its steady states have to be singular
w.r.t. the reference state $\rho$ (see, \eg \cite{AJPP1,Pi} for details).

In the case $\balpha=(0, {\bf 0}, {\bf 0})$, the operator  $L_{\frac1\balpha}$ reduces to the
$L^\infty$-Liouvillean (or $C$-Liouvillean) $L_\infty$ introduced in \cite{JP3}.  
In this work the relaxation to a non-equilibrium steady state was proven by using the identity
$$
\omega_t(A)=\langle\Omega_{\rm SF}|\e^{\i tL_\infty}\pi_{\rm SF}(A)\Omega_{\rm SF}\rangle,
$$
and by a careful study of resonances and resonance eigenfunctions of the operator $L_\infty$. 
This approach was adapted to the Spin-Boson model in \cite{MMS2}.

For a different approach to the large deviation principle for the spin-fermion and the 
spin-boson model we refer the reader to \cite{DR}.

\section{Electronic black box model}
\label{sect:EBBM}
\def\R{{\rm R}}
\def\L{{\rm L}}

\subsection{Model}
\bindex{model!electronic black box}\index{EBB|see{model, electronic black box}}

\begin{figure}[htbp]
\begin{center}
    \includegraphics[width=\textwidth]{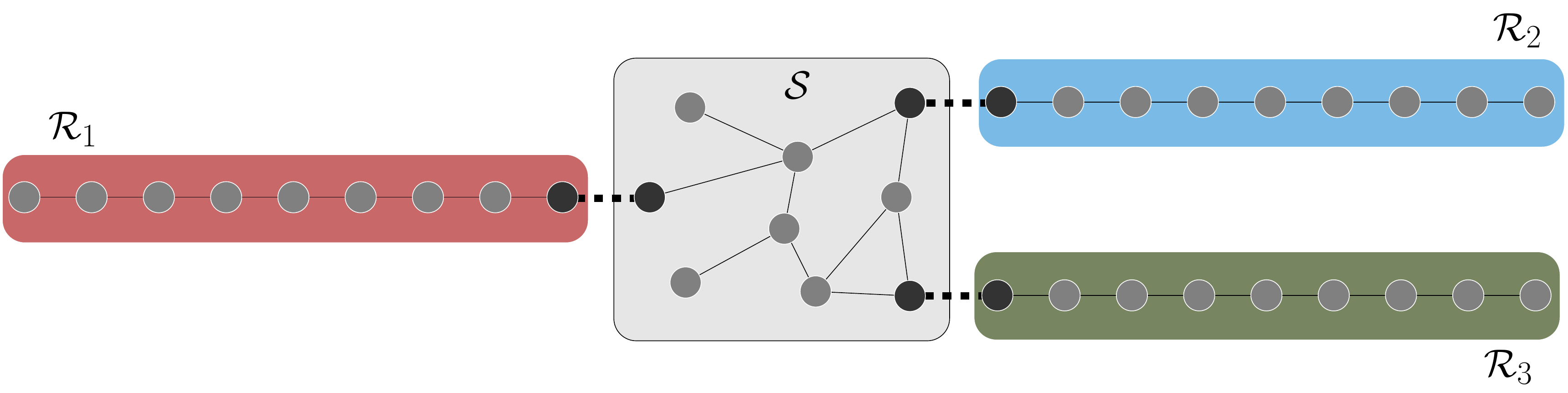}
    \caption{The EBB model with three leads.}
    \label{fig:EBBM}
\end{center}
\end{figure}

Let $\cS$ be a finite set and $h_\cS$ a one-particle Hamiltonian on $\cK_\cS=\ell^2(\cS)$.
We think of $\cS$ as a ``black box'' representing some electronic device (\eg a quantum dot).
To feed this device, we connect it to several, say $n$, reservoirs $\cR_1,\ldots,\cR_n$.
For simplicity, each reservoir $\cR_j$ is a finite lead described, in the tight binding approximation,
by a box $\Lambda=[0,M]$ in $\zz$ (see Figure \ref{fig:EBBM}). 
The  one-particle Hilbert space of a finite lead is  $\cK_j=\ell ^2(\Lambda)$  and its one-particle Hamiltonian is  
$h_j=-\frac{1}{2}\Delta_\Lambda$, where $\Delta_\Lambda$ denotes the discrete Laplacian on 
$\Lambda$ with Dirichlet boundary conditions (see Section \ref{sect:HarmoChainDef}).
\index{Laplacian!discrete Dirichlet ($\Delta_\Lambda$)}
The Electronic Black Box (EBB) model is a free Fermi gas with single particle Hilbert space
\[ 
\cK=\cK_\cS\oplus\left(\oplus_{j=1}^n\cK_j\right).
\]
In the following, we identify $\cK_\cS$ and $\cK_j$ with the corresponding subspaces of $\cK$
and we denote by $\one_\cS$ and $\one_j$ the orthogonal projections of $\cK$ on these subspaces.
In the absence of coupling between $\cS$ and the reservoirs, the Hamiltonian \index{Hamiltonian}
of the EBB model is
\[ 
H_0=\d\Gamma(h_0),
\]
where
\[ 
h_0=h_\cS \oplus\left(\oplus_{j=1}^nh_j\right).
\]
The reference state of the system, denoted $\omega_0$,  is the quasi-free state associated 
to the density 
\[ 
T_0=T_\cS\oplus\left(\oplus_{j=1}^nT_j\right),
\]
where $T_\cS>0$ is a density operator on $\cK_\cS$ which commutes with $h_\cS$ and
$$
T_j=(\one +\e^{\beta_j(h_j-\mu_j\one)})^{-1},
$$
is the Fermi-Dirac density describing the thermal equilibrium of the $j$-th reservoir at inverse
temperature $\beta_j$ and chemical potential $\mu_j$.

The coupling of the black box $\cS$ to the $j$-th reservoir is described as follows. 
Let $\chi_j\in\cK_\cS$ be a unit vector and let $\delta_0^{(j)}\in\cK_j$ be the 
Dirac delta function at site $0\in\Lambda$, both identified with elements of $\cK$. Set
$v_j=|\chi_j\rangle\langle\delta_0^{(j)}|+|\delta_0^{(j)}\rangle\langle\chi_j|$. The single  particle Hamiltonian
of the coupled EBB model is
\[
h_\lambda = h_0+\lambda\sum_{j=1}^nv_j,
\]
where $\lambda\in\rr$ is a coupling constant. Denoting by $a^\#$ the creation/annihilation operators on 
$\Gamma(\cK)$ and using Part (8) of Proposition \ref{midterm1}  we see that the full Hamiltonian of the coupled 
EBB model is
\[
H_\lambda=\d\Gamma(h_\lambda)=H_0
+\lambda\sum_{j=1}^n\left[ a^\ast(\chi_j)a(\delta_0^{(j)})+a^\ast(\delta_0^{(j)})a(\chi_j)\right],
\]
and that the induced dynamics on the CAR algebra over $\cK$ is completely determined by
$$
\tau_\lambda^t(a^\#(\psi))=\e^{\i tH_\lambda}a^\#(\psi)\e^{-\i tH_\lambda}=a^\#(\e^{\i th_\lambda}\psi).
$$
Assume that the black box $\cS$ is TRI, \ie that there exists an anti-unitary involution 
$\theta_\cS$ on $\cK_\cS$ such that $\theta_\cS h_\cS\theta_\cS^\ast=h_\cS$ and
$\theta_\cS T_\cS\theta_\cS^\ast=T_\cS$. If $\theta_\cS\chi_j=\chi_j$ for 
$j=1,\ldots,n$, then one easily shows that the EBB model is TRI, with \index{time reversal invariance}
the time reversal
$$
\Theta=\Gamma(\theta),\qquad \theta=\theta_\cS\oplus(\oplus_{j=1}^n\theta_j),
$$
where $\theta_j$ denotes the complex conjugation on $\cK_j=\ell^2(\Lambda)$.

\subsection{Fluxes} The energy operator of the $j$-th reservoir is $H_j=\d\Gamma(h_j)$.
Applying Equ. \eqref{FluxFormuLa}, using Relation \eqref{GammaLie} and Part (8) 
of Proposition \ref{midterm1},  we see that the energy flux observables 
\index{flux} are given by
\begin{align}
\Phi_j&=-\i[H_\lambda,H_j]=-\d\Gamma(\i[h_\lambda,h_j])=\lambda\,\d\Gamma(\i[h_j,v_j])\label{QFPhi}\\
&=\i\lambda\left(a^\ast(h_j\delta_0^{(j)})a(\chi_j)-a^\ast(\chi_j)a(h_j\delta_0^{(j)})\right).\nonumber
\end{align}
The charge operator of $\cS$ is $N_\cS=\d\Gamma(\one_\cS)$ and $N_j=\d\Gamma(\one_j)$ 
is the charge operator of $\cR_j$. Note that the total charge \index{charge}
$N=N_\cS+\sum_{j=1}^nN_j=\d\Gamma(\one)$  commutes with $H$. The charge flux observables are 
\begin{align}
{\cal J}_j&=-\i[H_\lambda,N_j]=-\d\Gamma(\i[h_\lambda,\one_j])=\lambda\,\d\Gamma(\i[\one_j,v_j])\label{QFJ}\\
&=\i\lambda\left(a^\ast(\delta_0^{(j)})a(\chi_j)-a^\ast(\chi_j)a(\delta_0^{(j)})\right).\nonumber
\end{align}
It follows from Part (6) of  Proposition \ref{midterm1} and Part (1) of Proposition
\ref{QuasiProps} that the heat and charge fluxes at time $t$ are
\begin{align*}
\omega_0(\tau_\lambda^t(\Phi_j))
&=2\lambda\,\Im\langle\e^{\i t h_\lambda}h_j\delta_0^{(j)}|T_0\e^{\i t h_\lambda}\chi_j\rangle,\\
\omega_0(\tau_\lambda^t({\cal J}_j))
&=2\lambda\,\Im\langle\e^{\i t h_\lambda}\delta_0^{(j)}|T_0\e^{\i t h_\lambda}\chi_j\rangle.
\end{align*}

\subsection{Entropy production} One easily concludes from Part (1) of proposition \ref{QuasiProps}
that $\omega_t=\omega_0\circ\tau_\lambda^t$ is the quasi-free state with density 
$T_t=\e^{-\i th_\lambda}T_0\e^{\i th_\lambda}$. We set
$$
k_0=\log\left(T_0(\one-T_0)^{-1}\right)
=\log\left(T_\cS(\one_\cS-T_\cS)^{-1}\right)\oplus
\left(\oplus_{j=1}^n\left[-\beta_j(h_j-\mu_j\one_j)\right]\right),
$$
so that
$$
k_t=\log\left(T_t(\one-T_t)^{-1}\right)=\e^{-\i th_\lambda}k_0\e^{\i th_\lambda}.
$$
Proposition \ref{AWModular} allows us to write the relative Hamiltonian of $\omega_t$ w.r.t. 
$\omega_0$ as\index{Hamiltonian!relative}
$$
\ell_{\omega_t|\omega_0}=\d\Gamma(k_t-k_0).
$$
It follows that the entropy production observable is\index{entropy!production}
\begin{equation}
\sigma=\left.\frac{\d\ }{\d t}\ell_{\omega_t|\omega_0}\right|_{t=0}
=\d\Gamma(-\i[h_\lambda,k_0])
=-\i[H_\lambda,Q_\cS]-\sum_{j=1}^n\beta_j(\Phi_j-\mu_j{\cal J}_j),
\label{EBBsigmaform}
\end{equation}
where $Q_\cS=\d\Gamma(\log(T_\cS(\one_\cS-T_\cS)^{-1}))$
(compare this expression with Equ. \eqref{SigmaExpro}).
The entropy balance equation thus reads
\begin{equation}
S(\omega_t|\omega_0)=\omega_0(\tau_\lambda^t(Q_\cS)-Q_\cS)
+\sum_{j=1}^n\beta_j\int_0^t\omega_s(\Phi_j-\mu_j{\cal J}_j)\,\d s.
\label{EBBBalance}
\end{equation}

\subsection{Entropic  pressure functionals} Not surprisingly, 
these functionals can be expressed in terms of one-particle quantities.
For $p\in[1,\infty[$ one has, by Lemma \ref{DetLemma},\index{entropic pressure}
\begin{align}
e_{p,t}(\alpha)
&=\log\tr\left[\left(\omega_0^{(1-\alpha)/p} \omega_t^{2\alpha/p}
\omega_0^{(1-\alpha)/p}\right)^{p/2}\right]\nonumber\\
&=\log\tr\left[
\frac1{Z_{T_0}}\Gamma\left(\left(\e^{k_0(1-\alpha)/p}\e^{k_t2\alpha/p}\e^{k_0(1-\alpha)/p}
\right)^{p/2}\right)\right]\nonumber\\
&=-\log Z_{T_0}+\log\det\left(\one+\left(\e^{k_0(1-\alpha)/p}\e^{k_t2\alpha/p}\e^{k_0(1-\alpha)/p}
\right)^{p/2}\right)\nonumber\\
&=\log\left[\frac{\det\left(\one+\left(\e^{k_0(1-\alpha)/p}\e^{k_t2\alpha/p}\e^{k_0(1-\alpha)/p}
\right)^{p/2}\right)}
{\det\left(\one+\e^{k_0}\right)}\right].\label{eptEBMone}
\end{align}
After some elementary algebra, one gets
$$
e_{p,t}(\alpha)=\log\det\left[\one+T_0\left(\e^{-k_0}\left(\e^{k_0(1-\alpha)/p}\e^{k_t2\alpha/p}\e^{k_0(1-\alpha)/p}
\right)^{p/2}-\one\right)\right].
$$
In particular, for $p=2$,
$$
e_{2,t}(\alpha)=\log\det\left(\one+T_0(\e^{-\alpha k_0}\e^{\alpha k_t}-\one)\right),
$$
and for $p=\infty$ we obtain
$$
e_{\infty,t}(\alpha)=\lim_{p\to\infty}e_{p,t}(\alpha)
=\log\det\left(\one+T_0(\e^{-k_0}\e^{(1-\alpha)k_0+\alpha k_t}-\one)\right).
$$

\medskip
\begin{exo}
\label{Exo:GenFMultiEBBM}
The multi-parameter formalism of Section \ref{sect:MultiSys} is easily adapted to the EBB model.
Indeed, one has
$$
\log\omega_0=(Q_\cS-\log Z_{T_0})-\sum_{j=1}^n\beta_j H_j+\sum_{j=1}^n\beta_j\mu_jN_j,
$$
and the $n+1$ terms in this sum form a commuting family (the scalar term $-\log Z_T$ plays no
role in the following, we can pack it with the $Q_\cS$ term which will turn out to become
irrelevant in the large time limit). Following Exercise 
\ref{Exo:AbelianOmega}, define
$$
\omega_0^\balpha=\e^{\alpha_\cS (Q_\cS-\log Z_{T_0})-\sum_{j=1}^n\alpha_j\beta_j H_j
+\sum_{j=1}^n\alpha_{n+j}\beta_j\mu_jN_j},\qquad
\omega_t^\balpha=\e^{-\i tH_\lambda}\omega_0^\balpha\e^{\i tH_\lambda},
$$
for $\balpha=(\alpha_\cS,\alpha_1,\ldots,\alpha_{2n})\in\rr^{2n+1}$.

\exop Show that the generating functional for multi-parameter full counting statistics
 \index{full counting statistics} is given by
$$
e_{2,t}(\balpha)=\log\tr\left(\omega_0^{{\bf 1}-\balpha}\omega_t^\balpha\right)
=\log\det\left(\one+T_0(\e^{-k(\balpha)}\e^{k_t(\balpha)}-\one)\right)
$$
where
$$
k(\balpha)=\alpha_\cS\log(T_\cS(\one-T_\cS)^{-1})-\sum_{j=1}^n\alpha_j\beta_j h_j
+\sum_{j=1}^n\alpha_{n+j}\beta_j\mu_j\one_j,
$$
and $k_t(\balpha)=\e^{-\i th_\lambda}k(\balpha)\e^{\i th_\lambda}$.

\exop Show that the ``naive'' generating function \eqref{naiveedef} is given by 
$$
e_{{\rm naive},t}(\balpha)=\log\det\left(\one+T_0(\e^{k_{-t}(\balpha)-k(\balpha)}-\one)\right).
$$
\end{exo}
\begin{exo}
\label{Exo:GenFEBBM}
Following Section \ref{sect:LinearResponse}, introduce the control parameters $X_j=\beta_\eq-\beta_j$ and 
$X_{n+j}=\beta_\eq\mu_\eq-\beta_j\mu_j$, where $\beta_\eq$ and $\mu_\eq$ are some equilibrium 
values of the inverse temperature and chemical potential. Denote by $\omega_X$ the quasi-free
state on the CAR algebra over $\cK$ with density
$$
T_X=\left(\one+\e^{\beta_\eq(h_\lambda-\mu_\eq\one)-\sum_{j=1}^n(X_jh_j+X_{n+j}\one_j)}\right)^{-1},
$$
and set $k_X=\log\left(T_X(\one-T_X)^{-1}\right)
=-\beta_\eq(h_\lambda-\mu_\eq\one)+\sum_{j=1}^n(X_jh_j+X_{n+j}\one_j)$.

\exop Show that 
$$
\sigma_X=\d\Gamma(-\i[h_\lambda,k_X])=\sum_{j=1}^nX_j\Phi_j+X_{n+j}{\cal J}_j,
$$
where the individual fluxes are given by \eqref{QFPhi} and \eqref{QFJ}.
\exop Show that the generalized entropic pressure is given by
$$
e_t(X,Y)=\log\det\left(\one+T_X\left(\e^{-k_X}\e^{k_{X-Y}+k_{Y,t}-k_0}-\one\right)\right),
$$
where $k_{Y,t}=\e^{-\i th_\lambda}k_Y\e^{\i th_\lambda}$.
\exop Develop the finite time linear response theory of the EBB model.
\end{exo}

\medskip

\subsection{Thermodynamic limit} 
The thermodynamic limit of the EBB model is achieved by letting $M\rightarrow \infty$, keeping the
system $\cS$ untouched. We shall not enter into a detailed description of this step which is completely 
analogous to the thermodynamic limit of the classical harmonic chain discussed in Section 
\ref{sect:ClassicTD} (see Exercise \ref{Exo:EBBTDL} below). \index{thermodynamic limit}
The one particle Hilbert space of the reservoir $\cR_j$ becomes 
$\cK_j=\ell^2(\nn)$ and its one particle Hamiltonian $h_j=-\frac12\Delta$, where $\Delta$ is the 
discrete Laplacian on $\nn$ with Dirichlet boundary condition. Using the discrete Fourier transform
$$
\widehat\psi(\xi)=\sqrt{\frac2\pi}\,\sum_{x\in\nn}\psi(x)\sin(\xi(x+1)),
$$
we can identify $\cK_j$ with $L^2([0,\pi],\d\xi)$ and $h_j$ becomes the operator of multiplication
by $\varepsilon(\xi)=1-\cos\xi$. In particular, the spectrum of $h_j$ is purely absolutely continuous
and fills the interval $[0,2]$ with constant multiplicity one. Thus, the spectrum of the decoupled
Hamiltonian $h_0$ consists of an absolutely continuous part filling the same interval with
constant multiplicity $n$ and of a discrete part given by the eigenvalues of $h_\cS$.
We denote by $\one_\cR=\one-\one_\cS=\sum_{j=1}^n\one_j$ the projection on the absolutely 
continuous part of $h_0$. In the momentum representation one has
$\cK_\cR=\Ran\one_\cR=L^2([0,\pi])\otimes\cc^n$.  Denoting by $1_j=|e_j\rangle\langle e_j|$ 
the orthogonal projection
of $\cc^n$ onto the subspace generated by the $j$-th vector of its standard basis, we have
$\one_j=\one\otimes 1_j$ and $h_j=\varepsilon(\xi)\otimes 1_j$.

\medskip
\begin{exo}\label{Exo:EBBTDL}
Denote by the subscript ${(\cdot)}_M$ the dependence on the parameter $M$ of the various objects
associated to the EBB model, \eg $\omega_{M,0}$ is the reference state with density 
$T_{M,0}=T_\cS\oplus(\oplus_{j=1}^nT_{M,j})$, etc.

\exop Show that
$$
\lim_{M\to\infty}T_{M,0}(\e^{-\alpha k_{M,0}}\e^{\alpha k_{M,t}}-\one)
=T_0(\e^{-\alpha k_0}\e^{\alpha k_t}-\one),
$$
holds in trace norm, where $T_0=\slim_{M\to\infty}T_{M,0}$, $k_0=\log(T_0(\one-T_0))$,
$k_t=\e^{-\i th_\lambda}k_0\e^{\i th_\lambda}$ and $h_\lambda=\slim_{M\to\infty}h_{M,\lambda}$.

\noindent{\sl Hint}: write $\e^{-\alpha k_{M,0}}\e^{\alpha k_{M,t}}-\one$ as the integral of its derivative w.r.t. $t$
and observe that $[h_{M,\lambda},k_{M,0}]$ is a finite rank operator that does not depend on $M$.

\exop Show that, for any $\alpha,t\in\rr$,
$$
\lim_{M\to\infty}e_{M,2,t}(\alpha)
=\log\det\left(\one+T_0(\e^{-\alpha k_0}\e^{\alpha k_t}-\one)\right).
$$

\noindent{\sl Hint}: recall that 
$\det(\one+T_{M,0}(\e^{-\alpha k_{M,0}}\e^{\alpha k_{M,t}}-\one))
=\omega_{M,0}(\Gamma(\e^{-\alpha k_{M,0}/2}\e^{\alpha k_{M,t}}\e^{-\alpha k_{M,0}/2}))>0$.
\newline
Remark. The implications of this exercise are described in Proposition \ref{paris-implications}.
\end{exo}
\begin{exo}
Let ${\rm P}_{M,t}$ denote the spectral measure of $\log(\Delta_{\omega_{M,t}|\omega_{M,0}})$ 
and $\xi_{\omega_{M,0}}$. Through the following steps, show that the spectral measure 
${\rm P}_{t}$ of $\log(\Delta_{\omega_{t}|\omega_{0}})$  and $\xi_{\omega_{0}}$ is the weak 
limit of the sequence $\{{\rm P}_{M,t}\}$. (Up to a rescaling, ${\rm P}_{M,t}$ is the
full counting statistics of the finite EBB model.)

\exop Show that, for all $\alpha\in\rr$, the characteristic function of ${\rm P}_{M,t}$,
\begin{align*}
\chi_{M,t}(\alpha)&=\int\e^{\i\alpha x}\d{\rm P}_{M,t}(x)
=(\xi_{\omega_{M,0}}|\Delta_{\omega_{M,t}|\omega_{M,0}}^{\i\alpha}\xi_{\omega_{M,0}})\\
&=\tr\left(\omega_{M,0}^{1-\i\alpha}\,\omega_{M,t}^{\i\alpha}\right)\\
&=\det\left(\one+T_{M,0}(\e^{\i\alpha k_{M,t}}\e^{-\i\alpha k_{M,0}}-\one)\right),
\end{align*}
converges, as $M\to\infty$, towards
$$
\chi_{t}(\alpha)=\det\left(\one+T_0(\e^{\i\alpha k_t}\e^{-\i\alpha k_0}-\one)\right)
=\omega_0\left(\Gamma(\e^{\i\alpha k_t}\e^{-\i\alpha k_0})\right).
$$

\exop In the Araki-Wyss representation associated to the state $\omega_0$, show that
$$
(\xi_{\omega_0}|\Delta_{\omega_t|\omega_0}^{\i\alpha}\xi_{\omega_0})=
(\xi_{\omega_0}|\Gamma_t(\alpha)\xi_{\omega_0}),
$$
where the cocycle \index{cocycle}
$\Gamma_t(\alpha)=\Delta_{\omega_t|\omega_0}^{\i\alpha}\Delta_{\omega_0}^{-\i\alpha}$
satisfies the Cauchy problem
$$
\frac{\d\ }{\d\alpha}\Gamma_t(\alpha)
=\i\,\Gamma_t(\alpha)\left(
\Delta_{\omega_0}^{\i\alpha}\pi_{\rm AW}(\ell_{\omega_t|\omega_0})\Delta_{\omega_0}^{-\i\alpha}
\right),
\qquad \Gamma_t(0)=\one.
$$
\exop Show that $\Gamma_t(\alpha)=\pi_{\rm AW}(\gamma_t(\alpha))$ where
$$
\frac{\d\ }{\d\alpha}\gamma_t(\alpha)
=\i\,\gamma_t(\alpha)
\left(\e^{\i\alpha\d\Gamma(k_0)}\ell_{\omega_t|\omega_0}\e^{-\i\alpha\d\Gamma(k_0)}\right),
\qquad \gamma_t(0)=\one.
$$
Conclude that
$(\xi_{\omega_0}|\Delta_{\omega_t|\omega}^{\i\alpha}\xi_{\omega_0})=\omega_0(\gamma_t(\alpha))$.
\exop Show that
$$
\gamma_t(\alpha)=[D\omega_t:D\omega_0]^\alpha
=\e^{\i\alpha\d\Gamma(k_t)}\e^{-\i\alpha\d\Gamma(k_0)}
=\Gamma(\e^{\i\alpha k_t}\e^{-\i\alpha k_0}),
$$
and conclude that
$$
\chi_{t}(\alpha)=\int\e^{\i\alpha x}\d{\rm P}_t(x).
$$ 

\exop Invoke the L\'evy-Cram\'er continuity theorem (Theorem 7.6 in \cite{Bi1})
to conclude that ${\rm P}_{M,t}$ converges weakly towards ${\rm P}_t$.
\end{exo}

\medskip
\subsection{Large time limit} Let us briefly discuss the limit $t\to\infty$. For simplicity, we shall
assume that the one particle Hamiltonian $h_\lambda$ has purely absolutely continuous spectrum.
This is the generic situation for small coupling $\lambda$ in the fully resonant case where 
$\sp(h_\cS)\subset]0,2[$. Since $h_\lambda-h_0=v=\sum_{j=1}^nv_j$ is finite rank,
the wave operators \index{wave operator}
$$
w_\pm=\slim_{t\to\pm\infty}\e^{\i th_\lambda}\e^{-\i th_0}\one_\cR,
$$
exist and are complete, $w_\pm w_\pm^\ast=\one$, $w_\pm^\ast w_\pm=\one_\cR$. The
scattering matrix $s=w_+^\ast w_-$ is unitary on $\cK_\cR$. \index{scattering matrix}
It acts as the operator of multiplication 
by a unitary $n\times n$ matrix $s(\xi)=[s_{jk}(\xi)]$. Since $[h_0,T_0]=0$, one has
\begin{align*}
\lim_{t\to\infty}\langle\psi|T_t\phi\rangle
&=\lim_{t\to\infty}\langle\e^{\i th_\lambda}\psi|T_0\e^{\i th_\lambda}\phi\rangle\\
&=\lim_{t\to\infty}\langle\e^{-\i th_0}\e^{\i th_\lambda}\psi|T_0\e^{-\i th_0}\e^{\i th_\lambda}\phi\rangle\\
&=\langle w_-^\ast\psi|T_0w_-^\ast\phi\rangle=\langle\psi|T_+\phi\rangle,
\end{align*}
whith $T_+=w_-T_0w_-^\ast$. It follows that for any polynomial $A$ in the creation/annihilation
operators on $\Gamma(\cK)$, one has
$$
\lim_{t\to\infty}\omega_0\circ\tau_\lambda^t(A)=\omega_+(A),
$$
where $\omega_+$ is the quasi-free state with density $T_+$. We conclude that the NESS $\omega_+$
of the EBB model is the quasi-free state with density\index{state!non-equilibrium steady}
\begin{equation}
T_+=w_-T_0w_-^\ast.
\label{Tplusform}
\end{equation}

The large time asymptotics of the entropic pressure functionals can be obtained along the same 
line as in Section \ref{sect:GF}. We shall only consider the case $p=2$
and leave the general case as an exercise.\index{entropic pressure}

Starting with \eqref{eptEBMone} and using the result of Exercise \ref{Exo:Dlog}, we can write
\begin{align*}
\frac{\d\ }{\d\alpha}\,e_{2,t}(\alpha)
&=\frac{\d\ }{\d\alpha}\,\tr\log\left(\one+\e^{(1-\alpha)k_0}\e^{\alpha k_t}\right)\\
&=\tr\left((\one+\e^{(1-\alpha)k_0}\e^{\alpha k_t})^{-1}\e^{(1-\alpha)k_0}(k_t-k_0)\e^{\alpha k_t}\right)\\
&=\tr\left((\one+\e^{-(1-\alpha)k_0}\e^{-\alpha k_t})^{-1}(k_t-k_0)\right)\\
&=-t\int_0^1\tr\left((\one+\e^{-(1-\alpha)k_0}\e^{-\alpha k_t})^{-1}
\e^{-\i tuh_\lambda}\i[h_\lambda,k_0]\e^{\i tuh_\lambda}\right)\d u\\
&=-t\int_0^1\tr\left(\e^{\i tuh_\lambda}
(\one+\e^{-(1-\alpha)k_0}\e^{-\alpha k_t})^{-1}\e^{-\i tuh_\lambda}\i[h_\lambda,k_0]\right)\d u\\
&=-t\int_0^1\tr\left(
(\one+\e^{-(1-\alpha)k_{-tu}}\e^{-\alpha k_{t(1-u)}})^{-1}
\i[h_\lambda,k_0]\right)\d u.
\end{align*}
The final relation 
\[
\frac{\d\ }{\d\alpha}\,e_{2,t}(\alpha)=-t\int_0^1\tr\left(
(\one+\e^{-(1-\alpha)k_{-tu}}\e^{-\alpha k_{t(1-u)}})^{-1}
\i[h_\lambda,k_0]\right)\d u
\]
remains valid after the thermodynamic limit is taken.
Since $k_0$ is a bounded operator commuting with $h_0$, one easily shows that
$$
\slim_{t\to\pm\infty}k_t=k_\pm=w_\mp k_0w_\mp^\ast,
$$
which leads to
\begin{align*}
\slim_{t\to\infty}(\one+\e^{-(1-\alpha)k_{-ts}}\e^{-\alpha k_{t(1-s)}})^{-1}
&=(\one+\e^{-(1-\alpha)k_-}\e^{-\alpha k_+})^{-1}\\
&=(\one+w_+\e^{-(1-\alpha)k_0}w_+^\ast w_-\e^{-\alpha k_0}w_-^\ast)^{-1}\\
&=(\one+w_-s^\ast\e^{-(1-\alpha)k_0}s\e^{-\alpha k_0}w_-^\ast)^{-1}\\
&=w_-(\one+s^\ast\e^{-(1-\alpha)k_0}s\e^{-\alpha k_0})^{-1}w_-^\ast.
\end{align*}
Since $\i[h_\lambda,k_0]$ is finite rank, it follows that
\begin{align*}
\lim_{t\to\infty}\tr_{\cK}&\left((\one+\e^{-(1-\alpha)k_{-tu}}\e^{-\alpha k_{t(1-u)}})^{-1}\i[h_\lambda,k_0]\right)\\
&=\tr_{\cK_\cR}\left((\one+s^\ast\e^{-(1-\alpha)k_0}s\e^{-\alpha k_0})^{-1}{\cal T}\right),
\end{align*}
where ${\cal T}=w_-^\ast\i[h_\lambda,k_0]w_-$. Since $e_{2,t}(0)=0$, we can write
$$
e_{2,+}(\alpha)=\lim_{t\to\infty}\frac1t\,e_{2,t}(\alpha)
=\lim_{t\to\infty}\frac1t\,\int_0^\alpha\frac{\d\ }{\d\gamma}\,e_{2,t}(\gamma)\,\d\gamma,
$$
and the dominated convergence theorem yields
\begin{align*}
e_{2,+}(\alpha)
&=-\int_0^\alpha\int_0^1
\tr\left((\one+s^\ast\e^{-(1-\gamma)k_0}s\e^{-\gamma k_0})^{-1}{\cal T}\right)\d u\d\gamma\\
&=-\int_0^\alpha
\tr\left((\one+s^\ast\e^{-(1-\gamma)k_0}s\e^{-\gamma k_0})^{-1}{\cal T}\right)\d\gamma.
\end{align*}
The trace class operator $\cal T$ on $\cK_\cR$ has an integral kernel ${\cal T}(\xi,\xi')$ in the 
momentum representation. Following the argument leading to \eqref{Tmatrixelt}, one shows that its 
diagonal is given by
\begin{equation}
{\cal T}(\xi,\xi)=\frac{\varepsilon'(\xi)}{2\pi}\left(s^\ast(\xi)k(\xi)s(\xi)-k(\xi)\right),
\label{DiagTform}
\end{equation}
where $k(\xi)$ is the operator on $\cc^n$ defined by
$$
k(\xi)=-\sum_{j=1}^n\beta_j(\varepsilon(\xi)-\mu_j)1_j.
$$
Thus, one has
\begin{align*}
&\tr_{\cK_\cR}\left((\one+s^\ast\e^{-(1-\alpha)k_0}s\e^{-\alpha k_0})^{-1}{\cal T}\right)\\
&=-\int_0^\pi\tr_{\cc^n}\left((\one+s^\ast(\xi)\e^{-(1-\alpha)k(\xi)}s(\xi)\e^{-\alpha k(\xi)})^{-1}
\left(k(\xi)-s^\ast(\xi)k(\xi)s(\xi)\right)\right)\varepsilon'(\xi)
\frac{\d\xi}{2\pi}\\
&=-\frac{\d\ }{\d\alpha}\int_0^\pi
\tr_{\cc^n}\left(\log(\one+s^\ast(\xi)\e^{(1-\alpha)k(\xi)}s(\xi)\e^{\alpha k(\xi)})\right)\varepsilon'(\xi)
\frac{\d\xi}{2\pi},
\end{align*}
and we conclude that
$$
e_{2,+}(\alpha)=\int_0^\pi\log\left[
\frac{\det\left(\one+\e^{(1-\alpha)k(\xi)}s(\xi)\e^{\alpha k(\xi)}s^\ast(\xi)\right)}
{\det\left(\one+\e^{k(\xi)}\right)}
\right]\frac{\d\varepsilon(\xi)}{2\pi}.
$$
After a simple algebraic manipulation, this can be rewritten as
\begin{equation}
e_{2,+}(\alpha)=\int_0^\pi\log
\det\left(\one+T(\xi)(\e^{-\alpha k(\xi)}s(\xi)\e^{\alpha k(\xi)}s^\ast(\xi)-\one)
\right)\frac{\d\varepsilon(\xi)}{2\pi},
\label{etwoplusformone}
\end{equation}
where $T(\xi)=(\one+\e^{-k(\xi)})^{-1}$.
In the following exercise, this calculation is extended to various other entropic functionals.

\medskip
\begin{exo}\label{Ex:final}
\exop Show that for $p\in[1,\infty[$ one has
\begin{align*}
&e_{p,+}(\alpha)=\lim_{t\to\infty}\frac1t\,e_{p,t}(\alpha)\\
&=\int_{0}^\pi\log\det\left[\one+T(\xi)\left(\e^{-k(\xi)}
\left(\e^{k(\xi)(1-\alpha)/p}s(\xi)\e^{k(\xi)2\alpha/p}s^\ast(\xi)\e^{k(\xi)(1-\alpha)/p}\right)^{p/2}-\one
\right)\right]\frac{\d\varepsilon(\xi)}{2\pi}.
\end{align*}
\exop Show that
\begin{align*}
e_{\infty,+}(\alpha)&=\lim_{t\to\infty}\frac1t\,e_{\infty,t}(\alpha)\\
&=\int_0^\pi\log\det\left(\one+T(\xi)
(\e^{-k(\xi)}\e^{(1-\alpha)k(\xi)+\alpha s(\xi)k(\xi)s(\xi)^\ast}-\one)\right)\frac{\d\varepsilon(\xi)}{2\pi}.
\end{align*}

\exop Compute 
\[
 e_{{\rm naive}, +}(\alpha)=\lim_{t \rightarrow \infty}\frac{1}{t}e_{{\rm naive}, t}(\alpha).
\]

\exop Show that the large time asymptotics of the multi-parameter functional of Exercise 
\ref{Exo:GenFMultiEBBM} is given by
\begin{align*}
e_{2,+}(\balpha)&=\lim_{t\to\infty}\frac1t\,e_{2,t}(\balpha)\\
&=\int_0^\pi\log\det\left(\one+T(\xi)
(\e^{-k(\balpha,\xi)}s(\xi)\e^{k(\balpha,\xi)}s^\ast(\xi)-\one)\right)\frac{\d\varepsilon(\xi)}{2\pi},
\end{align*}
where 
$$
k(\balpha,\xi)=-\sum_{j=1}^n\beta_j\left(\alpha_j\varepsilon(\xi)-\alpha_{n+j}\mu_j\right)1_j.
$$
Note in particular that $e_{2,+}(\balpha)$ does not depend on the first component $\alpha_\cS$
of $\balpha$.
\exop Show that the large time asymptotics of the generalized functional of Exercise 
\ref{Exo:GenFEBBM} is given by
\begin{align*}
e_{+}(X,Y)&=\lim_{t\to\infty}\frac1t\,e_{t}(X,Y)\\
&=\int_0^\pi\log\det\left(\one+T_X(\xi)
(\e^{-k_X(\xi)}\e^{k_{X-Y}(\xi)+s(\xi)k_Y(\xi)s^\ast(\xi)-k_0(\xi)}-\one)\right)\frac{\d\varepsilon(\xi)}{2\pi},
\end{align*}
where $k_X(\xi)$ is the diagonal $n\times n$ matrix with
entries $-(\beta_\eq-X_j)\varepsilon(\xi)+(\beta_\eq\mu_\eq+X_{n+j})$ and 
$T_X(\xi)=(\one+\e^{-k_X(\xi)})^{-1}$.
\exop Develop  the linear response theory of the EBB model.\index{linear response}
\end{exo}

\bigskip
For $\xi\in[0,\pi]$, denote by $\omega_\xi$ the density matrix
$$
\omega_\xi=\frac{\Gamma(\e^{k(\xi)})}{\tr_{\Gamma(\cc^n)}(\Gamma(\e^{k(\xi)}))},
$$
on $\Gamma(\cc^n)$. Clearly, $\omega_\xi$ defines a state on $\Gamma(\cc^n)$ which is
quasi-free with density $T(\xi)$. By Part (3) of Proposition \ref{QuasiProps}, the R\'enyi relative
entropy of the state $\Gamma(s(\xi))\omega_\xi\Gamma(s(\xi))^\ast$ w.r.t. $\omega_\xi$
is given by \index{entropy!R\'enyi}
\begin{align*}
S_\alpha(\Gamma(s(\xi))\omega_\xi\Gamma(s(\xi))^\ast|\omega_\xi)
&=\log\tr\left(\omega_\xi^{1-\alpha}\Gamma(s(\xi))\omega_\xi^\alpha\Gamma(s(\xi))^\ast\right)\\
&=\log\omega_\xi\left(\Gamma(\e^{-\alpha k(\xi)}s(\xi)\e^{\alpha k(\xi)}s^\ast(\xi))\right)\\
&=\log\det\left(\one+T(\xi)(\e^{-\alpha k(\xi)}s(\xi)\e^{\alpha k(\xi)}s^\ast(\xi)-\one)\right).
\end{align*}
Thus, we can rewrite Formula \eqref{etwoplusformone} as
$$
e_{2,+}(\alpha)
=\int_0^\pi S_\alpha(\Gamma(s(\xi))\omega_\xi\Gamma(s(\xi))^\ast|\omega_\xi)
\frac{\d\varepsilon(\xi)}{2\pi}.
$$
Using the second identity in \eqref{SDerivSalpha}, we deduce
$$
\left.\frac{\d\ }{\d\alpha}e_{2,+}(\alpha)\right|_{\alpha=1}
=-\int_0^\pi S(\Gamma(s(\xi))\omega_\xi\Gamma(s(\xi))^\ast|\omega_\xi)
\frac{\d\varepsilon(\xi)}{2\pi}.
$$
Since $\log\omega_\xi=\d\Gamma(k(\xi))-\log\det(\one+\e^{k(\xi)})$,
Relation \eqref{GammadGammaGamma} and Part (4) of Proposition \ref{QuasiProps} yield
\begin{align*}
S(\Gamma(s(\xi))\omega_\xi\Gamma(s(\xi))^\ast|\omega_\xi)&=
\tr\left[\Gamma(s(\xi))\omega_\xi\Gamma(s(\xi))^\ast
\left(\log\omega_\xi-\Gamma(s(\xi))\log\omega_\xi\Gamma(s(\xi))^\ast\right)\right]\\
&=\tr\left[\omega_\xi(\Gamma(s(\xi))^\ast\log\omega_\xi\Gamma(s(\xi))-\log\omega_\xi)\right]\\
&=\omega_\xi\left(\d\Gamma(s^\ast(\xi)k(\xi)s(\xi)-k(\xi))\right)\\
&=\tr\left(T(\xi)(s^\ast(\xi)k(\xi)s(\xi)-k(\xi))\right).
\end{align*}
Hence, it follows from \eqref{DiagTform}and \eqref{Tplusform} that
\begin{align*}
\int_0^\pi S(\Gamma(s(\xi))\omega_\xi\Gamma(s(\xi))^\ast|\omega_\xi)\frac{\d\varepsilon(\xi)}{2\pi}
&=-\int_0^\pi\tr\left(T(\xi){\cal T}(\xi,\xi)\right)\d\xi\\
&=-\tr(T_0{\cal T})\\
&=-\tr(T_0w_-^\ast\i[h_\lambda,k_0]w_-)\\
&=-\tr(T_+\i[h_\lambda,k_0]).
\end{align*}
Finally, \eqref{EBBsigmaform} allows us to write
$$
-\tr\left(T_+\i[h_\lambda,k_0]\right)=\omega_+\left(\d\Gamma(-\i[h_\lambda,k_0])\right)
=\omega_+(\sigma).
$$
Thus, we have shown that
$$
\left.\frac{\d\ }{\d\alpha}e_{2,+}(\alpha)\right|_{\alpha=1}=
\omega_+(\sigma)=-\int_0^\pi S(\Gamma(s(\xi))\omega_\xi\Gamma(s(\xi))^\ast|\omega_\xi)
\frac{\d\varepsilon(\xi)}{2\pi}.
$$
Invoking Part (1) of Proposition \ref{PropSprops} we observe that if $\omega_+(\sigma)=0$ then
we must have 
$$
S(\Gamma(s(\xi))\omega_\xi\Gamma(s(\xi))^\ast|\omega_\xi)=0,
$$
for almost all 
$\xi\in[0,\pi]$ which in turn implies that $\Gamma(s(\xi))\omega_\xi\Gamma(s(\xi))^\ast=\omega_\xi$,
\ie that $[k(\xi),s(\xi)]=0$ for almost all $\xi\in[0,\pi]$. The last condition can be written as
$$
\left[(\beta_k-\beta_j)\varepsilon(\xi)-(\beta_k\mu_k-\beta_j\mu_j)\right]s_{jk}(\xi)=0,
$$
for all $j,k\in\{1,\ldots,n\}$, and we conclude that if there exists $j,k\in\{1,\ldots,n\}$ and a set $\Omega\subset[0,\pi]$ of 
positive Lebesgue measure such  that $j\not=k$, $s_{jk}(\xi)\not=0$ for $\xi\in\Omega$ and
$(\beta_j,\mu_j)\not=(\beta_k,\mu_k)$, then $\omega_+(\sigma)>0$. In more physical terms,
if there is an open scattering channel between two leads $\cR_j$ and $\cR_k$ which are not in
mutual thermal equilibrium, then entropy production in the NESS is strictly positive. \index{entropy!production}

Note that since \eqref{EBBsigmaform} implies
$$
\omega_+(\sigma)=-\sum_{j=1}^n\beta_j\left(\omega_+(\Phi_j)-\mu_j\omega_+({\cal J}_j)\right),
$$
the expected currents $\omega_+(\Phi_j)$, $\omega_+({\cal J}_j)$ can not all vanish if entropy 
production is strictly positive.

\medskip
\begin{exo}
Deduce from Relation \eqref{EBBBalance} that
$$
-\lim_{t\to\infty}\frac1t S(\omega_t|\omega_0)=\omega_+(\sigma).
$$
Thus, if $\omega_+(\sigma)>0$ then the entropy of $\omega_t$ w.r.t. $\omega_0$ 
diverges as $t\to\infty$.
\end{exo}
\begin{exo}
\label{Exo:LBFormula}
Derive the Landauer-B\"uttiker formulas for the expected energy and charge currents in the
steady state $\omega_+$,\index{formula!Landauer-B\"uttiker}
\begin{align*}
\omega_+(\Phi_j)&=\sum_{k=1}^n\int_0^\pi t_{jk}(\xi)(\varrho_j(\xi)-\varrho_k(\xi))\varepsilon(\xi)
\frac{\d\varepsilon(\xi)}{2\pi},\\
\omega_+({\cal J}_j)&=\sum_{k=1}^n\int_0^\pi t_{jk}(\xi)(\varrho_j(\xi)-\varrho_k(\xi))
\frac{\d\varepsilon(\xi)}{2\pi},
\end{align*}
where $\varrho_j(\xi)=(1+\e^{\beta_j(\varepsilon(\xi)-\mu_j)})^{-1}$ is the Fermi-Dirac
density of the $j$-th reservoir and
$$
t_{jk}(\xi)=\left|s_{jk}(\xi)-\delta_{jk}\right|^2.
$$
{\sl Hint}: start with 
$\omega_+(\Phi_j)=-\tr\left(T_+\i[h_\lambda,h_j]\right)=-\tr\left(T_0{\cal T}_j\right)$ where
${\cal T}_j=w_-^\ast\i[h_\lambda,h_j]w_-$, and deduce from \eqref{DiagTform} that
the diagonal part of the integral kernel of ${\cal T}_j$ is given by
$$
{\cal T}_j(\xi,\xi)
=\frac{\varepsilon'(\xi)}{2\pi}\varepsilon(\xi)\left(s^\ast(\xi)1_js(\xi)-1_j\right).
$$
Proceed in a similar way for the charge currents. (For more information on the Landauer-B\"uttiker
formalism, see \cite{Da,Im}. More general mathematical 
derivations can be found in \cite{AJPP2,Ne,BS}.
\end{exo}
\begin{exo} 
Starting with  the Landauer-B\"uttiker formulas develop the linear response theory of  the EBB model.
\index{linear response} 
\end{exo}
\begin{exo} 
Consider the full counting statistics of charge transport in the framework of\index{full counting statistics}
Section \ref{sect:MultiFCS}. Let ${\mathbb P}^c_t({\bf q})$, ${\bf q}=(q_1,\ldots,q_n)$, denote the 
probability for the results, $\mathbf n$ and $\mathbf n'$, of two successive joint measurements of 
${\mathbf N}=(N_1,\ldots,N_n)$, at time $0$ and $t$, to be such that
${\mathbf n}'-{\mathbf n}=t{\mathbf q}$. Loosely speaking, ${\mathbb P}^c_t(q_1,\ldots,q_n)$ is
the probability for the charge (number of fermions) of the reservoir $\cR_j$ to increase by $tq_j$
($j=1,\ldots,n$) during the time interval $[0,t]$.
Denote by
$$
\chi_t({\boldsymbol\nu})=
\sum_{\bf q}{\mathbb P}^c_t({\bf q})\e^{-t{\boldsymbol\nu}\cdot{\bf q}},
$$
the Laplace transform of this distribution (that is, the moment generating function of
${\mathbb P}^c_t$).
\exop Show that the logarithm of $\chi_t({\boldsymbol\nu})$ is related to the functional 
$e_{2,t}(\balpha)$ of Exercise \ref{Exo:GenFMultiEBBM} by
$$
\log\chi_t({\boldsymbol\nu})=e_{2,t}({\bf 1}-\balpha),
$$
provided ${\boldsymbol\nu}=(\nu_1,\ldots,\nu_n)$ is related to 
$\balpha=(\alpha_\cS,\alpha_1,\ldots,\alpha_{2n})$ according to
$$
\alpha_j=\alpha_\cS=0,\quad
\nu_j=-\alpha_{n+j}\beta_j\mu_j, \qquad(j=1,\ldots,n).
$$
\exop Show that in the thermodynamic limit
$$
\chi_t({\boldsymbol\nu})
=\det\left(\one+T_0(\e^{q_t({\boldsymbol\nu})}\e^{-q({\boldsymbol\nu})}-\one)\right),
$$
where
$$
q({\boldsymbol\nu})=\sum_{j=1}^n\nu_j\one_j,
$$
and $q_t({\boldsymbol\nu})=\e^{-\i th_\lambda}q({\boldsymbol\nu})\e^{\i th_\lambda}$.

\noindent{\sl Hint}: combine Part 1 with the result of Exercise \ref{Exo:GenFMultiEBBM}.
\exop Derive the Levitov-Lesovik formula \bindex{formula!Levitov-Lesovik}
$$
\lim_{t\to\infty}\frac1t\log\chi_t({\boldsymbol\nu})=\int_0^\pi
\log\det\left(\one+T(\xi)(s^\ast(\xi)s^{\boldsymbol\nu}(\xi)-\one)\right)
\frac{\d\varepsilon(\xi)}{2\pi},
$$
where the matrix $s^{\boldsymbol\nu}(\xi)=[s^{\boldsymbol\nu}_{jk}(\xi)]$ is defined by
$$
s^{\boldsymbol\nu}_{jk}(\xi)=s_{jk}(\xi)\e^{\nu_k-\nu_j}.
$$
(See \cite{LL}, where the Fourier transform of the probability distribution
${\mathbb P}_t^c$ is considered instead of its Laplace transform. See also 
\cite{ABGK}.)
\end{exo}
\begin{exo}\label{believe} 
Consider EBB model with two reservoirs. Prove that the following statements are equivalent.
\exop $e_{p, +}(\alpha)$ does not depend on $p$. 

\exop $s_{11}(\xi)=s_{22}(\xi)=0$  for Lebesgue a.e. $\xi \in [0, \pi]$. 

\exop The fluctuation relation $e_{{\rm naive}, +}(\alpha)=e_{{\rm naive}, +}(1-\alpha)$ holds.

\exop $e_{{\rm naive}, +}(\alpha) = e_{\infty, +}(\alpha)$.

\end{exo}
\begin{exo} \label{big-trampo}Consider the following variant of the EBB model. ${\cal S}$ is a box 
$\Lambda = [-l, l]$ in $\zz$ and $h_\cS=-\frac{1}{2}\Delta_{\Lambda}$ is the discrete Laplacian on 
$\Lambda$ with Dirichlet boundary condition. \index{Laplacian!discrete Dirichlet ($\Delta_\Lambda$)}
The box ${\cal S}$ is connected to the left and right lead which, before the  thermodynamical limit 
is taken, are described by the  boxes $\Lambda_L=]-M, -l-1]$, $\Lambda_R=[l+1, M[$, where $l\ll  M$,
and after the thermodynamic limit is taken, by the boxes $\Lambda_L=]-\infty, -l-1]$, 
$\Lambda_R=[l+1, \infty[$. The one particle Hamiltonians are $h_L =-\frac{1}{2}\Delta_{\Lambda_L}$, 
$h_R=-\frac{1}{2} \Delta_{\Lambda_R}$, where, as usual, $\Delta_{\Lambda_L}$ and 
$\Delta_{\Lambda_R}$ are the discrete Laplacians on $\Lambda_L$ and $\Lambda_R$ with 
Dirichlet boundary condition. The corresponding EBB model  is a free Fermi gas with  single particle 
Hilbert space
\[
\ell^2(\Lambda_L)\oplus \ell^2(\Lambda) \oplus \ell^2(\Lambda_R)
=\ell^2(\Lambda_L \cup \Lambda\cup \Lambda_R).
\] 
In the absence of coupling its Hamiltonian is $H_0=\d\Gamma(h_0)$, where 
$h_0= h_L \oplus h_\cS\oplus h_R$. The Hamiltonian of the joint system is $H=\d\Gamma(h)$, 
where $h =-\frac{1}{2}\Delta_{\Lambda_L\cup\Lambda\cup\Lambda_R}$ and 
$\Delta_{\Lambda _L \cup \Lambda \cup \Lambda_R}$ is the discrete Laplacian on 
$\Lambda_L \cup \Lambda \cup \Lambda_R$ with Dirichlet boundary condition. The reference 
state of the system is a quasi free state with  density 
\[ 
T_0=T_L \oplus T_\cS\oplus T_R,
\]
where $T_\cS >0$ is a density operator on $\ell^2(\Lambda)$ that commutes with $h_\cS$ and 
\[
T_L =(\one +\e^{\beta_L (h_L-\mu_L\one)})^{-1}, \qquad
  T_R=(\one +\e^{\beta_R (h_R-\mu_R\one)})^{-1},
\]
are the Fermi-Dirac densities of the left and right reservoir. 
\exop Discuss in detail the thermodynamic limit $M\rightarrow \infty$ and compare the 
model with the classical harmonic chain discussed in Section  \ref{chap:HarmoChain}.

The remaining parts of this exercise concern the  infinitely extended model. 

\exop Using the discrete Fourier transform 
\[
\ell^2 (\Lambda_L)\oplus\ell^2( \Lambda_R)\ni \psi_L \oplus \psi_R \mapsto
\widehat \psi_L \oplus \widehat \psi_R \in L^2([0, \pi], \d\xi)\oplus L^2([0, \pi], \d\xi),
\]
\[ \widehat \psi_L(\xi)=\sqrt{\frac{2}{\pi}}\sum_{x\in {\Lambda _L}}\psi_L(x)\sin (\xi(x-1)), \qquad 
\widehat \psi_R(\xi)=\sqrt{\frac{2}{\pi}}\sum_{x\in {\Lambda _R}}\psi_R(x)\sin (\xi(x+1)),
\]
identify  $h_L \oplus h_R$ with the operator of multiplication by $(1-\cos\xi)\oplus (1-\cos \xi)$ on 
$L^2([0, \pi], \d\xi)\oplus L^2([0, \pi], \d\xi)$. 
The wave operators \index{wave operator}
\[
w_{\pm}=\slim_{t\rightarrow \pm \infty}\e^{\i t h}\e^{-\i t h_0}\one_{\cR},
\]
exist and  are complete ($\one_{\cR}$ is the orthogonal projection  onto 
$\ell^2(\Lambda_L)\oplus \ell^2(\Lambda_R)$).  The scattering matrix \index{scattering matrix} 
\[
s= w_+^\ast w_-: \ell^2(\Lambda_L)\oplus \ell^2(\Lambda_R)
\rightarrow \ell^2(\Lambda_L)\oplus \ell^2(\Lambda_R),
\]
is a unitary operator commuting with $h_L\oplus h_R$. Following computations in Section 
\ref{sect:ClassicScattering1} verify that in the Fourier representation $s$ acts as the operator 
of multiplication by the unitary matrix 
\[
s(\xi)=\e^{2\i l \xi}\left[ 
\begin{matrix}0 &1\\1&0
\end{matrix}
\right].
\]
\exop Show that for $p\in [1, \infty]$, 
\begin{equation}
e_{p, +}(\alpha)
=\frac{1}{2\pi}\int_0^2 \log\left(1
-\frac{\sinh \frac{\alpha(\beta_R(\varepsilon-\mu_R)-\beta_L(\varepsilon-\mu_L))}{2}
\sinh \frac{(1-\alpha)(\beta_R(\varepsilon -\mu_R) -\beta_L(\varepsilon - \mu_L)}{2}}
{\cosh\frac{\beta_L(\varepsilon -\mu_L)}{2}
\cosh\frac{\beta_R
(\varepsilon -\mu_R)}{2}}\right)\d \varepsilon.
\label{monday-end}
\end{equation}
Note that, in accordance with Exercise \ref{believe}, $e_{p,+}(\alpha)$ does not depend on $p$. 
The function \eqref{monday-end}  can be expressed in terms of Euler dilogarithm, see the end 
of Section \ref{sec-open-xy}.

\exop Verify directly that $e_{{\rm naive}, +}(\alpha)= e_{p, +}(\alpha)$.  
\exop (Recall Exercise \ref{Ex:final}). Show that
\begin{equation}
e_{2, +}(\balpha)=\frac{1}{2\pi}
\int_0^2 \log \left( 
1 +{\cal D}(\varepsilon)\right)\d\varepsilon,
\label{trampo}
\end{equation}
where 
\[ 
{\cal D}(\varepsilon)=\frac{\sinh\frac{\beta_L(\alpha_1\varepsilon -\alpha_3\mu_L)
-\beta_R(\alpha_2\varepsilon-\alpha_4\mu_L)}{2}
\sinh\frac{\beta_R((1-\alpha_2)\varepsilon - (1-\alpha_4)\mu_R) 
-\beta_L((1-\alpha_1)\varepsilon -(1-\alpha_3)\mu_L)}{2}}
{\cosh\frac{\beta_L(\varepsilon -\mu_L)}{2}
\cosh\frac{\beta_R
(\varepsilon -\mu_R)}{2}}.
\]

\exop Using \eqref{trampo} show that the steady state charge 
and heat fluxes out of the left reservoir are 
\begin{align*}
\omega_+({\cal J}_L) &=\frac{1}{2\pi}\int_{0}^2 \left[ \frac{1}{1+ \e^{\beta_L(\varepsilon -\mu_L)}} 
- \frac{1}{1+ \e^{\beta_R(\varepsilon -\mu_R)}}\right]\d \varepsilon,\\[3mm]
\omega_+(\Phi_L) &=\frac{1}{2\pi}\int_{0}^2\varepsilon \left[\frac{1}{1+ 
\e^{\beta_L(\varepsilon -\mu_L)}} - \frac{1}{1+ \e^{\beta_R(\varepsilon -\mu_R)}}\right]\d \varepsilon,
\end{align*}
and that $\omega_+({\cal J}_R)=-\omega_+({\cal J}_L)$,  $\omega_+(\Phi_R)=-\omega_+(\Phi_L)$.
\end{exo}
\begin{exo} \label{bruneau} This exercise is intended for technically  advanced reader. Consider 
an infinitely extended EBB model with two reservoirs except that now we keep the single particle 
Hilbert spaces  $\cK_j$ and  Hamiltonians $h_j$  general. The coupling is defined in the 
same way as previously except that now $\delta_0^{(j)}$ is just a given vector in $\cK_j$. We absorb 
$\lambda$ in $\delta_0^{(j)}$ and denote by $h$ the single particle Hamiltonian of the joint system. 
We shall suppose that the spectral measure $\nu_j$ for $h_j$  and $\delta_0^{(j)}$ is purely 
absolutely continuous and  denote by $\d \nu_j(\varepsilon)/\d\varepsilon$ its Radon-Nikodym
\index{Radon-Nikodym derivative}
derivative w.r.t. the Lebesgue measure. We also suppose that $h$ has purely absolutely continuous 
spectrum. Since $h$ preserves the cyclic subspace spanned by 
$\{ \cK_\cS, \delta_0^{(1)}, \delta_0^{(2)}\}$ and $h_0$, without loss of generality we may assume 
that $\cK_j=L^2(\rr, \d\nu_j)$ and that $h_j$ is the  operator of multiplication by $\varepsilon$. 
\exop Show that the scattering matrix is given by 
\[
s( \varepsilon) = \one +2\i\pi\left[\begin{array}{cc} \langle \chi_1| (h-\varepsilon +\i 0)^{-1}\chi_1\rangle 
\frac{\d\nu_1(\varepsilon)}{\d \varepsilon} & \langle \chi_1| (h-\varepsilon +\i 0)^{-1}\chi_2\rangle
 \sqrt{\frac{\d\nu_1(\varepsilon)}{\d \varepsilon}\frac{\d\nu_2(\varepsilon)}
{\d \varepsilon}} \\ \langle \chi_2| (h-\varepsilon +\i 0)^{-1}\chi_1\rangle \sqrt{\frac{\d\nu_1(\varepsilon)}{\d \varepsilon}\frac{\d\nu_2(\varepsilon)}{\d 
\varepsilon}} & \langle \chi_2| (h-\varepsilon +\i 0)^{-1}\chi_2\rangle  \frac{\d\nu_2(\varepsilon)}{\d \varepsilon} \end{array} \right].
\]
\exop Compute $e_{p, +}(\alpha)$ for $p\in [1, \infty]$. 

\exop Verify that Exercise \ref{believe} applies  to  this more general 
model. Classify the examples  for which  $e_{p, +}(\alpha)$ does not depend on $p$.

\exop Compute $e_{{\rm naive}, +}(\alpha)$.

\exop Compute $e_{2, +}(\balpha)$ and derive the formulas for the steady state charge 
and heat fluxes.

\exop Verify  the results  by comparing them with Exercise \ref{big-trampo}.
\end{exo}

\noindent {\bf Remark.} For more information about the Exercises \ref{believe}, \ref{big-trampo},
and \ref{bruneau} we refer the reader to \cite {BJP}.
\medskip

\subsection{Local interactions} One can easily modify the EBB model to allow for 
interactions between fermions in the device $\cS$. For example, let $q$ be a pair interaction on
$\cS$, \ie a self-adjoint operator on $\Gamma_2(\cK)$ acting like
$$
(q\psi)(x_1,x_2)=\left\{
\begin{array}{ll}
q(x_1,x_2)\psi(x_1,x_2)&\text{if }x_1,x_2\in\cS,\\[10pt]
0&\text{otherwise}.
\end{array}
\right.
$$
Then the operator
$$
Q=\frac12\sum_{x,y\in\cS}q(x,y)a^\ast(\delta_x)a^\ast(\delta_y)a(\delta_y)a(\delta_x),
$$
is self-adjoint on $\Gamma(\cK)$ and leaves all the $\Gamma_k(\cK)$ invariant.
It vanishes on $\Gamma_0(\cK)$ and $\Gamma_1(\cK)$ and acts like
$$
(Q\psi)(x_1,\ldots,x_k)=\left(
\frac12\sum_{\atop{x,y\in\{x_1,\ldots,x_k\}\cap\cS}{x\not=y}}q(x,y)\right)\psi(x_1,\ldots,x_k),
$$
on $\Gamma_k(\cK)$ for $k\ge2$. For $\kappa\in\rr$, the Hamiltonian
$$
H_{\lambda,\kappa}=H_\lambda+\kappa Q,
$$
is self-adjoint on $\Gamma(\cK)$ and defines a dynamics $\tau_{\lambda,\kappa}^t$ on the CAR 
algebra over $\cK$. It is easy to perform the thermodynamic limit of this locally interacting EBB 
model, the interaction term $Q$ being confined to the finite sample $\cS$. The large time limit 
is a more delicate problem. Hilbert space scattering techniques are no more adapted to this problem
and one has to deal with the much harder $C^\ast$-scattering theory, \eg the existence of
the limit
$$
\gamma_\pm(A)=\lim_{t\to\pm\infty}\tau_\lambda^{-t}\circ\tau_{\lambda,\kappa}^t(A).
$$
Such problems first appeared in the works of Hepp \cite{He} and Robinson \cite{Ro}. 
In the specific context of non-equilibrium statistical mechanics, the scattering approach was
advocated by Ruelle \cite{Ru1} (see also \cite{Ru2,Ru3}). A systematic approach
to the scattering problem for local perturbations of free Fermi gases has been developed 
by Botvich and Malyshev \cite{BM}, Aizenstadt and Malyshev \cite{AMa} and 
Malyshev \cite{Ma}.\index{expansion!Dyson}
It relies on the well known Cook argument and a uniform (in $t$) control of the Dyson expansion
\begin{align*}
\tau_{\lambda,\kappa}^t(A)&=\tau_{\lambda}^t(A)\\
&+\sum_{k\ge1}(\i\kappa)^k
\int_{0\le s_k\le\cdots\le s_1\le t}[\tau_\lambda^{s_k}(Q),[\cdots
[\tau_\lambda^{s_1}(Q),\tau_\lambda^{t}(A)]\cdots]]\d s_1\cdots\d s_k.
\end{align*}
Optimal bounds for the uniform convergence of such expansions have been obtained by
Maassen and Botvich \cite{MB}. The interested reader should consult
\cite{FMU,JOP2,JPP} and references therein.

\section{The XY-spin chain}
\label{Sect:XYChain}

In this section, we describe a simple example of extended quantum spin system on a 1D-lattice.
We shall follow closely the approach of Chapter \ref{chap:HarmoChain}, starting from the 
standard quantum mechanical description of a finite sub-lattice.

\subsection{Finite spin systems}

Let $\Lambda$ be a finite set. A spin $\frac 12$ system on $\Lambda$ is a finite quantum system
obtained by attaching to each site $x\in\Lambda$ a spin $\frac12$. Thus, the Hilbert space of such
a spin system is given by\index{spin system}
$$
\cH_\Lambda=\bigotimes_{x\in\Lambda}\cH_x,
$$
where each $\cH_x$ is a copy of $\cc^2$. The corresponding $\ast$-algebra is
$$
\cO_\Lambda=\bigotimes_{x\in\Lambda}\cO_x,
$$
where $\cO_x={\rm M}_2(\cc)$ is the algebra of $2\times2$ complex matrices.
Together with the identity $\one_x\in\cO_x$, the Pauli matrices
$$
\sigma_x^{(1)}=\left[\begin{array}{cc}0&1\\1&0\end{array}\right],\quad
\sigma_x^{(2)}=\left[\begin{array}{cc}0&-\i\\\i&0\end{array}\right],\quad
\sigma_x^{(3)}=\left[\begin{array}{cc}1&0\\0&-1\end{array}\right],
$$
form a basis of $\cO_x$ satisfying the well known relations
$$
\sigma_x^{(j)}\sigma_x^{(k)}=\delta_{jk}\one_x+\i\varepsilon^{jkl}\sigma_x^{(l)}.
$$
For $D\subset\Lambda$ we set $\one_D=\otimes_{x\in D}\one_x$.
We shall identify $T_x\in\cO_x$ with the element
$T_x\otimes\one_{\Lambda\setminus\{x\}}$
of $\cO_\Lambda$. With this convention, one has the relations
\begin{equation}
\sigma_x^{(j)}\sigma_x^{(k)}=\delta_{jk}\one_\Lambda+\i\varepsilon^{jkl}\sigma_x^{(l)},\qquad
[\sigma_x^{(j)},\sigma_y^{(k)}]=2\i\delta_{xy}\varepsilon^{jkl}\sigma_x^{(l)}.
\label{SUtwo}
\end{equation}
Moreover, any element of $\cO_\Lambda$ can be written as a finite sum
$$
\sum_{a}\prod_{x\in\Lambda}T_x^{(a)},
$$
with $T_x^{(a)}\in\{\one_\Lambda,\sigma_x^{(1)},\sigma_x^{(2)},\sigma_x^{(3)}\}$.
Since $\one_\Lambda=\sigma_x^{(j)2}$, it follows that the smallest $\ast$-subalgebra of
$\cO_\Lambda$ containing the set 
${\mathfrak S}_\Lambda=\{\sigma_x^{(j)}\,|\,x\in\Lambda,j=1,2,3\}$ is $\cO_\Lambda$.
By von Neumann's bicommutant theorem (Theorem \ref{BiCommThm}), we conclude that
${\mathfrak S}_\Lambda''=\cO_\Lambda$ and hence ${\mathfrak S}_\Lambda'=\cc\one_\Lambda$.

The dynamics of a spin chain is completely determined by its Hamiltonian $H_\Lambda$,
a self-adjoint element of $\cO_\Lambda$. The equilibrium state of the system at inverse temperature
$\beta$ is given by the density matrix
$$
\omega_{\beta\Lambda}=\frac{\e^{-\beta H_\Lambda}}{\tr\left(\e^{-\beta H_\Lambda}\right)}.
$$

The particular example we shall consider in the remaining part  of this section is the
XY-chain on the finite 1D-lattice $\Lambda=[A,B]\subset\zz$. It is defined by the
XY-Hamiltonian\index{Hamiltonian!-XY}
\begin{equation}
H_\Lambda=-\frac14\sum_{x\in[A,B[}J
\left(\sigma_x^{(1)}\sigma_{x+1}^{(1)}+\sigma_x^{(2)}\sigma_{x+1}^{(2)}\right)
-\frac12\sum_{x\in[A,B]}\lambda\sigma_x^{(3)},
\label{XYHamilton}
\end{equation}
where $J\in\rr$ is the nearest-neighbor coupling constant and $\lambda\in\rr$ is  the strength of
an external magnetic field in direction $(3)$\footnote{The name XY comes from the coupling between components
$(1)=(X)$ and $(2)=(Y)$ of the spins.}. The case $J>0$ corresponds to a ferromagnetic coupling
while $J<0$ describes an anti-ferromagnetic system.

\subsection{The Jordan-Wigner representation}
\label{Sect:JWrepres}

The natural ``spin'' interpretation of the $\ast$-algebra $\cO_\Lambda$ described in the previous
section is not very convenient for computational purposes. In this section, following Jordan and
Wigner \cite{JW}, we shall see that $\cO_\Lambda$ also carries an irreducible representation 
of a CAR algebra. Moreover, it turns out that the XY Hamiltonian \eqref{XYHamilton} takes a 
particularly simple form in this representation. In fact, we shall see that the XY-spin chain can be 
mapped to a free Fermi gas. \index{representation!Jordan-Wigner}

Let $\sigma_x^{(\pm)}=(\sigma_x^{(1)}\pm\i\sigma_x^{(2)})/2$ denote the spin raising/lowering 
operators at $x\in\Lambda$. Note that $\sigma_x^{(-)}$ and $\sigma_x^{(+)}=\sigma_x^{(-)\ast}$
satisfy the anti-commutation relations
$$
\{\sigma_x^{(+)},\sigma_x^{(+)}\}=\{\sigma_x^{(-)},\sigma_x^{(-)}\}=0\qquad
\{\sigma_x^{(+)},\sigma_x^{(-)}\}=\one_{\cH_\Lambda}.
$$
Thus, if $\Lambda$ reduces to the singleton $\{x\}$, then the maps 
$\alpha\mapsto\alpha\sigma_x^{(+)}$
and $\alpha\mapsto\bar\alpha\sigma_x^{(-)}$ define a representation of the CAR over the
Hilbert space $\cc=\ell^2(\{x\})$ (and one easily checks that this representation is irreducible).
This does not directly generalize to larger $\Lambda$. Indeed, if $\Lambda$ contains two distinct 
sites $x\not=y$ one has
$$
[\sigma_x^{(+)},\sigma_y^{(+)}]=[\sigma_x^{(-)},\sigma_y^{(-)}]=0\qquad
[\sigma_x^{(+)},\sigma_y^{(-)}]=0,
$$
\ie operators at distinct sites commute whereas they should anti-commute to define
a representation of the CAR over $\ell^2(\Lambda)$.

To transform commutation at distinct sites into anti-commutation, we make the following observation:
for $T_x\in\cO_x$ and $S_y\in\cO_y$ one has
$$
\{\sigma_A^{(3)}\cdots\sigma_{x-1}^{(3)}T_x,\sigma_A^{(3)}\cdots\sigma_{y-1}^{(3)}S_y\}=
\left\{
\begin{array}{ll}
\{\sigma_{x}^{(3)},T_x\}\sigma_{x+1}^{(3)}\cdots\sigma_{y-1}^{(3)}S_y&\text{for } x<y,\\[6pt]
\{T_x,S_x\}&\text{for } x=y,\\[4pt]
\{\sigma_{y}^{(3)},S_y\}\sigma_{y+1}^{(3)}\cdots\sigma_{x-1}^{(3)}T_x&\text{for } x>y.
\end{array}
\right.
$$
Since $\{\sigma_x^{(3)},\sigma_x^{(\pm)}\}=0$, it follows that
the Jordan-Wigner operators
\begin{equation}
b_x=\sigma_A^{(3)}\cdots\sigma_{x-1}^{(3)}\sigma_x^{(-)},\qquad
b_x^\ast=\sigma_A^{(3)}\cdots\sigma_{x-1}^{(3)}\sigma_x^{(+)},
\label{JWoper}
\end{equation}
satisfy
$$
\{b_x,b_y\}=\{b_x^\ast,b_y^\ast\}=0,\qquad
\{b_x,b_y^\ast\}=\delta_{xy}\one_{\Lambda}.
$$
Hence, the maps $\ell^2(\Lambda)\ni\alpha\mapsto b^\ast(\alpha)=\sum_x\alpha_x b_x^\ast$ and
$\ell^2(\Lambda)\ni\alpha\mapsto b(\alpha)=\sum_x\bar\alpha_x b_x$ define a representation of
the CAR over $\ell^2(\Lambda)$ on the Hilbert space $\cH_\Lambda$. We shall call it
the Jordan-Wigner representation.\index{representation!of CAR}

One easily inverts Relations \eqref{JWoper} to express the spin operators in terms of the 
Jordan-Wigner operators:
\begin{equation}
\sigma_x^{(1)}=V_x(b_x+b_x^\ast),\qquad
\sigma_x^{(2)}=\i V_x(b_x-b_x^\ast),\qquad
\sigma_x^{(3)}=2b_x^\ast b_x-\one_\Lambda,
\label{JWtrafo}
\end{equation}

where
\[
V_x=\left\{
\begin{array}{ll}
\one_\Lambda&\text{if }x=A,\\[6pt]
\prod_{y\in[A,x[}(2b_y^\ast b_y-1)&\text{otherwise.}
\end{array}
\right.
\]
If follows in particular that ${\mathfrak B}_\Lambda=\{b^\#_x\,|\,x\in\Lambda\}$ satisfies
${\mathfrak B}_\Lambda'={\mathfrak S}_\Lambda'=\cc\one_\Lambda$.
Hence, the Jordan-Wigner representation is irreducible. By Proposition
\ref{UniqueIredCAR}, there exists a unitary operator $U:\Gamma(\ell^2(\Lambda))\to\cH_\Lambda$
such that $b^\#(\alpha)=Ua^\#(\alpha)U^\ast$, where the $a^\#$ are the usual
creation/annihilation operators on the fermionic Fock space $\Gamma(\ell^2(\Lambda))$.

A simple calculation shows that 
$$
\sigma_x^{(1)}\sigma_{x+1}^{(1)}+\sigma_x^{(2)}\sigma_{x+1}^{(2)}
=-2(b_{x+1}^\ast b_x+b_x^\ast b_{x+1}),
$$
so that we can rewrite the XY-Hamiltonian as
$$
H_\Lambda=\frac J2\sum_{x\in[A,B[}(b_{x+1}^\ast b_x+b_x^\ast b_{x+1})
-\frac\lambda2\sum_{x\in[A,B]}(2b_x^\ast b_x-\one).
$$
By Part (8) of Proposition \ref{midterm1} we thus have $H_\Lambda=U\d\Gamma(h_\Lambda)U^\ast$,
up to an irrelevant additive constant, where the one-particle Hamiltonian $h_\Lambda$ is the
self-adjoint operator on $\ell^2(\Lambda)$ given by
$$
h_\Lambda=\frac J2\sum_{x\in[A,B[}(|\delta_{x+1}\rangle\langle\delta_x|
+|\delta_x\rangle\langle\delta_{x+1}|)
-\lambda\sum_{x\in[A,B]}|\delta_x\rangle\langle\delta_x|=\frac J2\Delta_\Lambda+(J-\lambda)\one,
$$
$\Delta_\Lambda$ being the discrete Laplacian on $\Lambda$ with Dirichlet boundary conditions
\eqref{DeltaLambda}. Thus, the unitary map $U$ provides an equivalence between the XY-chain on 
$\Lambda$ and the free Fermi gas with one particle Hamiltonian $h_\Lambda$. In particular, it maps
the equilibrium state $\omega_{\beta\Lambda}$ to the quasi-free state on the CAR 
algebra over $\ell^2(\Lambda)$ with density
$$
T_{\beta\Lambda}=(\one+\e^{\beta h_\Lambda})^{-1}.
$$

\begin{exo}
\exop Use the Jordan-Wigner representation of the XY-chain to show that, for all $x\in\Lambda$,
$$
\omega_{\beta\Lambda}(\sigma_x^{(1)})=\omega_{\beta\Lambda}(\sigma_x^{(2)})=0,\qquad
\frac12\omega_{\beta\Lambda}(\one_\Lambda+\sigma_x^{(3)})
=\frac2{|\Lambda|}\sum_{\xi\in\Lambda^\ast}\frac{\sin^2(\xi(x-A+1))}{1+\e^{\beta(J\cos\xi-\lambda)}},
$$
where $\Lambda^\ast=\{n\pi/(|\Lambda|+1)\,|\,n=1,\ldots,|\Lambda|\}$, $|\Lambda|=B-A+1$.

\exop Show that the mean magnetization per spin is given by
$$
m_\Lambda(\beta,J,\lambda)=
\frac1{|\Lambda|}\sum_{x\in\Lambda}\omega_{\beta\Lambda}(\sigma_x^{(3)})=
\frac1{|\Lambda|}\sum_{\xi\in\Lambda^\ast}{\tanh(\beta(\lambda-J\cos\xi)/2)}.
$$

\exop Show that, in the thermodynamic limit,
$$
\lim_{\Lambda\to\zz}m_\Lambda(\beta,J,\lambda)=\frac{2\sinh(\beta\lambda/2)}{\pi}\int_0^\pi
\frac{\d\xi}{\cosh(\beta\lambda/2)+\cosh(\beta(J\cos\xi-\lambda/2))}.
$$
\noindent{\sl Hint}: use the discrete Fourier transform to diagonalize the Laplacian $\Delta_\Lambda$.
\end{exo}

\subsection{The open XY-chain}\label{sec-open-xy}

\begin{figure}[htbp]
\begin{center}
    \includegraphics[width=\textwidth]{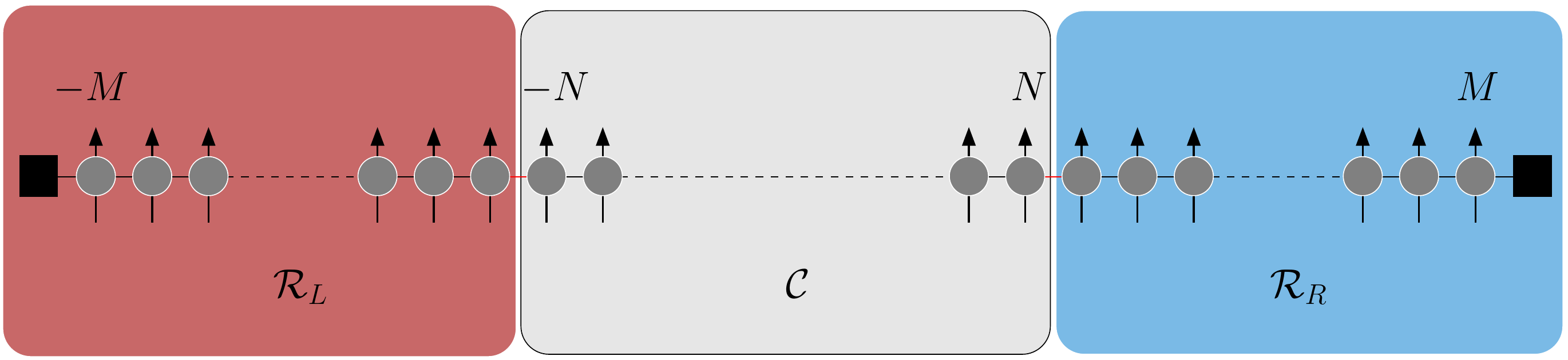}
    \caption{The XY-chain $\cC$ coupled at its left and right ends to the reservoirs $\cR_L$ and 
    $\cR_R$.}
    \label{fig:XYchain_coupled}
\end{center}
\end{figure}

To construct a model of open XY-chain we shall consider the same geometry as in the
classical harmonic chain of Chapter \ref{chap:HarmoChain}: a finite system
$\cC$, consisting of the XY-chain on $\Lambda=[-N,N]$, is coupled at its left and right ends
to two reservoirs $\cR_L$ and $\cR_R$ which are themselves XY-chains on
$\Lambda_L=[-M,-N-1]$ and $\Lambda_R=[N+1,M]$ (see Figure \ref{fig:XYchain_coupled}).
The size $N$ will be kept fixed and we shall discuss the thermodynamic limit $M\to\infty$. 

The Hamiltonian of the decoupled joint system $\cR_L+\cC+\cR_R$ is given by
$$
H_0=H_{\Lambda_L}+H_\Lambda+H_{\Lambda_R}.
$$
The coupled Hamiltonian is 
$$
H=H_{\Lambda_L\cup\Lambda\cup\Lambda_R}=H_0+V_L+V_R,
$$
with the coupling terms
$$
V_L=-\frac J4\left(\sigma_{-N-1}^{(1)}\sigma_{-N}^{(1)}+\sigma_{-N-1}^{(2)}\sigma_{-N}^{(2)}\right),
\qquad
V_R=-\frac J4\left(\sigma_{N}^{(1)}\sigma_{N+1}^{(1)}+\sigma_{N}^{(2)}\sigma_{N+1}^{(2)}\right).
$$
We consider the family of initial states
\begin{equation}
\omega_X=\frac{\e^{-\beta H+X_L H_{\Lambda_L}+X_RH_{\Lambda_R}}}
{\tr(\e^{-\beta H+X_L H_{\Lambda_L}+X_RH_{\Lambda_R}})},
\label{xy-initial}
\end{equation}
with control parameter $X=(X_L,X_R)\in\rr^2$. The entropy production observable is\index{entropy!production}
$$
\sigma_X=X_L\Phi_L+X_R\Phi_R,
$$
where the heat fluxes from $\cR_{L/R}$ to $\cC$ are easily computed
using the commutation relations \eqref{SUtwo},
\begin{align*}
\Phi_L&=-\i[H,H_{\Lambda_L}]
=\frac{J^2}8\left(\sigma_{-N-2}^{(2)}\sigma_{-N}^{(1)}
-\sigma_{-N-2}^{(1)}\sigma_{-N}^{(2)}\right)\sigma_{-N-1}^{(3)}
+\frac{\lambda J}4\left(\sigma_{-N-1}^{(1)}\sigma_{-N}^{(2)}-\sigma_{-N-1}^{(2)}\sigma_{-N}^{(1)}\right),\\
\Phi_R&=-\i[H,H_{\Lambda_R}]
=\frac{J^2}8\left(\sigma_{N}^{(1)}\sigma_{N+2}^{(2)}
-\sigma_{N}^{(2)}\sigma_{N+2}^{(1)}\right)\sigma_{N+1}^{(3)}
+\frac{\lambda J}4\left(\sigma_{N+1}^{(1)}\sigma_{N}^{(2)}-\sigma_{N+1}^{(2)}\sigma_{N}^{(1)}\right).
\end{align*}

In the Jordan-Wigner representation, the decoupled system is a free Fermi gas with one particle
Hilbert space $\ell^2(\Lambda_L\cup\Lambda\cup\Lambda_R)
=\ell^2(\Lambda_L)\oplus\ell^2(\Lambda)\oplus\ell^2(\Lambda_R)$ and one particle
Hamiltonian
$$
h_0=h_{\Lambda_L}\oplus h_\Lambda\oplus h_{\Lambda_R}.
$$
The one particle Hamiltonian of the coupled system is
$$
h=h_{\Lambda_L\cup\Lambda\cup\Lambda_R}=h_0+v_L+v_R,
$$
where the coupling terms
$$
v_L=\frac J2\left(|\delta_{-N-1}\rangle\langle\delta_{-N}|+|\delta_{-N}\rangle\langle\delta_{-N-1}|\right),
\qquad
v_R=\frac J2\left(|\delta_N\rangle\langle\delta_{N+1}|+|\delta_{N+1}\rangle\langle\delta_N|\right),
$$
are finite rank operators. The initial state $\omega_X$ is quasi-free with density
$$
T_X=\left(\one+\e^{-k_X}\right)^{-1},
$$
where
$$
k_X=-\beta h+X_Lh_{\Lambda_L}+X_Rh_{\Lambda_R}
=-\beta(h_\Lambda+v_L+v_R)-(\beta-X_L)h_{\Lambda_L}-(\beta-X_R)h_{\Lambda_R}.
$$

\begin{figure}[htbp]
\begin{center}
    \includegraphics[width=0.6\textwidth]{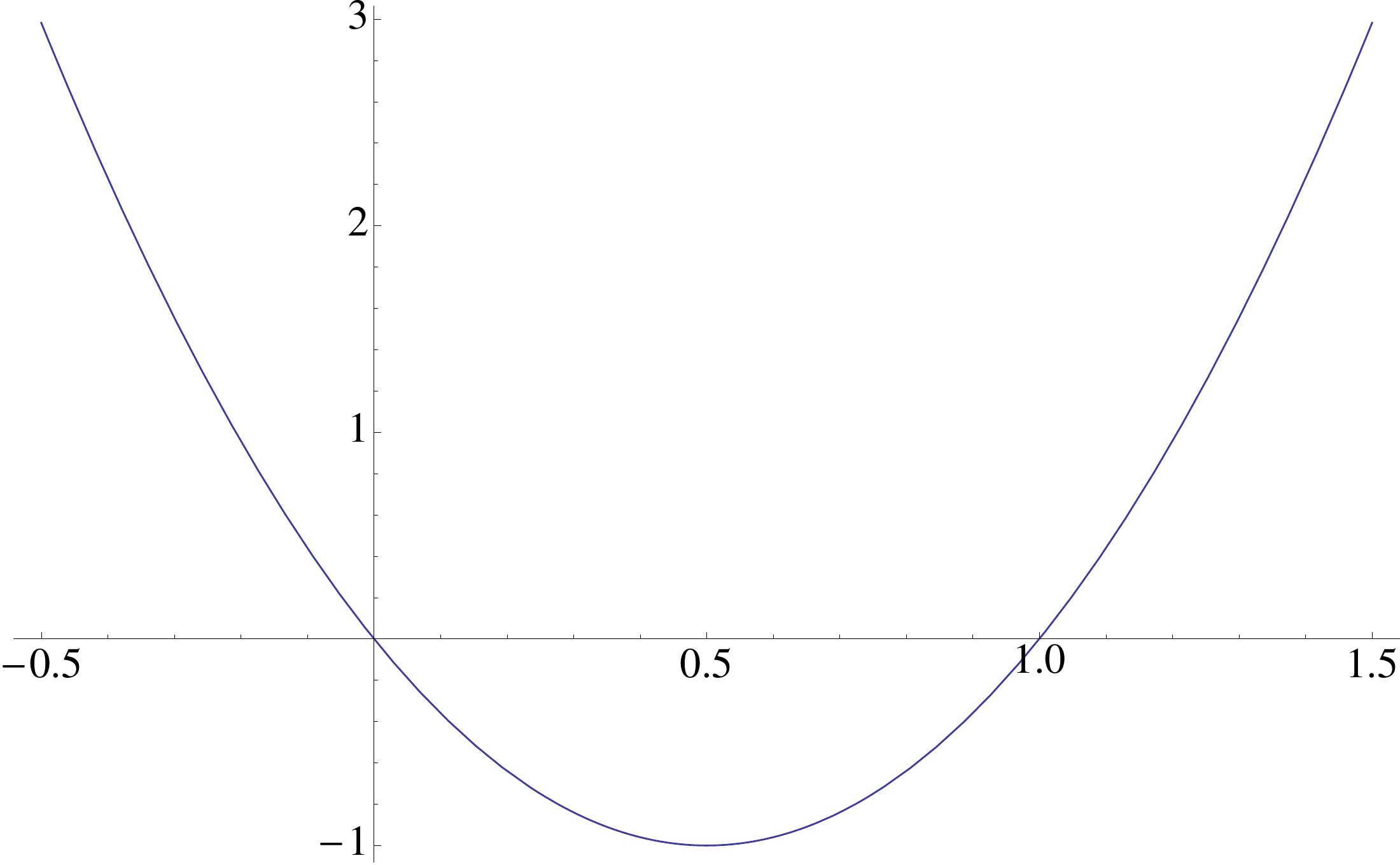}
    \caption{The entropic functional $e_{p,+}(\alpha)$ of the open XY-chain.}
    \label{fig:XYep}
\end{center}
\end{figure}

It is now apparent that the results of Section \ref{sect:EBBM} apply to the open XY-chain.
By Part (2) of Exercise \ref{Exo:GenFEBBM}, the generalized entropic pressure is given by\index{entropic pressure!generalized}
$$
e_t(X,Y)=\log\det\left(\one+T_X\left(\e^{-k_X}\e^{k_{X-Y}+k_{Y,t}-k_0}-\one\right)\right),
$$
where $k_{X,t}=\e^{-\i th}k_X\e^{\i th}$. The same formula holds in the thermodynamic limit,
provided $k_X$ is replaced by its strong limit. The large time limit follows from Part (5) of Exercise
\ref{Ex:final},
\begin{align*}
e_{+}(X,Y)&=\lim_{t\to\infty}\frac1t\,e_{t}(X,Y)\\
&=\int_0^\pi\log\det\left(\one+T_X(\xi)
(\e^{-k_X(\xi)}\e^{k_{X-Y}(\xi)+s(\xi)k_Y(\xi)s^\ast(\xi)-k_0(\xi)}-\one)\right)\frac{\d\varepsilon(\xi)}{2\pi},
\end{align*}
where $\varepsilon(\xi)=1-\cos\xi$, $k_X(\xi)$ is the diagonal $2\times 2$ matrix with entries 
$(\beta-X_j)(\lambda-J\cos(\xi))$ and $T_X(\xi)=(\one+\e^{-k_X(\xi)})^{-1}$. Using the explicit
form 
$$
s(\xi)=\e^{\pm2\i N\xi}\left[\begin{array}{cc}0&1\\1&0\end{array}\right],
$$
of the scattering matrix (see Section \ref{sect:ClassicScattering1}, the sign $\pm$ is opposite to the sign of the coupling constant $J$),
 we obtain 
 $$
e_+(X,Y)=\frac1{J\pi}\int_{u_-}^{u_+}
\log\left(1-\frac{\sinh(u\Delta Y)\sinh(u(\Delta X-\Delta Y))}
{\cosh(u(\beta-X_L))\cosh(u(\beta-X_R))}\right)\d u,
$$
where we have set $\Delta X=X_R-X_L$, $\Delta Y=Y_R-Y_L$ and $u_\pm=(\lambda\pm J)/2$. 
The steady heat current through the chain is given by
$$
\langle\Phi_L\rangle_+=\lim_{t\to\infty}\omega_{X,t}\left(\Phi_L\right)
=\left.-\partial_{Y_L}e_+(X,Y)\right|_{Y=0}=\frac1{J\pi}\int_{u_-}^{u_+}
u\left(\tanh(\beta_L u)-\tanh(\beta_R u)\right)\d u,
$$
where $\beta_{L/R}=\beta-X_{L/R}$. It follows that the entropy production \index{entropy!production}
$$
\langle\sigma\rangle_+=\frac1{J\pi}\int_{u_-}^{u_+}
(\beta_Lu-\beta_Ru)\left(\tanh(\beta_L u)-\tanh(\beta_R u)\right)\d u,
$$
is strictly positive iff $\beta_L\not=\beta_R$ and $J\not=0$.
\begin{exo}
Develop the linear response theory of the open XY-chain.
\end{exo}
\begin{exo} Instead of \eqref{xy-initial} consider the reference state 
\[
\omega=\frac{\e^{-\beta_L H_{\Lambda_L} -\beta H_\Lambda - \beta_R H_{\Lambda_R}}}
{\tr(\e^{-\beta_L H_{\Lambda_L} -\beta H_\Lambda - \beta_R H_{\Lambda_R}})}.
\]
In this case, up to irrelevant scaling, the Jordan-Wigner transformation maps the XY-chain to the  EBB model considered 
in  Exercise \ref{big-trampo}.  Show that for $p\in[1,\infty]$, 
\begin{equation}
e_{{\rm naive},  +}(\alpha)=e_{p,+}(\alpha)=\frac1{J\pi}\int_{u_-}^{u_+}
\log\left(1-\frac{\sinh(\alpha u\Delta\beta)\sinh((1-\alpha)u\Delta\beta)}
{\cosh(u\beta_L)\cosh(u\beta_R)}\right)\d u,
\label{xy-entropic}
\end{equation}
where $\Delta\beta=\beta_R-\beta_L$ (see Figure \ref{fig:XYep}).
Note that $e_{p, +}(\alpha)=e_{+}(X, \alpha X)$. 
\end{exo}

The formula \eqref{xy-entropic} can be rewritten in terms of Euler's dilogarithm
$$
{\rm Li}_2(z)=-\int_0^z\frac{\log(1-w)}{w}\,\d w,
$$
an analytic function on the cut plane $\cc\setminus[1,\infty[$ with a branching point
at $z=1$ (see \cite{Le}). More precisely, one has
$$
e_{p,+}(\alpha)=G(\bar\beta+(\alpha-1/2)\Delta\beta)+G(\bar\beta-(\alpha-1/2)\Delta\beta)
-G(\beta_L)-G(\beta_R),
$$
where $\bar\beta=(\beta_L+\beta_R)/2$ and
$$
G(x)=\frac{{\rm Li_2}\left(-\e^{2xu_+}\right)-{\rm Li_2}\left(-\e^{2xu_-}\right)}{\pi x(u_+-u_-)}.
$$
It follows that $e_{p,+}(\alpha)$ is analytic on the strip $|\Im\alpha|<\frac\pi{(|\lambda|+|J|)|\Delta\beta|}$.

{\noindent\bf Remark.} We were able to compute the TD and large time limits of the entropic 
functionals of the XY-chain  thanks to its Fermi-gas representation. We note however that the 
operator
$$
V_x=(2b^\ast_{-M}b_{-M}-\one)\cdots(2b^\ast_{x-1}b_{x-1}-\one),
$$
has no limit in the CAR algebra over $\ell^2(\zz)$ as $M\to\infty$, and the Jordan-Wigner 
transformation \eqref{JWtrafo} does not survive  the TD limit. In fact, to recover the full spin 
algebra in the TD limit, one needs to enlarge the CAR algebra over $\ell^2(\zz)$ with an element
$V$ formally equal to
$$
\lim_{M\to\infty}(2b^\ast_{-M}b_{-M}-\one)\cdots(2b^\ast_{-1}b_{-1}-\one).
$$
We refer to Araki \cite{A} for a complete exposition of this construction.
An alternative resolution  of the TD limit/Jordan-Wigner transformation conflict goes as follows.  

We set $\Lambda_M=[-M,M]\subset\zz$.
The operator $W=\sigma_{-M}^{(3)}\cdots\sigma_{M}^{(3)}\in\cO_{\Lambda_M}$ satisfies
$W=W^\ast=W^{-1}$. It implements the rotation by an angle $\pi$ around the $(3)$-axis of all the 
spins of the chain,
$$
W\sigma_x^{(j)}W^\ast=\left\{
\begin{array}{cl}
-\sigma_x^{(j)}&\text{for } j=1 \text{ or }j=2,\\[3pt]
\sigma_x^{(j)}&\text{for }j=3.
\end{array}
\right.
$$
Thus, $\theta(A)=WAW^\ast$ defines an involutive $\ast$-automorphism of $\cO_{\Lambda_M}$. 
In the fermionic picture, $\theta$ is completely characterized by $\theta(b_x)=-b_x$.

Since $\theta$ is  a linear involution on the vector space $\cO_{\Lambda_M}$, it
follows that $\cO_{\Lambda_M}=\cO_{\Lambda_M+}\oplus\cO_{\Lambda_M-}$, where
$$
\cO_{\Lambda_M\pm}=\{A\in\cO_{\Lambda_M}\,|\,\theta(A)=\pm A\},
$$
are vector subspaces. Note  that $\cO_{\Lambda_M+}$ is a $\ast$-subalgebra of 
$\cO_{\Lambda_M}$.  Since $H_\Lambda\in\cO_{\Lambda_M+}$, the dynamics 
$\tau_\Lambda^t(A)=\e^{\i tH_\Lambda}A\e^{-\i tH_\Lambda}$ satisfies
$\tau_\Lambda^t\circ\theta=\theta\circ\tau_\Lambda^t$ and, in particular, it preserves
both subspaces $\cO_{\Lambda_M\pm}$. Moreover, our initial state satisfies
$\omega_X\circ\theta=\omega_X$ which implies that $\omega_X|_{\cO_{\Lambda_M-}}=0$.
Thus, observables with non-trivial expectation belong to the subalgebra $\cO_{\Lambda_M+}$
and we may restrict ourselves to such observables. 

In the fermionic picture, $\cO_{\Lambda_M+}$ is the $\ast$-algebra of all polynomials in
the $b_x^\#$ which contain only monomials of even degree. In the spin picture, it
is generated by the operators $\sigma_x^{(3)}$ and $\sigma_x^{(s)}\sigma_{x'}^{(s')}$ with
$s,s'\in\{-,+\}$ and $x<x'$, which have a Jordan-Wigner representation surviving the TD limit,
\eg
$$
\sigma_x^{(-)}\sigma_y^{(+)}=b_x(2b_{x+1}^\ast b_{x+1}-\one)\cdots
(2b_{y-1}^\ast b_{y-1}-\one)b_y^\ast.
$$
Thus, at the price of restricting the dynamical system to the even subalgebra $\cO_{\Lambda_M+}$,
the XY-chain remains equivalent to a free Fermi gas in the TD limit. This  fact is  a starting 
point in the construction of  the NESS of the XY-chain. We refer the reader to \cite{AH,AP}
for the details of this construction and to \cite {AB1,AB2} for additional information about the 
NESS of the XY-chain.

\appendixsection{Appendix A: Large deviations}
\renewcommand{\thechapter}{A}
\refstepcounter{chapter}
\label{appx:LDP}

In this first appendix, we formulate some well known large deviation results that were used
in these lecture notes. We provide a proof in the simplest case of scalar random variables.

\appendixsubsection{A.1 Fenchel-Legendre transform}
\refstepcounter{section}
\label{sect:FLT}

In this section, we shall use freely some well known properties of convex real functions of a real 
variable, see, \eg \cite{RV}.

Let $I=[a,b]\subset\rr$ be a closed finite interval, denote by $\interior(I)=]a,b[$ its  interior, and let
$e:I\to\rr$ be a continuous convex function. Then $e$ admits finite left and right derivatives
\[
D^{\pm}e(s) =\lim_{h\downarrow 0}\frac{e(s\pm h)- e(s)}{\pm h},
\]
at every $s\in\interior(I)$. $D^+e(a)$ and $D^{-}e(b)$ exist, although they may be respectively
$-\infty$ and $+\infty$. By convention, we set $D^-e(a)=-\infty$ and $D^+e(b)=+\infty$.
The functions $D^\pm e(s)$ are increasing on $I$ and satisfy $D^{-}e(s)\leq D^{+}e(s)$.
Moreover, $D^-e(s)=D^+e(s)=e'(s)$ outside a countable set  in $\interior(I)$. If $e^\prime(s)$ exists 
for all $s\in\interior(I)$, then it is continuous on $\interior(I)$ and
\[
\lim_{s \downarrow a}e^\prime(s)= D^+e(a), \qquad \lim_{s\uparrow b} e^\prime(s)= D^-e(b).
\]
The subdifferential of $e$ at $s_0\in I$, denoted $\partial e(s_0)$, is the set of $\theta\in\rr$
such that the affine function $\ubar e(s)=e(s_0)+\theta(s-s_0)$ satisfies $e(s)\ge\ubar e(s)$ for all 
$s\in I$, \ie the graph of $\ubar e$ is tangent to the graph of $e$ at the point $(s_0,e(s_0))$. 
For any $s_0\in I$, one has $\partial e(s_0)=[D^-e(s_0), D^+e(s_0)]\cap\rr$.

It is convenient to extend the function $e$ to $\rr$ by setting $e(s)=+\infty$ for $s\not\in I$.
Then the function $e(s)$ is convex and lower semi-continuous on $\rr$, \ie 
$$
e(s_0)=\liminf_{s\to s_0} e(s),
$$
holds for all $s_0\in\rr$. The subdifferential of $e$ is naturally extended by setting 
$\partial e(s)=\emptyset$ for $s\not\in I$.

The function \index{Legendre transform}\index{Fenchel-Legendre transform}
\begin{equation}
\varphi (\theta)= \sup_{s\in I}(\theta s - e(s))=\sup_{s\in \rr} \,(\theta s -e(s))
\label{local-fenchel}
\end{equation}
is called the  Fenchel-Legendre transform of $e(s)$.
$\varphi(\theta)$ is finite and convex (hence continuous) on $\rr$.  Obviously,  if $a\geq 0$ then
$\varphi(\theta)$ is increasing and if $b \leq 0$ then $\varphi(\theta)$ is decreasing. The subdifferential
of $\varphi$ at $\theta\in\rr$ is $\partial \varphi(\theta)=[D^-\varphi(\theta), D^+\varphi(\theta)]$. 
The basic properties of the pair $(e, \varphi)$ are summarized in:

\bet\label{prop-fenchel}
\ben 
\item $\theta s\leq e(s) +\varphi(\theta)$ for all $s,\theta\in\rr$.
\item $\theta s=e(s) +\varphi(\theta)$ $\Leftrightarrow$ $\theta \in \partial e(s)$.
\item $e(s)=\sup_{\theta \in \rr}(\theta s-\varphi(\theta))$.
\item $\theta \in \partial e(s)$ $\Leftrightarrow$  $s\in \partial \varphi(\theta)$.
\item If $0\in ]a, b[$, then $\varphi(\theta)$ is decreasing on $]-\infty, D^-e(0)]$, increasing on 
$[D^+e(0), \infty[$, $\varphi(\theta)=-e(0)$ for $\theta \in \partial e(0)$, and $\varphi(\theta)>-e(0)$ for 
$\theta \not\in \partial e(0)$. 
\een
\eet

\demo
(1) Follows directly from the definition of $\varphi$. 

\medskip\noindent(2) Combining the inequality (1) with the equality $\theta s_0=e(s_0) +\varphi(\theta)$
we obtain that $e(s)\ge e(s_0)+\theta(s-s_0)$ for all $s\in\rr$ which implies $\theta\in\partial e(s_0)$. 
Reciprocally,
if  $\theta\in\partial e(s_0)$ then $e(s)\ge e(s_0)+\theta(s-s_0)$ holds for all $s\in\rr$ and hence
$\theta s_0\ge e(s_0)+\sup_s(\theta s-e(s))=e(s_0)+\varphi(\theta)$. Combined with inequality (1),
this yields $\theta s_0=e(s_0) +\varphi(\theta)$.

\medskip\noindent(3) It follows from Exercise \ref{SalphaUSC} that the function 
$\tilde e(s)=\sup_{\theta\in\rr}(\theta s-\varphi(\theta))$ is lower semi-continuous on $\rr$.
(1) implies that $\tilde e(s)\le e(s)$ for any $s\in\rr$.
$\partial e(s)\not=\emptyset$ for $s\in]a,b[$ we conclude from (2)   that
$\tilde e(s)=e(s)$. 

Note that $-e(s)\le-\min_{u\in I}(-e(u))=\varphi(0)$. Thus, for $\theta>0$, 
we have $\varphi(\theta)=\sup_{s\in[a,b]}(\theta s-e(s))\le\theta b+\varphi(0)$
and hence $\theta s-\varphi(\theta)\ge \theta(s-b)-\varphi(0)$. It follows that
$\tilde e(s)=+\infty=e(s)$ for $s>b$. A similar argument applies to the case $s<a$. 

Consider now the case $s=a$. From our previous conclusions, we can write
$\tilde e(a)=\liminf_{s\to a}\tilde e(s)=\lim_{s\downarrow a}\tilde e(s)=\lim_{s\downarrow a}e(s)=e(a)$.
A similar argument applies to $s=b$.

\medskip\noindent(4) By (2), $\theta_0\in\partial e(s)$ is equivalent to the equality 
$s\theta_0 =e(s) +\varphi(\theta_0)$ which, combined with the inequality (1) yields 
$\varphi(\theta)\ge\varphi(\theta_0)+s(\theta-\theta_0)$ for all $\theta\in\rr$ and hence 
$s\in\partial\varphi(\theta_0)$. Reciprocally, if $s\in\partial\varphi(\theta_0)$ then
$\varphi(\theta)\ge\varphi(\theta_0)+s(\theta-\theta_0)$ for all $\theta\in\rr$ and
we conclude from (3) that $e(s)\le\sup_\theta(\theta s-\varphi(\theta_0)-s(\theta-\theta_0))=
-\varphi(\theta_0)+s\theta_0$. Using (1) and (2), we conclude that $\theta_0\in\partial e(s)$.

\medskip\noindent(5) It follows from (4) that if $\theta_0\in\partial e(0)=[D^-e(0),D^+e(0)]$ then 
$0\in\partial\varphi(\theta_0)$, \ie $\varphi(\theta)\ge\varphi(\theta_0)$ for all $\theta\in\rr$.
Thus, $\varphi(\theta_0)=\min_\theta\varphi(\theta)=-e(0)$ and since $D^\pm\varphi(\theta)$ are 
increasing, $\varphi$ is decreasing for $\theta\le D^-e(0)$ and increasing for $\theta\ge D^+e(0)$.
\qed

\appendixsubsection{A.2 G\"artner-Ellis theorem in dimension $d=1$}
\refstepcounter{section}
\label{sect:GEoneD}
Let ${\cal I}\subset \rr_+$ be an unbounded  index set, $(M_t, {\cal F}_t, P_t)$, $t\in {\cal I}$, a family of measure spaces, and 
$X_t: M_t\rightarrow \rr$  a family of measurable  functions. We assume that the measures 
$P_t$ are finite for all $t$.  For $s\in \rr$ let
\[ e_t(s) =\log \int_{M_t}\e^{ sX_t}\d P_t.\]
$e_t(s)$ is a convex function taking values in $]-\infty, \infty]$. We make the following assumption: 

\medskip
{\noindent\bf (LD)}  For $s\in I=[a, b]$ the limit  
\[ 
e(s)=\lim_{t\rightarrow \infty}\frac{1}{t}e_t(s),
\]
exists and is finite. Moreover, the function $e(s)$ is continuous on $I$.

\medskip
Until the end of this section we shall assume that (LD) holds and set   $e(s)=\infty$ for $s\not\in I$.
The function $\varphi(\theta)$ is defined by \eqref{local-fenchel}.

\bep\label{LDP-up}
\ben 
\item Suppose that $0\in [a,b[$. Then 
$$
\limsup_{t\rightarrow \infty}\frac{1}{t}\log P_t(\{ x\in M_t\,|\, X_t(x) >t\theta\})\leq 
\begin{cases}
- \varphi(\theta)& \text{{\rm if} $\theta  \geq  D^+e(0)$}\\
e(0) &\text{{\rm if} $\theta <D^+e(0)$.}
\end{cases}
$$
\item Suppose that $0\in ]a,b]$. Then 
$$
\limsup_{t\rightarrow \infty}\frac{1}{t}\log P_t(\{ x\in M_t\,|\, X_t(x) <t\theta\})\leq 
\begin{cases}
- \varphi(\theta)& \text{{\rm if} $\theta  \leq  D^-e(0)$}\\
e(0) &\text{{\rm if} $\theta >D^-e(0)$.}
\end{cases}
$$
\een
\eep
\demo  We shall prove (1), the proof of (2) follows from (1) applied to $-X_t$ and $-\theta$. 
For $s\in]0,b]$, 
$$
P_t(\{ x\in M_t\,|\, X_t(x) >t\theta\})=P_t(\{ x\in M_t\,|\, \e^{sX_t(x)} >\e^{st\theta}\})
\leq \e^{-st\theta}\int_{M_t}\e^{sX_t}\d P_t,
$$
and so 
\[
\limsup_{t\rightarrow \infty}\frac{1}{t}\log P_t(\{ x\in M_t\,|\, X_t(x) >t\theta\})
\leq -\sup_{0\leq s \leq b}(\theta s -e(s)).
\]
For $\theta<D^+e(0)$ and $s\ge0$, one has $e(s)\ge e(0)+s D^+e(0)\ge e(0)+\theta s$, so that
$$
-e(0)\le\sup_{0\le s\le b}(\theta s -e(s))\le\sup_{0\le s\le b}(\theta s -e(0)-\theta s)=-e(0),
$$
and hence $\sup_{0\le s\le b}(\theta s -e(s))=-e(0)$. One shows in a similar way
that $\sup_{a\le s\le0}(\theta s -e(s))=-e(0)$ for $\theta\ge D^+e(0)$. It follows that
$$
\varphi(\theta)=\sup_{a\le s\le b}(\theta s -e(s))=\max\left(-e(0),
\sup_{0\le s\le b}(\theta s -e(s))\right)=\sup_{0\le s\le b}(\theta s -e(s)).
$$
The statement follows. \qed

\bep\label{LDP-basic}
Suppose that  $0\in ]a, b[$, $e(0)\leq 0$, and that  $e(s)$ is differentiable at $s=0$. Then 
for any $\delta >0$ there is $\gamma >0$ such that for $t$ large enough,
\[
P_t(\{ x\in M_t\, |\, |t^{-1}X_t(x) - e^\prime(0)|\geq \delta\})\leq  \e^{-\gamma t}.
\]
\eep
\demo
Part (2) of Theorem \ref{prop-fenchel} implies that $\varphi(e^\prime(0))=-e(0)$.
By Part (5) of the same theorem, one has $\varphi(\theta)>\varphi(e^\prime(0))\geq 0$ for 
$\theta\not=e^\prime(0)$. Since
\begin{align*}
P_t(\{ x\in M_t\, |\, |t^{-1}X_t(x) - e^\prime(0)|\geq \delta\})\leq 
&P_t(\{ x\in M_t\, |\, |X_t(x)\le t(e'(0)-\delta)\})\\
+&P_t(\{ x\in M_t\, |\, |X_t(x)\ge t(e'(0)+\delta)\}),
\end{align*}
Proposition \ref{LDP-up} implies
\[
\limsup_{t\rightarrow \infty}\frac{1}{t}\log P_t(\{ x\in M_t\,|\, |t^{-1}X_t(x)-e^\prime(0)|\geq \delta\})
\leq -\min\{\varphi(e^\prime(0)+\delta), \varphi(e^\prime(0)-\delta)\},
\]
and the statement follows. 
\qed

\bep\label{GE-THM}
Suppose that  $0\in]a,b[$ and  $e(s)$ is differentiable on $]a,b[$.  Then
$$
\liminf_{t\rightarrow \infty}\frac{1}{t}\log P_t(\{ x\in M_t\,|\, X_t(x) >t\theta)\}\geq -\varphi(\theta),
$$
for any $\theta \in ]D^+ e(a), D^-e(b)[$.
\eep

\demo Let $\theta \in ]D^+e(a), D^-e (b)[$ be given and let $\alpha $ and $\epsilon$ be such that 
$$
\theta < \alpha -\epsilon <\alpha <\alpha +\epsilon< D^-e(b).
$$
Let $s_\alpha\in ]a, b[$ be such that $e^\prime(s_\alpha)=\alpha$ 
(so $\varphi(\alpha)=\alpha s_\alpha - e(s_\alpha)$). Let 
\[\d\hat P_t =\e^{-e_t(s_\alpha)}\e^{s_\alpha X_t}\d P_t.\] Then $\hat P_t$ is a probability measure on 
$(M_t, {\cal F}_t)$ and 
\begin{align}
P_{t}(\{ x\in M_t\,|\, X_t(x)>t \theta\})
&\geq P_t(\{ x\in M_t\,|\, t^{-1}X_t(x)\in [\alpha-\epsilon, \alpha +\epsilon]\})\nonumber\\[3mm]
&=\e^{e_t(s_\alpha)}\int_{\{ t^{-1}X_t \in [\alpha -\epsilon, \alpha +\epsilon]\}}
\e^{-s_\alpha X_t}\d\hat P_t\label{LDP-long}\\[3mm]
&\geq \e^{e_t(s_\alpha)-s_\alpha t \alpha -|s_\alpha| t\epsilon}
\hat P_t(\{ x\in M_t\,|\, t^{-1}X_t \in [\alpha -\epsilon, \alpha +\epsilon]\}).\nonumber
\end{align}
Now, if $\hat e_t(s)=\log \int_{M_t}\e^{s X_t}\d\hat P_t$, then 
$\hat e_t(s)=e_t(s + s_\alpha)-e_t(s_\alpha)$ and so 
\[
\lim_{t\rightarrow \infty}\frac{1}{t}\hat e_t(s)=e(s+s_\alpha)-e(s_\alpha),
\]
for $s\in [a-s_\alpha, b-s_\alpha]$.  
Since $\hat e(0)=0$ and $\hat e^\prime(0)=e^\prime(s_\alpha)=\alpha$, it follows from 
Proposition \ref{LDP-basic} that 
\[
\lim_{t\rightarrow \infty}\frac{1}{t}\log 
\hat P_t(\{ x\in M_t\,|\, t^{-1}X_t(x) \in [\alpha -\epsilon, \alpha +\epsilon]\})=0,
\]
and \eqref{LDP-long} yields 
\[
\liminf_{t\rightarrow \infty}\frac{1}{t}\log P_t(\{ x\in M_t\,|\, X_t(x) >t\theta\})
\geq -s_\alpha \alpha +e(s_\alpha) -|s_\alpha|\epsilon = -\varphi(\alpha)-|s_\alpha|\epsilon.
\]
The statement follows by taking first  $\epsilon \downarrow 0$ and then $\alpha \downarrow \theta$.
\qed

\bigskip
The following local version of the G\"artner-Ellis theorem is a consequence of Propositions
\ref{LDP-up} and \ref{GE-THM}.\bindex{theorem!G\"artner-Ellis}

\bet\label{GEoneD}
If $e(s)$ is differentiable on $]a,b[$ and $0 \in ]a, b[$ then, for any open set 
${\mathbb J}\subset ]D^+e(a), D^- e(b)[$, 
\[
\lim_{t\rightarrow \infty}\frac{1}{t}\log P_t(\{ x\in M_t\,|\,t^{-1} X_t(x) \in {\mathbb J}\})
= -\inf_{\theta \in {\mathbb J}}\varphi(\theta).
\]
\eet
\demo
{\em Lower bound.} For any $\theta\in\mathbb J$ and $\delta>0$ such that
$]\theta-\delta,\theta+\delta[\subset\mathbb J$ one has
$$
P_t(\{ x\in M_t\,|\,t^{-1} X_t(x) \in {\mathbb J}\})
\ge P_t(\{ x\in M_t\,|\,t^{-1} X_t(x) \in]\theta-\delta,\theta+\delta[\}),
$$
and it follows from Proposition \ref{GE-THM} that
$$
\liminf_{t\to\infty}\frac1t\log P_t(\{ x\in M_t\,|\,t^{-1} X_t(x) \in {\mathbb J}\})
\ge -\varphi(\theta-\delta).
$$
Letting $\delta\downarrow0$ and optimizing over $\theta\in\mathbb J$, we obtain
\begin{equation}
\liminf_{t\to\infty}\frac1t\log P_t(\{ x\in M_t\,|\,t^{-1} X_t(x) \in {\mathbb J}\})
\ge -\inf_{\theta\in\mathbb J}\varphi(\theta).
\label{oneDlower}
\end{equation}

\medskip\noindent{\em Upper bound.} Note that $e(0)=0\in]a,b[$. By Part (5) of Proposition
\ref{prop-fenchel}, we have $\varphi(\theta)=0$ for $\theta=e'(0)$ and $\varphi(\theta)>0$
otherwise. Hence, if $e'(0)\in\closure(\mathbb J)$, then
$$
\limsup_{t\to\infty}\frac1t\log P_t(\{ x\in M_t\,|\,t^{-1} X_t(x) \in {\mathbb J}\})\le 0
=-\inf_{\theta\in\mathbb J}\varphi(\theta).
$$
In the case $e'(0)\not\in\closure(\mathbb J)$, there exist $\alpha,\beta\in\closure(\mathbb J)$ such
that $e'(0)\in]\alpha,\beta[\subset\rr\setminus\closure(\mathbb J)$. It follows that
\begin{align*}
P_t(&\{ x\in M_t\,|\,t^{-1} X_t(x) \in {\mathbb J}\})\\
&\le P_t(\{ x\in M_t\,|\,t^{-1} X_t(x)<\alpha\})
+P_t(\{ x\in M_t\,|\,t^{-1} X_t(x)>\beta\})\\
&\le 2\max\left(P_t(\{ x\in M_t\,|\,t^{-1} X_t(x)<\alpha\}),P_t(\{ x\in M_t\,|\,t^{-1} X_t(x)>\beta\})\right),
\end{align*}
and Proposition \ref{LDP-up} yields
\begin{align*}
\limsup_{t\to\infty}\frac1t\log P_t(\{ x\in M_t\,|\,t^{-1} X_t(x) \in {\mathbb J}\})\le
-\min(\varphi(\alpha),\varphi(\beta)).
\end{align*}
Finally, by Part (5) of Proposition \ref{prop-fenchel}, one has
$$
\inf_{\theta\in\mathbb J}\varphi(\theta)=\min(\varphi(\alpha),\varphi(\beta)),
$$
and therefore
\begin{equation}
\limsup_{t\to\infty}\frac1t\log P_t(\{ x\in M_t\,|\,t^{-1} X_t(x) \in {\mathbb J}\})\le
\inf_{\theta\in\mathbb J}\varphi(\theta),
\label{oneDupper}
\end{equation}
holds for any $\mathbb J\subset ]D^+e(a), D^- e(b)[$.
The result follows from the bounds \eqref{oneDlower} and \eqref{oneDupper}. 
\qed

\bigskip

\appendixsubsection{A.3 G\"artner-Ellis theorem in dimension $d>1$}
\refstepcounter{section}
\label{sect:GEmultiD}
Let ${\bf X}_t:M_t\to\rr^d$ be a family of measurable functions w.r.t. the probability  spaces
$(M_t,\cal F_t,P_t)$.  If $G\subset \rr^d$ is a Borel set, we denote by ${\rm int}(G)$ its interior, by ${\rm cl}(G)$ its closure, 
and by $\partial G$ its boundary. The following result is  a multi-dimensional version of the G\"artner-Ellis 
theorem.

\bet\label{GEmultidim}
Assume that  the limit
\beq
h({\bf Y})=\lim_{t \rightarrow \infty}\frac{1}{t}\log\int_{M_t}\e^{{\bf Y}\cdot{\bf X}_t}\,\d P_t,
\label{paris-end}
\eeq
exists in $[-\infty,+\infty]$ for all ${\bf Y}\in\rr^d$, that the function
$h({\bf Y})$ is lower semi-continuous on $\rr^d$, differentiable on the interior of the set 
$\mathcal D=\{{\bf Y}\in\rr^d\,|\,|h({\bf Y})|<\infty\}$ and satisfies
$$
\lim_{{\rm int} ({\mathcal D})\ni {\bf Y}\to {\bf Y}_0}|{\boldsymbol\nabla} h({\bf Y})|=\infty,
$$
for all ${\bf Y}_0\in\partial\mathcal D$. Suppose also that ${\bf 0}$ is an interior 
point of $\mathcal D$. Then, for all Borel sets $G \subset\rr^d$ we have
\begin{align*}
-\inf_{{\bf Z}\in {\rm int}({G})} I({\bf Z}) 
&\leq\liminf_{t \rightarrow \infty} 
\frac{1}{t} \log P_t\left(\left\{ x\in M_t\,|\, t^{-1}{\bf X}_t(x)\in G\right\}\right) \\
&\leq\limsup_{t \rightarrow \infty} 
\frac{1}{t} \log P_t\left(\left\{ x\in M\,|\, t^{-1}{\bf X}_t(x)\in G\right\}\right) \leq 
-\inf_{{\bf Z}\in {\rm cl}({G})} I({\bf Z}),
\end{align*}
where
\[
I({\bf Z}) = \sup_{{\bf Y}\in\rr^d }({\bf Y\cdot  Z}-h({\bf Y})).
\]
\eet
We now describe a local version of G\"artner-Ellis theorem in $d>1$. Set 
\begin{align*}
\bar h({\bf Y})&=\limsup_{t \rightarrow \infty}\frac{1}{t}\log\int_{M_t}\e^{{\bf Y}\cdot{\bf X}_t}\,\d P_t, \\[3mm]
\bar I({\bf Z}) &= \sup_{{\bf Y}\in\rr^d }({\bf Y\cdot  Z}-\bar h({\bf Y})).
\end{align*}
Let  $\bar {\mathcal D}= \{{\bf Y}\in\rr^d\,|\,\bar h({\bf Y})<\infty\}$ and let $\mathcal D$ be the set of all 
${\bf Y} \in \rr^d$ for which the  limit \eqref{paris-end} exists and is finite. Let $S\subset {\cal D}$ be the set of 
points at which $h({\bf Y})$ is differentiable and let ${\cal F}=\{ {\boldsymbol \nabla} h({\bf Y})\,|\, 
{\bf Y}\in S\}$. 

\bet Suppose that ${\bf 0}\in {\rm int}(\bar  {\mathcal D})$. Then
\ben 
\item 
For any  Borel set $G \subset \rr^d$, 
\[
\limsup_{t \rightarrow \infty} 
\frac{1}{t} \log P_t\left(\left\{ x\in M\,|\, t^{-1}{\bf X}_t(x)\in G\right\}\right) \leq 
-\inf_{{\bf Z}\in {\rm cl}({G})} \bar I({\bf Z}).
\]
\item For any Borel set $G \subset {\cal F}$, 
\[
\liminf_{t \rightarrow \infty} 
\frac{1}{t} \log P_t\left(\left\{ x\in M\,|\, t^{-1}{\bf X}_t(x)\in G\right\}\right) \geq 
-\inf_{{\bf Z}\in {\rm int}({G})} \bar I({\bf Z}).
\]
\een
\eet
We refer to \cite{DZ}  for  proofs and various extensions of these
fundamental results. 

\appendixsubsection{A.4 Central limit theorem}
\refstepcounter{section}
\label{sect:Bryc-CLT}

\index{theorem!central limit}
Bryc \cite{Br} has observed that under a  a suitable analyticity assumption the central limit theorem 
follows from the large deviation principle. In this appendix we state and prove Bryc's result. The 
setup is the same as in Appendix A.3. Let 
\[ 
h_t({\bf Y})=\frac{1}{t}\log\int_{M_t}\e^{{\bf Y}\cdot{\bf X}_t}\,\d P_t,
\]
and let $D_\epsilon$ be the open polydisk of $\cc^d$ of radius $\epsilon$
centered at ${\bf 0}$, \ie
$$
D_\epsilon=\{z=(z_1,\ldots,z_d)\in\cc^d\,|\,\max_j |z_j|<\epsilon\}.
$$
The analyticity assumption is:

\medskip
{\noindent\bf (A)}  For some $\epsilon >0$ and  all $t\in {\cal I}$ the function 
${\bf Y}\mapsto h_t({\bf Y})$ has an analytic continuation to 
the polydisc $D_\epsilon$ such that  
\[ 
\sup_{\atop{z\in D_\epsilon}{t\in {\cal I}}}|h_t(z)|<\infty.
\]
Moreover, for ${\bf Y}\in D_\epsilon$ real, the limit 
\[ h({\bf Y})=\lim_{t\rightarrow \infty} h_t({\bf Y})\]
exists.
\medskip

This assumption and Vitali's convergence theorem (see Appendix B below) imply that 
$h({\bf Y})$ has analytic extension to $D_\epsilon$ and  that all derivatives of $h_t(z)$ converge 
to corresponding derivatives of $h(z)$ as $t\rightarrow \infty$ uniformly on compact subsets of 
$D_\epsilon$. We denote 
\[{\bf m}_t={\boldsymbol \nabla} h_t({\bf Y})|_{{\bf Y}={\bf 0}}, \qquad {\bf m}={\boldsymbol \nabla} h({\bf Y})|_{{\bf Y}={\bf 0}}.\]
Clearly, 
Clearly, ${\bf m_t}$ is the expectation of ${\bf X}_t$ w.r.t. $P_t$ and 
\[
\lim_{t\rightarrow \infty}\frac{1}{t}{\bf m}_t = {\bf m}.
\]
Similarly, if ${\bf D}_t= [D_{jkt}]$ is the covariance of ${\bf X}_t$, then 
\[
\lim_{t\rightarrow \infty}\frac{1}{t} {\bf D}_t = {\bf D},
\]
where ${\bf D}=[D_{jk}]$ is given by 
\[ D_{jk}=\partial^2_{Y_jY_k}h({\bf Y)}|_{{\bf Y}={\bf 0}}.
\]

\bet Assumption {\rm (A)} implies the central limit theorem: for any Borel set 
$G \subset \rr^d$, 
\[
\lim_{t \rightarrow \infty}P_t\left(\left\{ x\in M_t\,\big|\, \frac{  {\bf X}_t(x) - {\bf m}_t}{\sqrt t}\in G\right\}\right)=\mu_{{\bf D}}(G),
\]
where $\mu_{{\bf D}}$ is centered Gaussian with variance ${\bf D}$.
\eet
\noindent{\bf Remark 1.} In general, the large deviation principle does not imply the central limit
theorem. In fact, assumption (A) cannot be significantly relaxed, see \cite{Br} for 
a discussion.
 
\noindent {\bf Remark 2.} Assumption (A) is typically difficult to check in practice. We emphasize, 
however, that a verification of assumptions of this type has played the central role in the works 
\cite{JOP1,JOP2,JOPP}.

\noindent{\bf Remark 3.} The proof below should be compared with Section \ref{sect:CLT}.

\demo By absorbing ${\bf m}_t$ into ${\bf X}_t$ we may assume that ${\bf m}_t={\bf 0}$. 
Let ${\bf k}=(k_1, \cdots, k_d)$, $k_j\geq 0$,  be a multi-index and 
\[ \chi_{\bf k}(t) =\frac{\partial^{k_1+\cdots +k_d}}{\partial Y_1^{k_1}\cdots \partial Y_d^{k_d}}\log \int_{M_t}\e^{\frac{{\bf Y}\cdot {\bf X}_t}{\sqrt t}}
\d P_t\big|_{{\bf Y}={\bf 0}},
\]
the ${\bf k}$-th cummulant of $t^{-1/2}{\bf X}_t$.

Set
$$
\Gamma_r=\{z=(z_1,\ldots,z_d)\in\cc^d\,|\, |z_j|=r\mbox{ for all } j\}.
$$

The Cauchy integral formula for polydisc yields
\begin{align*}
\frac{\partial^{k_1+\cdots+k_d}}{\partial z_1^{k_1}\cdots \partial z_d^{k_d}} h(z)\big|_{z=0}&= \frac{k_1!\cdots k_d!}{(2\pi\i)^d}
\oint _{\Gamma_{\frac{\epsilon}{2}}}\frac{h(z)}{z_1^{k_1+1}\cdots z_d^{k_d+1}}\,\d z_1 \cdots \d z_n\\[3mm]
&= 
\lim_{t \rightarrow \infty}\frac{k_1!\cdots k_d!}{(2\pi \i)^d}\oint_{\Gamma_{\frac{\epsilon}{2}}} \frac{h_t(z)}{z_1^{k_1+1}\cdots z_d^{k_d+1}}\,\d z_1\cdots \d z_n.
\end{align*}
Note that 
\begin{align*}
\oint_{\Gamma_{\frac\epsilon2}} \frac{h_t(z)}{z_1^{k_1+1}\cdots z_d^{k_d+1}}\d z&= 
\oint_{\Gamma_{\frac{\epsilon}{2\sqrt t}}} \frac{h_t(z)}{z_1^{k_1+1}\cdots z_d^{k_d+1}}\d z\\[3mm]
&=t^{\frac{k_1 +\cdots k_d}{2}}\oint_{\Gamma_{\frac{\epsilon}{2}}} \frac{h_t(t^{-1/2} z)}{z_1^{k_1+1}\cdots z_d^{k_d+1}}\d z,
\end{align*}
and so 
\begin{align*}
\frac{\partial^{k_1+\cdots+k_d}}{\partial z_1^{k_1}\cdots \partial z_d^{k_d}} h(z)\big|_{z=0}&= 
\lim_{t \rightarrow \infty}\frac{k_1!\cdots k_d!}{(2\pi \i)^d}t^{\frac{k_1+\cdots +k_d}{2}}\oint_{\Gamma_{\frac{\epsilon}{2}}} \frac{h_t(z)}{z_1^{k_1+1}\cdots z_d^{k_d+1}}\,\d z_1\cdots \d z_n.
\end{align*}

The Cauchy formula  implies  
\[\chi_{\bf k}(t)=t\frac{ k_1!\cdots k_d!}{(2\pi \i)^d}  \oint_{\Gamma_{\frac{\epsilon}{2}}} \frac{h_t(t^{-1/2} z)}{z_1^{k_1+1}\cdots z_d^{k_d+1}}\d z,\]
and we see that 
\[
\frac{\partial^{k_1+\cdots+k_d}}{\partial z_1^{k_1}\cdots \partial z_d^{k_d}} h(z)\big|_{z=0}= 
\lim_{t \rightarrow \infty}t^{\frac{k_1+\cdots +k_d}{2}-1}\chi_{\bf k}(t).
\]
Hence, if   $k_1+\cdots +k_d \geq 3$, then 
\[ \lim_{t\rightarrow \infty}\chi_{\bf k}(t)=0.\]
and  if $k_1+\cdots + k_d=2$ with  the  pair $k_{i}, k_j$ strictly positive, then 
\[ \lim_{t\rightarrow \infty}\chi_{\bf k}(t)=\frac{\partial^2}{\partial z_{k_i}\partial z_{k_j}}h(z)\big|_{z=0}.\]
Since the expectation of ${\bf X}_t$ is zero, we see that 
the cumulants of $t^{-1/2}{\bf X}_t$ converge to the  cumulants of the centered Gaussian on $\rr^d$ with covariance ${\bf D}$.
This implies that  the  moments of $t^{-1/2}{\bf X}_t$ converge to the  moments of the centered Gaussian 
with covariance ${\bf D}$, and theorem follows (see Section 30 in \cite{Bi2}).\qed

\appendixsection{Appendix B: Vitali convergence theorem}
\renewcommand{\thechapter}{B}
\refstepcounter{chapter}
\label{appx:Vitali}

For $\epsilon>0$ let  $D_\epsilon$ be the open polydisk of $\cc^n$ of radius $\epsilon$
centered at ${\bf 0}$, \ie
$$
D_\epsilon=\{z=(z_1,\ldots,z_n)\in\cc^n\,|\,\max_j|z_j|<\epsilon\}.
$$

\bet\label{PropVitali}
Let ${\cal I}\subset \rr_+$ be an unbounded  set  and let  $F_t:D_\epsilon\to\cc$, $t\in {\cal I}$,  be
analytic functions such that
$$
\sup_{\atop{z\in D_\epsilon}{t>0}}|F_t(z)|<\infty.
$$
Suppose that the limit 
\begin{equation}
\lim_{t\to\infty}F_t(z)=F(z),
\label{FtLimExist}
\end{equation}
exists for all $z\in D_\epsilon\cap\rr^n$. Then the limit \eqref{FtLimExist} exists for all $z\in D_\epsilon$
and is an analytic function on  $D_\epsilon$. Moreover, as $t\to\infty$, all derivatives of
$F_t$ converge uniformly on compact subsets of $D_\epsilon$ to the corresponding derivatives of 
$F$.
\eet

\bigskip
\demo Set
$$
\Gamma_r=\{z=(z_1,\ldots,z_n)\in\cc^n\,|\, |z_j|=r\mbox{ for all } j\}.
$$
For any $0<r<\epsilon$, the Cauchy integral formula for polydisks yields
\begin{equation}
\frac{\partial^{k_1+\cdots+k_n}F_t}{\partial z_1^{k_1}\cdots\partial z_n^{k_n}}(z)=
\frac{k_1!\cdots k_n!}{(2\pi\i)^n}\oint_{\Gamma_r}
\frac{F_t(w)}{(w_1-z_1)^{k_1+1}\cdots(w_n-z_n)^{k_n+1}}
\,\d w_1\cdots\d w_n,
\label{Cauchy}
\end{equation}
for all $z\in D_r$. It follows that the family of functions $\{ F_t\}_{t\in {\cal I}}$ is equicontinuous on 
$D_{r^\prime}$ for any $0<r^\prime <r$. By the Arzela-Ascoli theorem, the set $\{F_t\}$ is precompact 
in the Banach space $C(\closure(D_{r^\prime}))$ of all bounded continuous functions on 
$\closure(D_{r^\prime})$ equipped with the sup norm. The Cauchy integral formula \eqref{Cauchy}, 
where now $z\in D_{r^\prime}$ and the integral is over $\Gamma_{r^\prime}$, yields that any limit 
point of the net $\{ F_t\}_{t \in {\cal I}}$ (as $t\rightarrow \infty$)  in $C(\closure(D_{r^\prime}))$ is an 
analytic function on $D_{r^\prime}$. By the assumption, any two limit functions coincide for $z$  real, 
and hence they are identical. This yields the first part of the theorem. The convergence of the partial 
derivatives of $F_t(z)$ is an immediate consequence of the  Cauchy integral formula.
\qed

\vfill\eject
\refstepcounter{chapter}
\addcontentsline{toc}{chapter}{Notations}
\printindex[notation]
\refstepcounter{chapter}
\addcontentsline{toc}{chapter}{Index}
\printindex
\end{document}